%% file: HIG-19-009_temp.tex
\pdfoutput=1
\documentclass[11pt,twoside,a4paper,cmspaper,final,collab]{cms-tdr}
\def\svnVersion{b4b97ec-D}\def\svnDate{2021/08/01}\def\cmsCernNoTag{CERN-EP-2021-054}\def\cmsCernDate{\today}\def\cmsMessage{"Published in Physical Review D as \href{http://dx.doi.org/10.1103/PhysRevD.104.052004}{\doi{10.1103/PhysRevD.104.052004}.}"}
\begin{document}\cmsNoteHeader{HIG-19-009}

\newlength\cmsFigWidth
\ifthenelse{\boolean{cms@external}}{\setlength\cmsFigWidth{0.35\textwidth}}{\setlength\cmsFigWidth{0.5\textwidth}}
\ifthenelse{\boolean{cms@external}}{\providecommand{\CL}{C.L.\xspace}}{\providecommand{\CL}{CL\xspace}}
\newlength\cmsTabSkip\setlength\cmsTabSkip{1ex}
\newcommand\sss{}
\newcommand{\ROOT}{\textsc{root}\xspace}
\newcommand{\cPe}{\Pe\xspace}
\newcommand{\cPm}{\PGm\xspace}
\newcommand{\X}{\ensuremath{\cmsSymbolFace{X}}\xspace}
\newcommand{\jhugen}{\textsc{JHUGen}\xspace}
\newcommand{\minlo}{\textsc{MINLO}\xspace}
\newcommand{\mela}{\textsc{mela}\xspace}
\newcommand{\VV}{\ensuremath{\PV\PV}\xspace}
\newcommand{\WW}{\ensuremath{\PW\PW}\xspace}
\newcommand{\ZZ}{\ensuremath{\PZ\PZ}\xspace}
\newcommand{\HVV}{\ensuremath{\PH\PV\PV}\xspace}
\newcommand{\Hff}{\ensuremath{\PH\text{ff}}\xspace}
\newcommand{\Htt}{\ensuremath{\PH\PQt\PQt}\xspace}
\newcommand{\Hgg}{\ensuremath{\PH\Pg\Pg}\xspace}
\newcommand{\HWW}{\ensuremath{\PH\PW\PW}\xspace}
\newcommand{\HZZ}{\ensuremath{\PH\PZ\PZ}\xspace}
\newcommand{\VBF}{\ensuremath{\mathrm{VBF}}\xspace}
\newcommand{\VH}{\ensuremath{\PV\PH}\xspace}
\newcommand{\WH}{\ensuremath{\PW\PH}\xspace}
\newcommand{\ZH}{\ensuremath{\PZ\PH}\xspace}
\newcommand{\tqH}{\ensuremath{\PQt\PH}\xspace}
\newcommand{\ttH}{\ensuremath{\PQt\PAQt\PH}\xspace}
\newcommand{\bbH}{\ensuremath{\cPqb\PAQb\PH}\xspace}
\newcommand{\ggH}{\ensuremath{\Pg\Pg\PH}\xspace}
\newcommand{\ZX}{\ensuremath{\PZ+\X}\xspace}
\newcommand{\Hboson}{\PH boson\xspace}
\newcommand{\Hell}{\ensuremath{\PH\to4\ell}\xspace}
\newcommand{\qq}{\ensuremath{\Pq\Pq}\xspace}
\newcommand{\mH}{\ensuremath{m_{\PH}}\xspace}
\newcommand{\mell}{\ensuremath{m_{4\ell}}\xspace}
\newcommand{\bbar}{\ensuremath{\cPqb\PAQb}\xspace}
\newcommand{\muV}{\ensuremath{\mu_{\PV}}\xspace}
\newcommand{\Dbkg}{\ensuremath{{\mathcal{D}}_{\text{bkg}}}\xspace}
\newcommand{\DbkgEW}{\ensuremath{{\mathcal{D}}_{\text{bkg}}^{\mathrm{EW}}}\xspace}
\newcommand{\fB}{\ensuremath{f_{a2}}\xspace}
\newcommand{\fC}{\ensuremath{f_{a3}}\xspace}
\newcommand{\fL}{\ensuremath{f_{\Lambda1}}\xspace}
\newcommand{\fLZg}{\ensuremath{f_{\Lambda1}^{\PZ\gamma}}\xspace}
\newcommand{\fG}{\ensuremath{f_{a3}^{\Pg\Pg\PH}}\xspace}
\newcommand{\fT}{\ensuremath{f_{CP}^{\Htt}}\xspace}
\newcommand{\fF}{\ensuremath{f_{CP}^{\Hff}}\xspace}
\newcommand{\fai}{\ensuremath{f_{ai}}}
\newcommand{\onshell}{on-shell\xspace}
\newcommand{\specialcell}[2][c]{\begin{tabular}[#1]{@{}c@{}}#2\end{tabular}}
\ifthenelse{\boolean{cms@external}}{\providecommand{\cmsTable}[1]{#1}}{\providecommand{\cmsTable}[1]{\resizebox{\textwidth}{!}{#1}}}
\ifthenelse{\boolean{cms@external}}{\providecommand{\NA}{\ensuremath{\cdots}\xspace}}{\providecommand{\NA}{\ensuremath{\text{---}}\xspace}}
\ifthenelse{\boolean{cms@external}}{\providecommand{\toc}{\relax}}{\providecommand{\toc}{\tableofcontents}}
\cmsNoteHeader{HIG-19-009}
\title{Constraints on anomalous Higgs boson couplings to vector bosons and fermions in its production and decay using the four-lepton final state}
\date{\today}

\abstract{
Studies of $CP$ violation and anomalous couplings of the Higgs boson to vector bosons and fermions are presented.  
The data were acquired by the CMS experiment at the LHC and correspond to an integrated 
luminosity of $137\fbinv$ at a proton-proton collision energy of $13\TeV$.
The kinematic effects in the Higgs boson's four-lepton decay $\PH\to4\ell$ and its production in association with two jets, 
a vector boson, or top quarks are analyzed, using a full detector simulation and matrix element techniques to identify 
the production mechanisms and to increase sensitivity to the tensor structure of the Higgs boson interactions. A simultaneous 
measurement is performed of up to five Higgs boson couplings to electroweak vector bosons (\HVV), two couplings to gluons (\Hgg), 
and two couplings to top quarks (\Htt). The $CP$ measurement in the \Htt interaction is combined with the recent measurement 
in the $\PH\to\gamma\gamma$ channel. The results are presented in the framework of anomalous couplings 
and are also interpreted in the framework of effective field theory, including the first study of $CP$ properties of the \Htt and 
effective \Hgg couplings from a simultaneous analysis of the gluon fusion and top-associated processes. 
The results are consistent with the standard model of particle physics. 
}

\author{CMS Collaboration}

\hypersetup{
pdfauthor={CMS Collaboration},
pdftitle={Constraints on anomalous Higgs boson couplings to vector bosons and fermions in its production and decay using the four-lepton final state},
pdfsubject={CMS},
pdfkeywords={Higgs, width, spin, parity, CP, anomalous couplings}
}

\maketitle

\toc

\section{Introduction} 
\label{sec:intro}

The discovery by the ATLAS and CMS Collaborations of a Higgs boson ($\PH$) with a mass of 
$\approx$125\GeV~\cite{Aad:2012tfa,Chatrchyan:2012xdj,Chatrchyan:2013lba}
has confirmed the predictions of the standard model (SM) of particle physics~\cite{StandardModel67_1, Englert:1964et,
Higgs:1964ia,Higgs:1964pj,Guralnik:1964eu,StandardModel67_2,StandardModel67_3}.
The CMS~\cite{Chatrchyan:2012jja,Chatrchyan:2013mxa,Khachatryan:2014kca,Khachatryan:2015mma,Khachatryan:2016tnr,
Sirunyan:2017tqd,Sirunyan:2019twz,Sirunyan:2019nbs}
and ATLAS~\cite{Aad:2013xqa,Aad:2015mxa,Aad:2016nal,Aaboud:2017oem,Aaboud:2017vzb,Aaboud:2018xdt,Aad:2020mnm}
experiments have set constraints on the spin-parity properties of the \Hboson and anomalous \HVV couplings, 
where \PV stands for $\PW$, $\PZ$, and $\gamma$ electroweak (EW) gauge bosons, finding its quantum numbers 
to be consistent with $J^{PC}=0^{++}$, but leaving room for small anomalous \HVV couplings.
In theories beyond the SM (BSM), \Hboson interactions may generate several of them, 
which lead to new interaction tensor structures, both $CP$-even and $CP$-odd. 
The new anomalous tensor structures of the \Hboson interactions may also appear through loop corrections
in SM processes, but the size of their contributions is beyond the current experimental sensitivity. 
The $CP$-odd anomalous couplings between the \Hboson and either the top quark or new BSM particles, fermions or bosons,
contributing to the gluon fusion loop may generate $CP$ violation in \Hgg interactions (where \Pg stands for gluon). 
Possible $CP$ violation effects in couplings to fermions, \Hff, had not been experimentally probed until recently,
when the first constraints were reported by CMS~\cite{Sirunyan:2020sum} and ATLAS~\cite{Aad:2020ivc} in \ttH production
(where \PQt stands for top quark) using the $\PH\to\gamma\gamma$ channel. 

In this paper, we study the tensor structure of the \HVV, \Hgg, and \Htt 
interactions, and we search for several anomalous effects, 
including $CP$ violation, using the four-lepton final state $\PH\to\PV\PV\to4\ell$, where $\ell=\mu$ or $\Pe$. 
The \Hboson production processes considered in this paper include gluon fusion (\ggH), vector boson fusion (\VBF), 
associated production with a weak vector boson ($\PV\PH$, either $\PZ\PH$ or $\PW\PH$), 
with a top quark pair (\ttH), with a single top quark (\tqH), and with a bottom quark pair (\bbH). 
The Feynman diagrams for these processes are shown in Figs.~\ref{fig:Feyn_1}--\ref{fig:Feyn_5}
and discussed in detail in Section~\ref{sec:pheno}.
Kinematic effects in the \Hboson's decay and its production in association with two jets
generated in either the \VBF or \ggH processes, with a vector boson, or with top quarks are analyzed. 
Production and decay processes of the \Hboson are sensitive to certain anomalous contributions, 
or equivalently higher-dimensional operators in the effective field theory (EFT)~\cite{deFlorian:2016spz},
which modify the kinematic distributions of the \Hboson's decay products and of the particles produced in association 
with the \Hboson.
Prior measurements of EW processes limit the allowed values of certain EFT operators, and
the preferred EFT basis used in this paper is chosen to minimize the number of independent operators 
and their correlations. This allows us to reduce the number of operators to be measured. 
The results are also translated and reported in different, frequently used EFT bases.
The effects of EFT modifications in backgrounds are found to be negligible because of the high purity of the signal 
peak in the four-lepton invariant mass distribution and further constraints on the backgrounds from sidebands.

Each production process of the \Hboson is identified using its kinematic features, and events are assigned 
to corresponding categories. Two categorization schemes are employed in this analysis, one designed to study \Htt and \Hgg 
and the other designed to study \HVV anomalous couplings. Within each category, the matrix element likelihood approach 
(MELA)~\cite{Gao:2010qx,Bolognesi:2012mm,Anderson:2013afp,Gritsan:2016hjl,Gritsan:2020pib} is employed 
to construct observables that are optimal for the measurement of the chosen anomalous couplings,
or EFT operators, including $CP$-sensitive observables to search for $CP$-violating operators. 
In our approach, fully simulated Monte Carlo (MC) signal samples that include anomalous couplings allow 
the incorporation of detector effects into the likelihood analysis,
and observables explore all kinematic features of the events, including those sensitive to $CP$ violation
and to simultaneous anomalous effects in the production and decay of the \Hboson. 
These features distinguish our analysis from the recent measurements by the ATLAS Collaboration in the 
same $\PH\to4\ell$ decay channel~\cite{Aad:2020mkp,ATLAS:2020wny}, in which case the measurements 
are derived from the differential distributions based on either the simplified template cross sections 
(STXS)~\cite{deFlorian:2016spz} or unfolded fiducial measurements.

We follow the formalism used in the study of anomalous couplings in the earlier analyses of CMS data
in Refs.~\cite{Chatrchyan:2012jja,Chatrchyan:2013mxa,Khachatryan:2014kca,Khachatryan:2015mma,Khachatryan:2016tnr,
Sirunyan:2017tqd,Sirunyan:2019twz,Sirunyan:2019nbs,Sirunyan:2020sum}.
We focus on the measurements where the \Hboson is produced  \onshell; the extension to the off-shell region is considered in Ref.~\cite{Sirunyan:2019twz}, 
where joint constraints on the \Hboson width $\Gamma_{\PH}$ and its anomalous couplings are obtained using a partial data set. 
Some of the theoretical foundations relevant to the present analysis can be found in Refs.~\cite{Nelson:1986ki,Soni:1993jc,Chang:1993jy,Barger:1993wt,
Arens:1994wd,BarShalom:1995jb,Gunion:1996xu,Han:2000mi,Plehn:2001nj,Choi:2002jk,Buszello:2002uu,Hankele:2006ma,
Accomando:2006ga,Godbole:2007cn,Hagiwara:2009wt,Gao:2010qx,DeRujula:2010ys,Christensen:2010pf,
Gainer:2011xz,Bolognesi:2012mm,Ellis:2012xd,Chen:2012jy,Gainer:2013rxa,Artoisenet:2013puc,Anderson:2013afp,Chen:2013waa,
Chen:2013ejz,Gainer:2014hha,Gonzalez-Alonso:2014eva,Dolan:2014upa,Demartin:2014fia,Demartin:2014fia,Buckley:2015vsa,
Greljo:2015sla,Gritsan:2016hjl,deFlorian:2016spz,Brivio:2019myy,Gritsan:2020pib}.
The results are presented in a model-independent way, which allows interpretation in the scattering amplitude or 
effective Lagrangian approach, for example in the frameworks of the standard model effective field theory 
(SMEFT)~\cite{Weinberg:1979sa, Buchmuller:1985jz, Leung:1984ni, Dedes:2017zog} 
or pseudo observables (PO)~\cite{deFlorian:2016spz}.

Compared to our previous results on anomalous \HVV couplings~\cite{Sirunyan:2017tqd,Sirunyan:2019twz}, 
which used a subset of the data presented here, several substantial improvements have been introduced. 
First, as a result of the increased number of $\PH\to4\ell$  events, the expected  95\% confidence level (\CL)  constraints 
on \HVV couplings are now dominated by the tight limits from the analysis of kinematic distributions 
of particles produced in association with the \Hboson in the \VBF and \VH processes. 
Second, an improved fit implementation allows for the simultaneous measurement 
of up to five independent \HVV, two \Hgg, and two \Htt couplings using the single decay channel 
$\PH\to4\ell$. The couplings are parameterized with the signal strength and 
the fractional cross section contributions of the anomalous couplings. 
A direct constraint on $CP$ violation in the \Hgg coupling is obtained for the first time
by employing $CP$-sensitive observables.  
The $CP$ violation measurement in the \Htt coupling closely follows our recent measurement in
the $\PH\to\gamma\gamma$ channel~\cite{Sirunyan:2020sum}, and the results are combined. 
We also perform the first study of $CP$ properties in the \Htt and effective \Hgg couplings from a simultaneous 
analysis of the gluon fusion and top-associated processes. 
For all anomalous couplings, we interpret our results via the SMEFT framework in terms of \HVV, \Hgg, and \Htt operators.

The rest of this paper is organized as follows. The phenomenology of anomalous \HVV, \Hgg, and \Hff couplings 
and considerations in the EFT framework are discussed in Section~\ref{sec:pheno}. 
The data and MC simulation, event reconstruction and selection are discussed in Section~\ref{sec:cms}. 
The kinematic variables associated with the \Hboson's production and decay and its MELA analysis are introduced 
in Section~\ref{sec:kinematics}. The implementation of the maximum likelihood fit is shown in Section~\ref{sec:fit}.
The results are presented and discussed in Section~\ref{sec:results}. We conclude in Section~\ref{sec:Summary}.

\section{Parameterization of \texorpdfstring{\PH}{H} boson production and decay processes}
\label{sec:pheno}

\begin{figure*}[!tbp]
\centering
\includegraphics[width=0.4\textwidth]{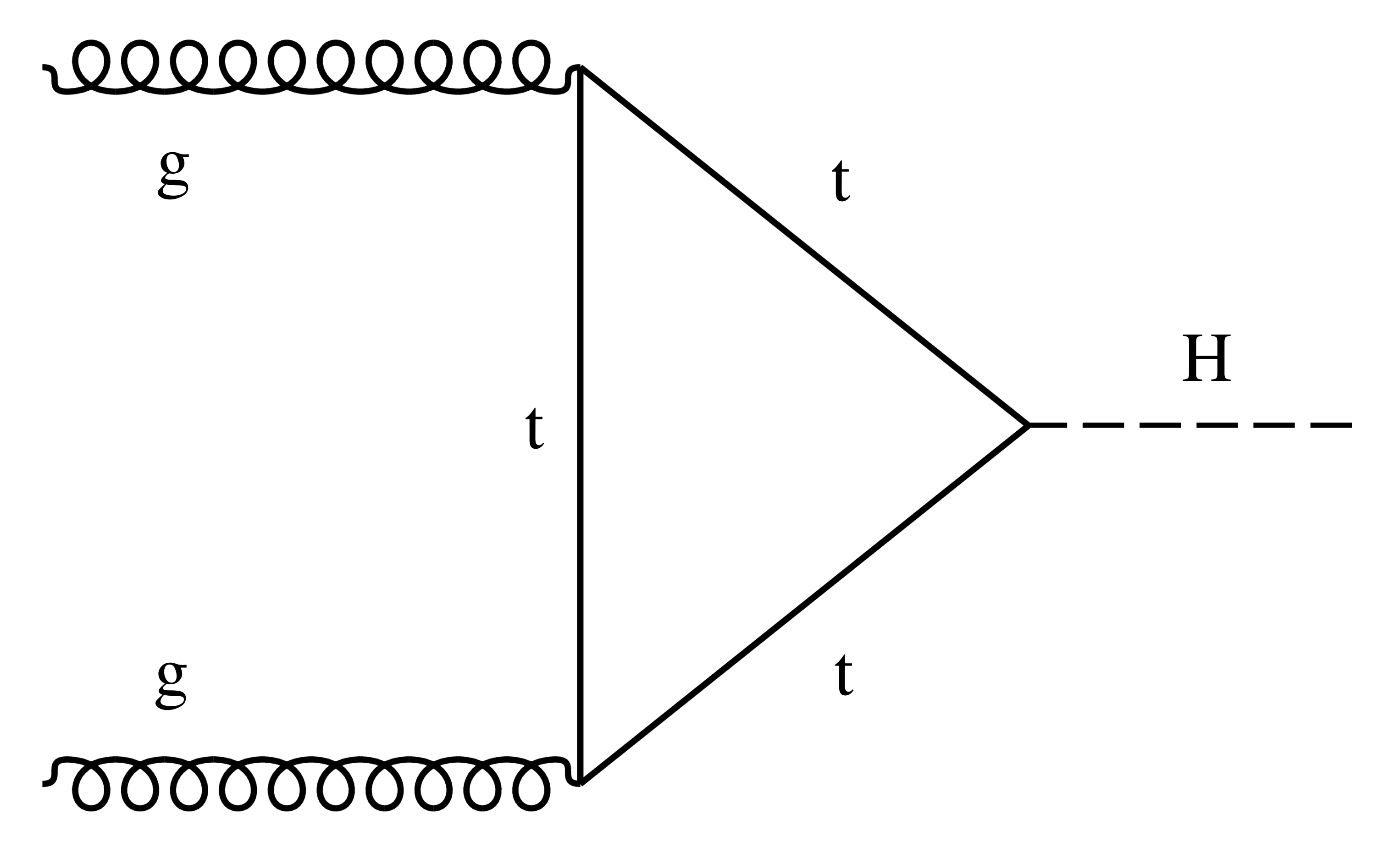}
~~~~~~
\includegraphics[width=0.4\textwidth]{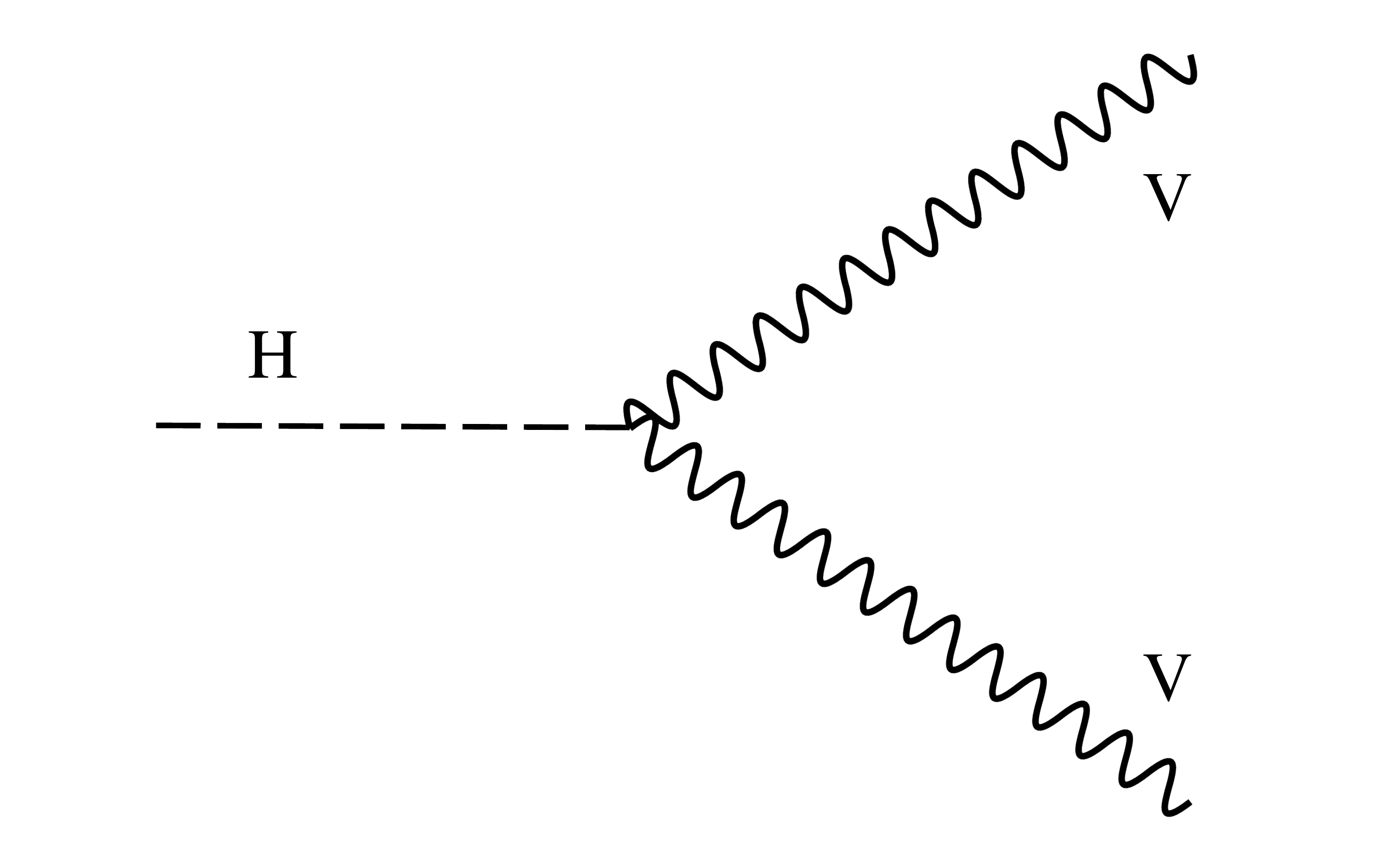}
\caption
{
Leading-order Feynman diagrams for the gluon fusion production mode (left) and $\PH\to\PV\PV$ decay (right). 
\label{fig:Feyn_1}
}
\end{figure*}

\begin{figure*}[!tbhp]
\centering
\includegraphics[width=0.4\textwidth]{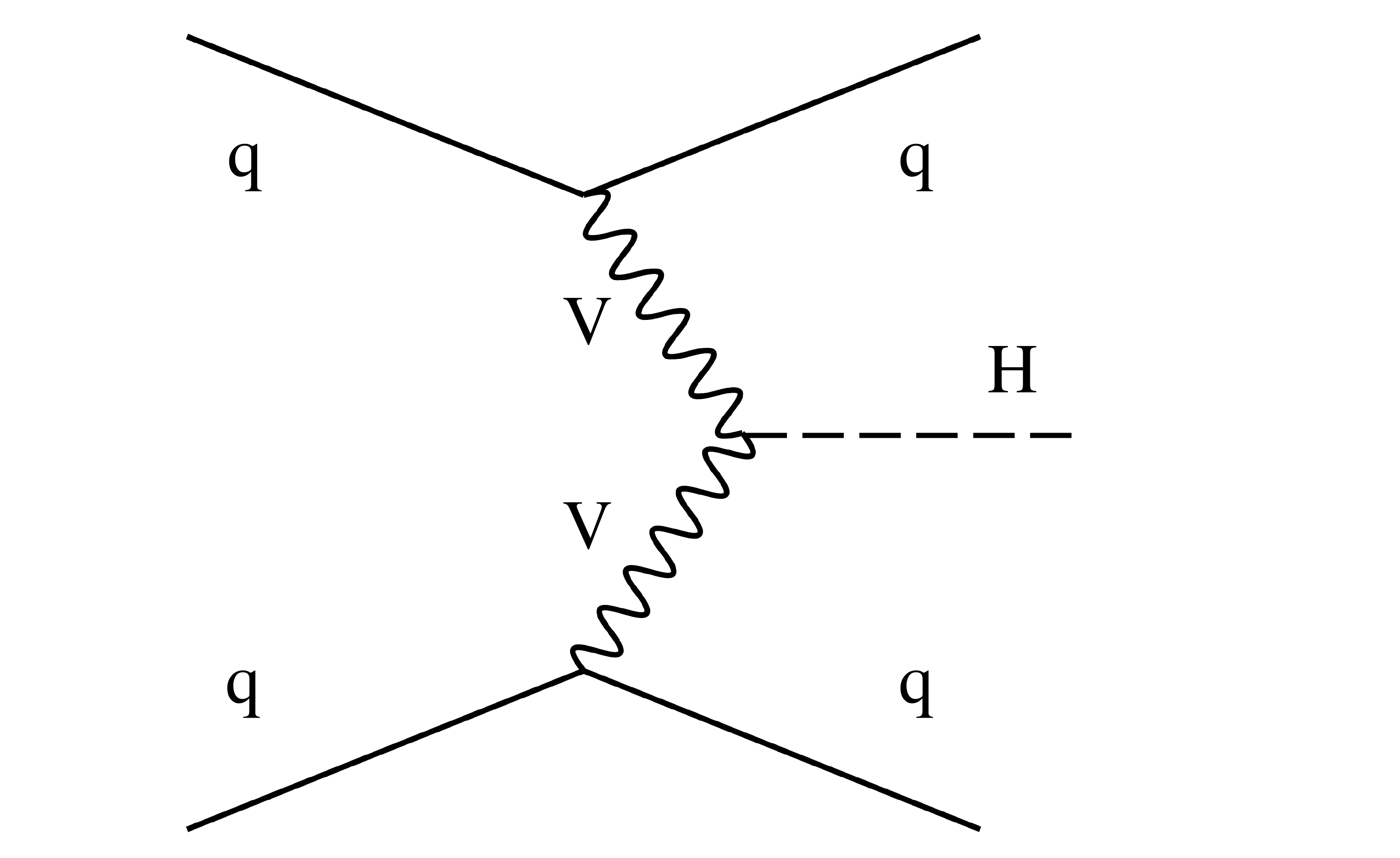}
~~~~~~
\includegraphics[width=0.4\textwidth]{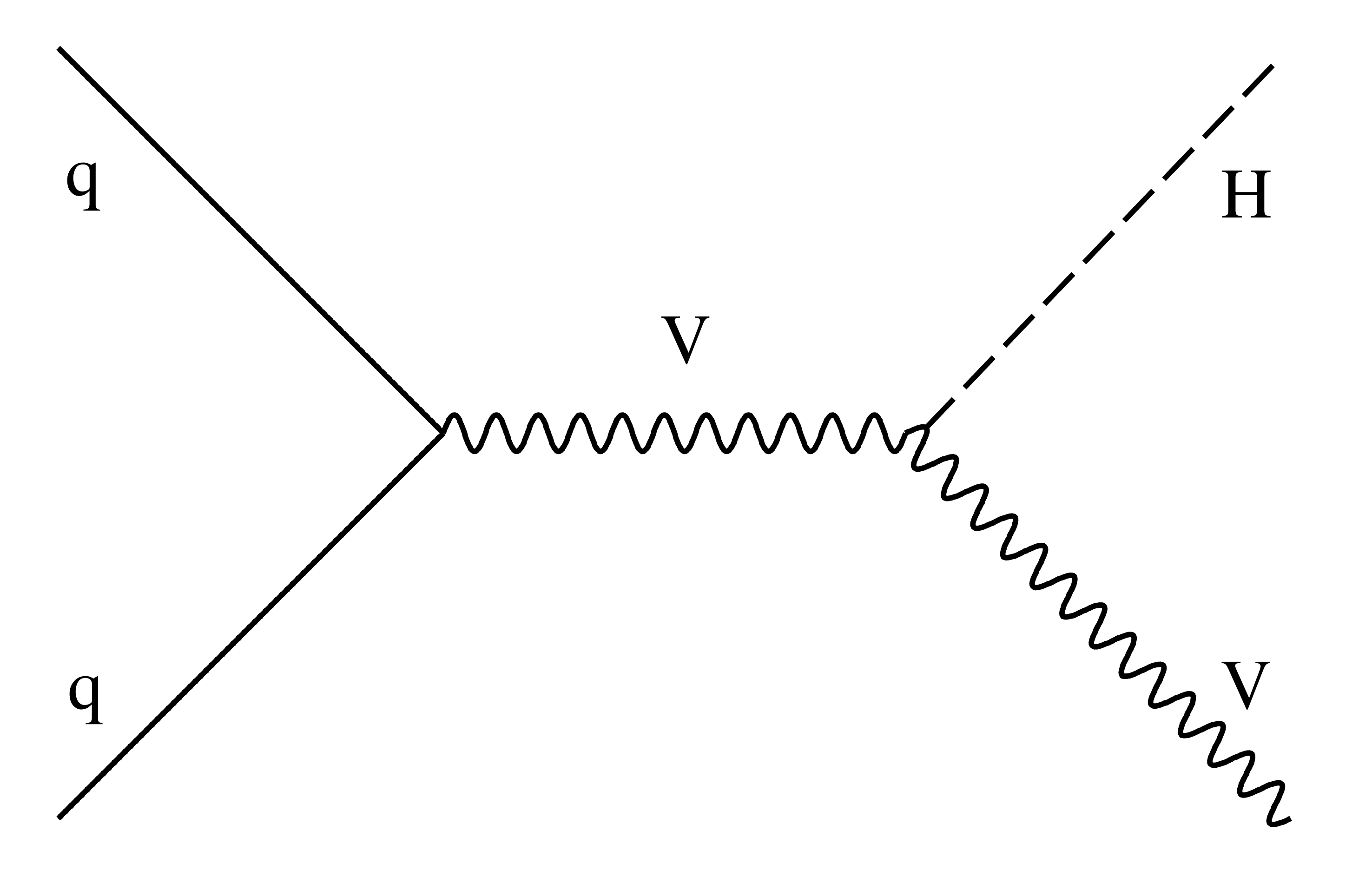}
\caption
{Leading-order Feynman diagrams for the \VBF (left) and \VH (right) production modes.
\label{fig:Feyn_2}
}
\end{figure*}

\begin{figure*}[!tbhp]
\centering
\includegraphics[width=0.35\textwidth]{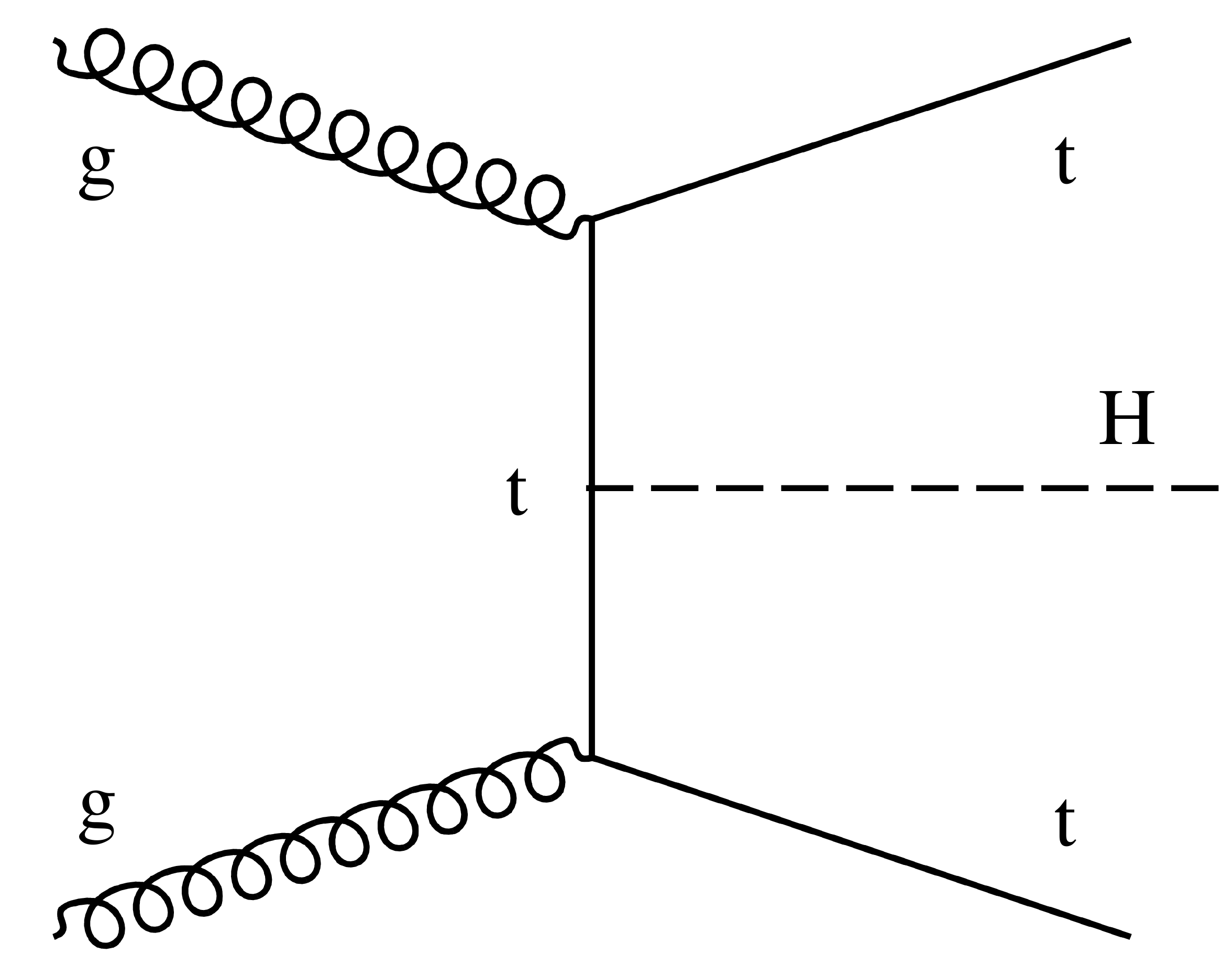}
~~~~~~
\includegraphics[width=0.35\textwidth]{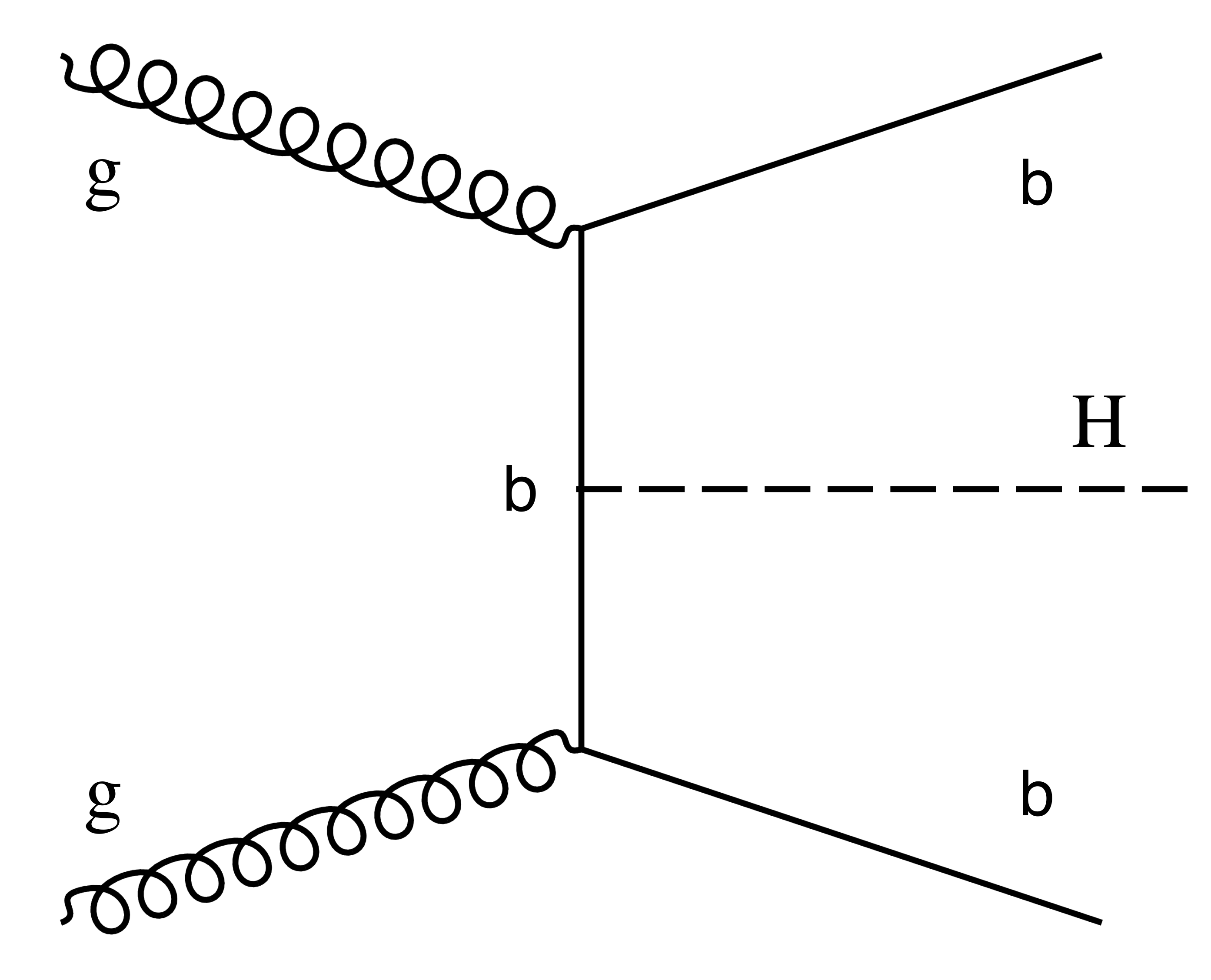}
\caption
{
Examples of leading-order Feynman diagrams for the $\ttH$ (left) and $\bbH$ (right) production modes. 
\label{fig:Feyn_3}
}
\end{figure*}

\begin{figure*}[!tbh]
\centering
\includegraphics[width=0.35\textwidth]{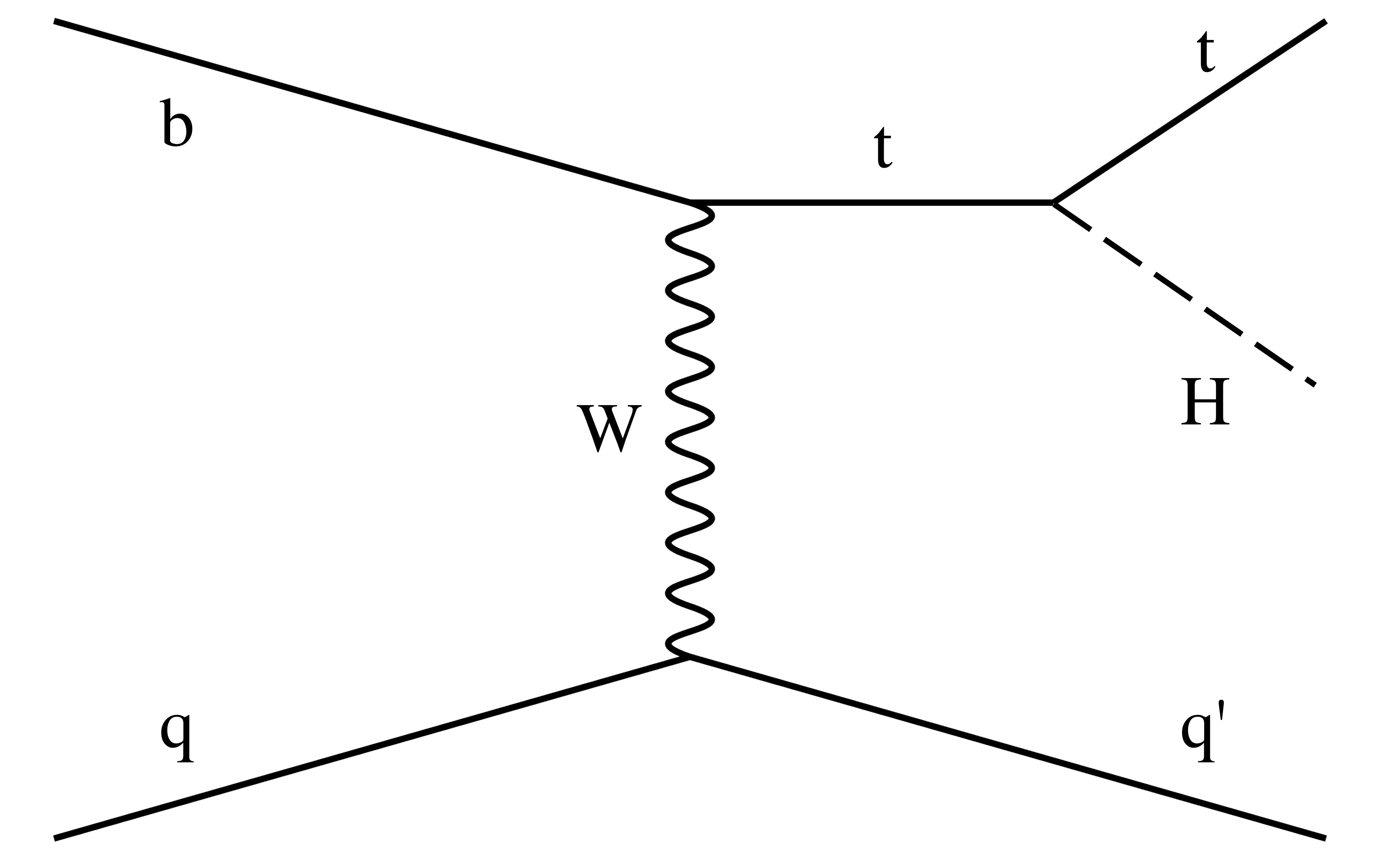}
~~~~~~
\includegraphics[width=0.35\textwidth]{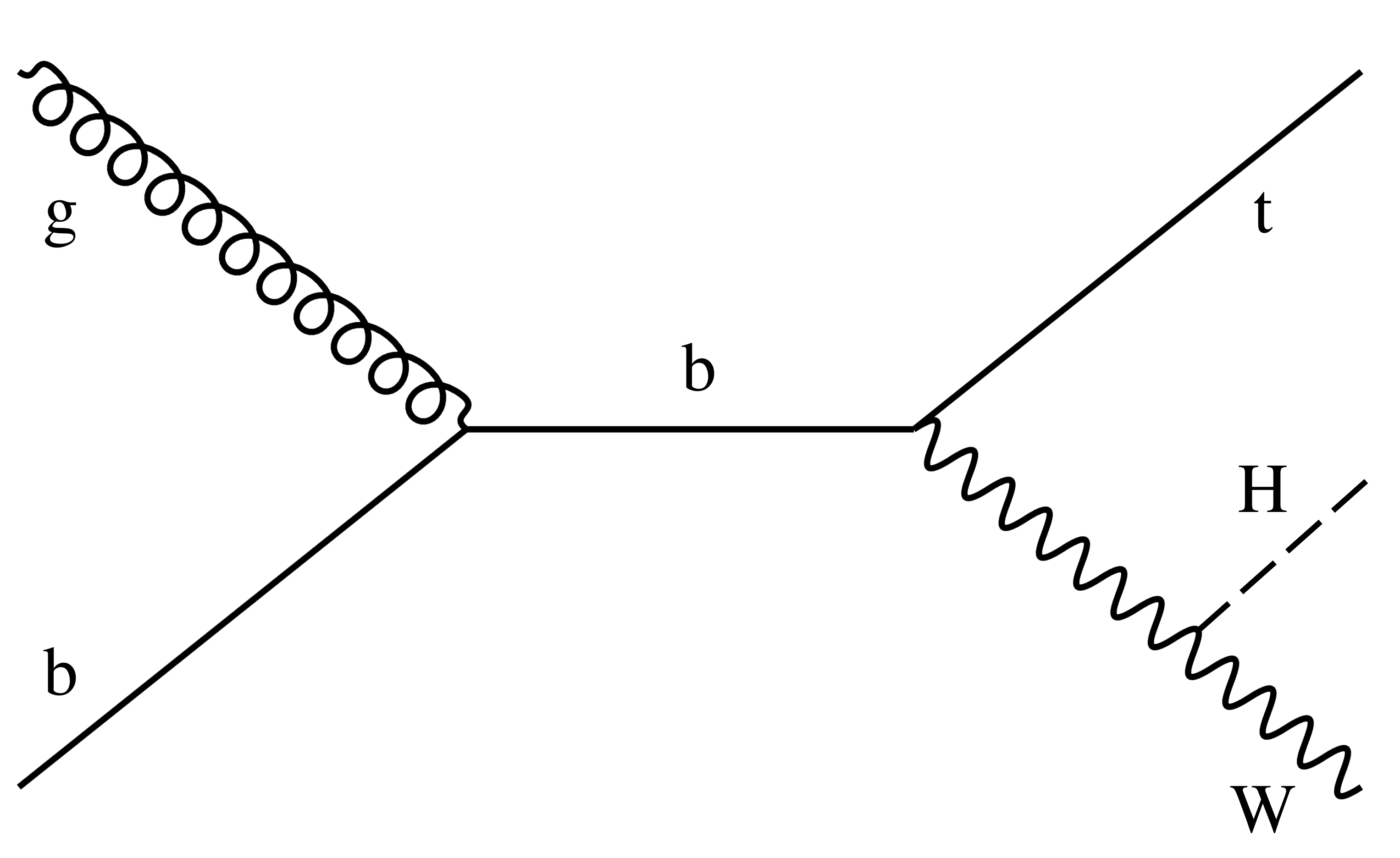}
\caption
{
Examples of leading-order Feynman diagrams for the $\tqH$ production mode.
\label{fig:Feyn_4}
}
\end{figure*}

\begin{figure*}[!tbh]
\centering
\includegraphics[width=0.4\textwidth]{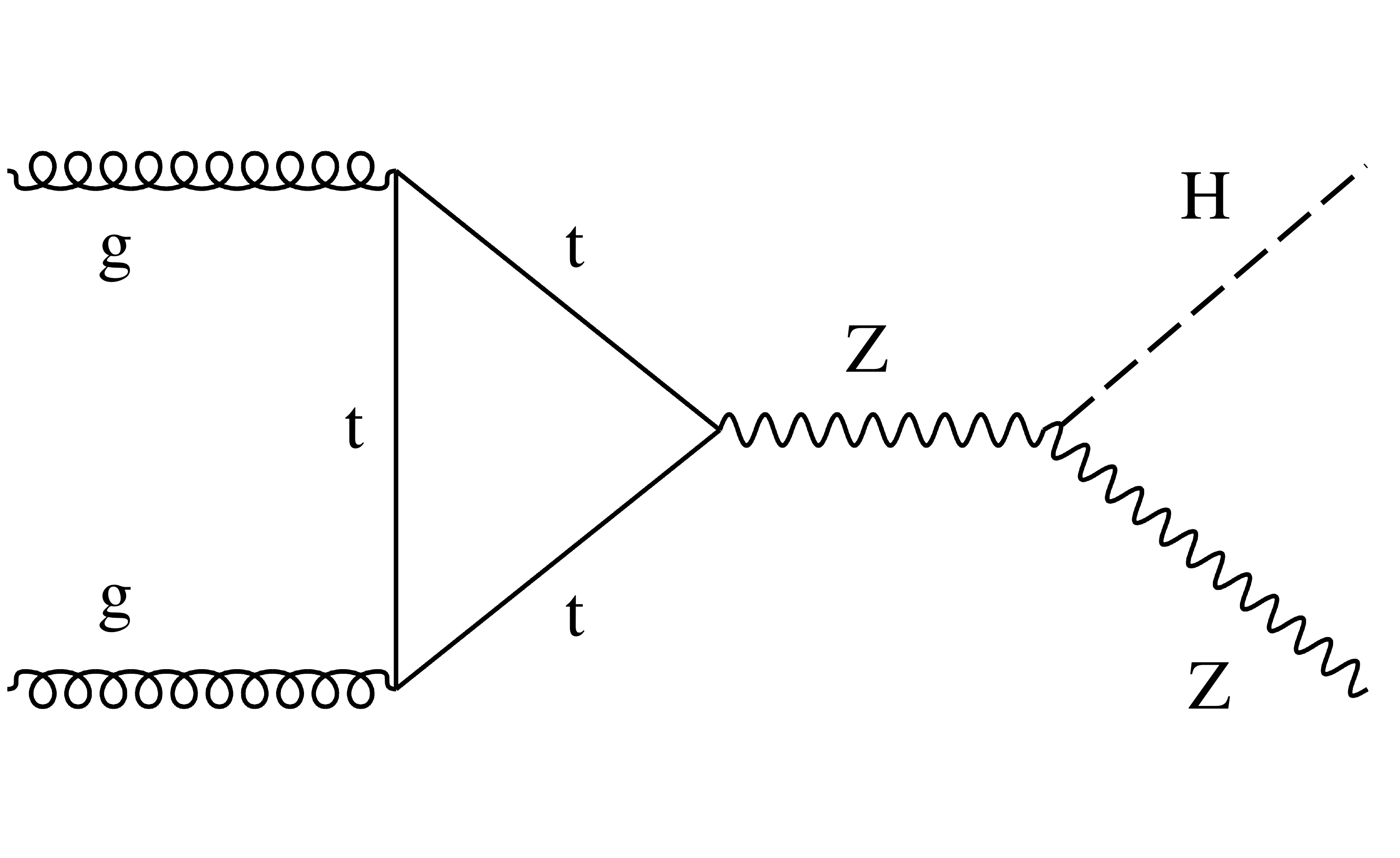}
~~~~~~
\includegraphics[width=0.4\textwidth]{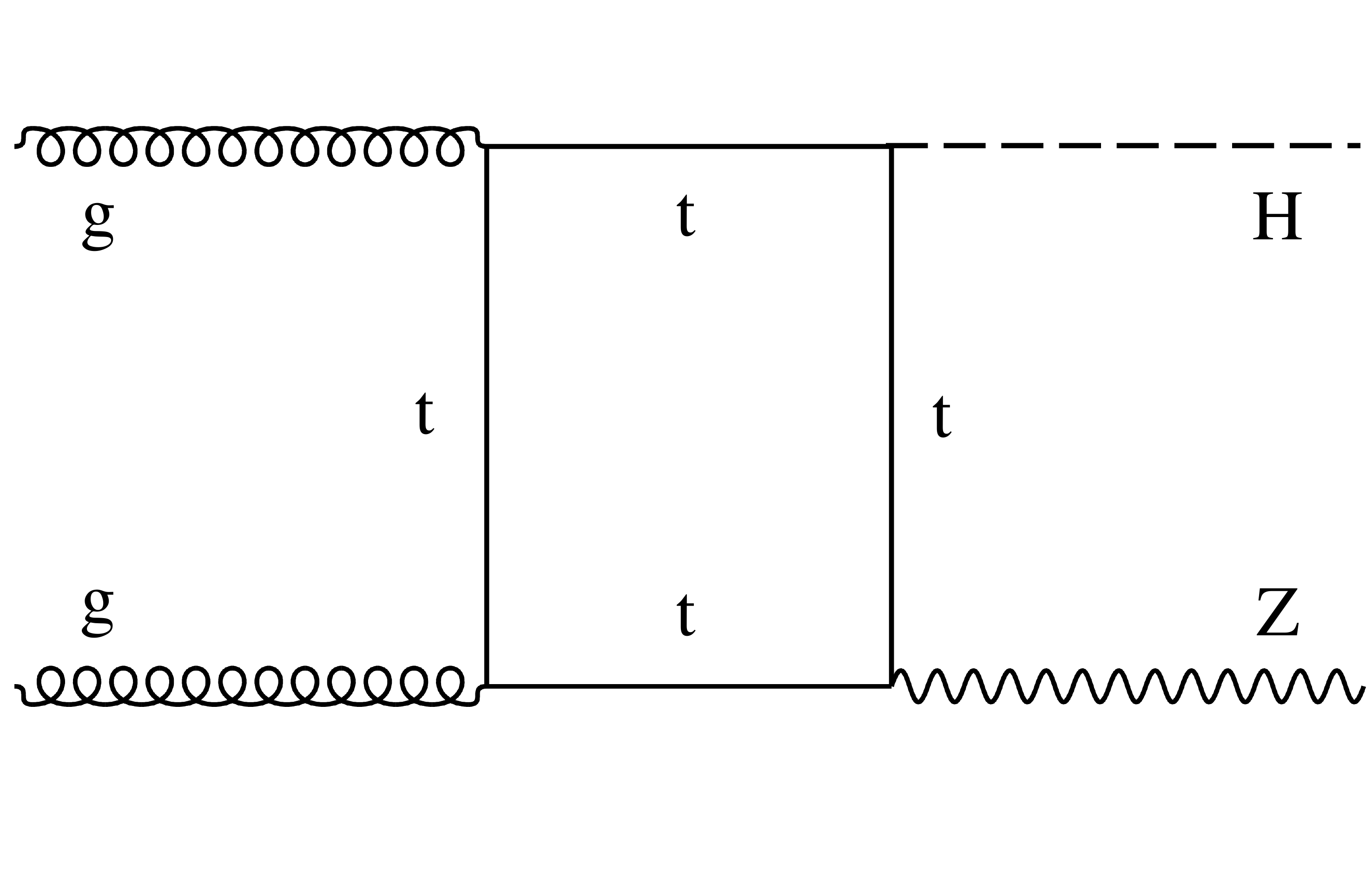}
\caption
{
Leading-order Feynman diagrams for the $\Pg\Pg\to\PZ\PH$ production mode.
\label{fig:Feyn_5}
}
\end{figure*}

The goal of this study is to search for $CP$ violation and, more generally, anomalous couplings of the \Hboson, 
in its interactions with fermions and vector bosons in the production and decay processes. 
These potential sources of $CP$ violation and anomalous tensor structures of interactions may arise from BSM effects,
including those considered in the EFT formulation. The dominant processes sensitive to such interactions 
are shown in Figs.~\ref{fig:Feyn_1}--\ref{fig:Feyn_5}~\cite{deFlorian:2016spz}. 

The main decay process considered in this paper is $\PH\to\PV\PV\to4\ell$, with the $\HVV$ vertex shown in Fig.~\ref{fig:Feyn_1}, right. 
The dominant \Hboson production mechanism is gluon fusion \ggH, shown in Fig.~\ref{fig:Feyn_1}, left. The dominant contribution 
to the gluon fusion loop comes from the top quark, with smaller contributions from the bottom quark and lighter quarks. 
However, contribution of new BSM states to the loop and variation of $CP$ properties of the \Hboson couplings to SM quarks
are also possible and are considered in this paper. 
The $\HVV$ vertex also appears in the vector boson fusion \VBF and  associated production with a weak vector 
boson $\PZ\PH$ or $\PW\PH$, shown in Fig.~\ref{fig:Feyn_2}, which are the next dominant production 
mechanisms of the \Hboson. 

The production of an \Hboson in association with a top quark pair \ttH is shown in Fig.~\ref{fig:Feyn_3}, left.
This is the main channel that allows to study the $CP$ property in the \Hboson coupling to fermions. 
We also combine our results with the recent \ttH measurements in the $\PH\to\gamma\gamma$ channel~\cite{Sirunyan:2020sum}. 
The production of an \Hboson in association with a single top quark \tqH is shown in Fig.~\ref{fig:Feyn_4}.
This production receives contributions from both $\HVV$ and $\Htt$ couplings, but its expected cross section 
is smaller than that of \ttH production. Both $\HVV$ and $\Htt$ couplings also contribute to the $\Pg\Pg\to\PZ\PH$ production mode,
shown in Fig.~\ref{fig:Feyn_5}. However, this gluon fusion production mode of $\PZ\PH$ is expected 
to contribute only about 5\% of the \VH process cross section shown in Fig.~\ref{fig:Feyn_2}, 
the dependence on anomalous $\HVV$ couplings is suppressed in this process~\cite{Gritsan:2020pib},
and for these reasons this production mode is neglected in this analysis. 
Finally, we also consider the \bbH production mode shown in Fig.~\ref{fig:Feyn_3}, right. 
However, this process does not provide kinematic features that could distinguish the $CP$ structure of interactions~\cite{Gritsan:2016hjl}
or the experimental signatures that would allow its isolation from the other more dominant production mechanisms. 

\subsection{Parameterization of production and decay amplitudes}

Anomalous effects in the \Hboson couplings to fermions, such as in the $\ttH$ and $\bbH$ production and 
partially in the $\tqH$ and $\Pg\Pg\to\PZ\PH$ production, can be parameterized with the amplitude 
\begin{align}
A(\Hff) = - \frac{m_\mathrm{f}}{v} \overline{\psi}_\mathrm{f} \left ( \kappa_\mathrm{f} + \mathrm{i}  \tilde\kappa_\mathrm{f} \gamma_5 \right ) {\psi}_\mathrm{f}, \label{eq:ampl-spin0-qq}
\end{align}
defined for each fermion type $\mathrm{f}$,
where $\overline{\psi}_\mathrm{f}$ and ${\psi}_\mathrm{f}$ are the fermions' Dirac spinors, 
$\kappa_\mathrm{f}$ and $\tilde\kappa_\mathrm{f}$ are the corresponding coupling strengths, 
$m_\mathrm{f}$ is the fermion mass, and $v$ is the SM Higgs field vacuum expectation value.
In the SM, the coupling strengths are $\kappa_\mathrm{f}=1$ and $\tilde\kappa_\mathrm{f}=0$. 
The presence of both $CP$-even $\kappa_\mathrm{f}$ and $CP$-odd $\tilde\kappa_\mathrm{f}$ couplings will lead to $CP$ violation. 
In an experimental analysis of the \bbH process it is not possible to resolve the 
$\kappa_\cPqb$ and $\tilde\kappa_\cPqb$ couplings~\cite{Gritsan:2016hjl}, 
but it is possible to resolve the $\kappa_\cPqt$ and $\tilde\kappa_\cPqt$ couplings
in the \ttH and \tqH processes, which we explore in this paper. 

Anomalous effects in EW \Hboson production (\VBF, $\ZH$, and $\WH$), \ggH production, 
$\PH\to\PV\PV$ decay, and partially in the $\tqH$ and $\Pg\Pg\to\PZ\PH$ production, are described by the $\PH\PV_1\PV_2$ couplings. 
The scattering amplitude describing the interaction between a spin-zero $\PH$ boson and two spin-one gauge bosons 
$\PV_1\PV_2$, such as $\ZZ$, $\PZ \gamma$, $\gamma\gamma$, $\PW\PW$, or $\Pg\Pg$, is written as
\begin{widetext}
\begin{align}
A(\PH\PV_1\PV_2) =
\frac{1}{v}
\left[ a_{1}^{\PV\PV}
+ \frac{\kappa_1^{\PV\PV}q_{\PV1}^2 + \kappa_2^{\PV\PV} q_{\PV2}^{2}}{\left(\Lambda_{1}^{\PV\PV} \right)^{2}} 
+ \frac{\kappa_3^{\PV\PV}(q_{\PV1} + q_{\PV2})^{2}}{\left(\Lambda_{Q}^{\PV\PV} \right)^{2}} \right]
m_{\PV1}^2 \epsilon_{\PV1}^* \epsilon_{\PV2}^*  
\nonumber \\
+ \frac{1}{v}a_{2}^{\PV\PV}  f_{\mu \nu}^{*(1)}f^{*(2),\mu\nu}
+ \frac{1}{v}a_{3}^{\PV\PV}   f^{*(1)}_{\mu \nu} {\tilde f}^{*(2),\mu\nu} ,
\label{eq:formfact-fullampl-spin0}
\end{align}
\end{widetext}
where $f^{(i){\mu \nu}} = \epsilon_{{\PV}i}^{\mu}q_{{\PV}i}^{\nu} - \epsilon_{{\PV}i}^\nu q_{{\PV}i}^{\mu}$,
${\tilde f}^{(i)}_{\mu \nu} = \frac{1}{2} \epsilon_{\mu\nu\rho\sigma} f^{(i),\rho\sigma}$, and
$\epsilon_{{\PV}i}$, $q_{{\PV}i}$, and $m_{{\PV}i}$ are the polarization vector, four-momentum, and pole mass 
of a gauge boson $i=1$ or 2. The constants $\Lambda_{1}$ and $\Lambda_{Q}$ are the scales 
of BSM physics necessary to keep the $\kappa_i^{\PV\PV}$ couplings unitless, 
and $a_1^{\PV\PV}$, $a_2^{\PV\PV}$, $a_3^{\PV\PV}$, $\kappa_1^{\PV\PV}$, $\kappa_2^{\PV\PV}$, and $\kappa_3^{\PV\PV}$ 
are real numbers that modify the corresponding amplitude terms.
Equation~(\ref{eq:formfact-fullampl-spin0}) describes couplings to both EW bosons and gluons, so $\PH\PV_1\PV_2$
 can stand for $\HVV$ or $\Hgg$. 

In Eq.~(\ref{eq:formfact-fullampl-spin0}),
the only nonzero tree-level contributions in the SM are $a_{1}^{\PZ\PZ}\ne 0$ and $a_{1}^{\PW\PW} \ne 0$.
In the SM, $a_{1}^{\PZ\PZ}=a_{1}^{\PW\PW}=2$.
The rest of the ${\PZ\PZ}$ and ${\PW\PW}$ couplings are considered to be anomalous contributions, 
which are either small contributions arising in the SM because of loop effects or new BSM contributions. 
Among the anomalous contributions, considerations of symmetry and gauge invariance require
$\kappa_1^{\PZ\PZ}=\kappa_2^{\PZ\PZ}$, $\kappa_1^{\PW\PW}=\kappa_2^{\PW\PW}$, and
$a_{1}^{\PZ\gamma}=a_{1}^{\gamma\gamma}=a_{1}^{\Pg\Pg}=\kappa_1^{\gamma\gamma}=\kappa_2^{\gamma\gamma}=
\kappa_1^{\Pg\Pg}=\kappa_2^{\Pg\Pg}=\kappa_1^{\PZ\gamma}=\kappa_3^{\PV\PV}=0$~\cite{Gritsan:2020pib}.
Therefore, there are a total of 13 independent parameters describing couplings of the \Hboson\ to EW gauge bosons
and two parameters describing couplings to gluons. 
The presence of any of the $CP$-odd couplings $a_{3}^{\PV\PV}$ together with any of the other couplings, which are all $CP$-even,
will lead to $CP$ violation in a given process. 

Since in our analysis it is not possible to disentangle the top quark, bottom quark, and any other heavy BSM particle 
contributions to the gluon fusion loop from kinematic features of the event, we parameterize the \Hgg coupling with only two 
parameters: $CP$-even $a_2^{\Pg\Pg}$ and $CP$-odd  $a_3^{\Pg\Pg}$, which absorb all SM and BSM loop contributions. 
However, when the gluon fusion process is analyzed jointly with the \ttH and \tqH processes, it may be possible 
to disentangle the top quark contributions in the loop from the relative rates of the processes, and we allow these 
contributions to be separated. 

\subsection{Symmetry considerations and SMEFT formulation}

The formulation in Eqs.~(\ref{eq:ampl-spin0-qq}) and~(\ref{eq:formfact-fullampl-spin0})
is presented in the approach of anomalous amplitude decomposition. However, it is fully equivalent to the
Lagrangian parameterization with dimension-4 operators, such as the $a_{1}^{\PV\PV}$ term in Eq.~(\ref{eq:formfact-fullampl-spin0}),
and dimension-6 operators, such as the other terms in Eq.~(\ref{eq:formfact-fullampl-spin0}), using the 
mass eigenstate basis~\cite{deFlorian:2016spz}. The dimension-8 and higher-dimension contributions are neglected. 
We apply additional symmetry considerations discussed below, which reduce the 
number of independent parameters to be measured. The chosen basis coincides with the Higgs basis~\cite{deFlorian:2016spz}
under SU(2)$\times$U(1) symmetry. The choice of the EFT operator basis is motivated
by the natural parameterization in terms of observable states and, as a consequence, allows more transparent 
construction of the data analysis and presentation of the results. However, we also present results in the 
Warsaw basis~\cite{deFlorian:2016spz}, using tools in Refs.~\cite{Falkowski:2015wza,Gritsan:2020pib} to perform the translation.
Our approach with SU(2)$\times$U(1) symmetry is equivalent to the SMEFT formulation~\cite{deFlorian:2016spz}.

The parameterization in Eq.~(\ref{eq:formfact-fullampl-spin0}) is the most general one, and we apply SU(2)$\times$U(1) 
symmetry in the relationships of anomalous couplings as follows~\cite{deFlorian:2016spz,Gritsan:2020pib}:
\begin{widetext}
\begin{align}
  a_1^{\WW} &= a_1^{\ZZ}  + \frac{\Delta m_{\PW}}{m_{\PW}},   \label{eq:EFT1}  \\
  a_2^{\WW} &= c_\mathrm{w}^2 a_2^{\ZZ} + s_\mathrm{w}^2 a_2^{\gamma\gamma} + 2 s_\mathrm{w} c_\mathrm{w} a_2^{\PZ\gamma}, \label{eq:EFT2} \\
  a_3^{\WW} &= c_\mathrm{w}^2 a_3^{\ZZ} + s_\mathrm{w}^2 a_3^{\gamma\gamma} + 2 s_\mathrm{w} c_\mathrm{w} a_3^{\PZ\gamma}, \label{eq:EFT3} \\
  \frac{\kappa_1^{\WW}}{(\Lambda_1^{\WW})^2} (c_\mathrm{w}^2-s_\mathrm{w}^2) &= \frac{\kappa_1^{\ZZ}}{(\Lambda_1^{\ZZ})^2}
                                                           +2 s_\mathrm{w}^2 \frac{a_2^{\gamma\gamma}-a_2^{\ZZ}}{m_\PZ^2} 
                                                          +2 \frac{s_\mathrm{w}}{c_\mathrm{w}} (c_\mathrm{w}^2-s_\mathrm{w}^2) \frac{a_2^{\PZ\gamma}}{m_\PZ^2},    \label{eq:EFT4} \\
  \frac{\kappa_2^{\PZ\gamma}}{(\Lambda_1^{\PZ\gamma})^2} (c_\mathrm{w}^2-s_\mathrm{w}^2) &= 2 s_\mathrm{w} c_\mathrm{w} \left( \frac{\kappa_1^{\ZZ}}{(\Lambda_1^{\ZZ})^2} 
                                                                              + \frac{ a_2^{\gamma\gamma} - a_2^{\ZZ}}{m_\PZ^2}  \right)
                                                                               +2 (c_\mathrm{w}^2-s_\mathrm{w}^2) \frac{a_2^{\PZ\gamma}}{m_\PZ^2},         \label{eq:EFT5}
\end{align}
\end{widetext}
where $c_\mathrm{w}=\cos\theta_W$, $s_\mathrm{w}=\sin\theta_W$, ${m}_{\PW}$ and ${m}_{\PZ}$ are the ${\PW}$ and 
${\PZ}$ boson masses, and $\Delta{m}_{\PW}$ is a shift in the ${\PW}$ mass. 
Since ${m}_{\PW}$ is measured to high precision~\cite{PDG2020}, we set $\Delta{m}_{\PW}=0$, leading 
to $a_{1}^{\PZ\PZ}=a_{1}^{\PW\PW}$. The latter relationship also appears under custodial symmetry~\cite{Sikivie:1980hm}.
Therefore, the set of $13+2$ independent parameters describing the $\HVV+\Hgg$ couplings 
can be reduced to $8+2$ with the above symmetry relationships. 

In our measurements, we further reduce the number of independent parameters in the following way. 
We assume that the four loop-induced couplings $a_{2,3}^{\gamma\gamma}$ and $a_{2,3}^{\PZ\gamma}$
are constrained to yield the SM rates of the direct decays $\PH\to\gamma\gamma$ and $\PZ\gamma$.
Therefore, in our analysis of EW production and $\PH\to 4\ell$ decay, we set these four couplings to zero 
because their allowed values are expected to have negligible effect in our coupling measurements. 
These four anomalous couplings have been tested in our earlier analysis of the $\PH\to\PV\PV\to4\ell$ process
with Run~1 data~\cite{Khachatryan:2014kca} and the obtained constraints were significantly looser than 
those from the direct decays with \onshell photons. 

We adopt two approaches to set the relationship between the $\HZZ$ and $\HWW$ couplings. 
The $\HWW$ couplings do not contribute to the $\PH\to4\ell$ decay, but they do contribute to the EW production. 
In this paper, the relationship between the $\HZZ$ and $\HWW$ couplings is mostly relevant for \VBF production with 
\ZZ and \WW fusion. There are no kinematic differences between these two processes and it is impossible 
to disentangle the \HZZ and \HWW couplings from these data. 
We used Approach~1 in our previous publications~\cite{Sirunyan:2017tqd,Sirunyan:2019twz,Sirunyan:2019nbs},
where we set $a_i^{\PW\PW}=a_i^{\PZ\PZ}$.
Approach~2 corresponds to the SMEFT formulation with SU(2)$\times$U(1) symmetry. 

In Approach~1, we set the \ZZ and \WW couplings to be equal, $a_i^{\PW\PW}=a_i^{\PZ\PZ}$. 
Formally, this could be considered as the relationship in Eqs.~(\ref{eq:EFT1}--\ref{eq:EFT4}) in the limiting case $c_\mathrm{w}=1$. 
As a result, we are left with four anomalous couplings to be measured, in addition to the SM-like couplings $a_1$:
$a_2$, $a_3$, $\kappa_1/(\Lambda_1)^2$, and $\kappa_2^{\PZ\gamma}/(\Lambda_1^{\PZ\gamma})^2$,
where we drop the \ZZ superscript from the couplings.
We adopt this approach both for its simplicity and to be able to relate the four anomalous couplings 
constrained with the $\PH\to4\ell$ decay to the equivalent four couplings in the PO approach.
This requires an independent measurement of the $\kappa_2^{\PZ\gamma}$ term, which would 
otherwise be eliminated by the relationship in Eq.~(\ref{eq:EFT5}). 
Therefore, this approach is slightly less restrictive than the SMEFT formulation adopted in Approach~2 discussed below. 

In Approach~2, we adopt the full set of SU(2)$\times$U(1) symmetry 
relationships in Eqs.~(\ref{eq:EFT1}--\ref{eq:EFT5}) with $s_\mathrm{w}^2 =0.23119$~\cite{PDG2020}. 
The number of independent $\HVV$ parameters is further reduced from five to four: 
$a_1$, $a_2$, $a_3$, and $\kappa_1/(\Lambda_1)^2$.
There is a linear one-to-one relationship between the amplitude couplings in Eq.~(\ref{eq:formfact-fullampl-spin0})
and the EFT couplings in the Higgs basis in the notation of Refs.~\cite{deFlorian:2016spz,Gritsan:2020pib}:
\begin{align}
  & \delta c_\mathrm{z}  = \frac12 a_1 - 1,   \label{eq:EFTpar1}  \\
  & c_{\mathrm{z} \Box} = \frac{m_\PZ^2 s_\mathrm{w}^2}{4\pi\alpha} \frac{\kappa_1}{(\Lambda_1)^2} ,   \label{eq:EFTpar2}  \\
  & c_\mathrm{zz} = -\frac{s_\mathrm{w}^2 c_\mathrm{w}^2}{2\pi\alpha} a_2 ,  \label{eq:EFTpar3}  \\
  & \tilde c_\mathrm{zz} = -\frac{s_\mathrm{w}^2 c_\mathrm{w}^2}{2\pi\alpha} a_3.  \label{eq:EFTpar4} 
\end{align}
Ignoring small loop-induced corrections, the above four parameters are expected to be zero in the SM.
Since we set $a_{2,3}^{\gamma\gamma}$ and $a_{2,3}^{\PZ\gamma}$ to zero, the four
corresponding parameters in the EFT Higgs basis 
$c_{\gamma\gamma}$, $c_{z\gamma}$, $\tilde{c}_{\gamma\gamma}$, and $\tilde{c}_{z\gamma}$
are also zero. 

In the case of $\Hgg$ couplings, the two EFT parameters are defined following the notation of Refs.~\cite{deFlorian:2016spz,Gritsan:2020pib} as:
\begin{align}
  &  c_{\Pg\Pg} = -\frac{1}{2\pi\alpS} a_2^{\Pg\Pg} ,  \label{eq:EFTpar5}  \\
  & \tilde c_{\Pg\Pg} = -\frac{1}{2\pi\alpS} a_3^{\Pg\Pg} ,   \label{eq:EFTpar6} 
\end{align}
where in the SM, $c_{\Pg\Pg} = 0$ and $\tilde c_{\Pg\Pg} =0$, and the SM process is generated by the quark loop
not accounted for in the $a_2^{\Pg\Pg}$ coupling. 
Finally, in the case of the $\Hff$ couplings, the $\kappa_\mathrm{f}$  and $\tilde\kappa_\mathrm{f}$
parameters in Eq.~(\ref{eq:ampl-spin0-qq}) can be treated as EFT parameters~\cite{deFlorian:2016spz},
where in the SM, $\kappa_\mathrm{f}=1$  and $\tilde\kappa_\mathrm{f}=0$.

\subsection{Parameterization of cross sections}
\label{sec:cross-section}

We use the narrow-width approximation and parameterize differential cross section of the \onshell\ \Hboson 
production process $j$ and decay to a final state $f$ following Refs.~\cite{deFlorian:2016spz,Gritsan:2020pib} as:
\ifthenelse{\boolean{cms@external}}
{
  \begin{multline}
    \sigma(j\to\PH\to f)\\
     \propto
    \frac{\left(\sum_{il}  \alpha_{il}^{(j)}a_ia_l\right)\left(\sum_{mn} \alpha_{mn}^{(f)}a_ma_n\right)}{\Gamma_{\PH}} ,
    \label{eq:diff-cross-section2}
  \end{multline}
}
{
  \begin{align}
    \sigma(j\to\PH\to f) \propto
    \frac{\left(\sum_{il}  \alpha_{il}^{(j)}a_ia_l\right)\left(\sum_{mn} \alpha_{mn}^{(f)}a_ma_n\right)}{\Gamma_{\PH}} ,
    \label{eq:diff-cross-section2}
  \end{align}
}
where $a_i$ are the real couplings describing the $\Hff$, $\Hgg$, or $\HVV$ interactions and include 
generically the $\kappa_i$ in Eqs.~(\ref{eq:ampl-spin0-qq}) and (\ref{eq:formfact-fullampl-spin0}).
The coefficients $\alpha^{(k)}_{il}$ are in general functions of kinematic observables for the differential 
cross section distributions and are modeled with simulation.
The total width $\Gamma_{\PH}$ depends on the couplings $a_i$ and potentially on the partial decay 
width to unobserved or invisible final states, a dependence that must be taken into account when interpreting
cross section measurements in terms of couplings. 

When we perform the amplitude analysis of the data, in the likelihood based on Eq.~(\ref{eq:diff-cross-section2}),
we keep all cross-terms in the expansion of powers of $\Lambda^{-N}$ with $N=0,2,4,6,8$, where formally each 
dimension-6 operator, or anomalous $a_i$ coupling, receives the $\Lambda^{-2}$ contribution in Eq.~(\ref{eq:formfact-fullampl-spin0}), 
even if not explicitly shown, while the SM tree-level coupling carries no such contribution. 
While this may create inconsistency in the EFT approach with the $\Lambda^{-N}$ terms
kept or neglected from the higher-dimension contributions, this allows us to keep the likelihood positive definite, which is an 
important consideration in the experimental analysis of the data discussed in Section~\ref{sec:fit}. Because interference 
contributions may become negative, dropping certain terms in the expansion may lead to negative probability. 
The importance of $N=4,6,8$ contributions may be considered as testing whether the current precision is sufficient 
to treat our results within the EFT approach, and we leave this test to the interpretation of the results. 
However, regardless of EFT validity, our results are presented in a fully self-consistent 
formulation of amplitude decomposition, which can either be translated to the EFT interpretation
or treated as a test of consistency of the data with the SM, including the search for new sources of $CP$ violation.

\subsection{Parameterization of the signal strength and cross section fractions}
\label{sec:pheno-fai}

We present the primary results in terms of cross sections, 
or equivalently, signal strengths $\mu_j=\sigma_j/\sigma_j^\mathrm{SM}$,
and the fractional contributions $f_{ai}$ of the couplings $a_i$ to cross sections 
$\left(\sum_{mn} \alpha_{mn} a_ma_n\right)$ of a given decay process.
The ratios of couplings entering Eq.~(\ref{eq:diff-cross-section2}) can be expressed through $f_{ai}$,
and the common factors, such as the total width $\Gamma_{\PH}$ and the SM-like coupling squared, 
are absorbed into the signal strength.
This formulation with $\mu_j$ and $f_{ai}$ allows the presentation of experimental results in the most direct way, 
with a minimal and complete set of parameters describing the given processes. 
This approach has several convenient features.
The cross sections and their ratios are invariant with respect to the coupling 
convention, such as the scaling in Eqs.~(\ref{eq:EFTpar1}--\ref{eq:EFTpar6}). 
The cross section fractions $f_{ai}$ reflect kinematic features in either production or decay in a direct way.
They are conveniently bound between $-1$ and $+1$, and most systematic uncertainties cancel in the ratio.

The cross section fraction for \Hff couplings is defined as 
\begin{align}
\fF = \frac{\abs{\tilde\kappa_\mathrm{f}}^2 }{\abs{\kappa_\mathrm{f}}^2 + \abs{\tilde\kappa_\mathrm{f}}^2}
\, \sign\left(\frac{\tilde\kappa_\mathrm{f}}{\kappa_\mathrm{f}} \right) .
\label{eq:fCP_definitions}
\end{align}
Similarly, the cross section fraction for \Hgg couplings is defined as 
\begin{align}
\fG = \frac{\abs{a_3^{\Pg\Pg}}^2 } {\abs{a_2^{\Pg\Pg}}^2  + \abs{a_3^{\Pg\Pg}}^2 } 
\, \sign\left(\frac{a_{3}^{\Pg\Pg}}{a_{2}^{\Pg\Pg}} \right) .
\label{eq:fggH_definitions}
\end{align}
Both definitions incorporate the relative sign of the possible BSM $CP$-odd and SM-like $CP$-even couplings. 
They are based on the observation that the cross sections of the $\PH\to\Pg\Pg$
process are equal for $a_2^{\Pg\Pg}=1$ and $a_3^{\Pg\Pg}=1$, as are the cross sections of the
$\PH\to\mathrm{f}\overline{\mathrm{f}}$ process for $\kappa_\mathrm{f}=1$ and 
$\tilde\kappa_\mathrm{f}=1$ in the limit of $m_\mathrm{f} \ll m_\PH$. 
We note that $\fG$ is defined following the convention that ${a_2^{\Pg\Pg}}$ and ${a_3^{\Pg\Pg}}$
absorb both point-like interactions and quark contributions to the loop. 
Following Ref.~\cite{Gritsan:2020pib}, the $\fG$ measurement can also be interpreted in terms of $\fF$ 
under the assumption that only the top and bottom quarks contribute to gluon fusion with 
$\kappa_\cPqt=\kappa_\cPqb$ and $\tilde\kappa_\cPqt=\tilde\kappa_\cPqb$:
\ifthenelse{\boolean{cms@external}}
{
  \begin{equation}
    \begin{aligned}
    \abs{\fF} &= \left(1 +2.38 \left[  \frac{1}{\abs{\fG}} -1  \right]  \right)^{-1}     
    &= \sin^2\alpha^{\Hff},
    \end{aligned}
    \label{eq:fai-relationship-hgg-tth}
  \end{equation}  
}
{\begin{align}
	\abs{\fF} = \left(1 +2.38 \left[  \frac{1}{\abs{\fG}} -1  \right]  \right)^{-1} = \sin^2\alpha^{\Hff},
\label{eq:fai-relationship-hgg-tth}
\end{align}
}
where the signs of \fF and \fG are equal, and $\alpha^{\Hff}$ is an effective parameter sometimes used
to describe the $CP$-odd contribution to the \Hboson Yukawa couplings. 
A more detailed analysis of the gluon fusion loop could be performed 
without the assumption that only the top and bottom quarks contribute. 

The cross section fractions in the \HVV couplings of the \Hboson to EW gauge bosons
require more parameters. Since in both of our approaches the \HWW couplings are expressed through 
other $a_i^{\PV\PV}$ couplings following Eqs.~(\ref{eq:EFT1}--\ref{eq:EFT4}),
and because we prefer that our definitions not depend on parton distribution functions (PDFs)
and other effects that involve measurement uncertainties,
we use the $\PH\to\PZ\PZ$ / $\PZ\gamma^*$ / $\gamma^*\gamma^* \to 2\Pe2\mu$ decay process
to define the cross section fractions as
 \begin{align}
 f_{ai}^{\PV\PV} =  \frac{\abs{a_i^{\PV\PV}}^2 \alpha_{ii}^{(2\Pe2\mu)} }{\sum_{j}{\abs{a_{j}^{\PV\PV}}^2 \alpha_{jj}^{(2\Pe2\mu)} } } 
 ~ \sign\left(\frac{a_i^{\PV\PV}}{a_{1}} \right),
\label{eq:fa_definitions_hvv}
\end{align}
where the $\alpha_{ii}^{(2\Pe2\mu)} $ coefficients are introduced in Eq.~(\ref{eq:diff-cross-section2}).
The numerical values of these coefficients are given in Table~\ref{tab:xsec_ratio}, where they are normalized
with respect to the $\alpha_{11}^{(2\Pe2\mu)} $ coefficient, corresponding to the cross section calculated
for $a_{1}=1$. The $\alpha_{ii}^{(2\Pe2\mu)} $ are the cross sections for $a_{i}^{\PV\PV}=1$, which are
different in the two approaches of the coupling relationship as a result of Eq.~(\ref{eq:EFT5}) adopted in Approach~2. 
The cross section fractions in Eq.~(\ref{eq:fa_definitions_hvv}) can be converted to coupling ratios as
\begin{align}
\frac{a_i^{\PV\PV}}{a_j^{\PV\PV}}=
\sqrt{\frac{\abs{f_{ai}^{\PV\PV}} \alpha_{jj}^{(2\Pe2\mu)} }{\abs{f_{aj}^{\PV\PV}} \alpha_{ii}^{(2\Pe2\mu)}} }
~ \sign\left(f_{ai}^{\PV\PV}f_{aj}^{\PV\PV}\right).
\label{eq:ai}
\end{align}

\begin{table}[!t]
\centering
\topcaption{
List of anomalous \HVV couplings $a_i^{\PV\PV}$ considered, the corresponding measured cross section fractions 
$f_{ai}^{\PV\PV}$ defined in Eq.~(\ref{eq:fa_definitions_hvv}), and the translation coefficients $\alpha_{ii}/\alpha_{11}$
in this definition with the relationship $a_i^{\PZ\PZ}=a_i^{\PW\PW}$ (Approach~1), and with the SMEFT relationship 
according to Eqs.~(\ref{eq:EFT1}--\ref{eq:EFT5}) (Approach~2). 
In the case of the $\kappa_1$ and $\kappa_2^{\PZ\gamma}$ couplings, the numerical values 
$\Lambda_{1}=\Lambda_{1}^{\PZ\gamma}=100\GeV$ are adopted in this calculation 
to make the coefficients have the same order of magnitude
and the negative sign indicates the convention in Eq.~(\ref{eq:fa_definitions_hvv}) adopted earlier~\cite{Khachatryan:2014kca}.
In Approach~2, $\kappa_2^{\PZ\gamma}$ is a dependent parameter expressed through Eq.~(\ref{eq:EFT5}) 
and does not require a translation coefficient. 
}
\label{tab:xsec_ratio}
\begin{scotch}{cccc}
\vspace{-0.3cm} \\
  \multicolumn{1}{c}{Coupling}         &  \multicolumn{1}{c}{Fraction}             &  \multicolumn{1}{c}{Approach 1} &  \multicolumn{1}{c}{Approach 2}  \\
  \multicolumn{1}{c}{$a_i^{\PV\PV}$}    &  \multicolumn{1}{c}{$f_{ai}^{\PV\PV}$}   & \multicolumn{1}{c}{$\alpha_{ii}/\alpha_{11}$} & \multicolumn{1}{c}{$\alpha_{ii}/\alpha_{11}$}  \\
[\cmsTabSkip]
\hline\\ [-1ex]
 $a_3$ & $f_{a3}$ & $0.153$ & $0.153$ \\
 $a_2$ & $f_{a2}$ & $0.361$ & $6.376$  \\
 $-\kappa_1$ & $f_{\Lambda1}$  & $0.682$ & $5.241$ \\
 $-\kappa_2^{\PZ\gamma}$ & $f_{\Lambda1}^{\PZ\gamma}$ & $1.746$ & \NA \\
[\cmsTabSkip]
\end{scotch}
\end{table}

The measured values of $\mu_j$ and $f_{ai}$ should be sufficient to adopt them in the fits for EFT parameters jointly 
with the data from other \Hboson, top quark, and EW measurements. They allow constraints on the $\kappa_i$ and $a_i$ 
couplings in Eqs.~(\ref{eq:ampl-spin0-qq}) and (\ref{eq:formfact-fullampl-spin0}). 
However, it is required to perform a simultaneous measurement of all production and decay 
channels of the \Hboson, including unobserved and invisible channels, as they contribute to the total width 
in Eq.~(\ref{eq:diff-cross-section2}). In this paper, we present only a limited interpretation of our data in 
terms of couplings by making certain assumptions about their relationship.  
We leave more extensive interpretation to a future combination with other channels.

\section{The CMS detector, data sets, and event reconstruction} 
\label{sec:cms}

The \Hell decay candidates are produced in proton-proton ($\Pp\Pp$) collisions at the LHC and
are collected and reconstructed in the CMS detector~\cite{Chatrchyan:2008zzk}. The data sample used in this analysis corresponds 
to integrated luminosities of $35.9\fbinv$ collected in 2016, $41.5\fbinv$ collected in 2017, and $59.7\fbinv$ collected in 2018, 
for a total of $137\fbinv$ collected during Run~2 at a $\Pp\Pp$ center-of-mass energy of 13\TeV.

The CMS detector comprises a silicon pixel and strip tracker, a lead tungstate crystal electromagnetic
calorimeter (ECAL), and a brass/scintillator hadron calorimeter, each composed of a barrel and two endcap sections,
all within a superconducting solenoid of 6\unit{m} internal diameter, providing a magnetic field of 3.8\unit{T}.
Extensive forward calorimetry complements the coverage provided by the barrel and endcap detectors.
Outside the solenoid are the gas-ionization detectors for muon measurements, which are embedded
in the steel flux-return yoke. A detailed description of the CMS detector can be found in Ref.~\cite{Chatrchyan:2008zzk}.

Events of interest are selected using a two-tiered trigger system. The first level, composed of custom hardware 
processors, uses information from the calorimeters and muon detectors to select events at a rate of around 100\unit{KHz} 
within a fixed latency of about 4\mus~\cite{Sirunyan:2020zal}. The second level, known as the high-level trigger, 
consists of a farm of processors running a version of the full event reconstruction software optimized for fast processing, 
and reduces the event rate to around 1\unit{KHz} before data storage~\cite{Khachatryan:2016bia}.

\subsection{Event reconstruction and selection}

The selection of $4\ell$ events and associated particles closely follows the methods used in the analyses
of the Run~1~\cite{Chatrchyan:2013mxa,Khachatryan:2014kca} 
and Run~2~\cite{Sirunyan:2017exp,Sirunyan:2017tqd,Sirunyan:2019twz,Sirunyan:2021rug} data sets.
The main triggers for the Run~2 analysis select either a pair of electrons or muons, or an electron and a muon, 
passing loose identification and isolation requirements.
The transverse momentum (\PT) for the leading electron or muon is required to be larger than 23 or 17\GeV,
while that of the subleading lepton is required to be larger than 12 or 8\GeV, respectively. 
To maximize the signal acceptance, triggers requiring three leptons with lower \pt thresholds and no isolation 
requirement are also used, as well as isolated single-electron and single-muon triggers with thresholds of 27 and 22\GeV
in 2016, or 35 and 27\GeV in 2017 and 2018, respectively. The overall trigger efficiency for
simulated signal events that pass the full selection chain of this analysis is larger than 99\%.

Event reconstruction is based on the particle-flow algorithm~\cite{Sirunyan:2017ulk},
which exploits information from all the CMS subdetectors to identify and reconstruct individual particles in the event.
The particle-flow candidates are classified as charged or neutral hadrons, photons, electrons, or muons, and they are
then used to build higher-level objects, such as jets, and to calculate the lepton isolation quantities.
Electrons or muons are reconstructed within the geometrical acceptance defined by a requirement
on the pseudorapidity $\abs{\eta} <2.5$ or 2.4 and $\PT >7$ or 5\GeV, 
with an algorithm that combines information from the tracker and the ECAL or muon system, respectively.
Muons are selected from a list of reconstructed muon track candidates by applying minimal requirements on the track in both
the muon system and inner tracker system, and taking into account compatibility with small energy deposits in the calorimeters.

To discriminate between leptons from prompt particle decays and those arising from hadron decays within jets, 
an isolation variable is calculated for electrons and muons~\cite{Sirunyan:2021rug}. 
An isolation requirement is imposed on the muons. 
Electrons are identified using a multivariate discriminant which includes observables sensitive to
the presence of bremsstrahlung along the electron trajectory, the geometrical and momentum-energy matching between the
electron trajectory and the associated cluster in the ECAL, the shape of the electromagnetic shower in the ECAL,
and variables that discriminate against electrons originating from photon conversions. This discriminant also includes 
the isolation to suppress electrons originating from EW decays of hadrons within jets~\cite{Sirunyan:2021rug}. 
A dedicated algorithm is used to collect the final-state radiation of leptons~\cite{Sirunyan:2017exp}.

The jets are clustered using the anti-\kt jet finding algorithm~\cite{Cacciari:2008gp,Cacciari:2011ma} with a distance 
parameter of 0.4. The jet momentum is determined as the vector sum of all particle momenta in the jet. 
Jets must satisfy $\pt>30\GeV$ and $\abs{\eta}<4.7$ and must be separated
from all selected lepton candidates and any selected final-state radiation photons with a requirement on the
parameter $\Delta R(\ell/\cPgg,{\mathrm{jet}})>0.4$, where $(\Delta R)^2 = (\Delta \eta)^2 + (\Delta \phi)^2$.
Jets are \cPqb-tagged using the DeepCSV algorithm~\cite{Sirunyan:2017ezt}, which combines information 
about impact parameter significance, the secondary vertex, and jet kinematics.

The reconstructed vertex with the largest value of summed physics-object $\pt^2$ is taken to be the
primary $\Pp\Pp$ interaction vertex. The physics objects are the jets, clustered using the jet finding algorithm~\cite{Cacciari:2008gp,Cacciari:2011ma}
 with the tracks assigned to candidate vertices as inputs, and the associated missing transverse
momentum, taken as the negative vector \pt sum of those jets.
In order to suppress muons from in-flight decays of hadrons and electrons from photon conversions, 
leptons are rejected if the ratio of their impact parameter in three dimensions, computed
with respect to the primary vertex position, to their uncertainty is greater than four.

\begin{figure}[!tbp]
	\centering
		\includegraphics[width=0.49\textwidth]{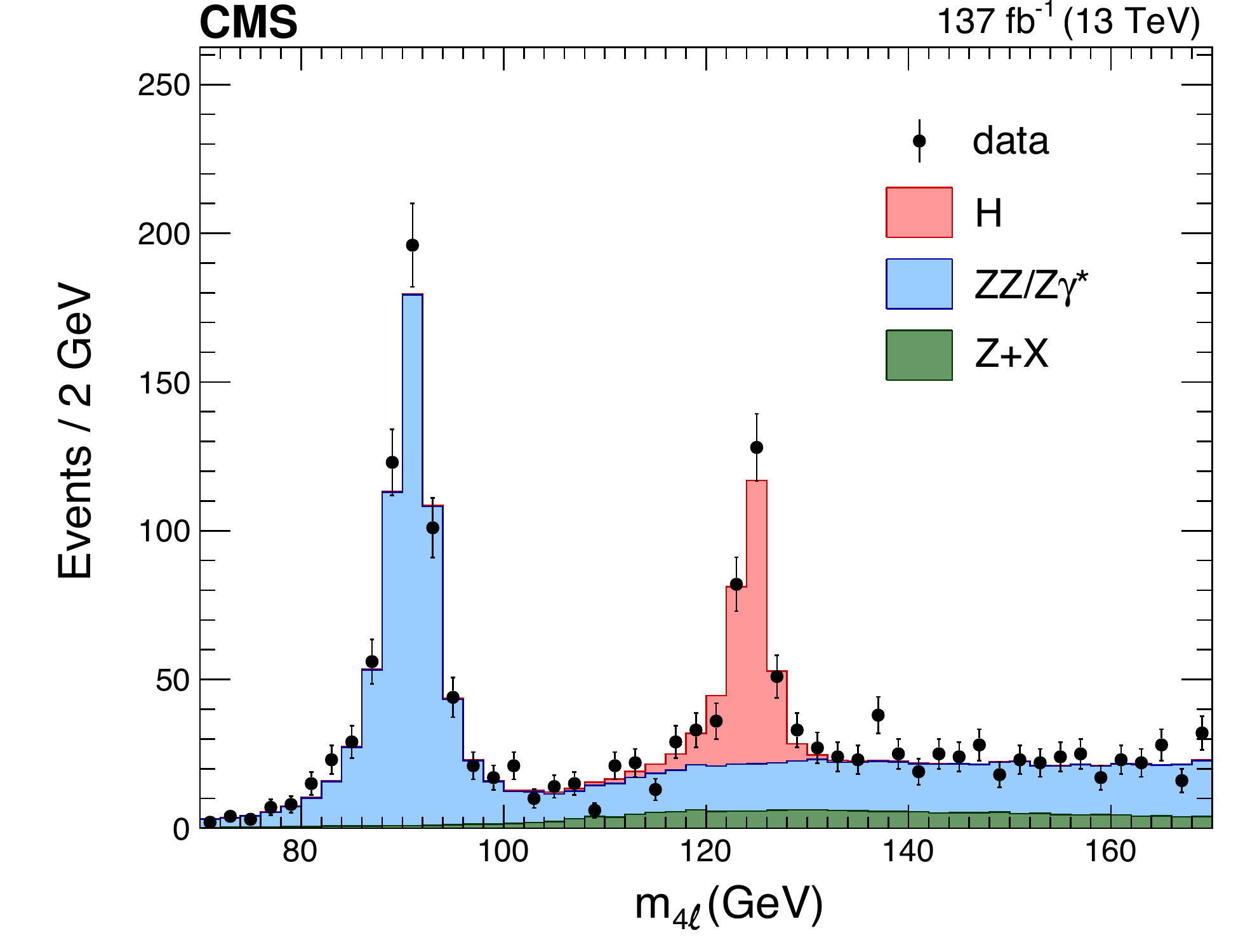}
		\caption{
		Four-lepton invariant mass distribution of observed events (data points) and expectation from MC simulation 
		or background estimates (histograms) in the region between 70 and 170\GeV~\cite{Sirunyan:2021rug}.
		The peaks from the $\PZ$ and $\PH\to4\ell$ decays are visible near 91 and 125\GeV, respectively. 
		\label{fig:combination}}
\end{figure}

We consider three mutually exclusive lepton flavor channels: $\PH\to\PV\PV\to 4\Pe$, $4\Pgm$, and $2\Pe 2\Pgm$.
At least two leptons are required to have $\pt >10\GeV$, and at least one is required to have $\pt >20\GeV$.
All four pairs of oppositely charged leptons that can be built with the four leptons are required
to satisfy $m_{\ell^{+}\ell^{-}} >4\GeV$, regardless of lepton flavor, to further suppress events with leptons 
originating from hadron decays in jet fragmentation or from the decay of low-mass resonances.
The $\PV$ candidates are formed with pairs of leptons of the same flavor and opposite charge
that pass the requirement $12 < m_{\ell^{+}\ell^{-}} < 120\GeV$, where $m_1$ is the invariant mass of the 
$\PV$ candidate that is closest to the nominal $\cPZ$ boson mass and $m_2$ is the mass of the other one. 
A value of $m_1>40\GeV$ is required. The reconstructed four-lepton invariant mass, \mell, distribution in the region 
between 70 and 170\GeV is shown in Fig.~\ref{fig:combination}. The \mell region between 105 and 140\GeV 
is considered in this analysis, which is wide enough to use sidebands for constraining the background 
normalization in the later fitting procedure~\cite{Sirunyan:2021rug}.

\subsection{Event categorization}

In order to perform a dedicated study of a particular kinematic topology, events are further split into several mutually 
exclusive categories based on the presence of other particles produced in association with the \Hboson candidate~\cite{Sirunyan:2021rug}.
Two independent categorization, Schemes~1 and~2, discussed below, are followed in this study. 
Scheme~1 targets \Htt and \Hgg anomalous couplings, while Scheme~2 targets \HVV anomalous couplings. 

\begin{figure*}[!tbp]
\centering
\includegraphics[width=0.32\textwidth]{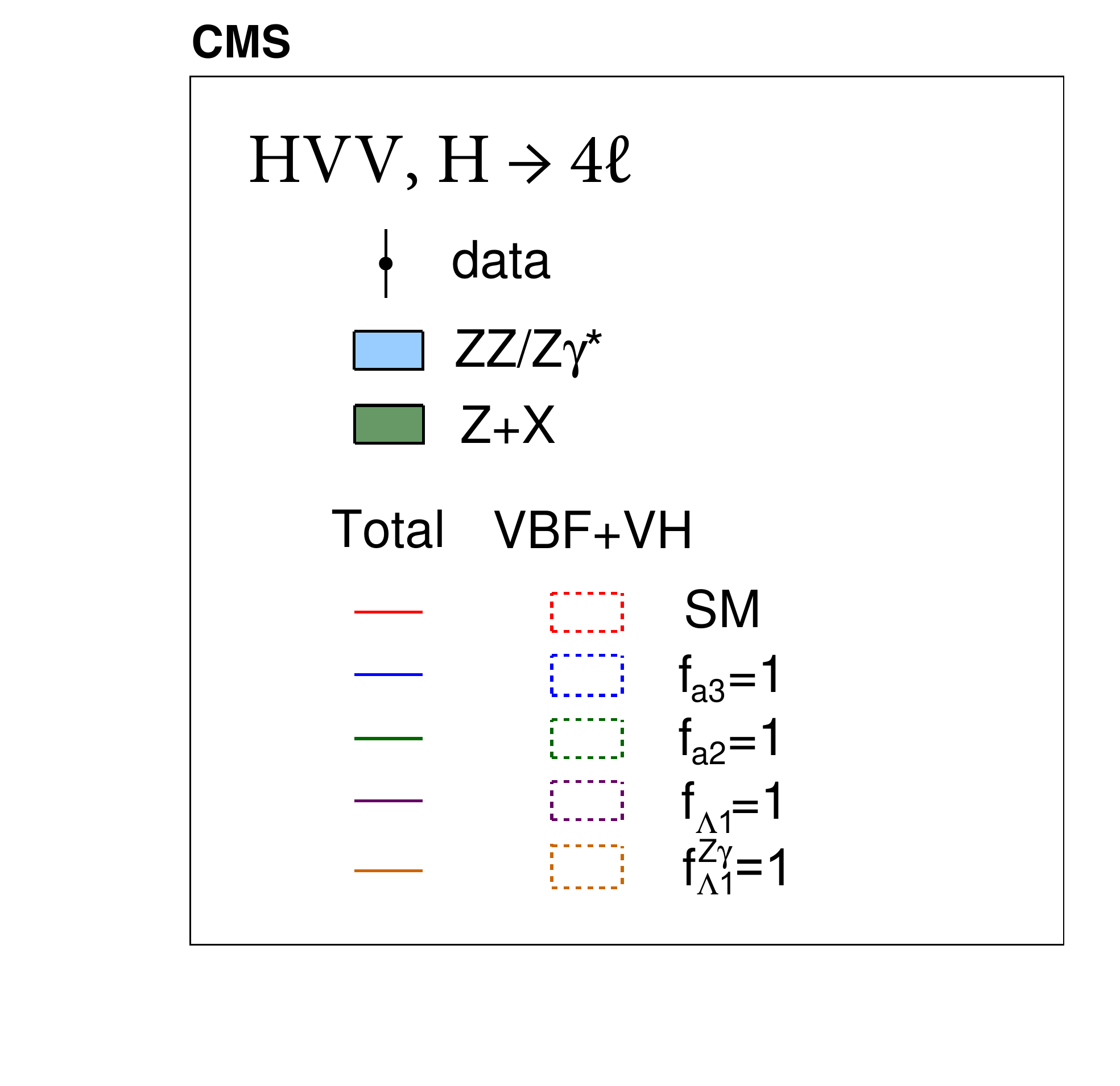}
\includegraphics[width=0.32\textwidth]{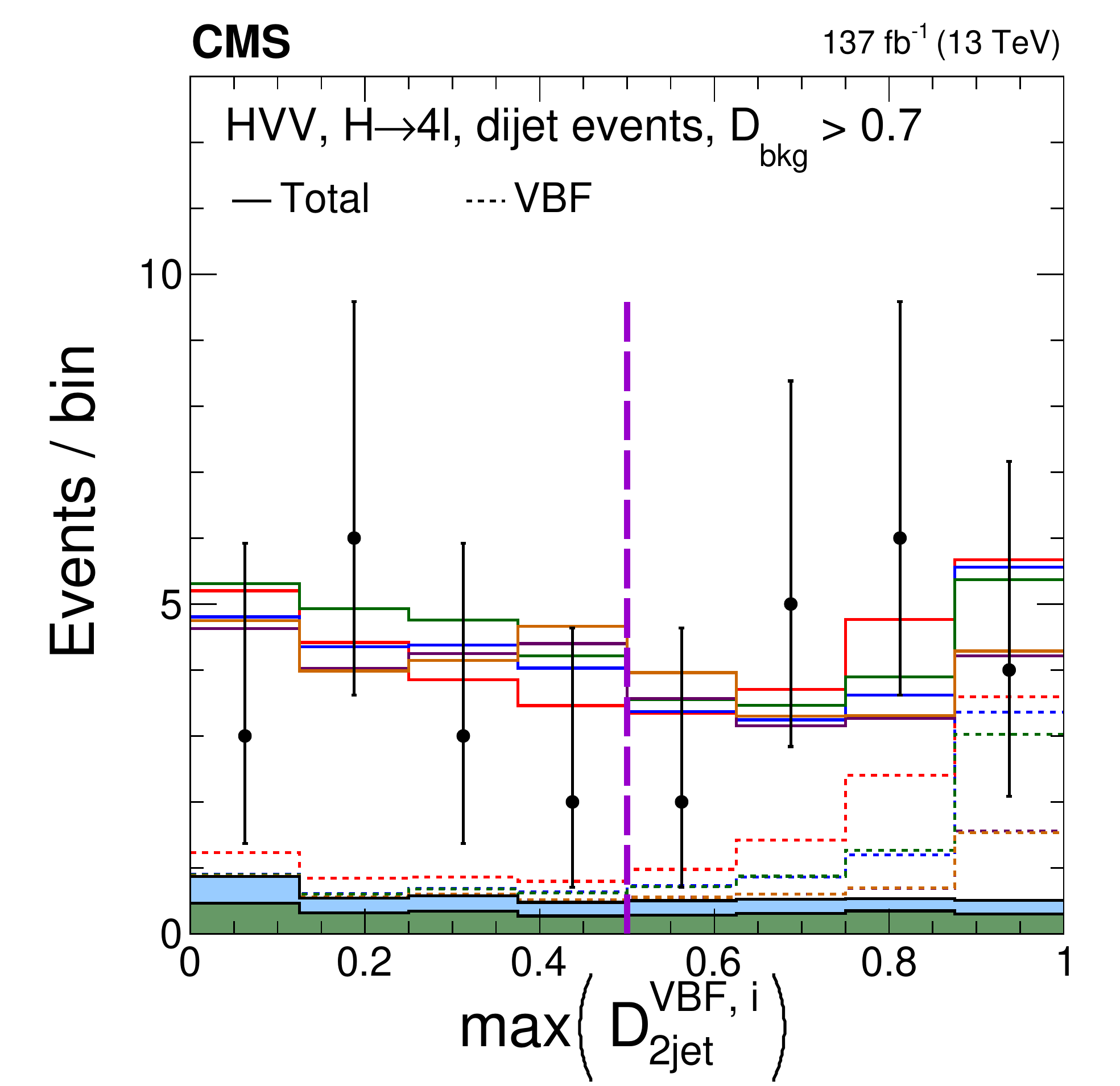}
\includegraphics[width=0.32\textwidth]{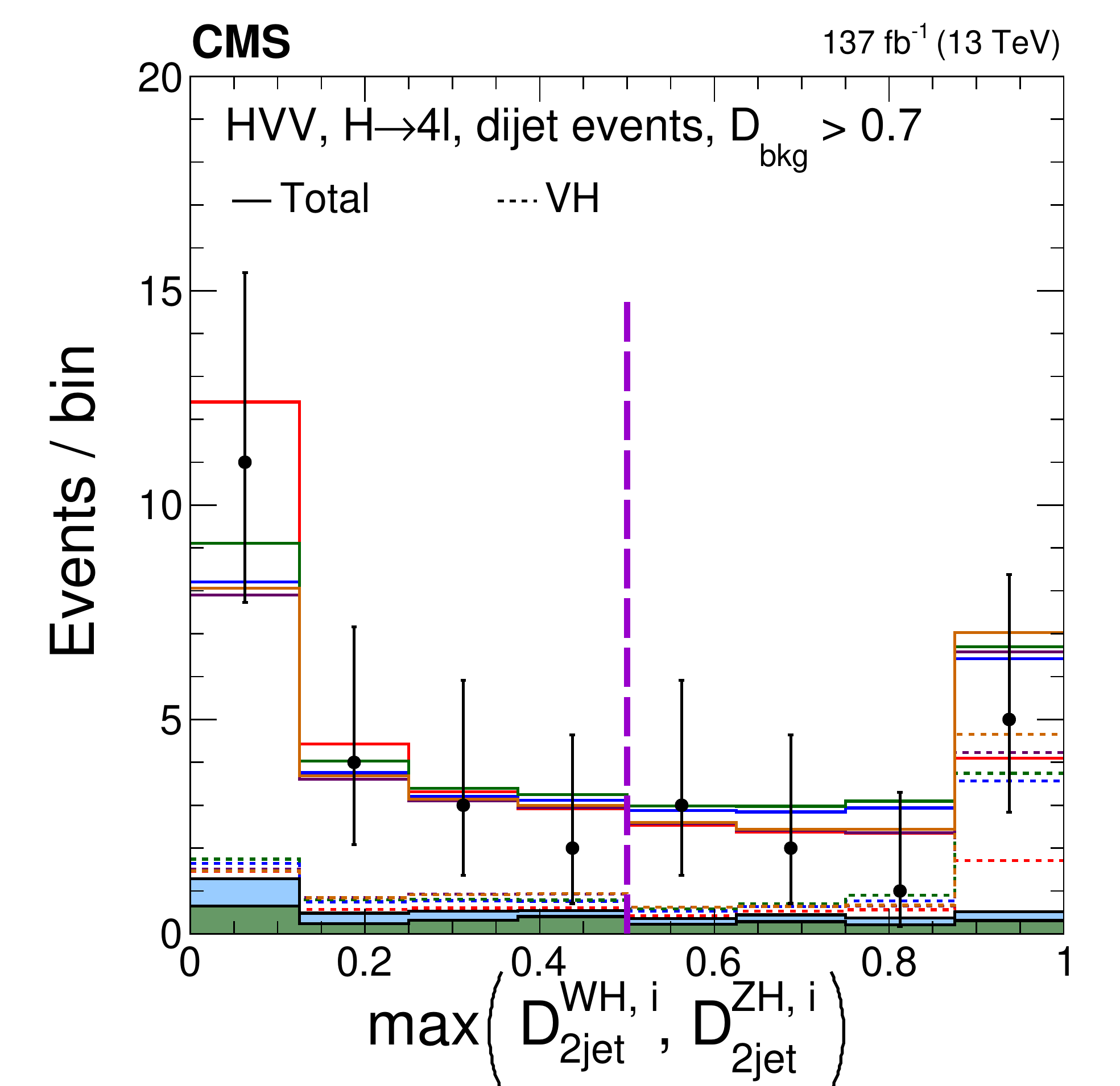}
\caption{
The distributions of observed events (data points) and expectation (histograms) 
for $\max\left(\mathcal{D}_\mathrm{2jet}^{\VBF,i} \right)$ (middle) and
$\max\left(\mathcal{D}_\mathrm{2jet}^{\WH,i},\mathcal{D}_\mathrm{2jet}^{\ZH,i} \right)$ (right),
where maximum is evaluated over the the SM and the four anomalous coupling hypotheses~$i$, described in the legend (left). 
Only events with at least two reconstructed jets are shown, and the requirement $\Dbkg>0.7$ is applied 
in order to enhance the signal contribution over the background, where \Dbkg is calculated using decay information only. 
The expectation is shown for the total distribution, including background and all production mechanisms of the \Hboson,
and for the \VBF (middle) and \VH (right) signals, which are enhanced in the region above $0.5$, indicated with the vertical dashed line. 
}
\label{fig:d2jet}
\end{figure*}

We use the values of kinematic discriminants and other selection requirements to perform the categorization. 
The definition of these discriminants can be found in Refs.~\cite{Sirunyan:2017exp,Sirunyan:2017tqd,Sirunyan:2019twz,Sirunyan:2021rug}
and is further discussed in Section~\ref{sec:kinematics}. They are calculated using the MELA approach 
while employing the matrix elements at leading order (LO) in quantum chromodynamics (QCD). 
These discriminants use full kinematic information from the \Hboson and from associated jet production and 
are labeled to indicate a specific topology (1jet, 2jet) and production mechanism (\VBF, \WH, \ZH), 
which is discriminated against the dominant gluon fusion process:
$\mathcal{D}_\mathrm{1jet}^{\VBF}$, $\mathcal{D}_\mathrm{2jet}^{\VBF}$, $\mathcal{D}_\mathrm{2jet}^{\PZ\PH}$,
and $\mathcal{D}_\mathrm{2jet}^{\PW\PH}$. 
The $\mathcal{D}_\mathrm{2jet}$ discriminants are calculated using both SM and anomalous coupling hypotheses, 
leading to a set $\mathcal{D}_\mathrm{2jet}^{i}$, all of which 
are tested in order to maintain high efficiency of \VBF and \VH categorization in the presence of anomalous couplings. 
The discriminants defined for the two-jet topology are illustrated in Fig.~\ref{fig:d2jet}, where the expected distributions 
are based on the MC signal simulation discussed in Section~\ref{sec:cms_mc} and the background estimate in Section~\ref{sec:cms_bkg}.
To enhance the signal to background ratio in this illustration in Fig.~\ref{fig:d2jet}, a selection of $\Dbkg>0.7$ is applied.
This observable uses information from the lepton kinematic distributions and does not use information from associated jets,
as also discussed in Section~\ref{sec:kinematics}. 

In categorization Scheme 1, the \Htt and \Hgg anomalous couplings are targeted.
The categories and selection criteria are identical to the first step of the categorization scheme 
in Ref.~\cite{Sirunyan:2021rug} and are optimized to measure the rates of \Hboson production modes.
Because anomalous \Htt and \Hgg couplings have only a small effect on the fractions 
of \ttH and \ggH events in each category, the optimization based on SM kinematic distributions used for the study 
in Ref.~\cite{Sirunyan:2021rug} remains optimal here. The sequential selection criteria in Scheme 1 are as follows:
\begin{itemize}
\item[--]  The {$\VBF$-2jet} category requires exactly four leptons. In addition, there must be
	either two or three jets of which at most one is \cPqb-tagged, or at least four jets and no \cPqb-tagged jets.
Finally, $\mathcal{D}_\text{2jet}^{\VBF}>0.5$ is required. 
\item[--]  The {$\VH$-hadronic} category requires exactly four leptons. In addition, there must be
either two or three jets, or at least four jets and no \cPqb-tagged jets.
Finally,  $\mathcal{D}_\text{2jet}^{\VH}=\max\left( \mathcal{D}_\text{2jet}^{\WH},\mathcal{D}_\text{2jet}^{\ZH} \right)>0.5$ is required. 
\item[--] The {\VH-leptonic} category requires no more than three jets and no \cPqb-tagged jets in the event,
and exactly one additional lepton or one additional pair of opposite-sign same-flavor leptons. 
This category also includes events with no jets and at least one additional lepton.
\item[--] The {\ttH-hadronic} category requires at least four jets, if one is a \cPqb-tagged jet, and no additional leptons.
\item[--] The {\ttH-leptonic} category requires at least one additional lepton in the event.
\item[--] The {\VBF-1jet} category requires exactly four leptons, exactly one jet and $\mathcal{D}_\mathrm{1jet}^{\VBF}>0.7$.
\item[--] The {Untagged} category consists of the remaining events.
\end{itemize}
The number of events expected from signal simulation and background estimation are shown along
with the observed number of events for each Scheme~1 category in Table~\ref{tab:Hffcategory}.

\begin{table*}[!t]
\centering
\topcaption{The numbers of events expected in the SM for different $\PH$ signal ({sig}) and background ({bkg}) contributions 
and the observed number of events in each category defined in Scheme 1 targeting \Hff and \Hgg anomalous couplings. 
The \ttH signal expectation is quoted for the SM and anomalous coupling 
($\kappa_{\PQt}=0$, $\tilde\kappa_{\PQt}=1.6$) scenario, both generated with the same cross section. 
\label{tab:Hffcategory}}
\begin{scotch}{cccccccc}
\vspace{-0.3cm} \\
	& {Untagged}  & \specialcell{VBF-\\1jet} & \specialcell{VBF-\\2jet} & \specialcell{$\PV\PH$-\\leptonic} & \specialcell{$\PV\PH$-\\hadronic}
	                                                                                                        & \specialcell{$\ttH$-\\leptonic}     & \specialcell{$\ttH$-\\hadronic}          \\
[\cmsTabSkip]
\hline\\[-1ex]
$\ggH$  {sig}            &182.98        &15.50     &6.70   &0.35   &4.68    &0.02    &0.18   \\
$\VBF$  {sig}            &7.23          &3.28      &7.23   &0.05   &0.28    &0.01    &0.05    \\
$\WH$   {sig}            &2.68          &0.22      &0.22   &1.07   &1.17    &0.03    &0.03     \\
$\ZH$    {sig}            &2.20 &0.14 &0.15 &0.26 &0.78 &0.02 &0.05  \\
$\bbH$    {sig}          &1.90          &0.13      &0.08   &0.03   &0.07    &0.00    &0.01     \\
$\ttH$     {sig}         &0.43          &0.00      &0.08   &0.14   &0.15    &0.68    &0.86    \\
($\tilde\kappa_{\PQt}=1.6$)     &(0.45)          &(0.00)      &(0.12)   &(0.15)   &(0.15)    &(0.87)    &(1.18)     \\
$\tqH$     {sig}         &0.14          &0.01      &0.10   &0.04   &0.03    &0.04    &0.03    \\
[\cmsTabSkip]
Total sig  &197.89        &19.31     &14.57  &2.00   &7.40    &0.80    &1.23     \\
[\cmsTabSkip]
$\qqbar\to4\ell$ {bkg}             &210.50        &6.93      &1.92   &2.23   &1.87    &0.08    &0.04   \\
$\Pg\Pg\to4\ell$ {bkg}            &19.79         &1.53      &0.56   &0.38   &0.24    &0.01    &0.01    \\
EW {bkg}               &3.43          &0.18      &1.37   &0.26   &0.57    &0.24    &1.07     \\
$\ZX$   {bkg}          &77.94         &2.46      &4.88   &1.20   &3.29    &0.21    &1.07    \\
[\cmsTabSkip]
Sig + bkg   &509.55        &30.41     &23.30  &6.05   &13.38   &1.33    &3.41     \\
[\cmsTabSkip]
Observed                 &539        &27     &20  &10  &12   &0    &2     \\
[\cmsTabSkip]
\end{scotch}
\end{table*}

\begin{table*}[!t]
\centering
\topcaption{The numbers of events expected in the SM for different $\PH$ signal ({sig}) and background ({bkg}) contributions 
and the observed number of events in each category defined in Scheme~2 targeting \HVV anomalous couplings. 
The EW (\VBF, \WH, and \ZH) signal expectation is quoted for the SM and four anomalous coupling 
($a_3/a_2/\kappa_1/\kappa_2^{\PZ\gamma}$) scenarios $f_{ai}=1$, all generated with the same total EW production cross section. 
\label{tab:HVVcategorization}}
\cmsTable{
\begin{scotch}{ccccccc}
\vspace{-0.3cm} \\
 & {Untagged} & {Boosted}  & \specialcell{VBF-\\1jet} & \specialcell{VBF-\\2jet}  & \specialcell{$\PV\PH$-\\leptonic} & \specialcell{$\PV\PH$-\\hadronic}\\
[\cmsTabSkip]
\hline\\[-1ex]
$\ggH$ {sig} & 171.46 & 6.48 & 15.15 & 10.44 & 0.35 & 5.99 \\
[\cmsTabSkip]
 \specialcell{$\VBF$ {sig} \\($a_3/a_2$/\\$\kappa_1/\kappa_2^{\PZ\gamma}$)}
  &  \specialcell{5.06\\(0.29/0.29/\\0.05/0.09) } & \specialcell{1.18\\(0.69/0.54/\\0.52/0.48)} &  \specialcell{2.64\\(0.12/0.09/\\0.03/0.05)} & \specialcell{8.60\\(6.10/4.95/\\1.91/1.83) }  &  \specialcell{0.06\\(0.03/0.02/\\0.01/0.01)} &  \specialcell{0.54\\(0.28/0.21/\\0.07/0.07) } \\
[\cmsTabSkip]
 \specialcell{$\WH$ {sig} \\($a_3/a_2$/\\$\kappa_1/\kappa_2^{\PZ\gamma}$)}
 & \specialcell{2.18\\(1.93/3.15/\\0.72/0.00)} & \specialcell{0.43\\(3.81/3.20/\\6.28/0.00)} & \specialcell{0.29\\(0.83/0.92/\\0.22/0.00)} & \specialcell{0.22\\(1.20/1.05/\\2.04/0.00) } & \specialcell{1.11\\ (2.75/2.86/\\3.47/0.00) } & \specialcell{1.20\\(3.43/3.33/\\2.93/0.00) } \\
[\cmsTabSkip]
 \specialcell{$\ZH$ {sig} \\($a_3/a_2$/\\$\kappa_1/\kappa_2^{\PZ\gamma}$)}
 & \specialcell{1.87\\(0.99/1.89/\\0.68/1.17)} & \specialcell{0.34\\(1.87/1.66/\\4.14/12.34)} & \specialcell{0.16\\(0.30/0.35/\\0.12/0.27)} & \specialcell{0.16\\(0.56/0.51/\\1.30/3.88)} & \specialcell{0.26\\(0.42/0.48/\\0.65/1.82)} & \specialcell{0.79\\(1.42/1.53/\\1.84/4.69) } \\
[\cmsTabSkip]
$\bbH$ {sig} & 1.84 & 0.04 & 0.13 & 0.09 & 0.03 & 0.09 \\
$\ttH$ {sig} & 1.65 & 0.04 & 0.00 & 0.32 & 0.13 & 0.19 \\
$\tqH$ {sig} & 0.13 & 0.02 & 0.01 & 0.12 & 0.04 & 0.05 \\
[\cmsTabSkip]
 \specialcell{Total sig  \\($a_3/a_2$/\\$\kappa_1/\kappa_2^{\PZ\gamma}$)}
  &  \specialcell{184.1\\(178.2/180.3/\\176.4/176.2)}  &  \specialcell{8.5\\(12.9/12.0/\\17.5/19.4) }  &  \specialcell{18.4\\(16.5/16.7/\\15.7/15.6)}  &  \specialcell{19.8\\(18.7/17.4/\\16.1/16.6)}  &  \specialcell{1.9\\(3.7/3.9/\\4.6/2.3) }  &  \specialcell{8.8\\(11.4/11.4/\\11.1/11.0) }  \\
[\cmsTabSkip]
$\qqbar\to4\ell$ {bkg} & 206.05 & 1.89 & 6.78 & 2.78 & 2.21 & 2.30 \\
$\Pg\Pg\to4\ell$ {bkg} & 19.05 & 0.38 & 1.52 & 0.76 & 0.37 & 0.31 \\
EW {bkg} & 3.50 & 0.66 & 0.20 & 1.98 & 0.23 & 0.85 \\
   $\ZX$  {bkg} & 69.87 & 3.73 & 2.46 & 9.70 & 1.20 & 4.10 \\
[\cmsTabSkip]
\specialcell{Sig + bkg  \\($a_3/a_2$/\\$\kappa_1/\kappa_2^{\PZ\gamma}$)}
&  \specialcell{481.3\\(475.4/477.5/\\473.6/473.4) } &  \specialcell{15.1\\(19.5/18.6/\\24.1/26.0) } &  \specialcell{29.3\\(27.4/27.6/\\26.6/26.5)} &  \specialcell{34.9\\(33.8/32.4/\\31.1/31.6)} &  \specialcell{5.9\\(7.7/7.9/\\8.6/6.3) } &  \specialcell{16.24\\(18.83/18.78/\\18.54/18.47)} \\
[\cmsTabSkip]
Observed                  & 512 & 18 & 27 & 30 & 10 & 13 \\
[\cmsTabSkip]
\end{scotch}
}
\end{table*}

In categorization Scheme 2, which targets anomalous \HVV couplings, the categorization sequence is modified in three ways in order 
to be more sensitive to the \HVV couplings. First, the $\mathcal{D}_\text{2jet}$ discriminants calculated using the SM hypothesis for \VBF 
or \VH are less sensitive to anomalous \VBF or \VH production, so the selection for the {\VBF-2jet} and {\VH-hadronic} 
categories is modified to be more efficient for BSM hypotheses. Second, the \ttH categories are dropped and these events are 
merged into the Untagged category; \ttH forms a small background to \VBF and \VH.  
Third, a Boosted category is added. This category, adapted from the second (and finer) categorization scheme
in Ref.~\cite{Sirunyan:2021rug}, is designed for events where not all associated particles are fully reconstructed,
so that the full kinematic information cannot be used to measure anomalous couplings.
After these modifications, Scheme 2 contains six categories, with sequential selection criteria as follows:
\begin{itemize}
\item[--]  The {$\VBF$-2jet} category requires exactly four leptons. In addition, there must be 
either two or three jets of which at most one is \cPqb-tagged, or at least four jets and no \cPqb-tagged jets. 
Finally, $\max\left(\mathcal{D}_\mathrm{2jet}^{\VBF,i} \right)>0.5$ using either the SM or any of the four BSM signal 
hypotheses ($i$) for the \VBF production is required. See Fig.~\ref{fig:d2jet} (middle) for illustration.
\item[--]  The {$\VH$-hadronic} category requires exactly four leptons. In addition, there must be
either two or three jets, or at least four jets and no \cPqb-tagged jets. \\
Finally, $\max\left(\mathcal{D}_\mathrm{2jet}^{\WH,i},\mathcal{D}_\mathrm{2jet}^{\ZH,i} \right)>0.5$
using either the SM or any of the four BSM signal  hypotheses ($i$) for the \VH production is required.
See Fig.~\ref{fig:d2jet} (right) for illustration. 
\item[--]  The {\VH-leptonic} category requires no more than three jets and no \cPqb-tagged jets in the event,
and exactly one additional lepton or one additional pair of opposite-sign same-flavor leptons. 
This category also includes events with no jets and at least one additional lepton.
\item[--]  The {\VBF-1jet} category requires exactly four leptons, exactly one jet and $\mathcal{D}_\mathrm{1jet}^{\VBF}>0.7$.
\item[--]  The {Boosted} category requires exactly four leptons, three or fewer jets, or at least four jets and no \cPqb-tagged jets,
and the transverse momentum of the four-lepton system $\PT^{4\ell}>120\GeV$.
\item[--]  The {Untagged} category consists of the remaining events.
\end{itemize}
The number of events expected from signal simulation and background estimation are shown along
with the observed number of events for each Scheme~2 category in Table~\ref{tab:HVVcategorization}.
 
The events in each category in either Scheme~1 or~2 are further characterized with several observables using the kinematic
features of the \Hboson decay and associated particles, as discussed in Section~\ref{sec:kinematics}.

\subsection{Monte Carlo simulation}
\label{sec:cms_mc}

Monte Carlo simulation is used to model signal processes, which involve the \Hboson,
and background processes in $\Pp\Pp$ interactions at the LHC and their reconstruction in the CMS detector. 
All MC samples are interfaced with \PYTHIA~8~\cite{Sjostrand:2014zea} to simulate parton
showering and multi-parton interactions, using version 8.230 for all years with the CUETP8M1 tune~\cite{Khachatryan:2015pea} 
for the simulation of the 2016 data-taking period, and the CP5 tune~\cite{Sirunyan:2019dfx} 
for the simulation of the 2017 and 2018 data taking periods.
The NNPDF~3.0 parton distribution functions are used~\cite{Ball:2011uy}.
Simulated events include the contribution from additional $\Pp\Pp$ interactions within the same or adjacent 
bunch crossings (pileup) and are weighted to reproduce the observed pileup distribution in data.
The MC samples are further processed through a dedicated simulation of the CMS detector based
on \GEANTfour~\cite{Agostinelli2003250}.

The \jhugen~7.3.0~\cite{Gao:2010qx,Bolognesi:2012mm,Anderson:2013afp,Gritsan:2016hjl,Gritsan:2020pib}
MC program is used to simulate all anomalous couplings in the \Hboson production and
$\PH\to \ZZ$ / $\PZ\gamma^*$ / $\gamma^*\gamma^*\to4\ell$ decay as discussed in Section~\ref{sec:pheno}. 
The \mela~\cite{Gao:2010qx,Bolognesi:2012mm,Anderson:2013afp,Gritsan:2016hjl,Gritsan:2020pib} 
package contains a library of matrix elements from \jhugen for the signal
and \MCFM~7.0.1~\cite{MCFM} for the background; these matrix elements are used to apply weights to events 
in any MC sample to model any other set of anomalous or SM couplings.

The SM production of the \Hboson through \VBF, in association with a $\PW$ or $\PZ$ boson, or with 
a \ttbar pair is simulated using both \jhugen at LO in QCD and 
\POWHEG~2~\cite{Frixione:2007vw,Bagnaschi:2011tu,Nason:2009ai,Luisoni:2013kna,Hartanto:2015uka} 
at next-to-leading order (NLO) in QCD.
Production in association with a \bbar pair or single top quark is simulated at LO in QCD via \jhugen.
In the \VBF, \VH, and \ttH production modes, the \jhugen and \POWHEG simulations are explicitly compared
after parton showering in the SM case, and no significant differences are found in kinematic observables.
Therefore, the \jhugen simulation is adopted to describe kinematic distributions in the \VBF, \VH, \ttH, \tqH, and \bbH 
production modes with anomalous couplings, with the expected yields scaled to match the SM theoretical 
predictions~\cite{deFlorian:2016spz} for inclusive cross sections and \POWHEG simulation for categorization 
of events based on associated particles in the SM. 
There are no observable anomalous effects in kinematic distributions of the \bbH process~\cite{Gritsan:2016hjl},
but we keep this process in modeling its event contribution. 
The considered \VH process does not include $\Pg\Pg\to\PZ\PH$ production, which is expected 
to contribute about 5\% of the \VH cross section and is therefore neglected in this analysis. 
This process has been studied with \jhugen, including anomalous $\HVV$ and $\Hff$ couplings, 
and it was found that the dependence on anomalous $\HVV$ couplings is suppressed~\cite{Gritsan:2020pib}. 

Gluon fusion production is simulated with the \POWHEG~2 event generator at NLO in QCD.
The kinematic features of events produced in gluon fusion with two associated jets are also modified by anomalous
$\PH\Pg\Pg$ couplings. These effects are studied using \MGvATNLO 2.6.0~\cite{Alwall:2014hca,Demartin:2014fia}
and \jhugen. Simulation with the \minlo~\cite{Hamilton:2012np} program at NLO in QCD is used for evaluation of 
systematic uncertainties related to the modeling of two associated jets. 
The relationship between the \Hff and \Hgg couplings follows \jhugen with the relative sign of $CP$-odd and $CP$-even 
coefficients opposite to that assumed in \MGvATNLO~2.6.0, as discussed in Ref.~\cite{Gritsan:2020pib}. 
The sign convention of the photon field in \jhugen~7.3.0 is opposite to that in \MGvATNLO~2.6.0,
which leads to the opposite sign of the $\PH\PZ\gamma$ couplings. This sign convention depends on 
the sign in front of the gauge fields in the covariant derivative. 

In all of the above cases, the subsequent decay $\PH\to \ZZ$ / $\PZ\gamma^*$ / $\gamma^*\gamma^*\to4\ell$
is modeled with \jhugen.
All signal processes have been generated under the assumption that the \Hboson mass is $\mH=125\GeV$.   
This value has been used in calculations in Sections~\ref{sec:pheno} and~\ref{sec:cms}. 
However, in the analysis of the data discussed in Sections~\ref{sec:kinematics} and~\ref{sec:results},
the $\mH=125.38\GeV$ value from Ref.~\cite{Sirunyan:2020xwk} is used. 
The $\mell$ parameterization, the cross sections and the branching fractions of all processes~\cite{deFlorian:2016spz} 
are adjusted according to $\mH=125.38\GeV$,  but the effect on other kinematic distributions of the \Hboson decay 
products and associated particles is neglected owing to the small difference between the two $\mH$ values.

\subsection{Background modeling}
\label{sec:cms_bkg}

The main background in this analysis, $\qqbar\to\ZZ/\PZ\gamma^*/\gamma^*\gamma^*\to 4\ell$, 
is estimated from NLO simulation with \POWHEG.
A fully differential cross section has been computed at next-to-next-to-leading order (NNLO) 
in QCD~\cite{Grazzini:2015hta} and the NNLO/NLO QCD correction is applied as a function of \mell.
The $\Pg\Pg\to\ZZ/\PZ\gamma^*/\gamma^*\gamma^*\to 4\ell$ background process is simulated with
\MCFM~7.0.1~\cite{MCFM,Campbell:2011bn,Campbell:2013una,Campbell:2015vwa} at LO in QCD.
The cross section of this background process is corrected with an NNLO K factor as 
a function of \mell~\cite{Catani:2007vq,Grazzini:2008tf,Grazzini:2013mca}, assuming that the signal 
and background processes have the same correction for higher orders in QCD. 
The EW background includes the vector boson scattering and $\PV\PZ\PZ$ processes, generated within the  
\jhugen framework by adopting the \MCFM matrix elements for the background processes. 
The EW background also incorporates other $\PV\PV\PV$, $\ttbar\PV\PV$, and $\ttbar\PV$ processes, 
which are generated with~\MGvATNLO.

Other background contributions are estimated using control samples in reconstructed data 
without relying on simulation. Different sources of leptons such as leptons originating from decays 
of heavy flavor quarks or light mesons may produce additional background to the \Hboson signal. 
We denote this background collectively as the \ZX background and employ a data-driven method
for its estimation. The same method has been used in the analyses of Run~1~\cite{Chatrchyan:2013mxa,Khachatryan:2014kca}
and Run~2~\cite{Sirunyan:2017exp,Sirunyan:2017tqd,Sirunyan:2019twz,Sirunyan:2021rug} data sets. 
The lepton misidentification rates are first derived using $\PZ+1\ell$ control 
regions with relaxed selection requirements on the third lepton, and the extracted rates are then applied 
in $\PZ+2\ell$ control regions, where the two additional leptons with relaxed selection requirements 
have the same lepton flavor of equal or opposite charge~\cite{Sirunyan:2021rug}.

\section{Kinematic effects in production and decay of the \texorpdfstring{\PH}{H} boson}
\label{sec:kinematics}

\begin{figure*}[!tbp]
\centering
\includegraphics[width=0.49\textwidth]{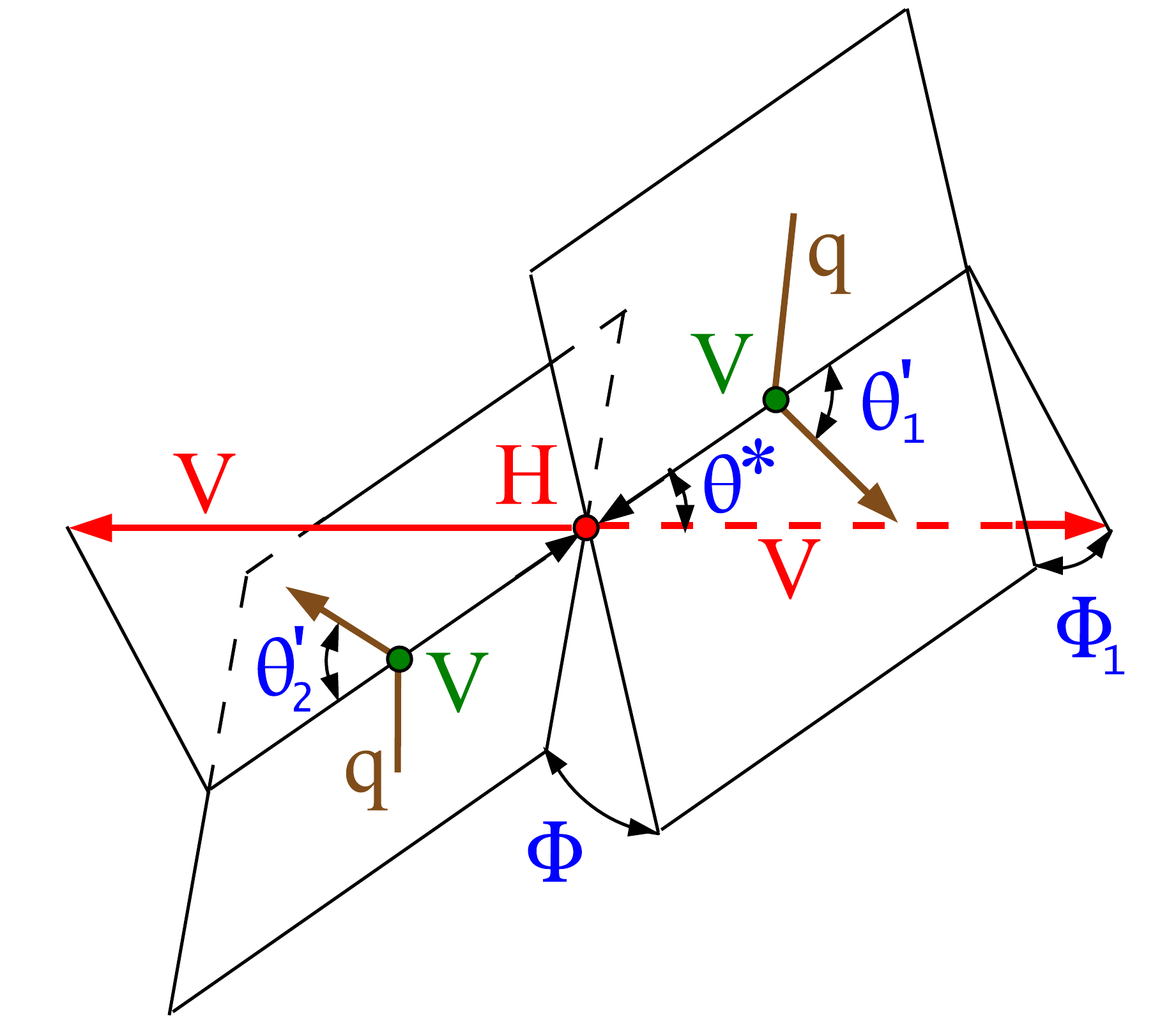}
\includegraphics[width=0.49\textwidth]{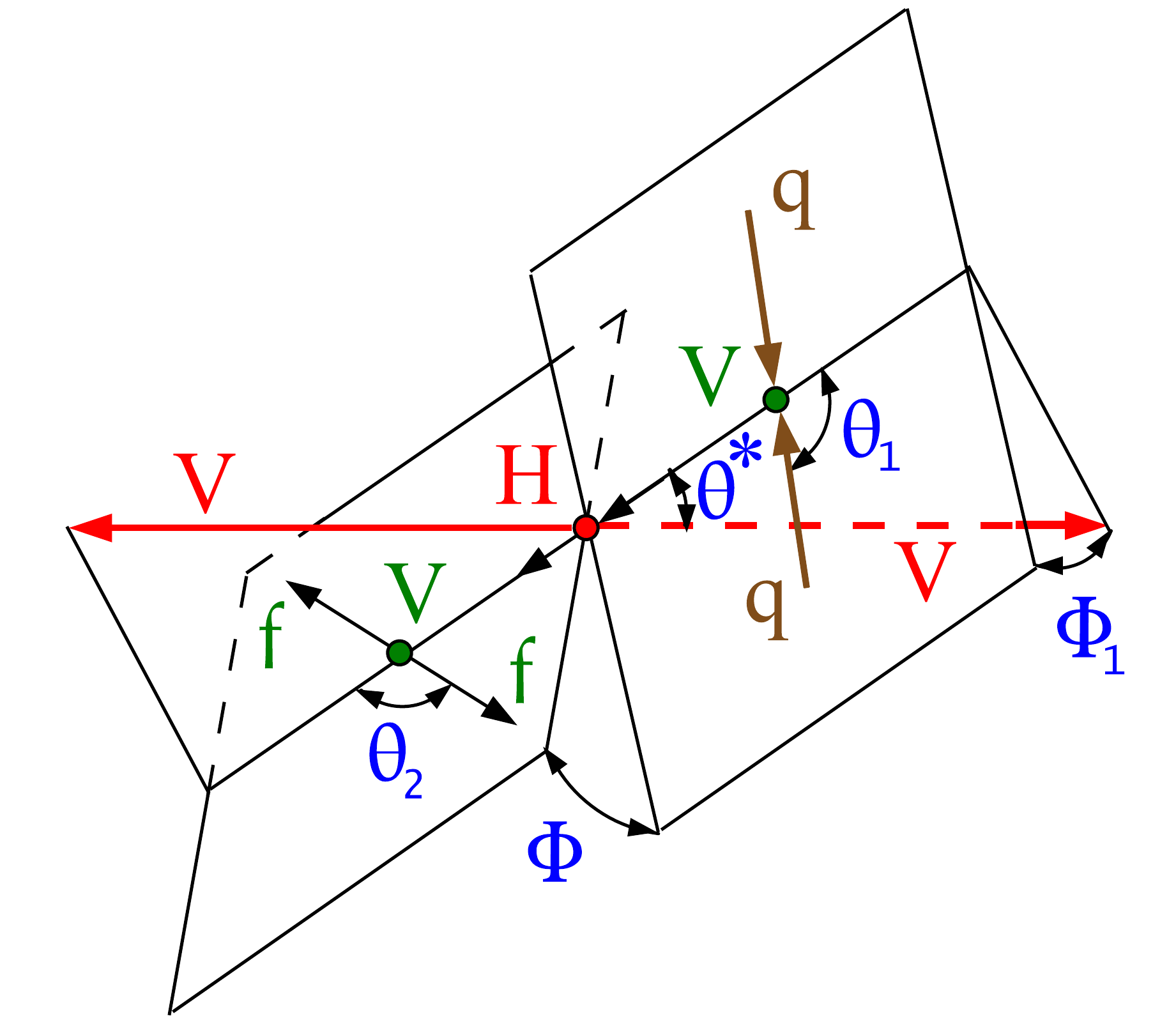} \\
\includegraphics[width=0.49\textwidth]{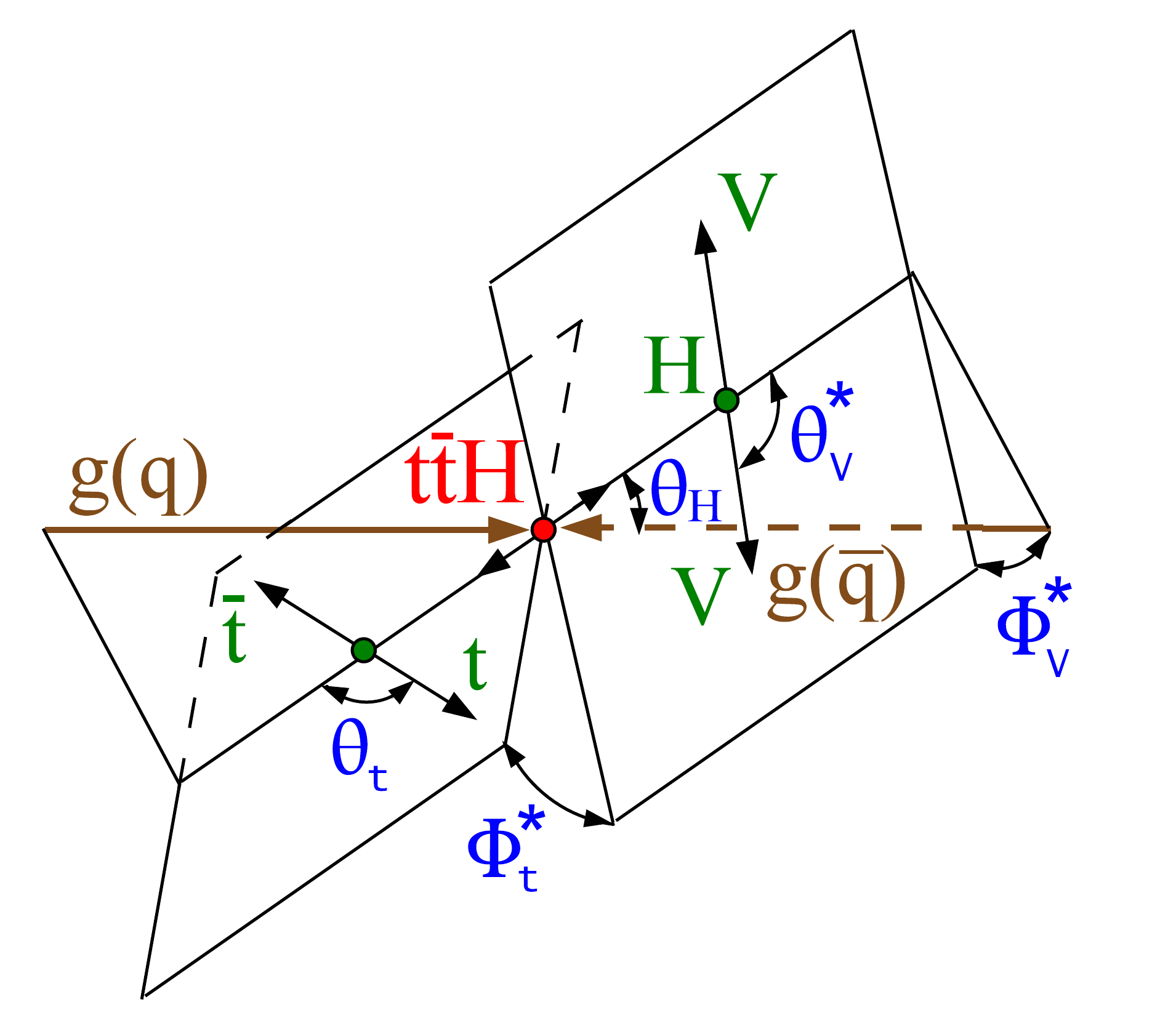}
\includegraphics[width=0.49\textwidth]{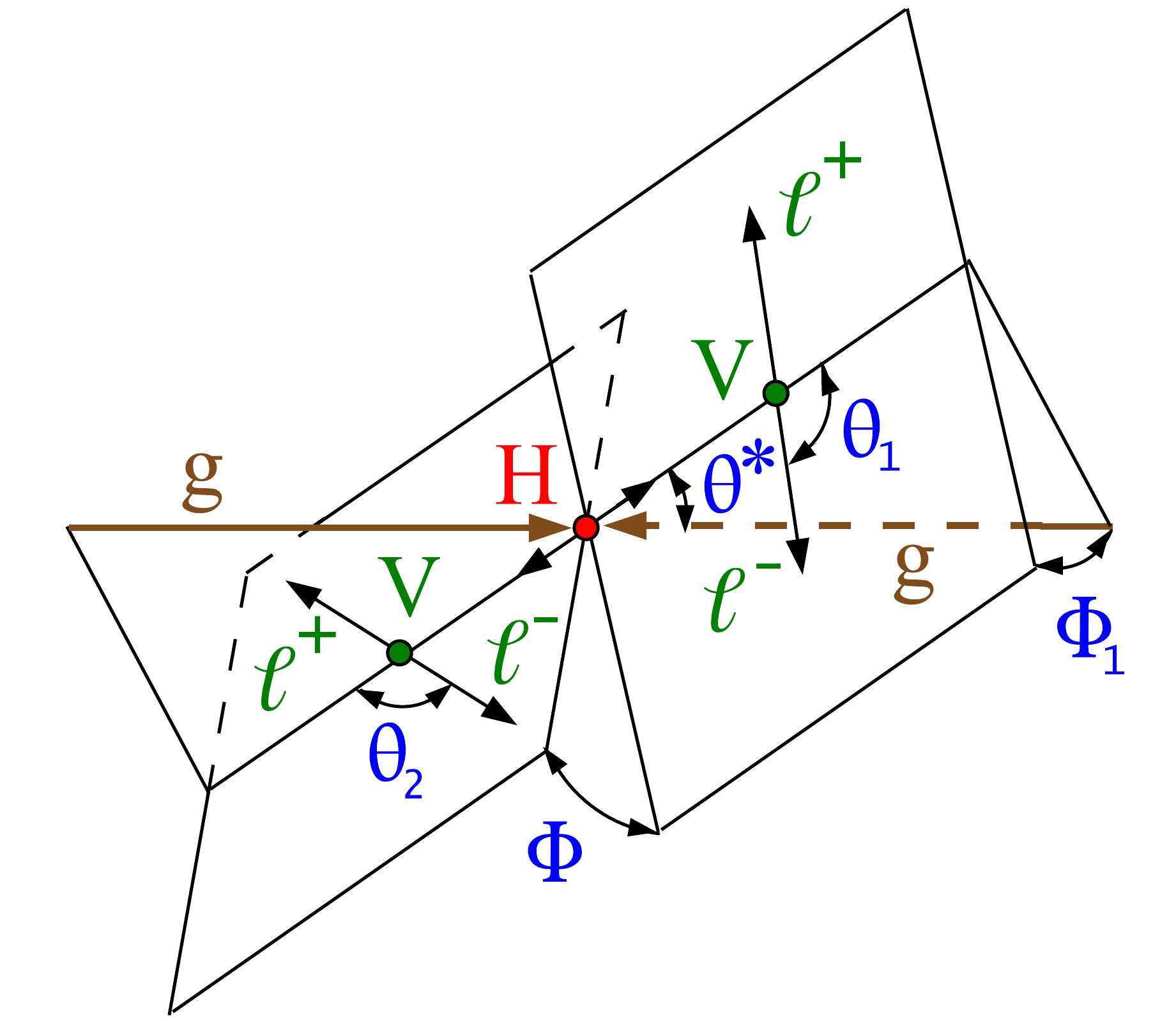}
\caption{
Four topologies of the \Hboson production and decay: 
gluon or EW vector boson fusion $\qq\to \PV_1\PV_2 (\qq) \to \PH (\qq) \to (\VV) (\qq)$ (upper left);
associated production $\qq\to \PV \to \PV\PH \to (\mathrm{ff})\ (\VV)$ (upper right); 
\Hboson production in association with the top quarks $\ttH$ or $\tqH$ (lower left);
and four-lepton decay $\PH \to \VV \to 4\ell$ where the incoming gluons $\Pg\Pg$ indicate the collision axis (lower right), 
and which proceeds either with or without associated particles.
The incoming partons are shown in brown and the intermediate or final-state particles are shown in red and green.
The angles characterizing kinematic distributions are shown in blue and are defined in the respective rest 
frames~\cite{Gao:2010qx,Anderson:2013afp,Gritsan:2016hjl}.
The subsequent top quark decay is not shown. See Ref.~\cite{Gritsan:2016hjl} for details. 
}
\label{fig:kinematics}
\end{figure*}

Kinematic distributions of particles produced in the \Hboson decay or in association with the \Hboson production
are sensitive to the quantum numbers and anomalous couplings of the \Hboson.
Four main production topologies are studied: \ggH, \VBF, \ZH or \WH, and \ttH or \tqH, as illustrated in Fig.~\ref{fig:kinematics}. 

In the $\PH\to\VV\to4\ell$ decay, shown in Fig.~\ref{fig:kinematics}, lower right, eight observables fully characterize 
the kinematic distributions of the decay products and the orientation of the decay frame with respect to the production axis
$\boldsymbol{\Omega}^{\text{dec}}=\{\theta_1, \theta_2, \Phi, \theta^*, \Phi_1, m_1, m_2, m_{4\ell} \}$~\cite{Gao:2010qx}.
Sets of observables $\boldsymbol{\Omega}^{\text{prod}}$ for the \ggH, \VBF, \VH, and \ttH production processes 
are defined in a similar way~\cite{Anderson:2013afp,Gritsan:2016hjl}, as shown in Fig.~\ref{fig:kinematics}.
As a result, 13 or more kinematic observables can be defined for the associated production process, with subsequent 
\Hboson decay to a four-fermion final state.
The MELA approach is designed to reduce the number of observables to a minimum, while retaining all essential information. 

\subsection{Kinematic discriminants}

Full kinematic information from each event, using either the $\PH\to\VV\to4\ell$ decay or associated particles 
in its production, is extracted using discriminants from matrix element calculations using the \mela package.
The discriminants use a complete set of mass and angular input observables 
$\boldsymbol{\Omega}$~\cite{Gao:2010qx,Anderson:2013afp,Gritsan:2016hjl} to describe kinematic distributions at LO in QCD.
Full reconstruction of the four-lepton decay chain and associated particles is employed in the matrix element 
calculation, following the selection chain discussed in Section~\ref{sec:cms}. Events with partial reconstruction
of associated particles are retained in the analysis by using other kinematic observables, such as the transverse momentum 
of the reconstructed \Hboson. In the case of the \ttH topology, where full reconstruction of the full decay chain of
the top quarks is a challenging task, an approximation to the matrix element approach is achieved with machine 
learning~\cite{Gritsan:2020pib}. 

Two types of discriminants are defined for either the production process, the decay process, or the full 
production + decay process. These discriminants are
\begin{align}
& \mathcal{D}_\mathrm{alt}\left(\boldsymbol{\Omega}\right) =  \frac{\mathcal{P}_\text{sig}\left(\boldsymbol{\Omega}\right) }
        {\mathcal{P}_\text{sig}\left(\boldsymbol{\Omega}\right) +\mathcal{P}_\mathrm{alt}\left(\boldsymbol{\Omega}\right) },
\label{eq:melaD} \\
& \mathcal{D}_\text{int}\left(\boldsymbol{\Omega}\right) = 
\frac{\mathcal{P}_\text{int}\left(\boldsymbol{\Omega}\right) }
{2 \ \sqrt{{\mathcal{P}_\text{sig}\left(\boldsymbol{\Omega}\right) \ \mathcal{P}_\mathrm{alt}\left(\boldsymbol{\Omega}\right) }}},
\label{eq:melaDint}
\end{align}
where the probability density $\mathcal{P}$ of a certain process is calculated using the full kinematic description characterized
by $\boldsymbol{\Omega}$ for the processes denoted as ``sig'' for a signal model and ``alt'' for an alternative model,
which could be an alternative \Hboson production mechanism (used to categorize events),
background (to isolate signal), or an alternative \Hboson coupling model (to measure coupling parameters).
The ``int'' label refers to the interference between the two model contributions.
The probability densities $\mathcal{P}$ are calculated from the matrix elements provided by the \mela package and
are normalized to give the same integrated cross section for both processes in the relevant phase space.
This normalization leads to a balanced distribution of events in the range between 0 and 1
for the $\mathcal{D}_\mathrm{alt}$ discriminants, or between $-1$ and 1 for $\mathcal{D}_\text{int}$.
In the special case where the $\mathcal{D}_\text{int}$ is calculated between $CP$-even and $CP$-odd models, 
it is denoted as $\mathcal{D}_{CP}$. 
The $\mathcal{D}_{CP}$ observable is $CP$-odd, and a forward-backward
asymmetry in its distribution would indicate $CP$ violation. This motivates the index ``CP".

When events are split into the \VBF-1/2jet and \VH-hadronic categories, 
a set of discriminants $\mathcal{D}_\text{1/2jet}$ is constructed, following Eq.~(\ref{eq:melaD}),
where $\mathcal{P}_\text{sig}$ corresponds to the signal probability density for the \VBF ($\WH$ or $\ZH$)
production hypothesis in the \VBF-tagged (\VH-tagged) category, and $\mathcal{P}_\mathrm{alt}$
corresponds to that of \Hboson production in association with two jets via gluon fusion.
When more than two jets pass the selection criteria, the two jets with the highest $\PT$ are chosen 
for the matrix element calculations. Thereby, the $\mathcal{D}_\text{1/2jet}$ discriminants separate the 
target production mode of each category from gluon fusion production,
in all cases using only the kinematic properties of the \Hboson and two associated jets.
The application of the $\mathcal{D}_\text{1/2jet}$ discriminants is described in Section~\ref{sec:cms},
where we introduce four types of discriminants
$\mathcal{D}_\mathrm{1jet}^{\VBF}$, $\mathcal{D}_\mathrm{2jet}^{\VBF,i}$, $\mathcal{D}_\mathrm{2jet}^{\ZH,i}$,
and $\mathcal{D}_\mathrm{2jet}^{\WH,i}$, with the SM and the four anomalous coupling hypotheses~$i$ considered 
in the signal model.

\begin{table*}[t!]
\centering
\topcaption{The list of kinematic observables used for category selection and fitting in categorization Schemes~1 and~2. 
Only the main features involving the kinematic discriminants in the category selection are listed, 
while complete details are given in Section~\ref{sec:cms}.
The Untagged category includes the events not selected in the other categories. 
\label{tab:Dobservables}
}
\cmsTable{
\begin{scotch}{lll}
\vspace{-0.3cm} \\
Category & Selection  & Observables $\vec{x}$ for fitting \\
[\cmsTabSkip]
\hline \\ [-1ex]
Scheme~1 \\
[\cmsTabSkip]  
{\VBF-1jet}  & $\mathcal{D}_\mathrm{1jet}^{\VBF}>0.7$  & $ \Dbkg $ \\
 {\VBF-2jet}  & $\mathcal{D}_\mathrm{2jet}^{\VBF}>0.5$ & $ \Dbkg, \mathcal{D}_\mathrm{2jet}^{\VBF}, \mathcal{D}_\mathrm{0-}^{\ggH}, \mathcal{D}_{CP}^{\ggH} $ \\
 {\VH-hadronic} & $\mathcal{D}_\mathrm{2jet}^{\VH}>0.5$ & $ \Dbkg $ \\
 {\VH-leptonic} & see Section~\ref{sec:cms} & $ \Dbkg $ \\
 {\ttH-hadronic} 	 & see Section~\ref{sec:cms} & $ \Dbkg, \mathcal{D}_\mathrm{0-}^{\ttH} $ \\
 {\ttH-leptonic} & see Section~\ref{sec:cms} & $ \Dbkg, \mathcal{D}_\mathrm{0-}^{\ttH} $ \\
 {Untagged} & none of the above  & $ \Dbkg $ \\
[\cmsTabSkip]
Scheme~2 \\
[\cmsTabSkip]
 {Boosted} & $\PT^{4\ell}>120\GeV$ & $ \Dbkg, \PT^{4\ell} $ \\
{\VBF-1jet}  & $\mathcal{D}_\mathrm{1jet}^{\VBF}>0.7$  & $ \Dbkg, \PT^{4\ell} $ \\
 {\VBF-2jet}  & $\mathcal{D}_\mathrm{2jet}^{\VBF}>0.5$ & $ \DbkgEW,
 \mathcal{D}^\mathrm{\VBF+dec}_\mathrm{0h+}, \mathcal{D}^\mathrm{\VBF+dec}_\mathrm{0-}, \mathcal{D}^\mathrm{\VBF+dec}_{\Lambda1}, 
 \mathcal{D}^{\PZ\gamma,\mathrm{\VBF+dec}}_{\Lambda1}, \mathcal{D}_{\text{int}}^{\VBF} , \mathcal{D}_{CP}^{\VBF} $ \\
{\VH-hadronic} & $\mathcal{D}_\mathrm{2jet}^{\VH}>0.5$ & $ \DbkgEW,
  \mathcal{D}^\mathrm{\VH+dec}_\mathrm{0h+}, \mathcal{D}^\mathrm{\VH+dec}_\mathrm{0-}, \mathcal{D}^\mathrm{\VH+dec}_{\Lambda1}, 
 \mathcal{D}^{\PZ\gamma,\mathrm{\VH+dec}}_{\Lambda1}, \mathcal{D}_{\text{int}}^{\VH}, \mathcal{D}_{CP}^{\VH}  $ \\
 {\VH-leptonic} & see Section~\ref{sec:cms} & $ \Dbkg, \PT^{4\ell} $ \\
{Untagged} & none of the above & $ \Dbkg,
  \mathcal{D}^\mathrm{dec}_\mathrm{0h+}, \mathcal{D}^\mathrm{dec}_\mathrm{0-}, \mathcal{D}^\mathrm{dec}_{\Lambda1}, 
 \mathcal{D}^{\PZ\gamma,\mathrm{dec}}_{\Lambda1}, \mathcal{D}_{\text{int}}^{\mathrm{dec}} , \mathcal{D}_{CP}^{\mathrm{dec}} $ \\
[\cmsTabSkip]
\end{scotch}
}
\end{table*}

Several arrays of observables $\vec{x}$ are defined in each category of events, uniquely targeting kinematic
features of each category, and are listed in Table~\ref{tab:Dobservables}.
One observable, \Dbkg, is common to most production categories in both Schemes~1 and~2. 
This observable is calculated using Eq.~(\ref{eq:melaD}) and
is designed to separate signal from the dominant background production of four leptons. 
The $\mathcal{P}_\text{alt}$ probability density is calculated for the dominant $\qqbar\to4\ell$ background process. 
The signal and background probability densities include both the matrix element probability based on the four-lepton 
kinematic properties from \mela and the empirical \mell probability density parameterization extracted 
from the simulation of detector effects. 
In the \VBF-2jet and \VH-hadronic categories in Scheme~2, the observable \DbkgEW is a modified version of \Dbkg
which includes the jet information. In this case, $\mathcal{P}_\text{sig}$ and $\mathcal{P}_\text{alt}$ still include
the \mell probability parameterization and four-lepton kinematic information, but they also include kinematic information 
for the two associated jets. The $\mathcal{P}_\text{alt}$ probability density represents the EW and QCD background 
processes $4\ell+2$\,jets, while $\mathcal{P}_\text{sig}$ represents the EW $\PH$ production processes summed together, 
\VBF, \WH, and \ZH. The \Dbkg or \DbkgEW calculation employs the SM hypothesis for the signal, 
while BSM kinematic information is incorporated in the observables discussed next.

\subsection{Observables targeting anomalous \texorpdfstring{\Htt}{Htt} and \texorpdfstring{\Hgg}{Hgg} couplings}

In Scheme~1, designed to study anomalous \Htt and \Hgg couplings, seven event categories are used.
In the Untagged, \VBF-1jet, \VH-leptonic, and \VH-hadronic categories, only one observable $\Dbkg$ is used. 
These categories do not provide additional information for separating $CP$-even and $CP$-odd contributions 
in the \Htt and \Hgg couplings, but are included in the fit in order to constrain the rates of the processes.
The probability density parameterization of $\Dbkg$ in these categories is not sensitive to the $CP$ structure 
of either \Hff or \HVV interactions.

\begin{figure*}[!tbp]
\hspace*{\fill}
\includegraphics[width=0.32\textwidth]{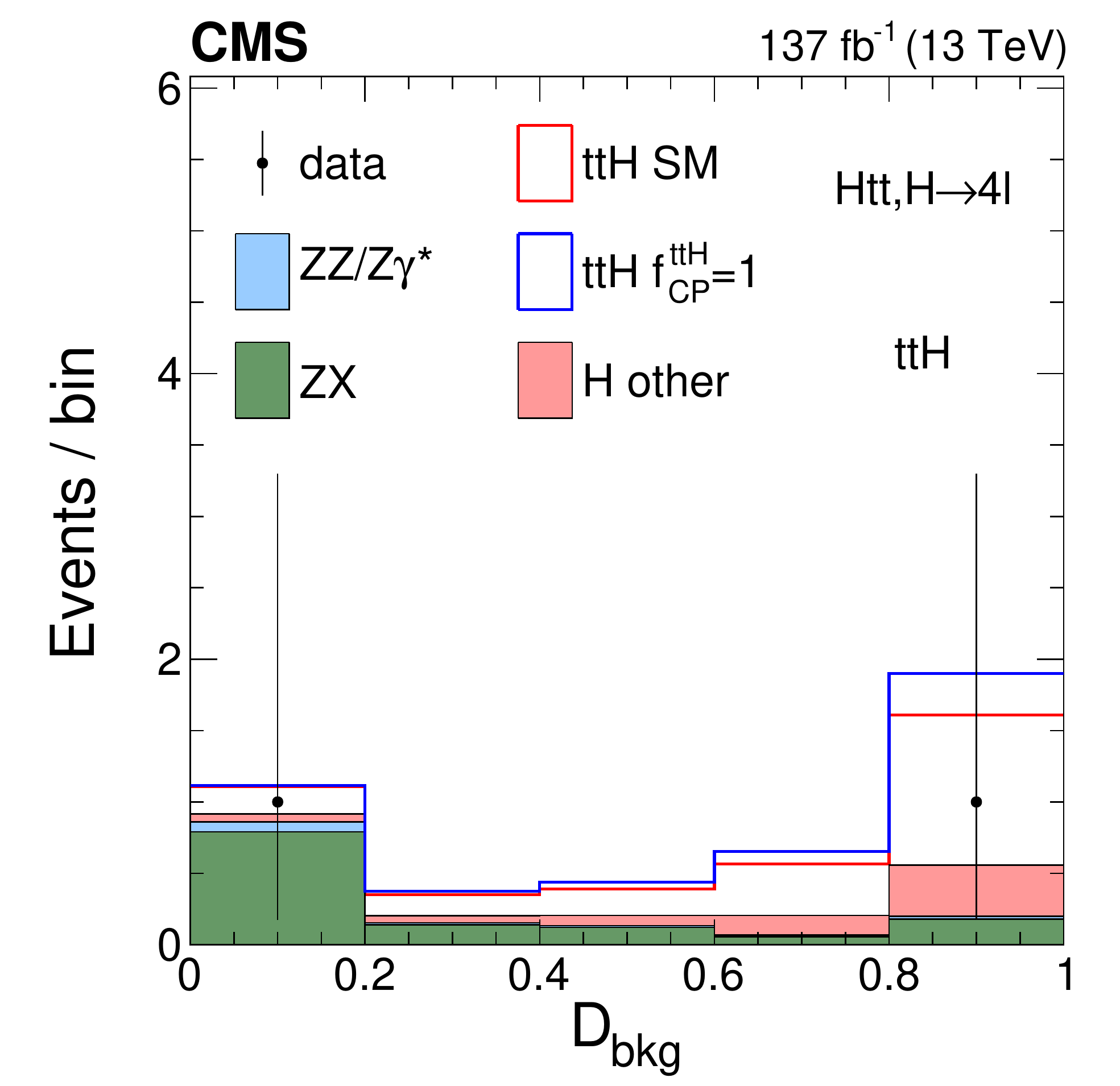}
\hspace*{\fill}
\includegraphics[width=0.32\textwidth]{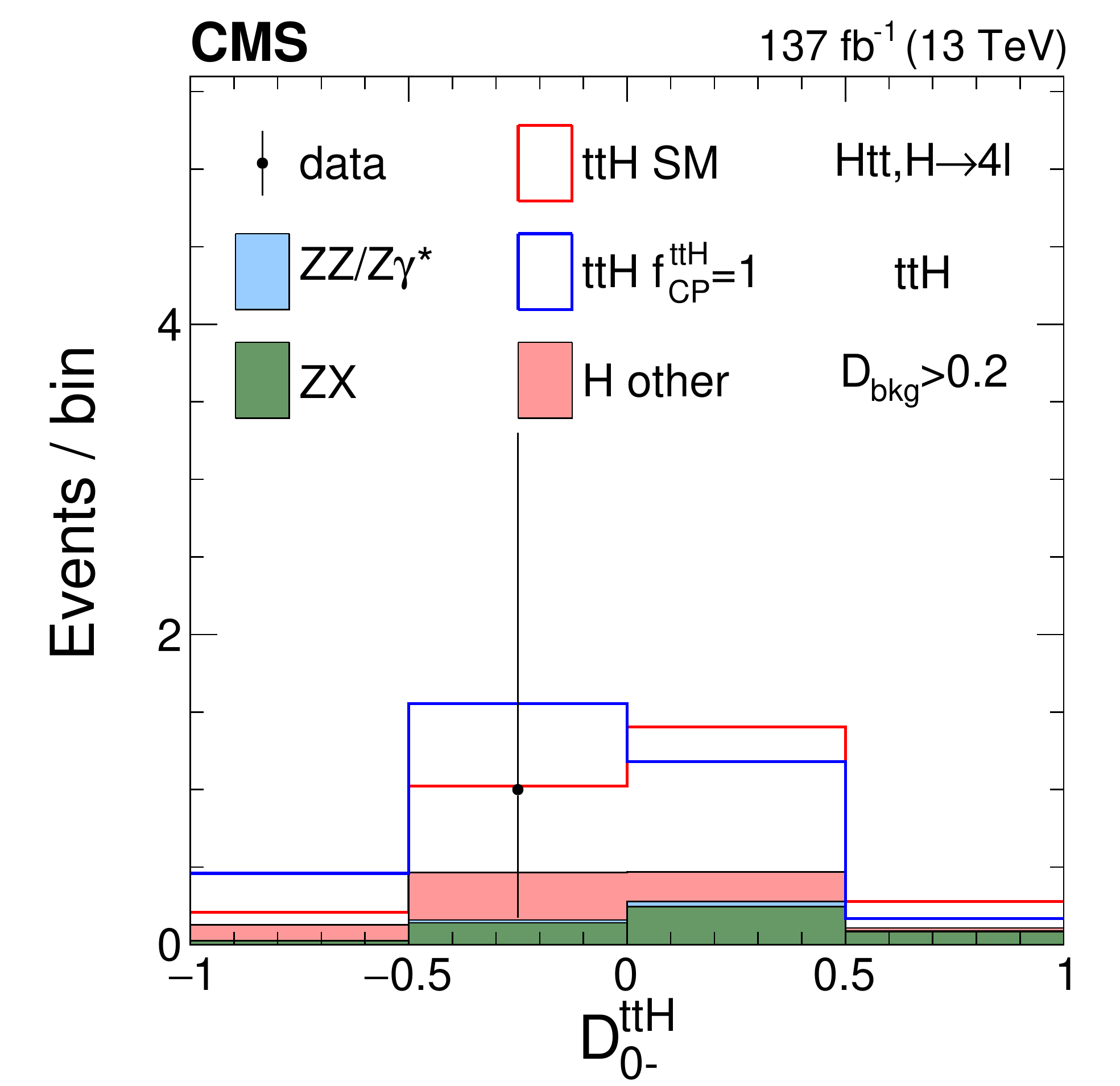}
\hspace*{\fill}
\caption
{
Distribution of the \Dbkg (left) and $\mathcal{D}_\text{0-}^\text{\ttH}$ (right), discriminants in the 
sum of the \ttH-leptonic and \ttH-hadronic categories in Scheme~1. The latter distribution is shown 
with the requirement $\Dbkg>0.2$ in order to enhance the signal over the background contribution.
\label{fig:ttH_BDT}
}
\end{figure*}

There is rich kinematic information in \ttH production because of the sequential decay of the top quarks, 
as discussed further in Ref.~\cite{Gritsan:2016hjl}. While it is possible to construct observables, as defined
in Eqs.~(\ref{eq:melaD}) and~(\ref{eq:melaDint}), with matrix element techniques~\cite{Gritsan:2016hjl},
we adopt a machine learning approach to account for partial reconstruction and possible permutations of the jets.
A boosted decision tree (BDT) classifier is trained to separate $CP$-even, corresponding to the $\kappa_{\PQt}$
coupling, and $CP$-odd, corresponding to the $\tilde\kappa_{\PQt}$ coupling, contributions independently in the 
\ttH-leptonic and \ttH-hadronic categories. The discriminant $\mathcal{D}_\mathrm{0-}^{\ttH}$ 
is obtained with this approach as the best approximation to Eq.~(\ref{eq:melaD}), provided the full kinematic 
information is made available in the calculation~\cite{Gritsan:2016hjl}. This technique still ensures that the 
maximal information is retained in the discriminant and is based on the same matrix element used in simulation. 

We achieve the full kinematic information in the $\mathcal{D}_\mathrm{0-}^{\ttH}$ calculation by including 
the following observables in the training: the four-momenta of the reconstructed \Hboson and of the six jets 
with the largest \pt, as well as the \cPqb-tagging scores of the six jets for resolving their permutation.
In addition, in the \ttH-leptonic category, the lepton multiplicity and the four-momentum of the highest-\pt 
lepton not originating from the \Hboson decay are used as input to the BDT classifier. 
It is not possible to construct the $\mathcal{D}_{CP}^{\ttH}$ discriminant, corresponding 
to Eq.~(\ref{eq:melaDint}), without tagging the flavors of the jets, including distinguishing 
quarks from antiquarks~\cite{Gritsan:2016hjl}. Alternatively, in the leptonic decay of both top quarks one 
could use the charges of the leptons, but the efficiency of such a method is very low. Therefore, in the two 
\ttH categories, two observables are used: $\vec{x} = \{ \Dbkg, \mathcal{D}_\text{0-}^{\ttH}\}$. 
The distributions of these two discriminants are shown in Fig.~\ref{fig:ttH_BDT}.

The analysis of gluon fusion production with associated jets is performed in the \VBF-2jet category. 
There are two discriminants that are sensitive to $CP$-even terms, corresponding to the $a_2^{\Pg\Pg}$ coupling, 
and to $CP$-odd terms, corresponding to the $a_3^{\Pg\Pg}$ coupling, $\mathcal{D}_\text{0-}^{\ggH}$ 
and $\mathcal{D}_{\text{CP}}^{\ggH}$, following Eqs.~(\ref{eq:melaD}) and~(\ref{eq:melaDint}), respectively. 
The matrix element for the gluon fusion \Hboson production in association with two jets includes three
possible initial states: quark-quark, quark-gluon, and gluon-gluon. Only the quark-quark initial state is used
to calculate these discriminants, because this configuration corresponds to the gluon scattering 
topology sensitive to $CP$ properties of the \Hgg coupling~\cite{Gritsan:2016hjl}, by analogy with the weak 
vector boson scattering process. The jets in the other configurations, quark-gluon and gluon-gluon, are more 
likely to be initiated from gluon radiation or splitting and are less likely to carry information about the $CP$ properties.
For similar reasons, we include the $\mathcal{D}_\text{2jet}^{\VBF}$ discriminant as one of the observables. 
This discriminant allows us to isolate the \VBF-like topology of the events, which is more characteristic of the 
quark-quark-initiated process. 

\begin{figure*}[!tbp]
\centering
\includegraphics[width=0.32\textwidth]{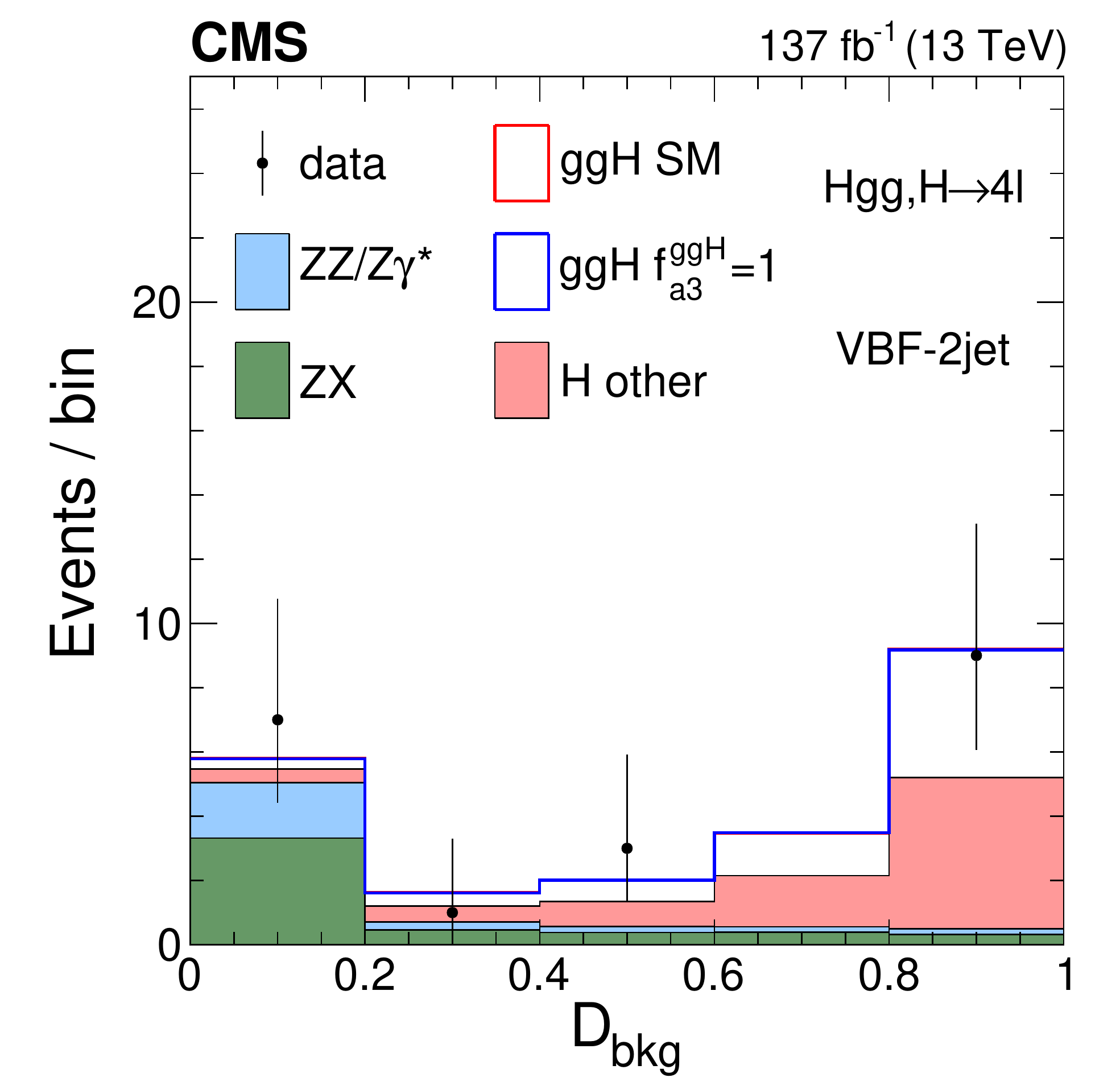}
\includegraphics[width=0.32\textwidth]{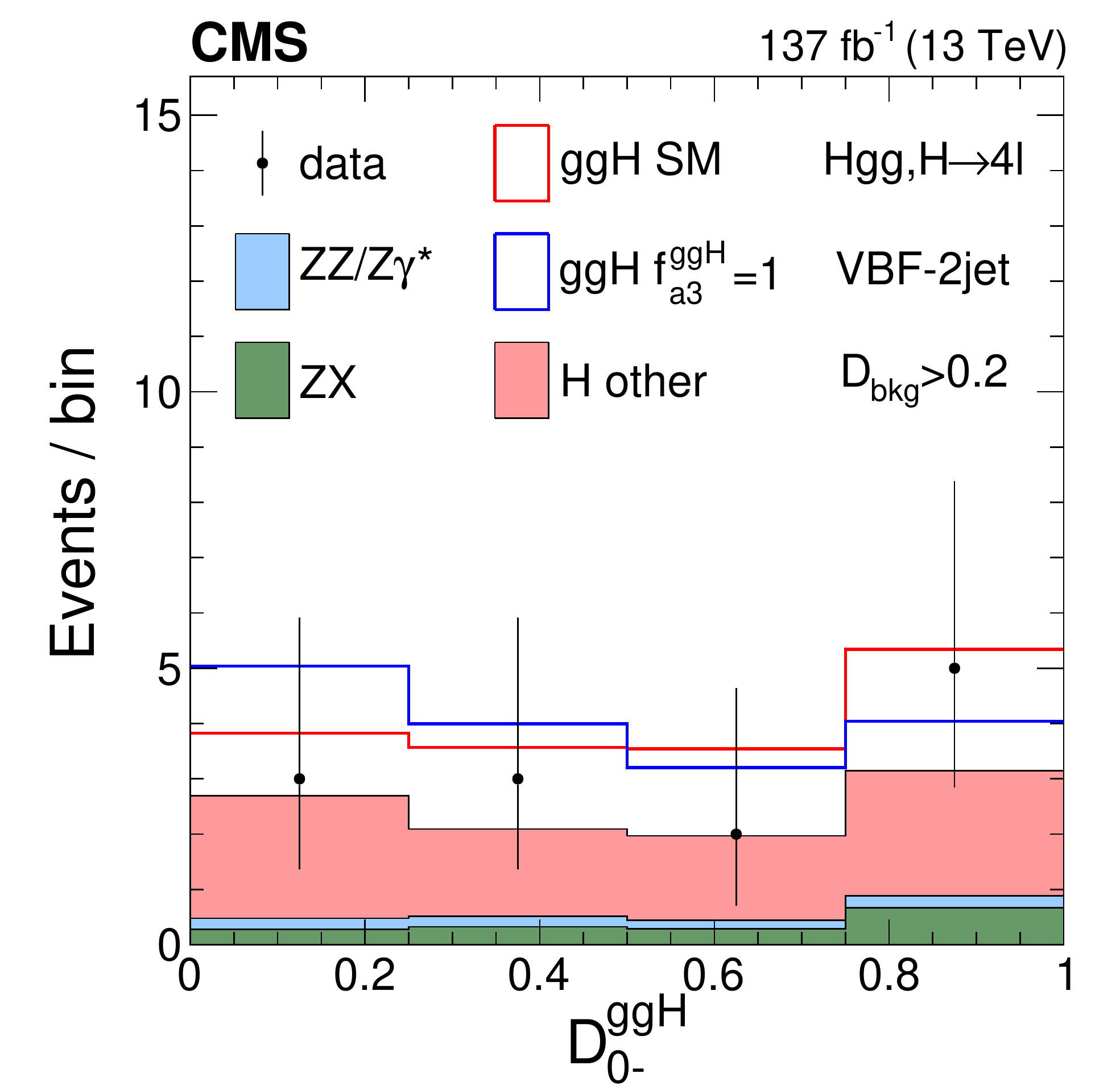}
\includegraphics[width=0.32\textwidth]{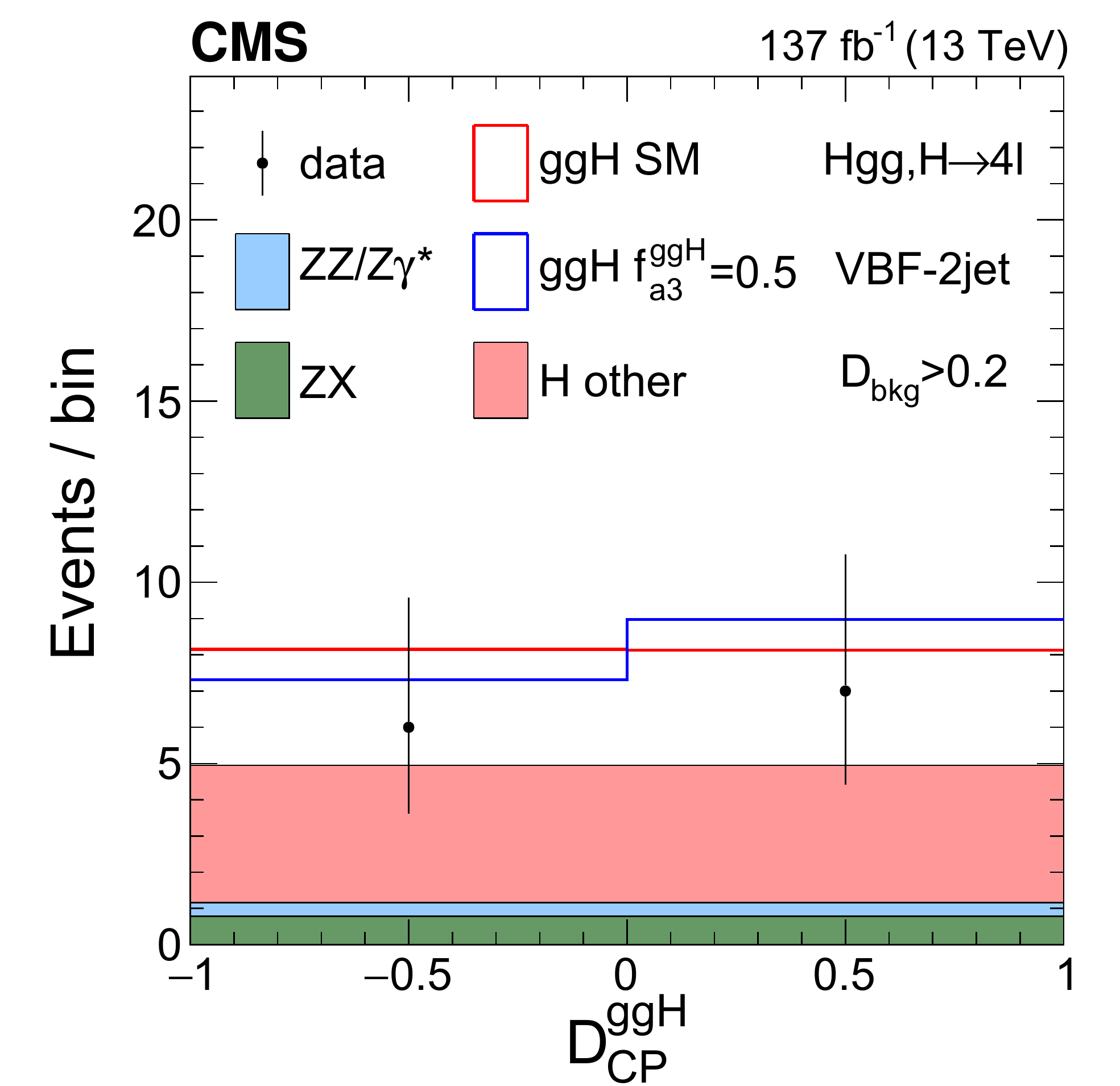}
\caption
{
Distribution of the \Dbkg (left), $\mathcal{D}_\text{0-}^{\ggH}$ (middle), and $\mathcal{D}_{\text{CP}}^{\ggH}$ (right) 
discriminants in the VBF-2jet category in Scheme~1. The latter two distributions are shown with the requirement 
$\Dbkg>0.2$ in order to enhance the signal over the background contribution.
\label{fig:ggH_D0mn}
}
\end{figure*}

As a result, in the VBF-2jet category in Scheme~1 the observables
$\vec{x} = \{ \Dbkg, \mathcal{D}_\text{2jet}^{\VBF}, \mathcal{D}_\text{0-}^{\ggH}, \mathcal{D}_{\text{CP}}^{\ggH} \}$
are used, as summarized in Table~\ref{tab:Dobservables}.
The distributions of the three observables $\Dbkg, \mathcal{D}_\text{0-}^{\ggH},$ and $\mathcal{D}_{\text{CP}}^{\ggH}$
in this category are shown in Fig.~\ref{fig:ggH_D0mn}, while the $\mathcal{D}_\text{2jet}^{\VBF}$ observable 
is shown in Fig.~\ref{fig:d2jet}. It has been shown~\cite{Plehn:2001nj,Gritsan:2020pib} that the azimuthal 
angle between the two jets carries information similar to $\mathcal{D}_\text{0-}^{\ggH}$ and 
$\mathcal{D}_{\text{CP}}^{\ggH}$, but the latter two are better in terms of performance and 
practical application to parameterization in the fit discussed in Section~\ref{sec:fit}. 

\subsection{Observables targeting anomalous \texorpdfstring{\HVV}{HVV} couplings}

In Scheme~2, designed to study anomalous \HVV couplings, six event categories are used. Two of these categories, 
\VBF-2jet and \VH-hadronic, target full reconstruction of the associated jets in EW production of the \Hboson.
Therefore, the full matrix element calculation using both production and decay information is employed, as 
discussed below and summarized in Table~\ref{tab:Dobservables}. Three other categories, Boosted, \VBF-1jet, and 
\VH-leptonic, also target  EW production, but without full reconstruction of the associated particles. Therefore, 
in these categories matrix element calculations are not employed, and instead the transverse momentum of the \Hboson
candidate $\PT^{4\ell}$ is used as the second observable. Anomalous couplings in EW production lead to 
a harder $\PT^{4\ell}$ spectrum. Finally, in the Untagged category, dominated by the \ggH events without 
two associated jets, matrix element calculations using $\PH \to \VV \to 4\ell$ decay information are employed.

Since we target four anomalous \HVV couplings appearing in Eq.~(\ref{eq:formfact-fullampl-spin0}), namely 
$a_2$, $a_3$, $\kappa_1/(\Lambda_1)^2$, and $\kappa_2^{\PZ\gamma}/(\Lambda_1^{\PZ\gamma})^2$,
optimal analysis of the $\PH \to \VV \to 4\ell$ decay requires four discriminants of the type given by 
Eq.~(\ref{eq:melaD}) and four discriminants of the type given by Eq.~(\ref{eq:melaDint}). 
In the Untagged category, the former four discriminants are defined as
$\mathcal{D}^\mathrm{dec}_\mathrm{0h+}$, $\mathcal{D}^\mathrm{dec}_\mathrm{0-}$,  $\mathcal{D}^\mathrm{dec}_{\Lambda1}$,
and $\mathcal{D}^{\PZ\gamma,\mathrm{dec}}_{\Lambda1}$, respectively, 
where the index ``dec" indicates that only the four-lepton decay information is used. 
Among the latter four interference discriminants, it was found that the two discriminants corresponding to
the $\kappa_1/(\Lambda_1)^2$ and $\kappa_2^{\PZ\gamma}/(\Lambda_1^{\PZ\gamma})^2$ couplings are strongly 
correlated with $\mathcal{D}^\mathrm{dec}_{\Lambda1}$ and $\mathcal{D}^{\PZ\gamma,\mathrm{dec}}_{\Lambda1}$,
and therefore these two interference discriminants are not used. This observation follows from the fact that the
$a_1$ and $\kappa_1/(\Lambda_1)^2$ or $\kappa_2^{\PZ\gamma}/(\Lambda_1^{\PZ\gamma})^2$
couplings correspond to the same tensor structure in Eq.~(\ref{eq:formfact-fullampl-spin0}) 
and differ only in the $q_{\PV}^2$ dependence. The remaining two interference 
discriminants, $\mathcal{D}_{\text{int}}^{\mathrm{dec}}$ and $\mathcal{D}_{CP}^{\mathrm{dec}}$,
corresponding to the $a_2$ and $a_3$ alternative couplings, respectively, are employed in the fit. 

In the \VBF-2jet and \VH-hadronic categories, the system of six discriminants discussed above is
extended to include both production and decay information, because these categories allow full
reconstruction of associated particles. 
The same four types of discriminants of the Untagged category following Eq.~(\ref{eq:melaD}) are used, 
namely $\mathcal{D}^\mathrm{\VBF+dec}_\mathrm{0h+}$, $\mathcal{D}^\mathrm{\VBF+dec}_\mathrm{0-}$, 
$\mathcal{D}^\mathrm{\VBF+dec}_{\Lambda1}$, and $\mathcal{D}^{\PZ\gamma,\mathrm{\VBF+dec}}_{\Lambda1}$
in the \VBF-2jet category, and
$\mathcal{D}^\mathrm{\VH+dec}_\mathrm{0h+}$, $\mathcal{D}^\mathrm{\VH+dec}_\mathrm{0-}$, 
$\mathcal{D}^\mathrm{\VH+dec}_{\Lambda1}$, and $\mathcal{D}^{\PZ\gamma,\mathrm{\VH+dec}}_{\Lambda1}$
in the VH-hadronic category. Here the index ``VBF+dec" or ``VH+dec" indicates that both
production and decay information is used, which means that the kinematic information from the 
associated jets and the four leptons are utilized in the \VBF or \VH matrix element calculations. 
In the case of the \VH process, the matrix elements of the \WH and \ZH processes are summed. 
There are more interference discriminants in cases where anomalous couplings appear both in production and decay. 
However, using interference discriminants with production information only following Eq.~(\ref{eq:melaDint}) 
is the better approach if one has to limit the number of discriminants. 
Therefore, $\mathcal{D}_{\text{int}}^{\VBF}$ and $\mathcal{D}_{CP}^{\VBF}$ 
are used in the \VBF-2jet category, and $\mathcal{D}_{\text{int}}^{\VH}$ and $\mathcal{D}_{CP}^{\VH}$ 
are used in the \VH-hadronic category.

\begin{figure*}[!tbp]
\centering
\includegraphics[width=0.32\textwidth]{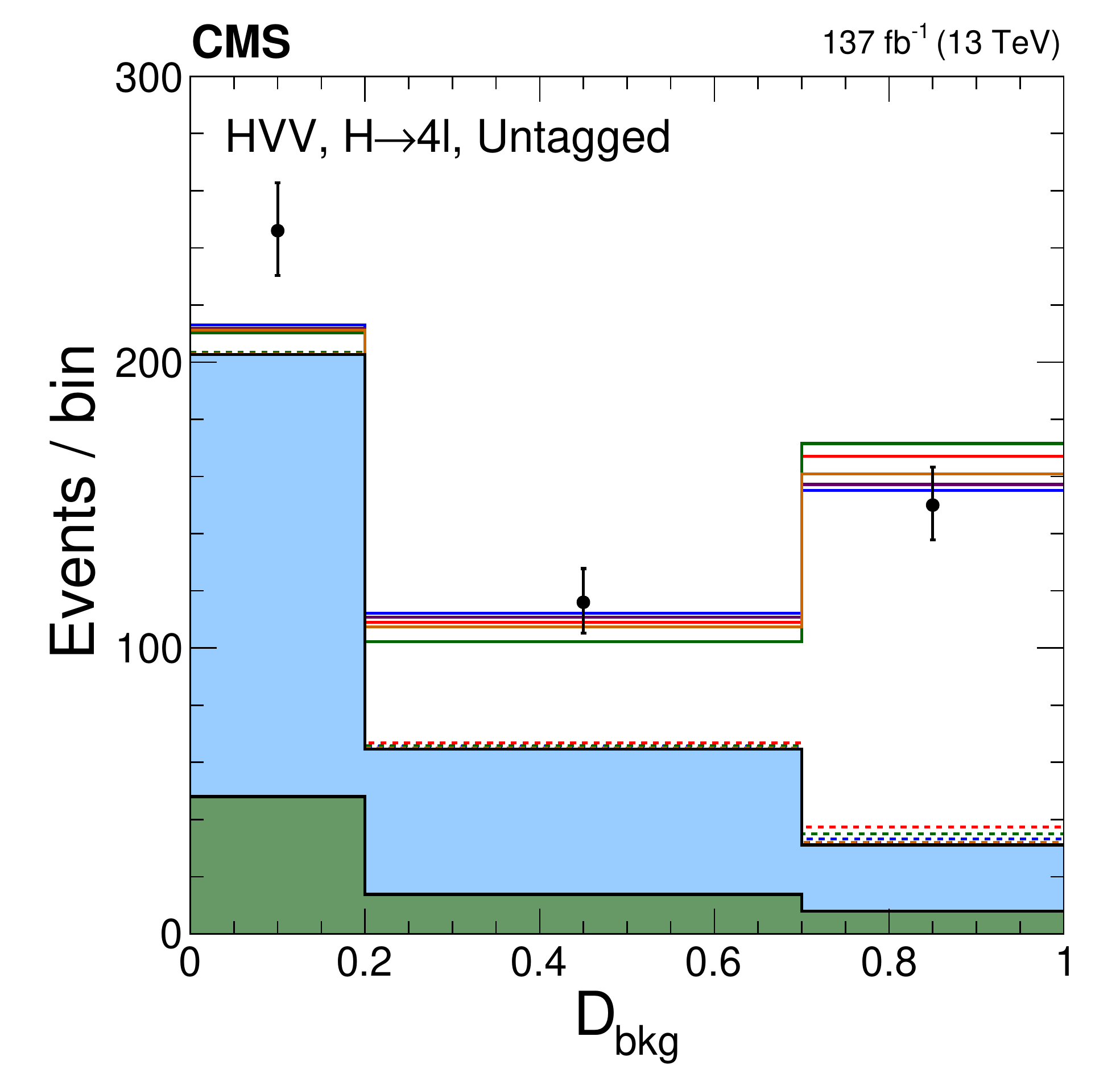}
\includegraphics[width=0.32\textwidth]{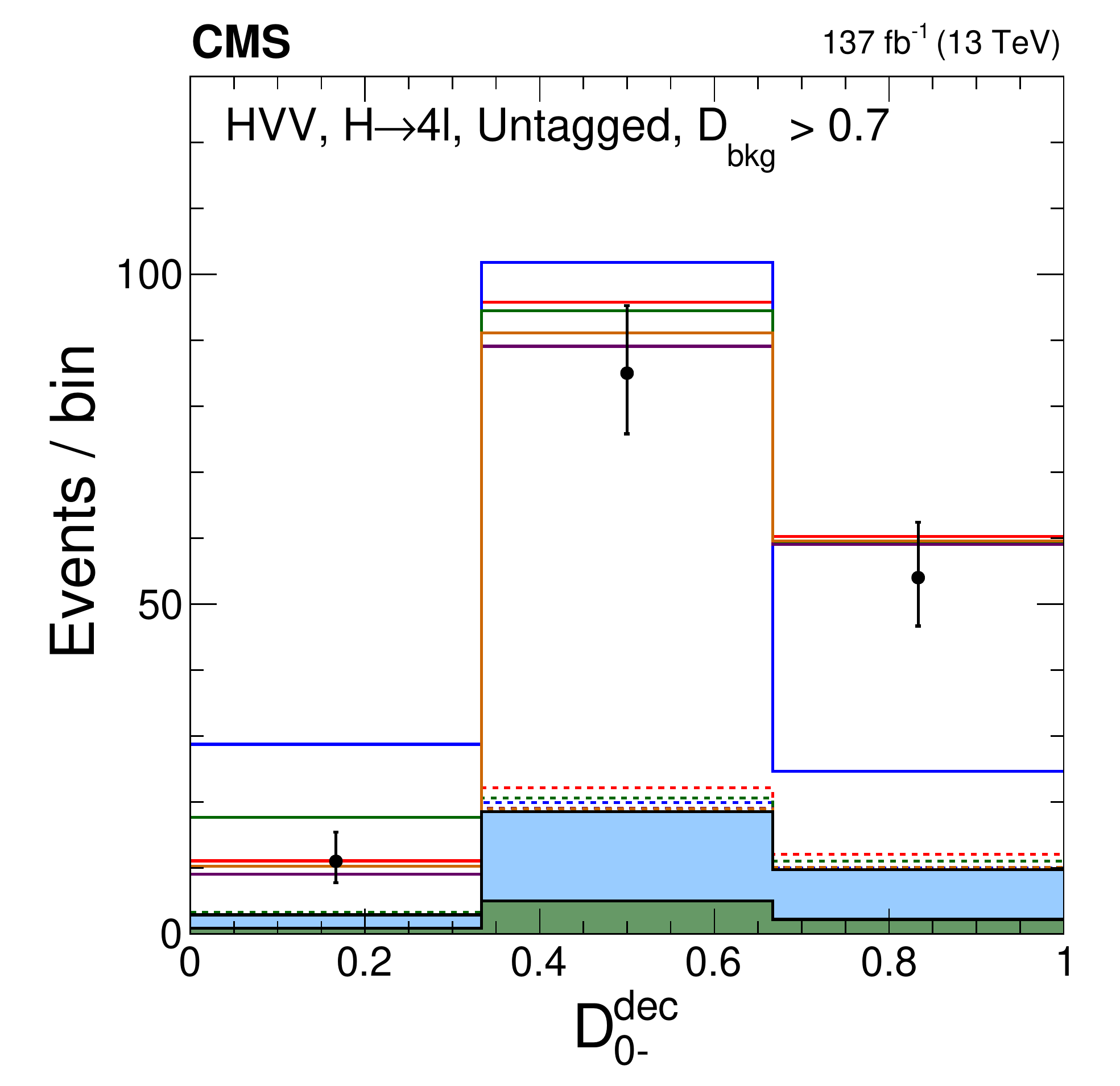}
\includegraphics[width=0.32\textwidth]{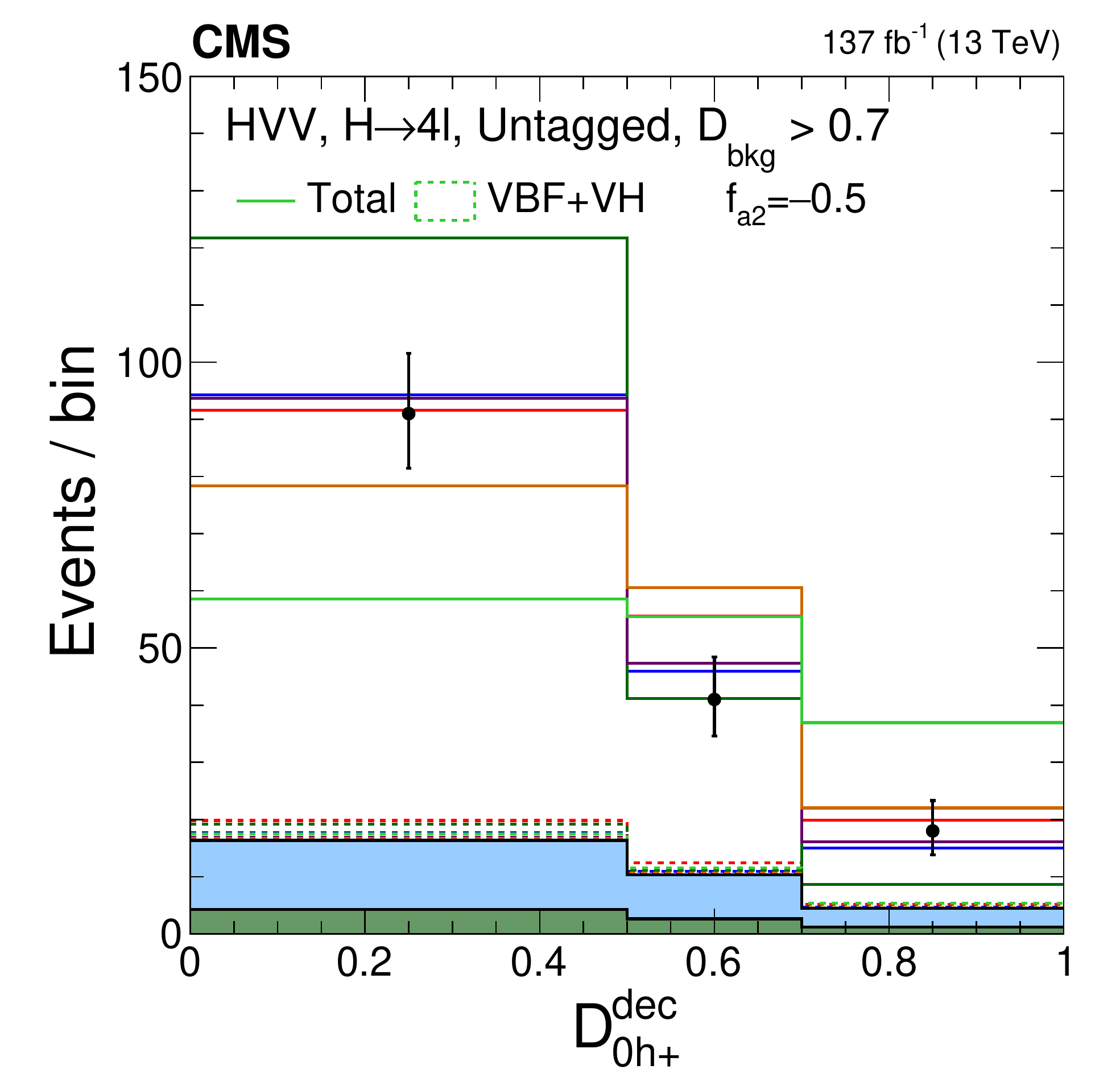} \\
\includegraphics[width=0.32\textwidth]{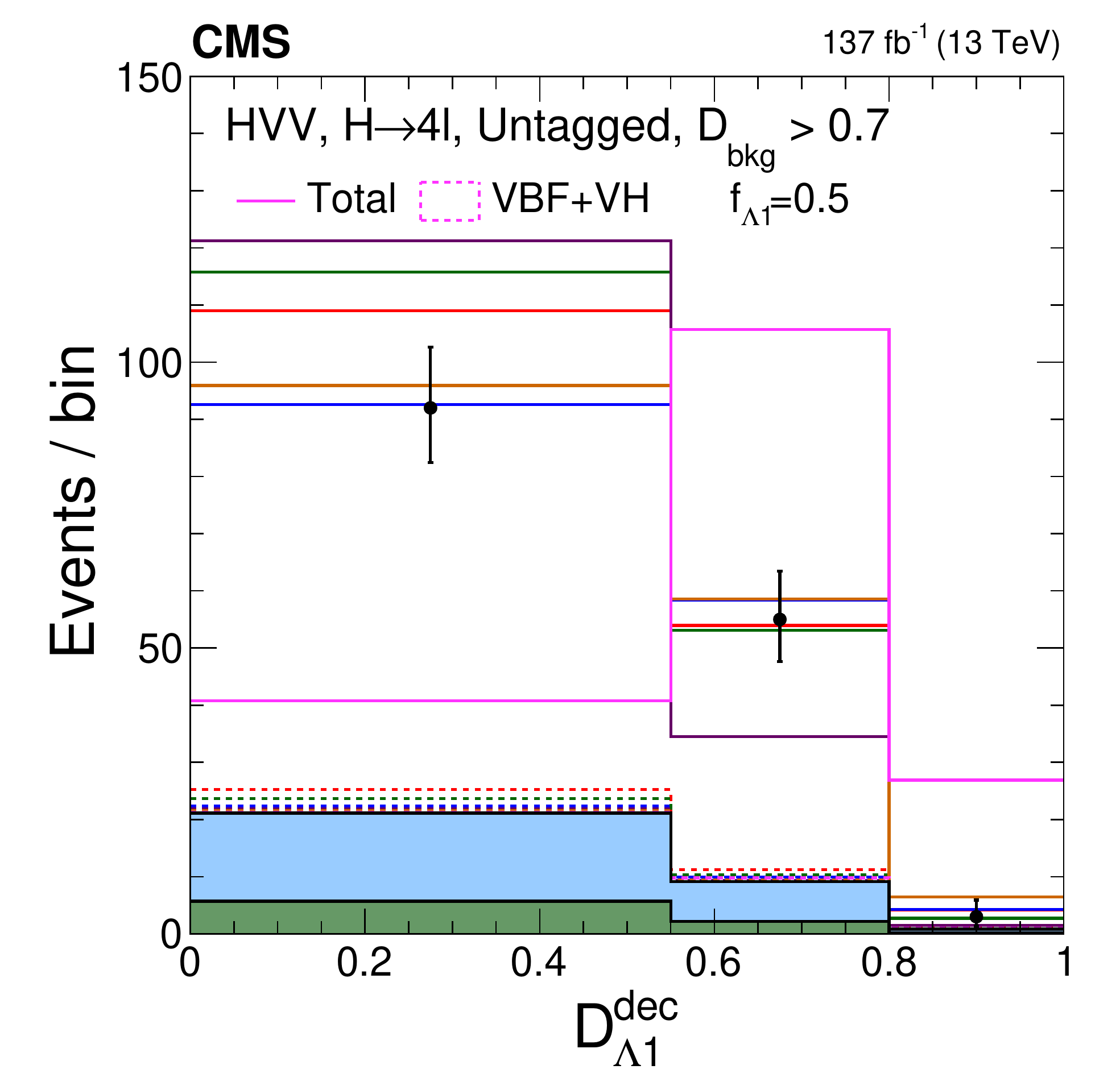}
\includegraphics[width=0.32\textwidth]{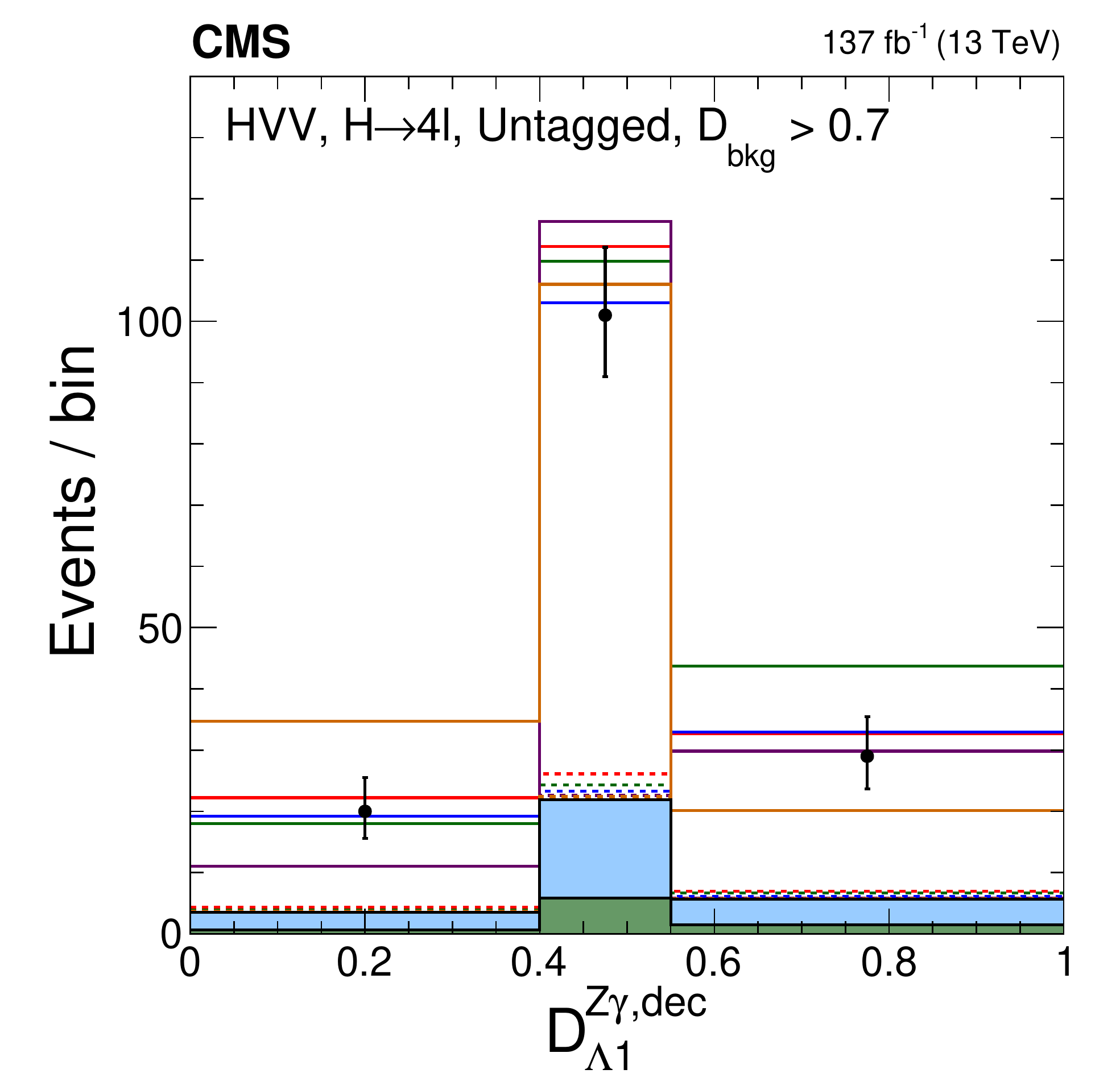}
\includegraphics[width=0.32\textwidth]{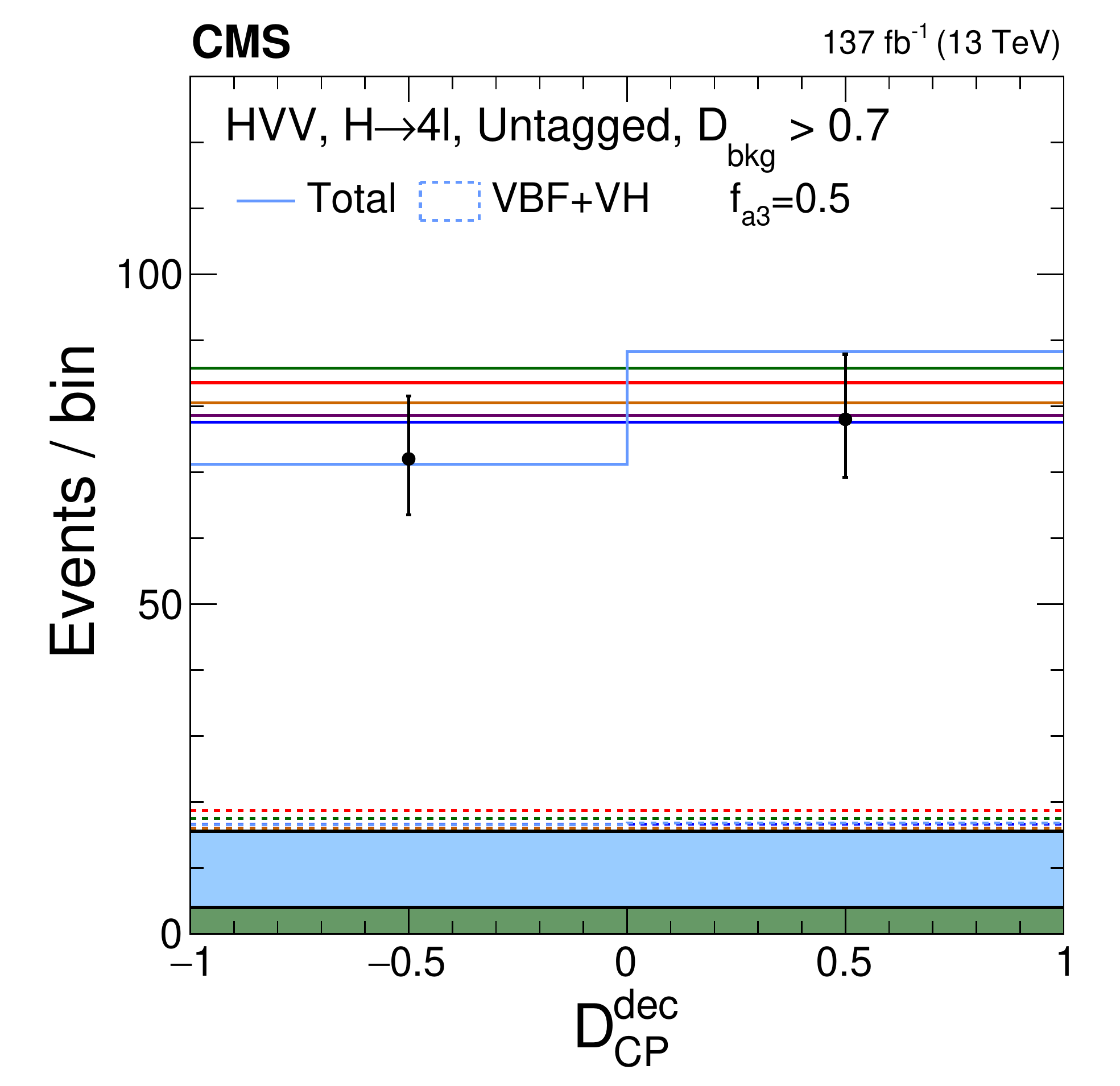} \\
\includegraphics[width=0.32\textwidth]{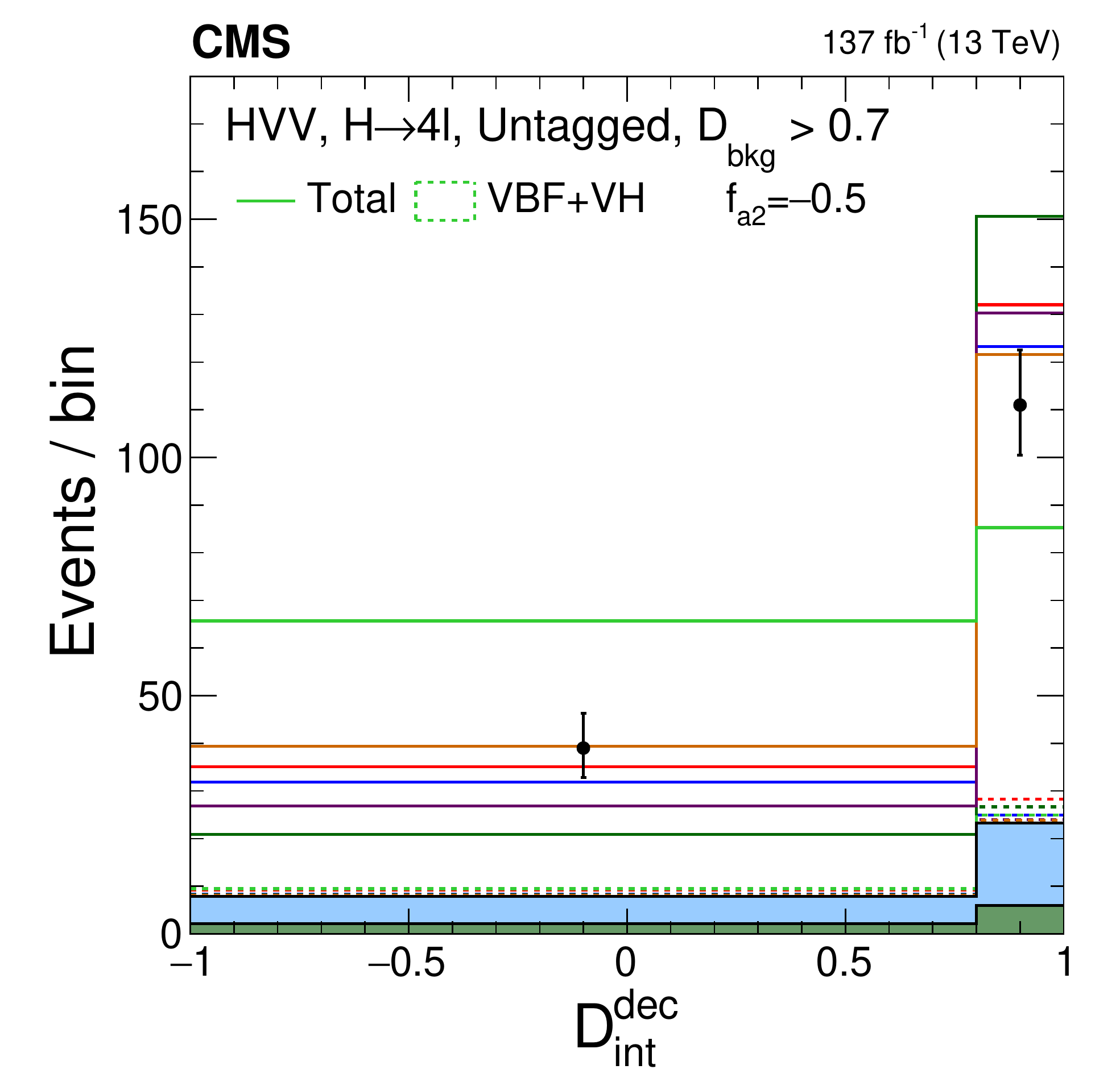}
\includegraphics[width=0.32\textwidth]{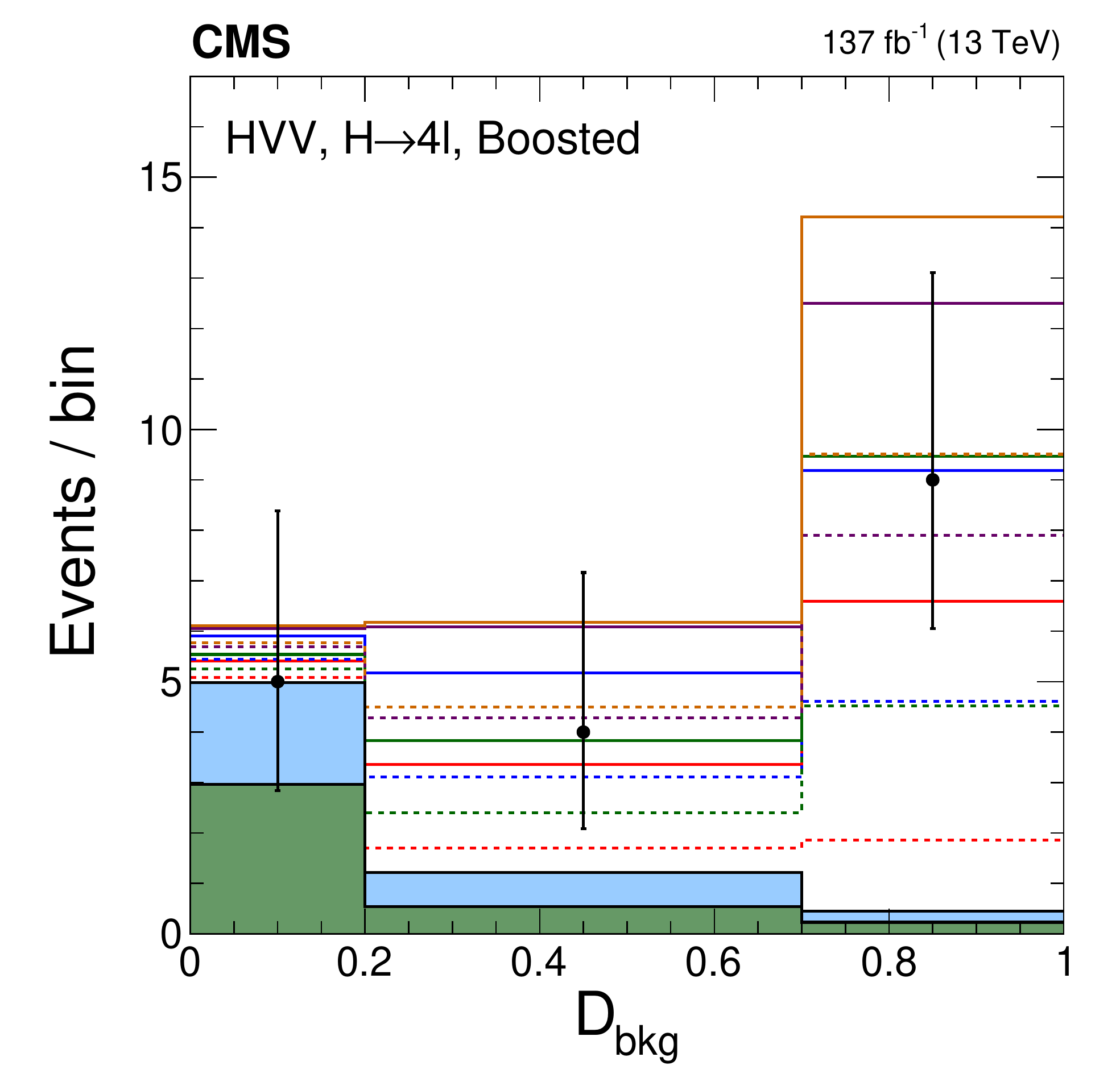}
\includegraphics[width=0.32\textwidth]{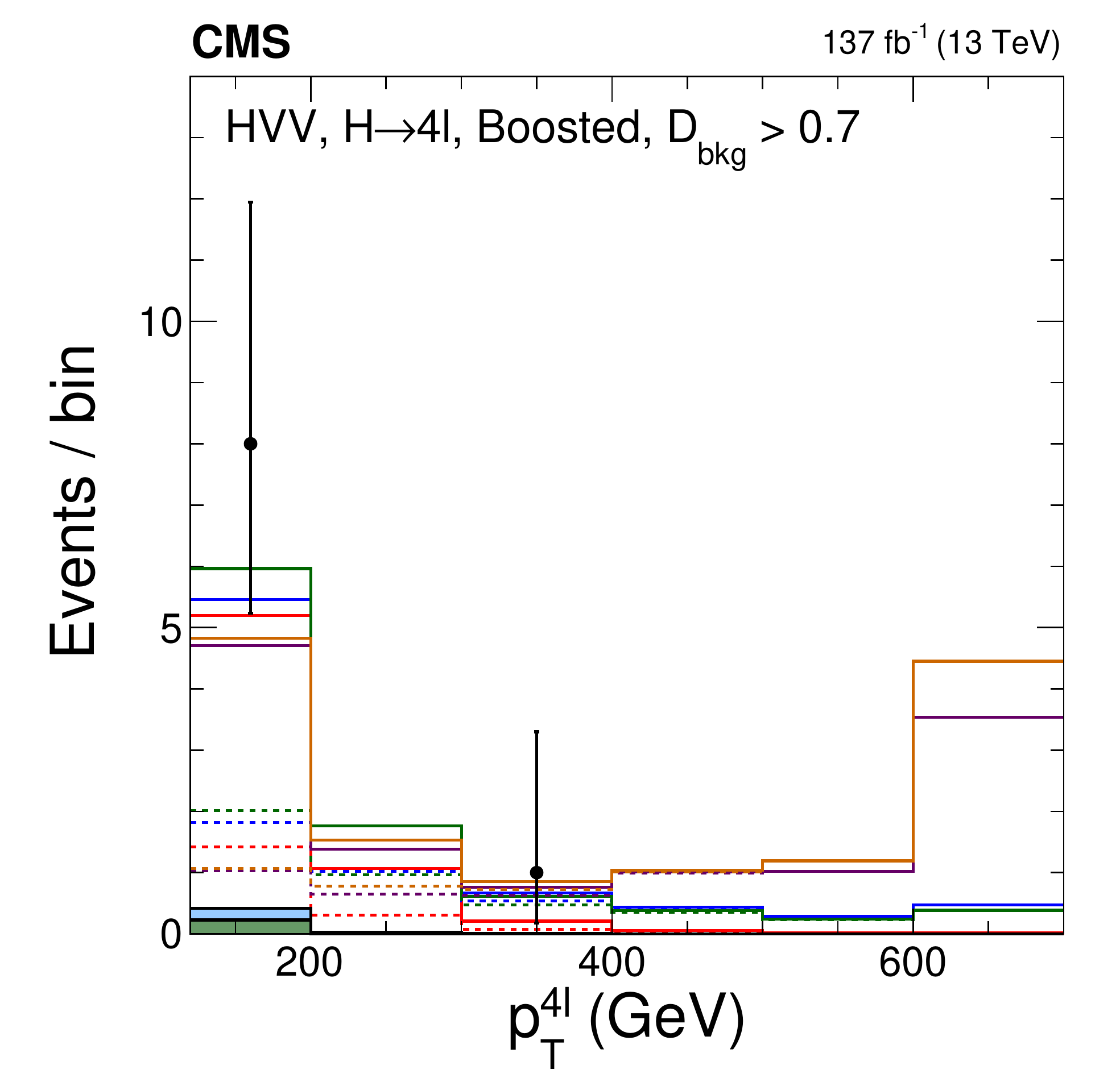}
\caption{
Distributions of events in the observables used in categorization Scheme~2.
The first seven plots are in the Untagged category:
The upper left plot shows \Dbkg.  The other distributions are shown with the requirement
$\Dbkg>0.7$ in order to enhance the signal over the background contribution:
$\mathcal{D}_{0-}^{\text{dec}}$ (upper middle),  $\mathcal{D}_{0h+}^{\text{dec}}$ (upper right),
$\mathcal{D}_{\Lambda1}^{\text{dec}}$ (middle left), $\mathcal{D}_{\Lambda1}^{\PZ\gamma, \text{dec}}$ (middle middle),
$\mathcal{D}_{CP}^{\text{dec}}$, and $\mathcal{D}_\text{int}^{\text{dec}}$.
The last two plots are shown in the Boosted category: \Dbkg (lower middle) and $\PT^{4\ell}$
with the requirement $\Dbkg>0.7$ and overflow events included in the last bin (lower right).
Observed data, background expectation, and five signal models are shown in the plots,
as indicated in the legend in Fig.~\ref{fig:d2jet} (left). 
In several cases, a sixth signal model with a mixture of the SM and BSM couplings 
is shown and is indicated in the legend explicitly.
}
\label{fig:stackPlotsOnshellUntaggedBoosted}
\end{figure*}

\begin{figure*}[!tbp]
\centering
\includegraphics[width=0.32\textwidth]{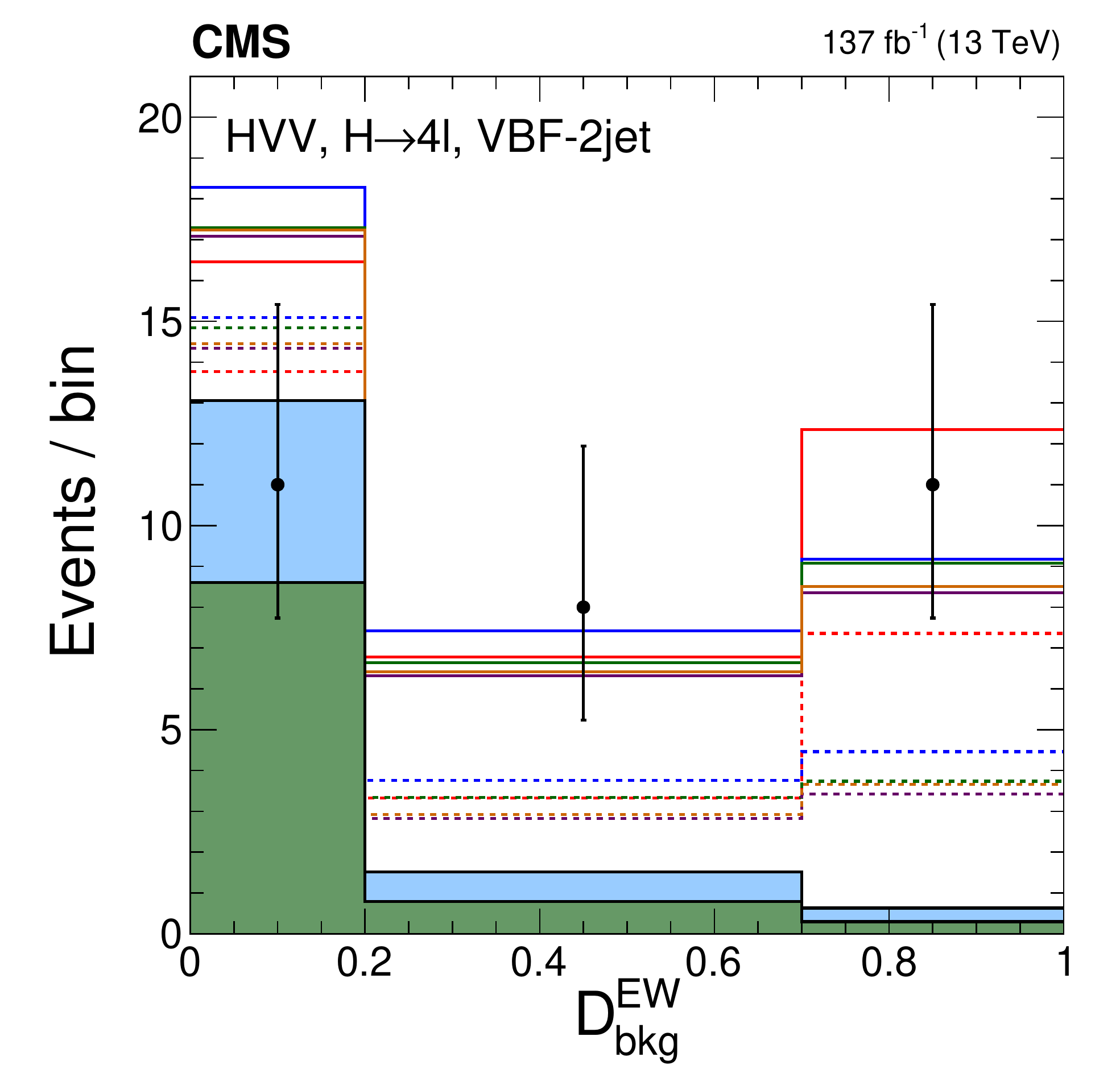}
\includegraphics[width=0.32\textwidth]{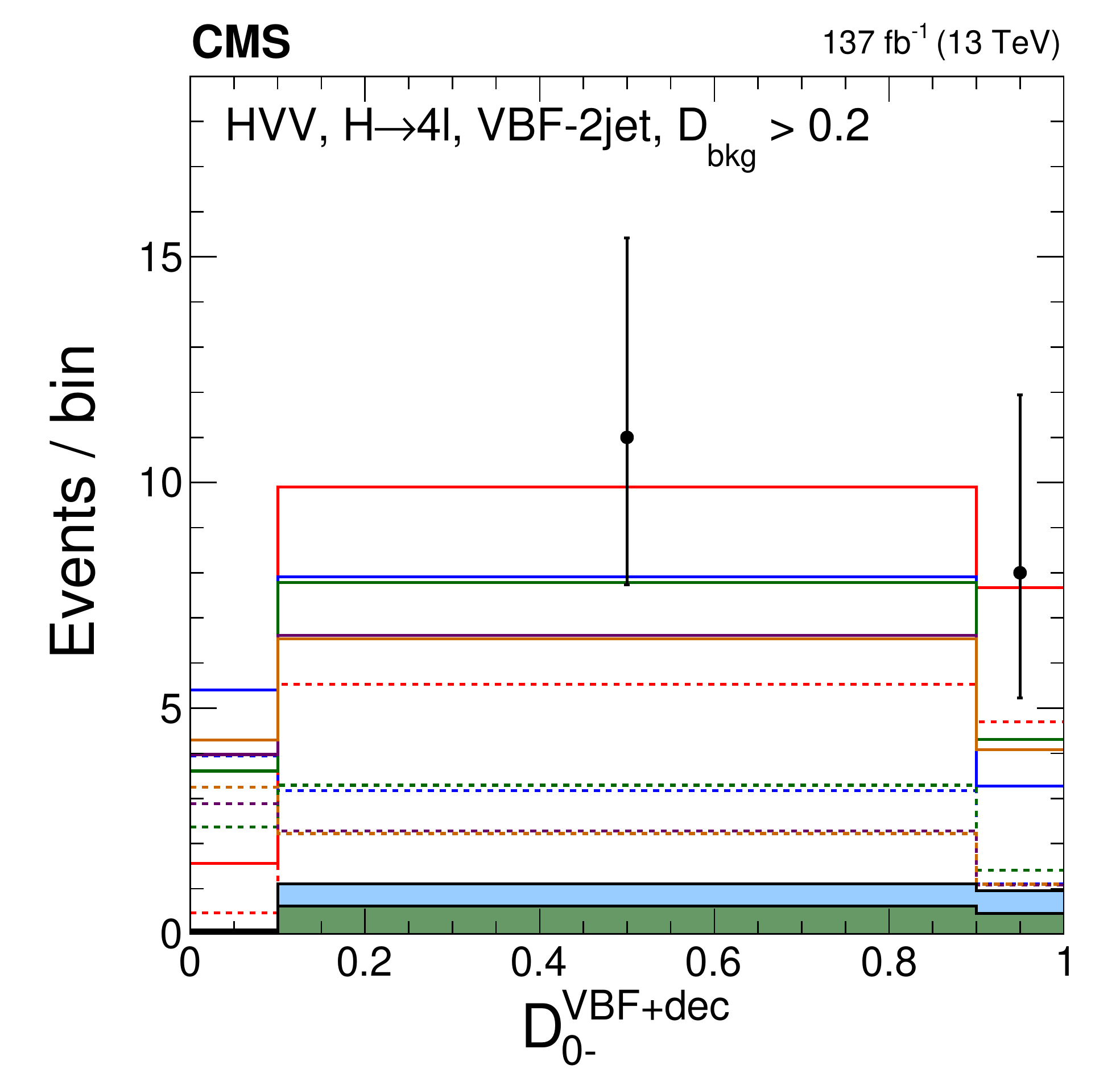}
\includegraphics[width=0.32\textwidth]{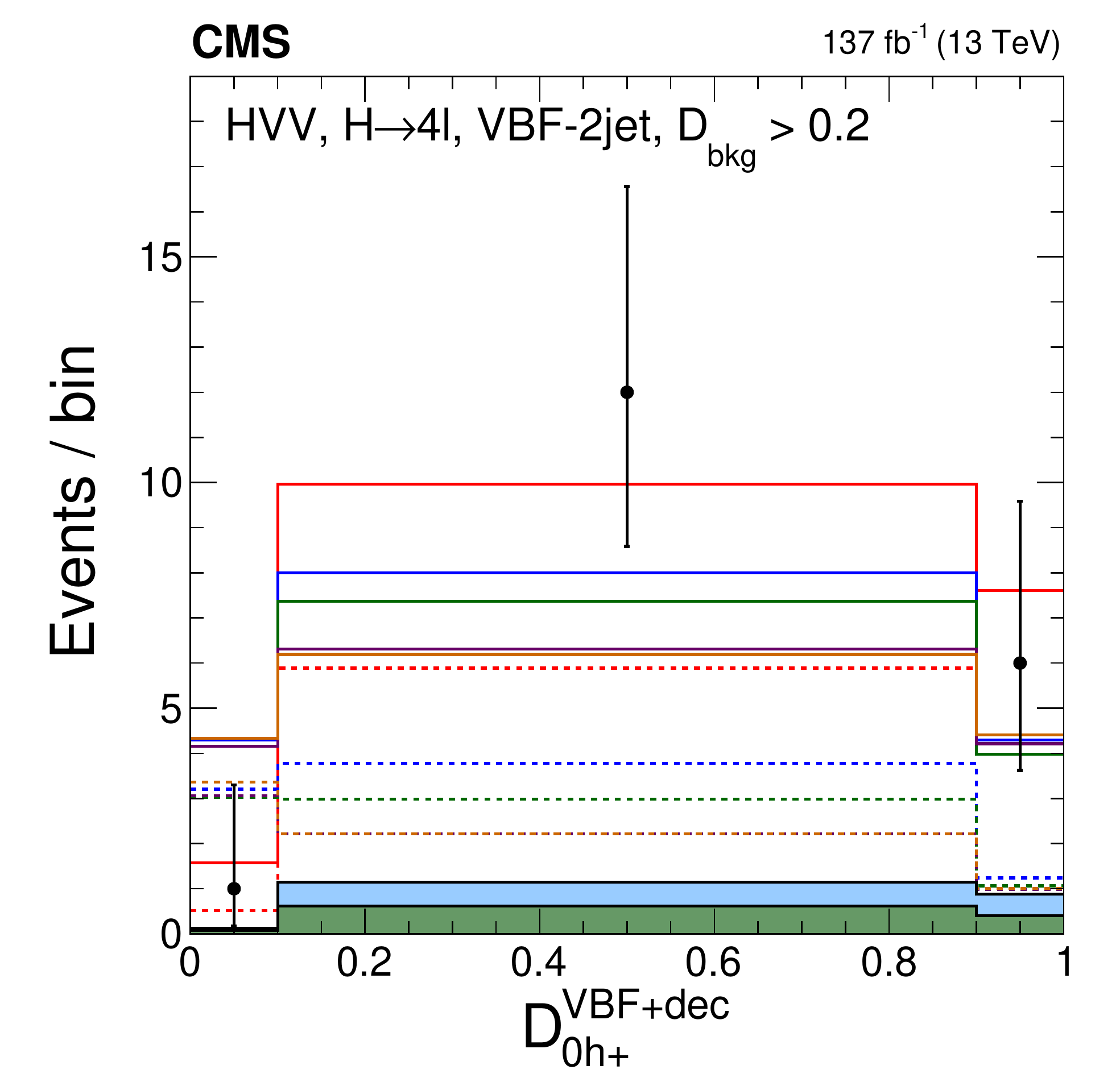} \\
\includegraphics[width=0.32\textwidth]{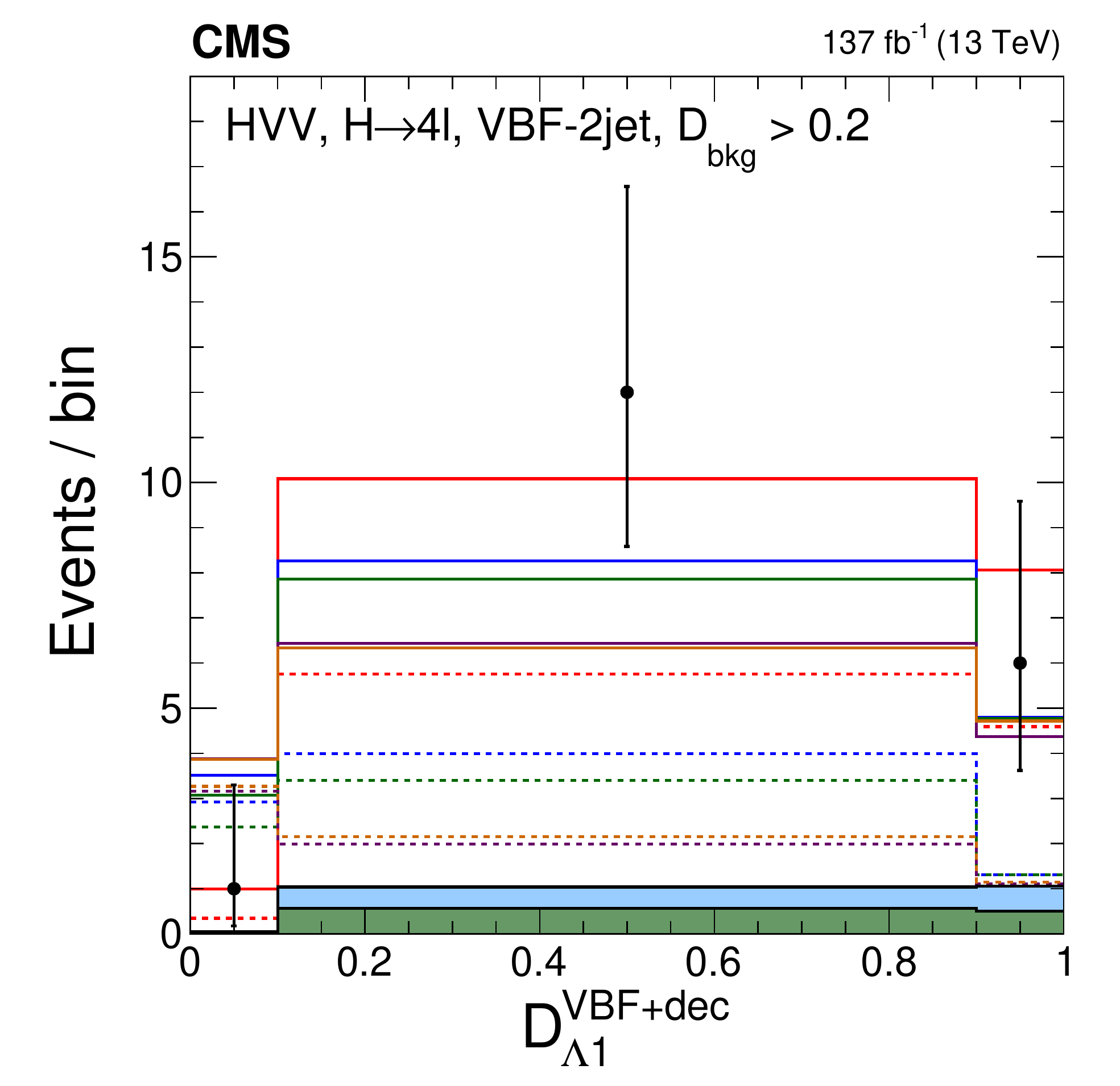}
\includegraphics[width=0.32\textwidth]{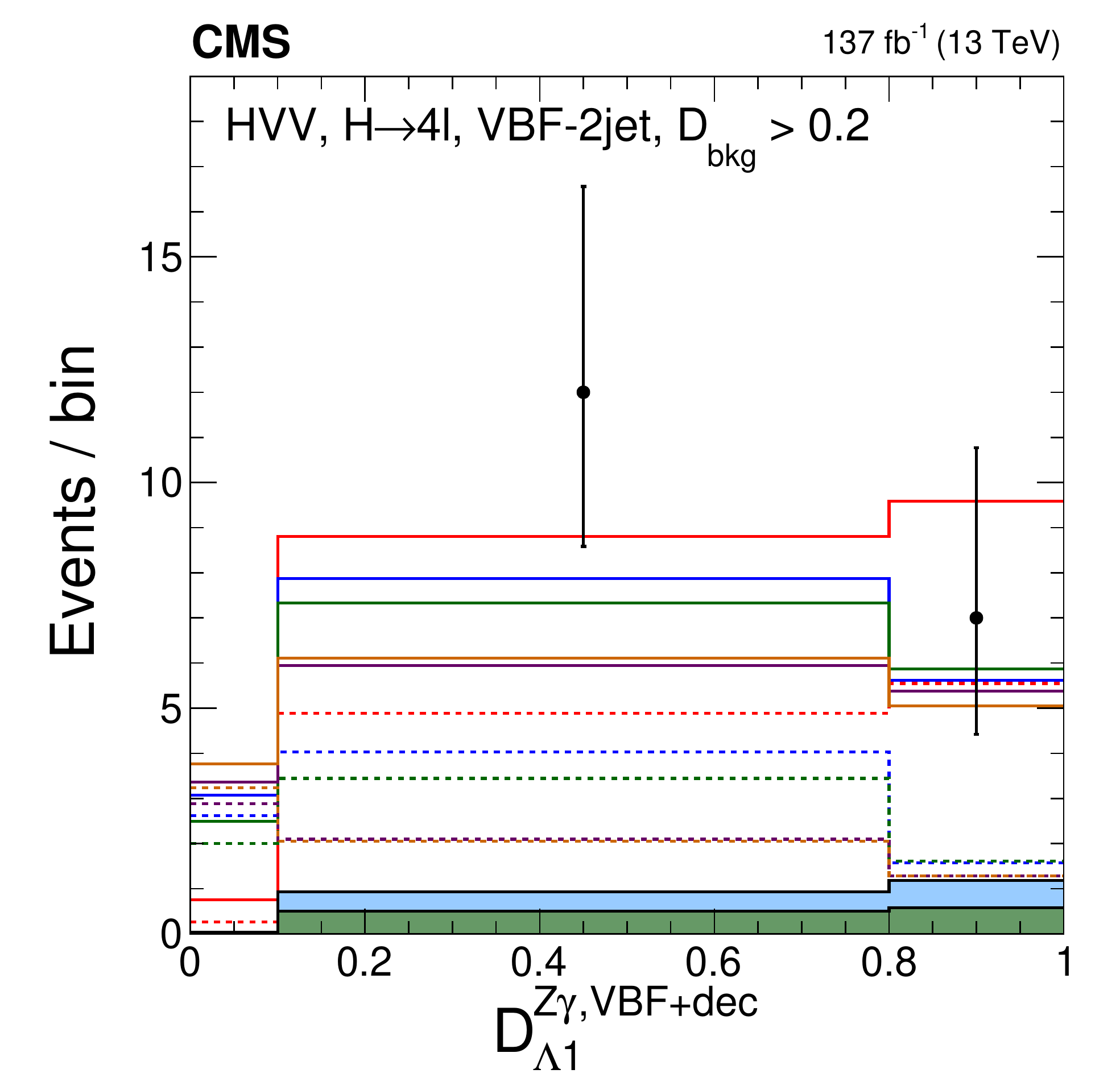}
\includegraphics[width=0.32\textwidth]{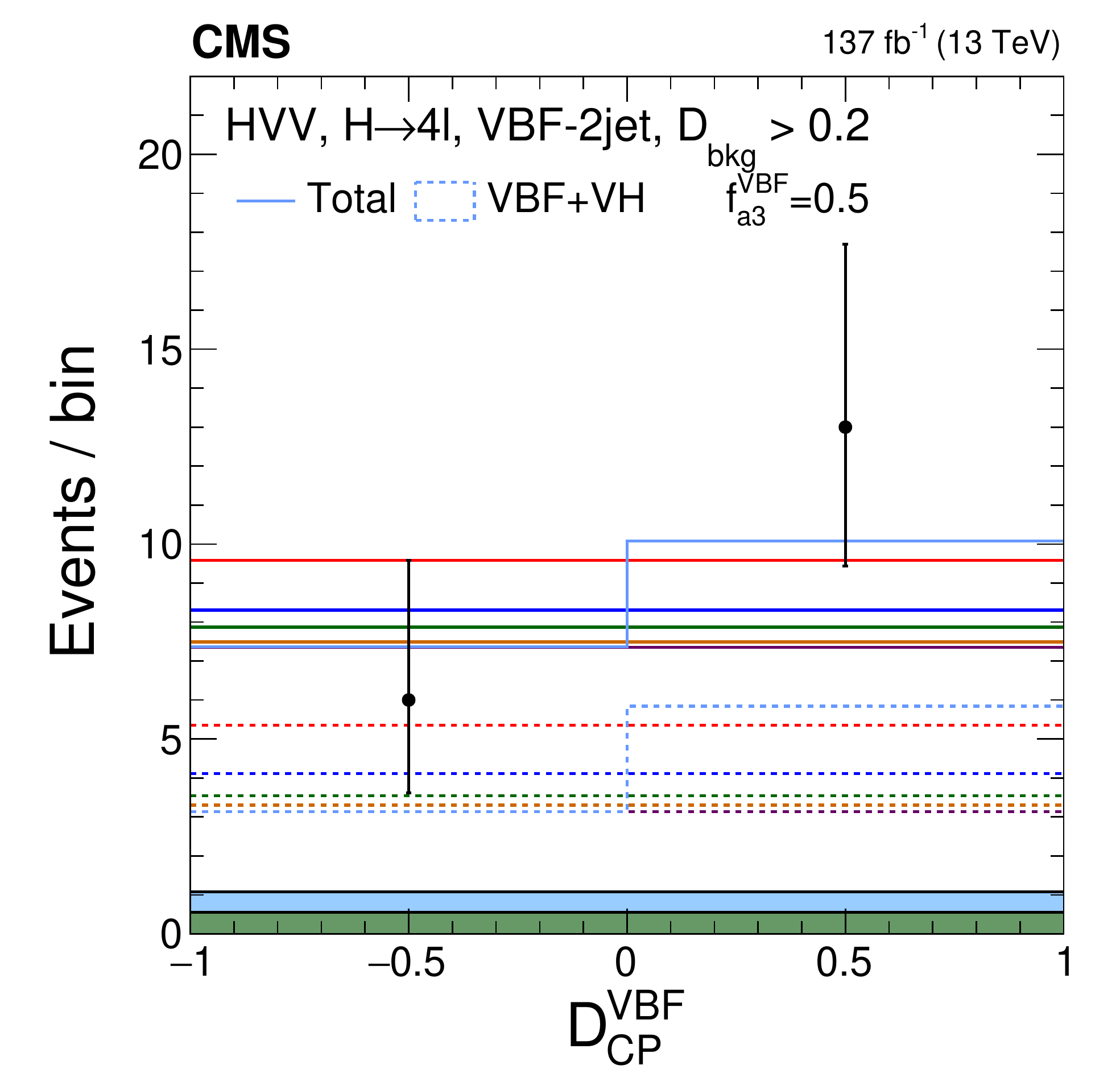} \\
\includegraphics[width=0.32\textwidth]{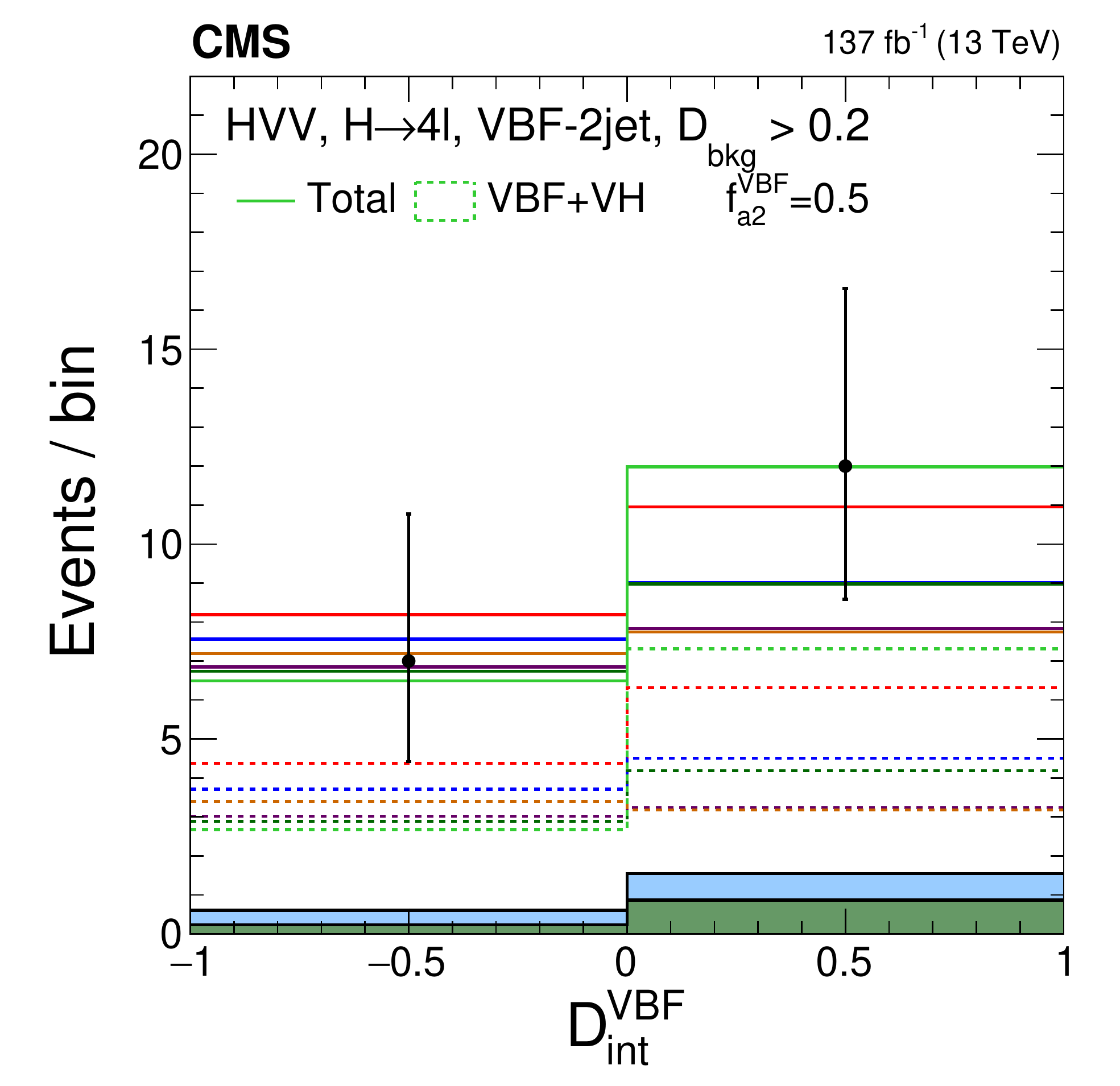}
\includegraphics[width=0.32\textwidth]{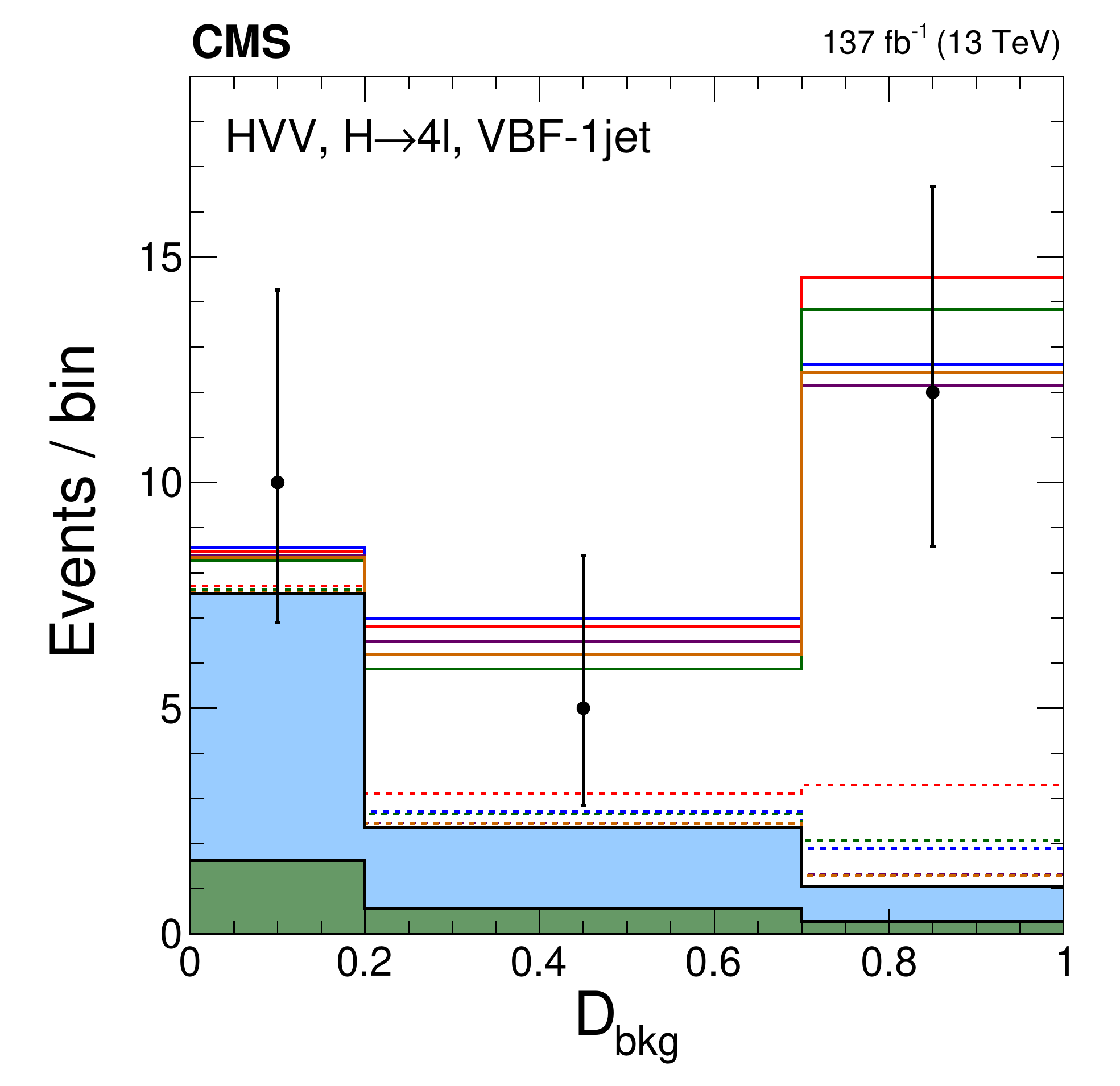}
\includegraphics[width=0.32\textwidth]{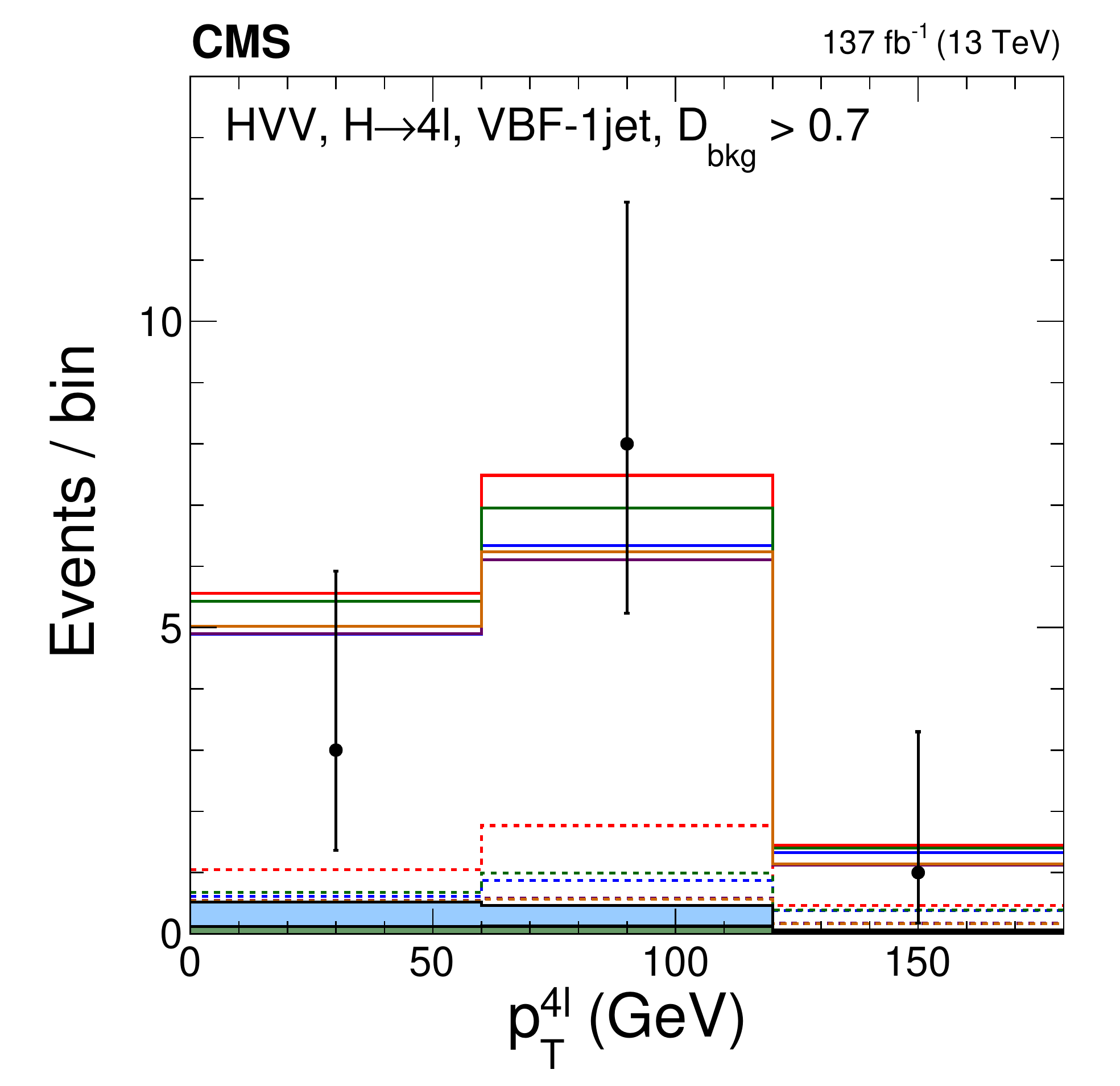}
\caption{
Distributions of events in the observables used in categorization Scheme~2.
The first seven plots are in the VBF-2jet category:
The upper left plot shows \DbkgEW.  The other distributions are shown with the requirement
$\DbkgEW>0.2$ in order to enhance the signal over the background contribution:
$\mathcal{D}_{0-}^{\text{VBF}+\text{dec}}$ (upper middle),  $\mathcal{D}_{0h+}^{\text{VBF}+\text{dec}}$ (upper right),
$\mathcal{D}_{\Lambda1}^{\text{VBF}+\text{dec}}$ (middle left), $\mathcal{D}_{\Lambda1}^{\PZ\gamma, \text{VBF}+\text{dec}}$ (middle middle),
$\mathcal{D}_{CP}^{\text{VBF}}$, and $\mathcal{D}_\text{int}^{\text{VBF}}$.
The last two plots are shown in the VBF-1jet category: \Dbkg (lower middle) and $\PT^{4\ell}$
with the requirement $\Dbkg>0.7$ and overflow events included in the last bin (lower right).
Observed data, background expectation, and five signal models are shown in the plots, 
as indicated in the legend in Fig.~\ref{fig:d2jet} (left). 
In several cases, a sixth signal model with a mixture of the SM and BSM couplings 
is shown and is indicated in the legend explicitly.
}
\label{fig:stackPlotsOnshellVBF}
\end{figure*}

\begin{figure*}[!tbp]
\centering
\includegraphics[width=0.32\textwidth]{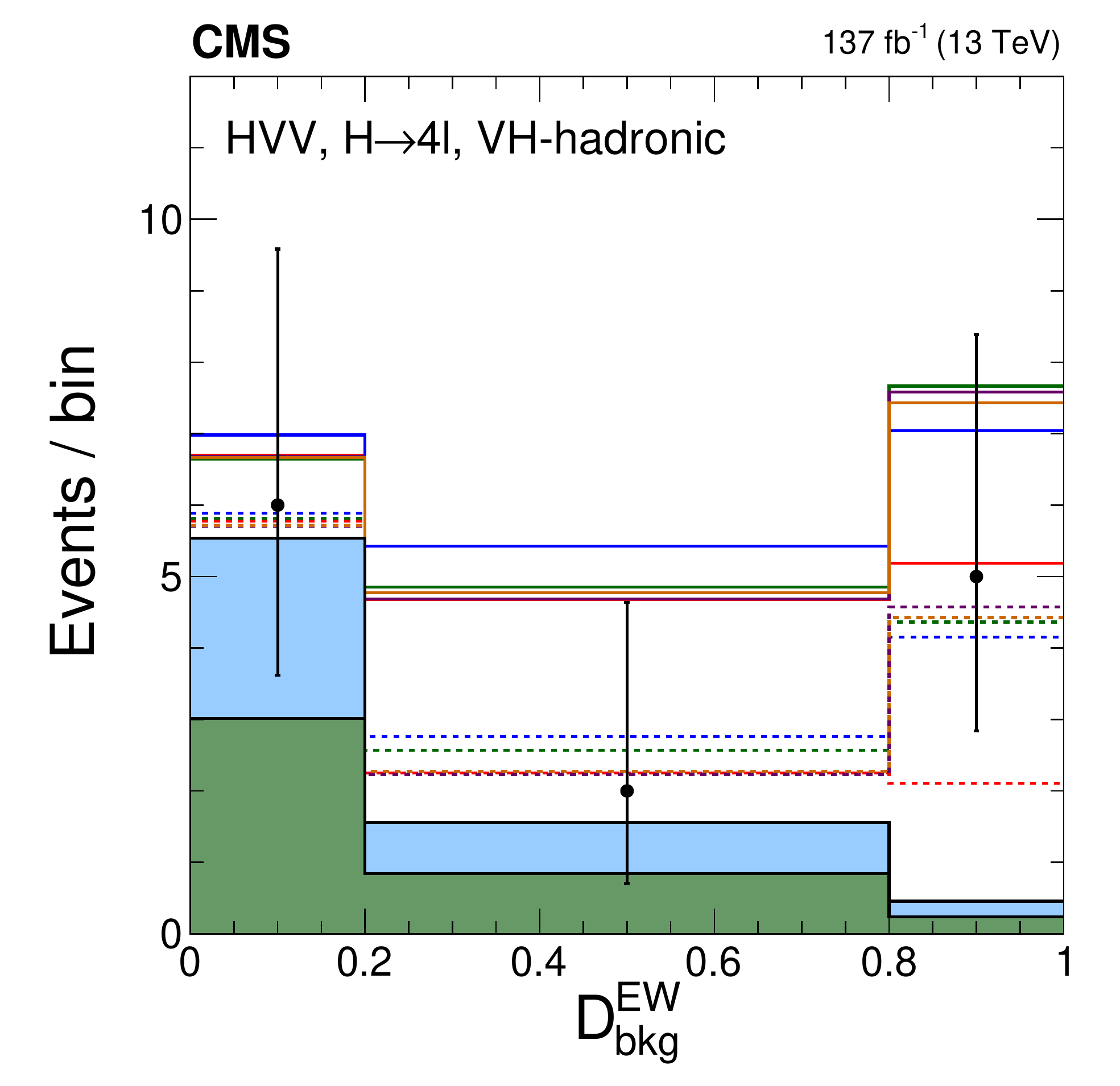}
\includegraphics[width=0.32\textwidth]{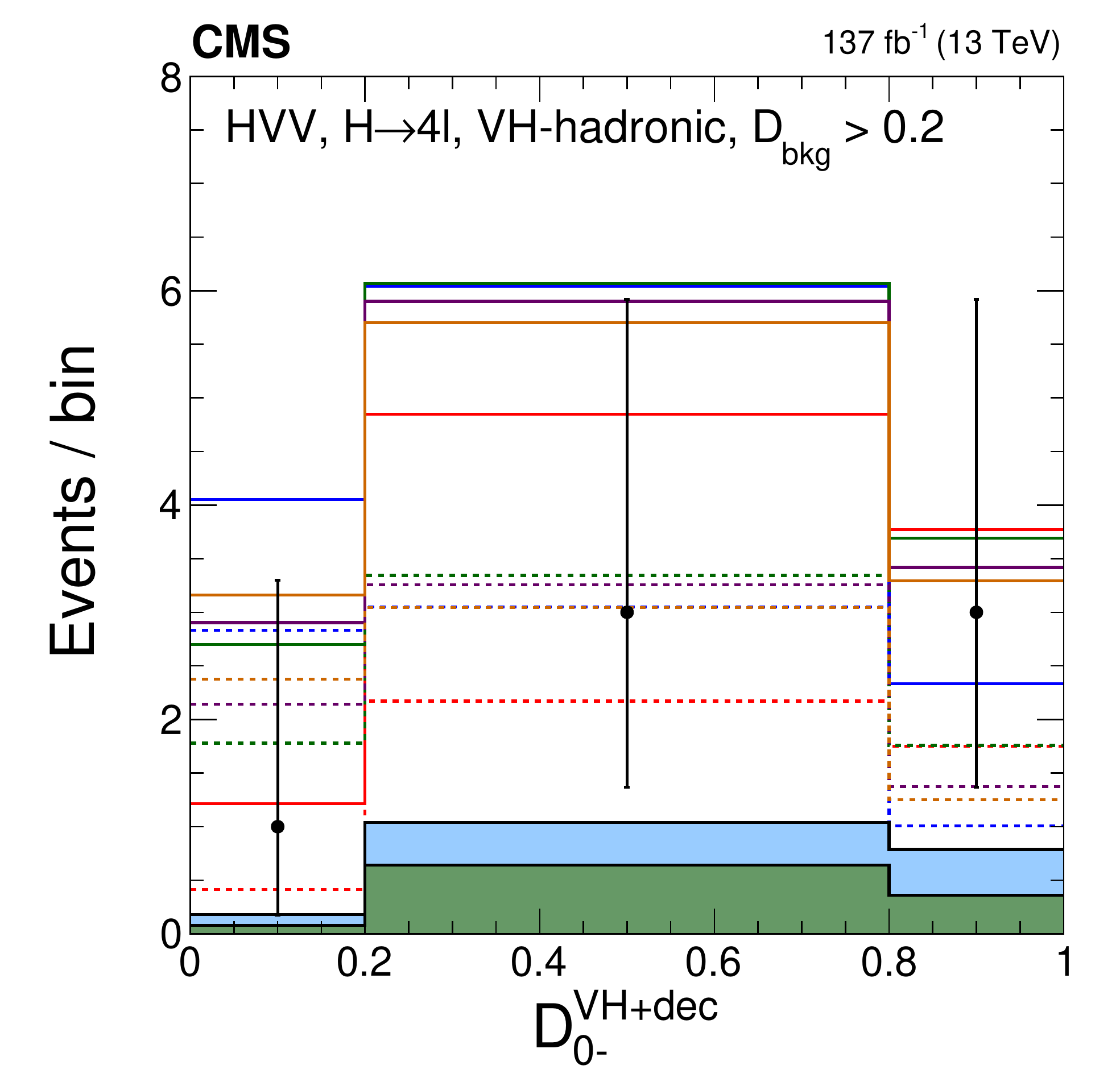}
\includegraphics[width=0.32\textwidth]{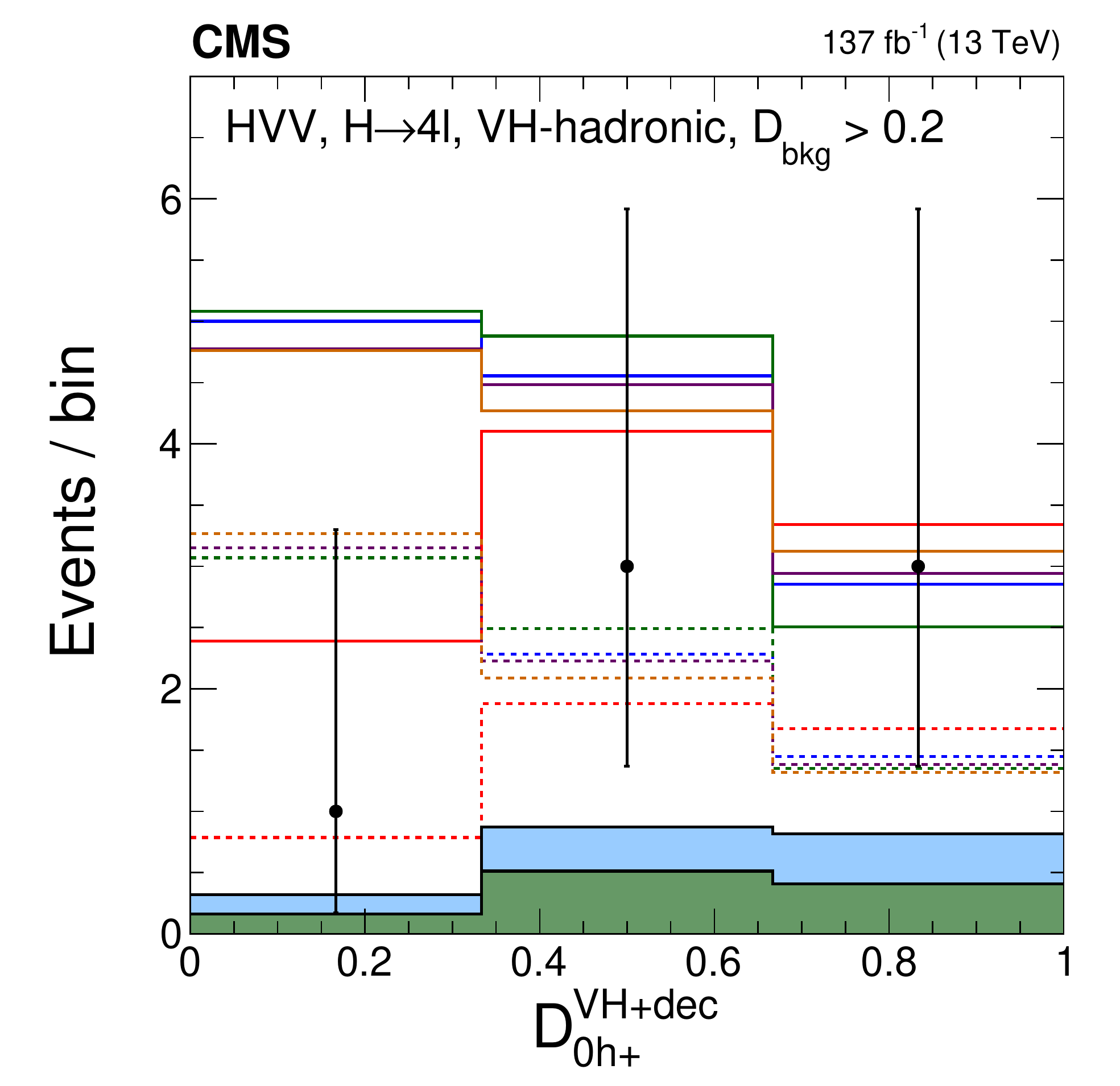} \\
\includegraphics[width=0.32\textwidth]{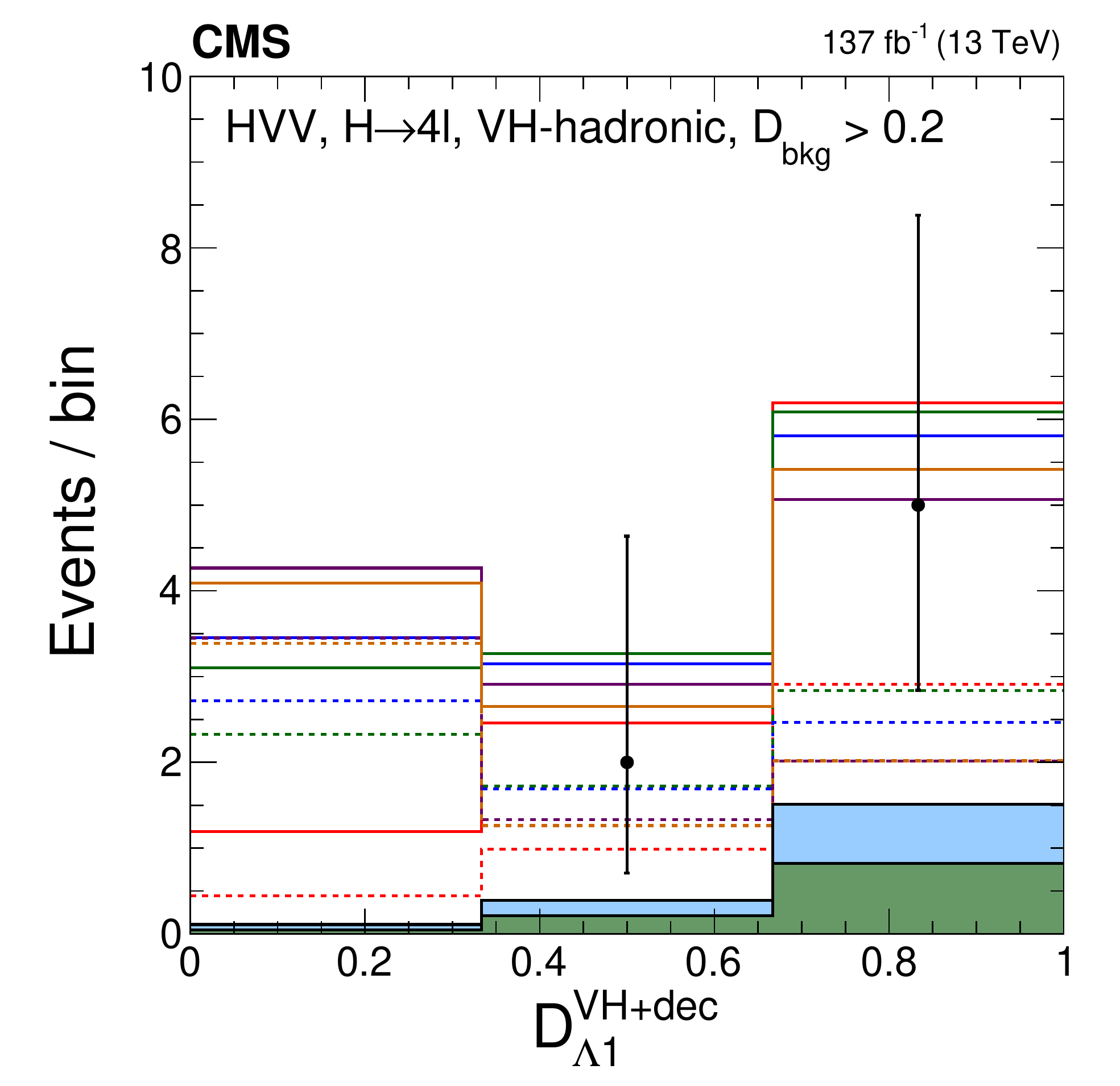}
\includegraphics[width=0.32\textwidth]{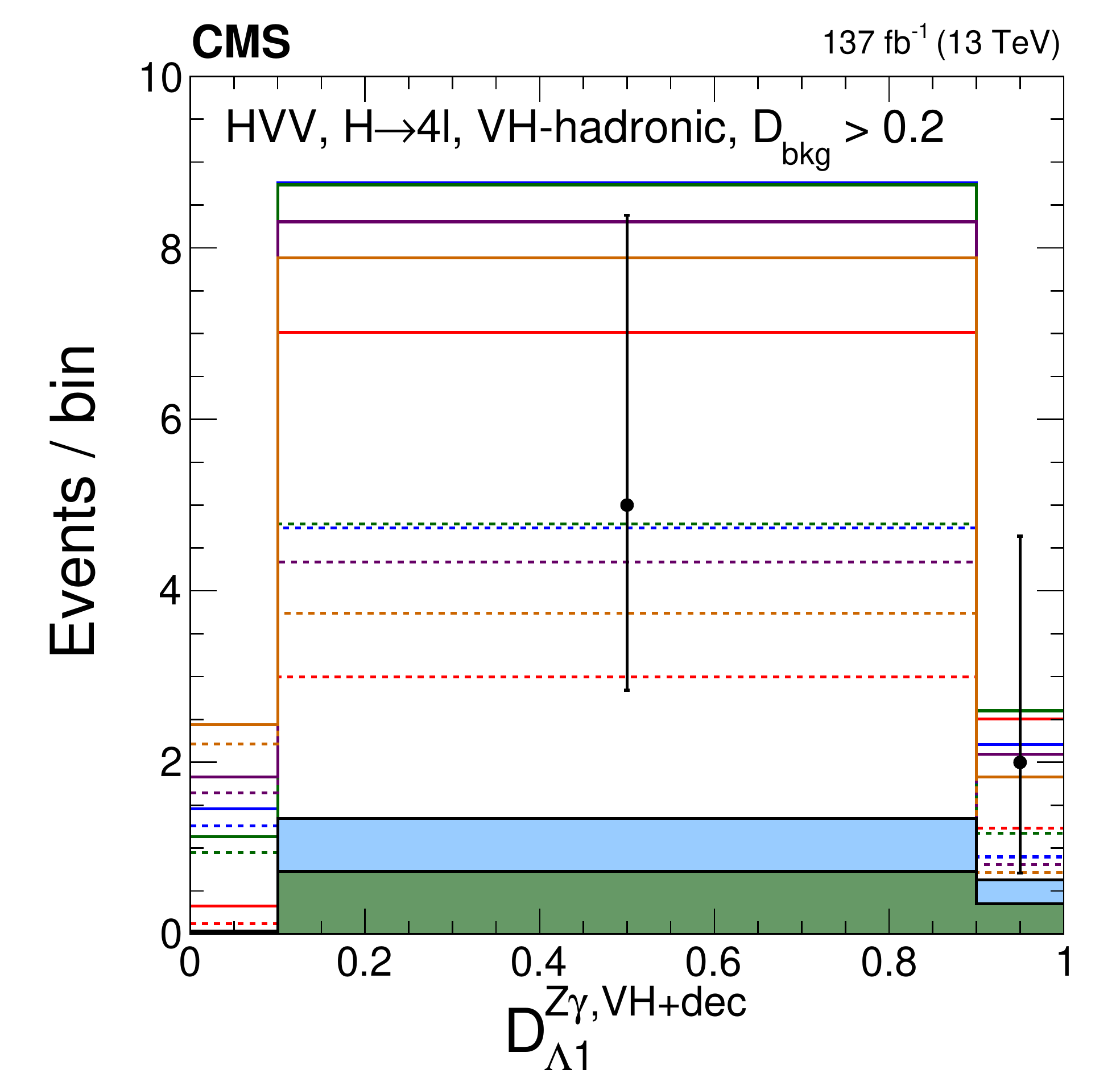}
\includegraphics[width=0.32\textwidth]{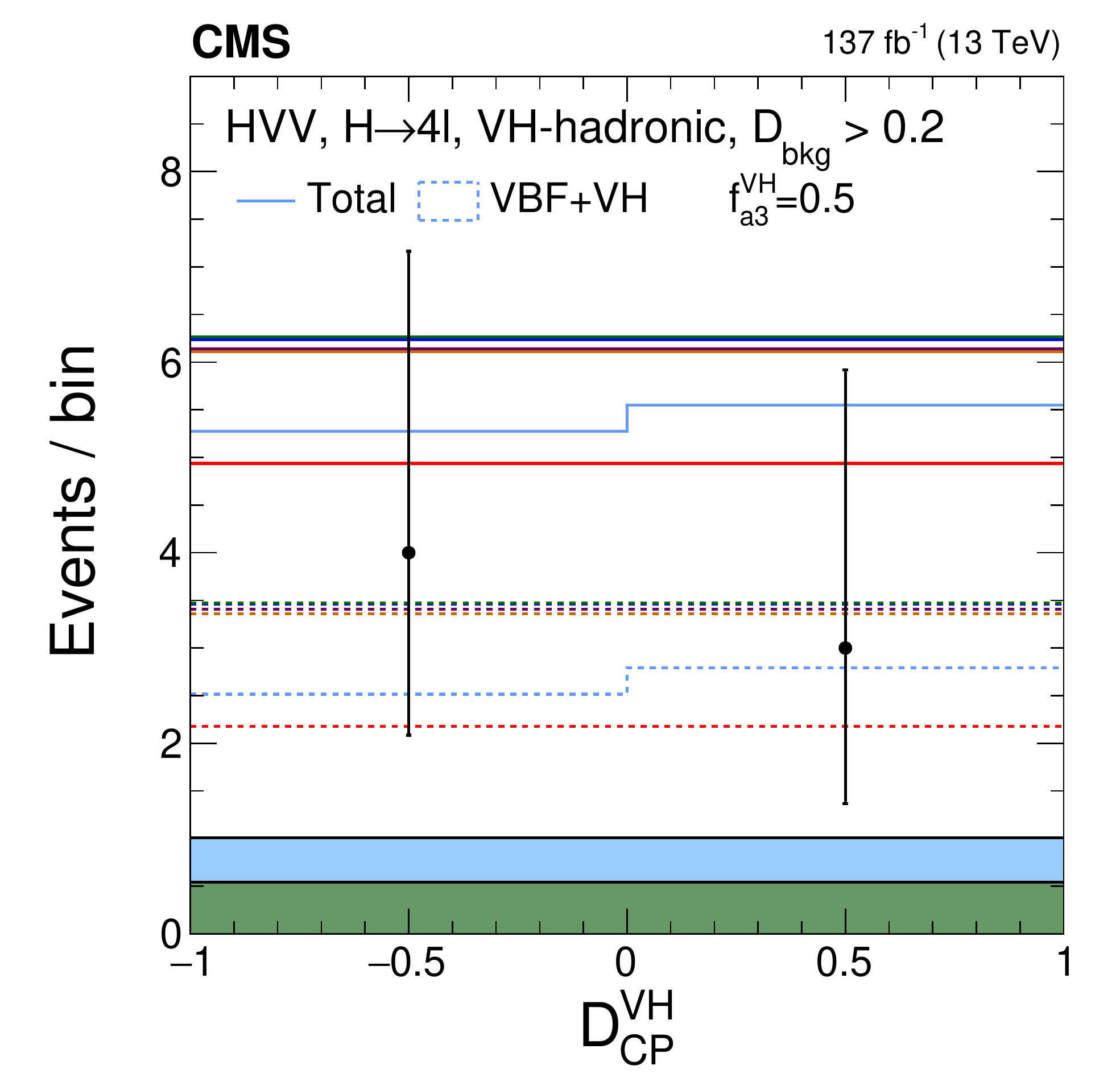} \\
\includegraphics[width=0.32\textwidth]{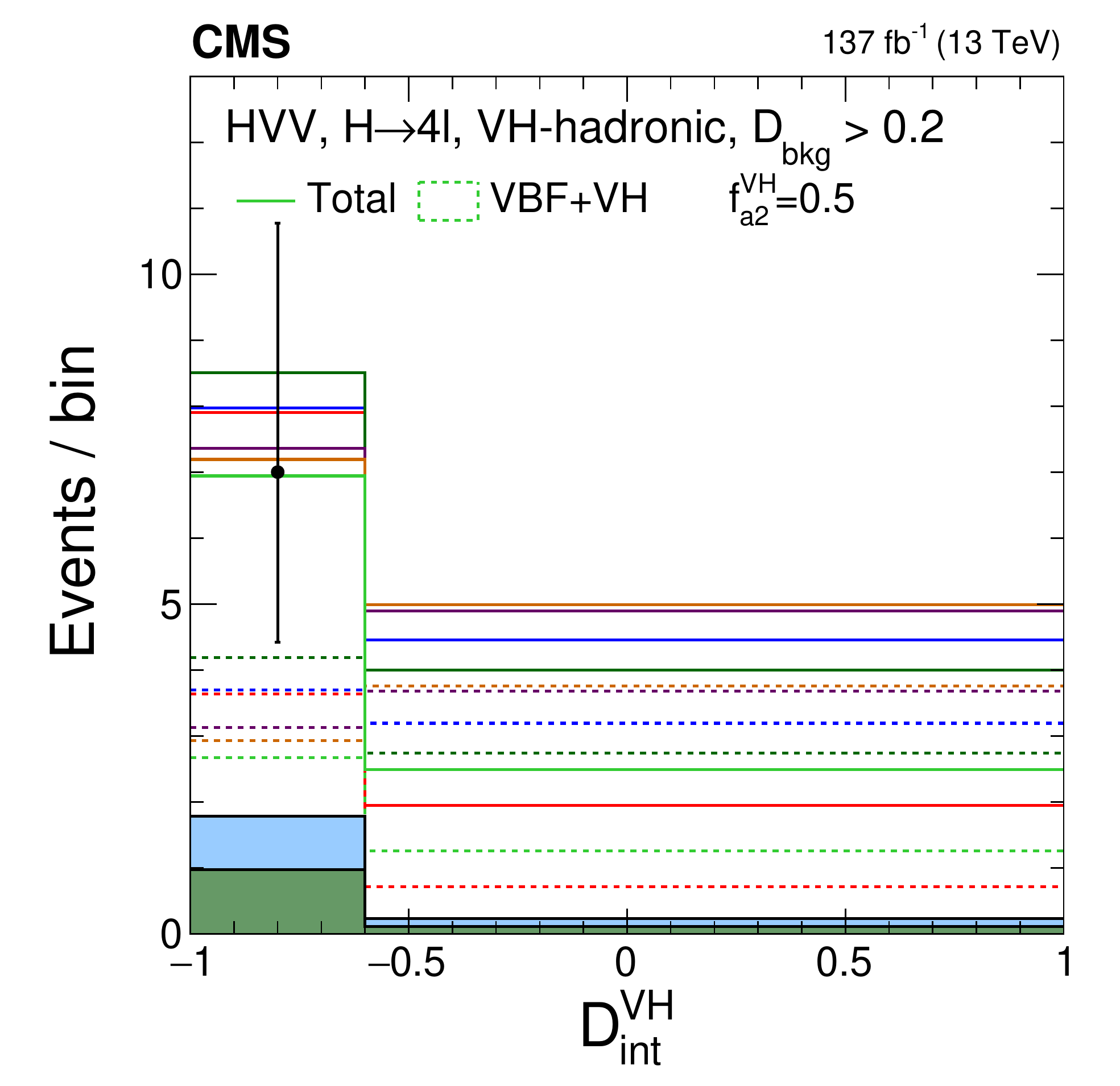}
\includegraphics[width=0.32\textwidth]{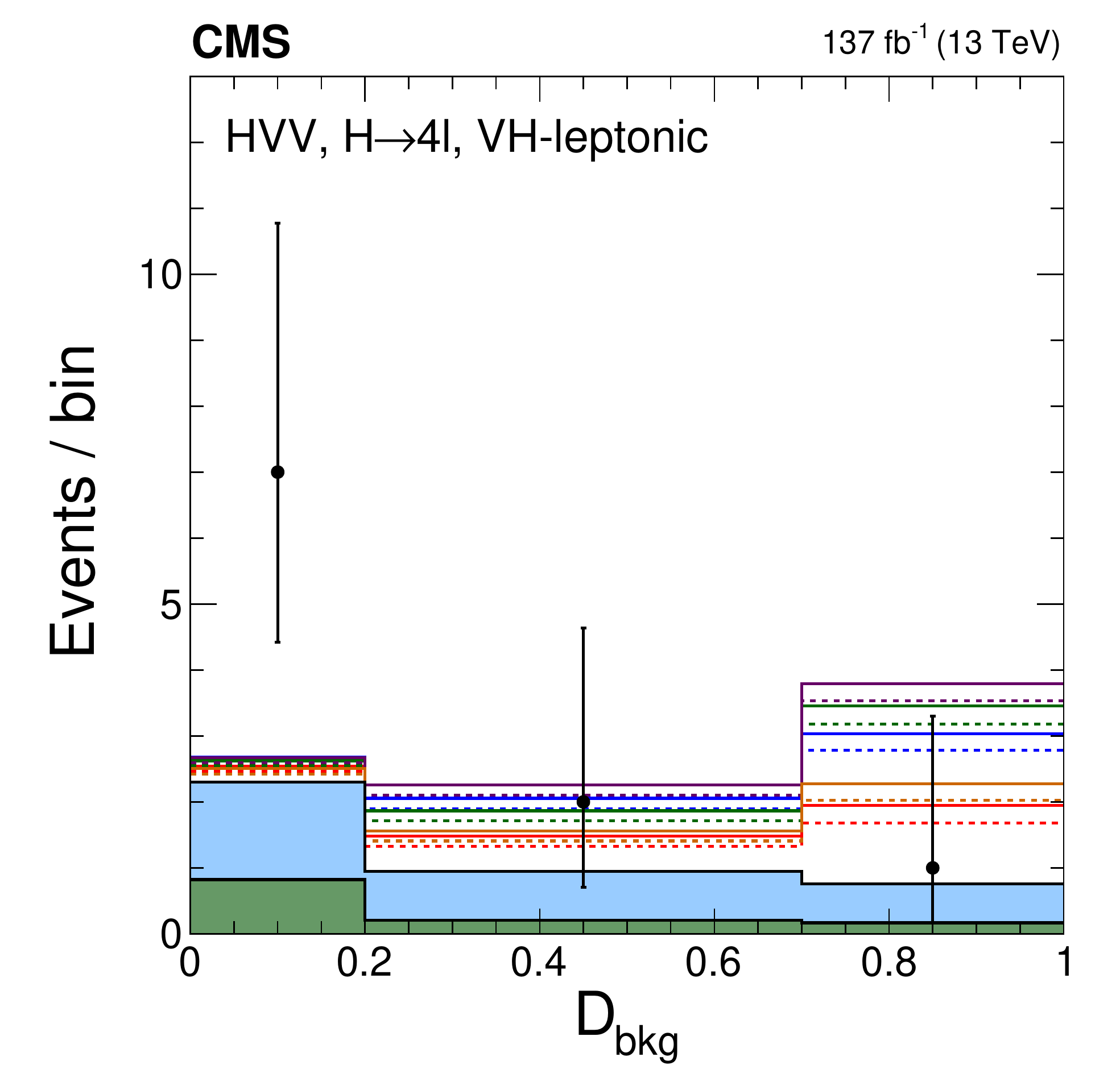}
\includegraphics[width=0.32\textwidth]{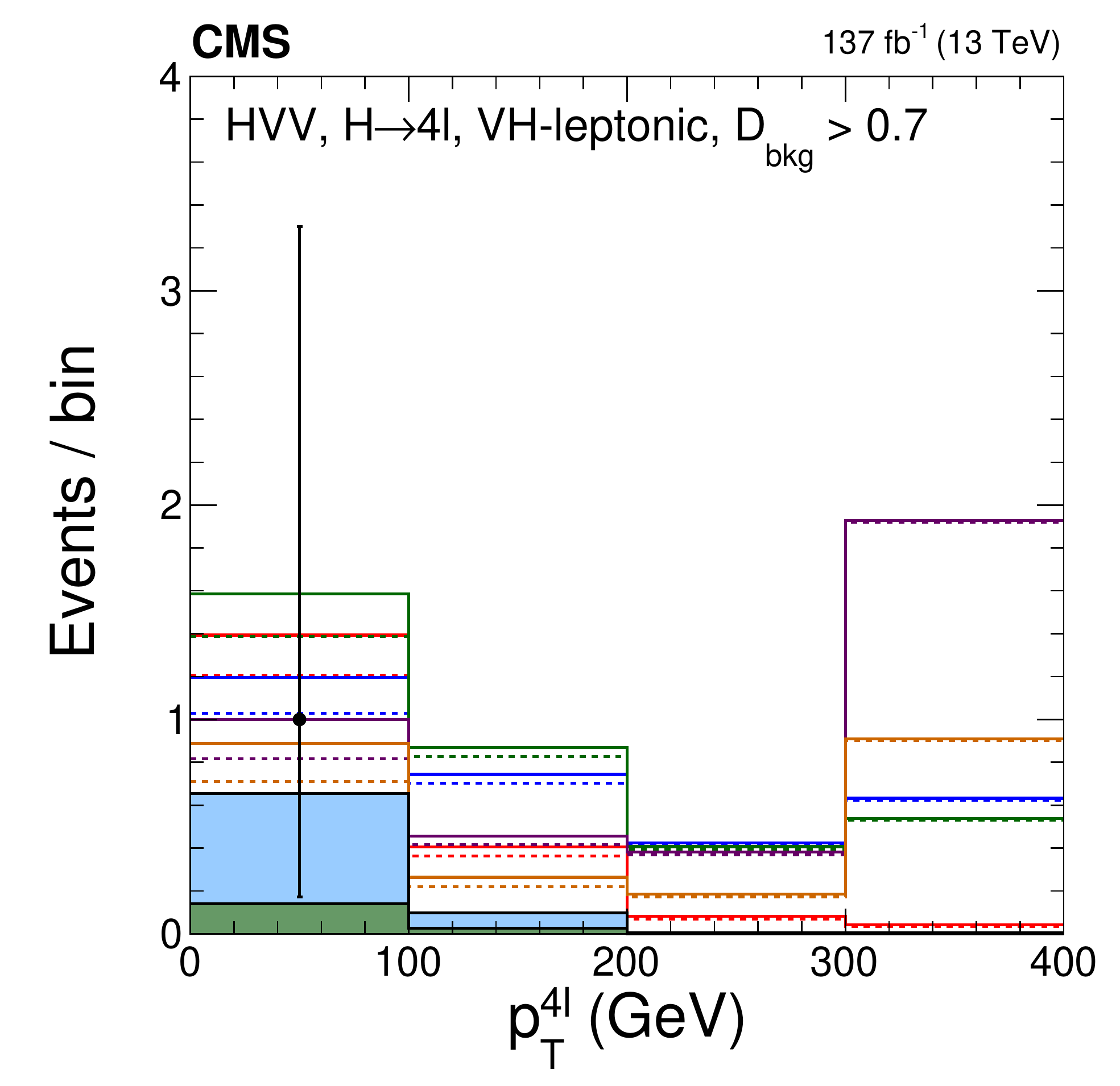}
\caption{
Distributions of events in the observables used in categorization Scheme~2.
The first seven plots are in the \VH-hadronic category:
The upper left plot shows \DbkgEW.  The other distributions are shown with the requirement
$\DbkgEW>0.2$ in order to enhance the signal over the background contribution:
$\mathcal{D}_{0-}^{\VH+\text{dec}}$ (upper middle);  $\mathcal{D}_{0h+}^{\VH+\text{dec}}$ (upper right);
$\mathcal{D}_{\Lambda1}^{\VH+\text{dec}}$ (middle left); $\mathcal{D}_{\Lambda1}^{\PZ\gamma, \VH+\text{dec}}$ (middle middle);
$\mathcal{D}_{CP}^{\VH}$, and $\mathcal{D}_\text{int}^{\VH}$.
The last two plots are shown in the \VH-leptonic category: \Dbkg (lower middle) and $\PT^{4\ell}$
with the requirement $\Dbkg>0.7$ and overflow events included in the last bin (lower right).
Observed data, background expectation, and five signal models are shown in the plots 
as indicated in the legend in Fig.~\ref{fig:d2jet} (left). 
In several cases, a sixth signal model with a mixture of the SM and BSM couplings 
is shown and is indicated in the legend explicitly.
}
\label{fig:stackPlotsOnshellVH}
\end{figure*}

Distributions of events for the observables $\vec{x}$ in Scheme~2 are illustrated in 
Figs.~\ref{fig:stackPlotsOnshellUntaggedBoosted}--\ref{fig:stackPlotsOnshellVH}.
Here and in Figs.~\ref{fig:ttH_BDT} and~\ref{fig:ggH_D0mn}
the expected distributions are based on signal MC simulation discussed in Section~\ref{sec:cms_mc} and 
the background estimate in Section~\ref{sec:cms_bkg}, where cross sections of all processes, 
including those with the BSM couplings, are set to the SM expectations.
The full list of kinematic observables employed in the fit in each category is summarized 
in the third column of~Table~\ref{tab:Dobservables}. 

\section{Implementation of fitting and associated uncertainties}
\label{sec:fit}

In the analysis of the $\Hff$, $\Hgg$, or $\HVV$ anomalous couplings, the events are split into a total of 63 
(in Scheme 1, designed to study $\Hff$ and $\Hgg$) or 54 (in Scheme 2, designed to study \HVV) categories
according to the seven or six production categories, three lepton flavor combinations 
($4\Pe$, $4\Pgm$, and $2\Pe 2\Pgm$), and three data periods (2016, 2017, and 2018).
Each event is characterized by its discrete category $k$ and set of input observables $\vec{x}$, 
as discussed in detail in Section~\ref{sec:kinematics} and summarized in Table~\ref{tab:Dobservables}. 
The observed distributions of events across the discriminants $\vec{x}$ are illustrated 
in Figs.~\ref{fig:ttH_BDT}--\ref{fig:stackPlotsOnshellVH} and 
are compared to the expected distributions in SM and BSM. However, quantitative
characterization of these distributions requires a careful analysis of the multidimensional 
space of observables $\vec{x}$ and categories $k$, which is discussed below. 
Preparation of this analysis was performed in a blind way, which means that 
observed distributions of events were not examined until all details of the fit discussed
below were finalized.

\subsection{Likelihood parameterization}

We perform an extended maximum likelihood fit~\cite{Barlow:1990vc} in which the
probability density is normalized to the total event yield in each category $k$ as a sum over 
all signal processes $j$ and background processes $i$ according to
\begin{align}
\mathcal{P}_{k}(\vec{x}) =
\sum_j \mu_j \mathcal{P}_{jk}^{\text{sig}} \left( \vec{x};\vec{\xi}_{jk},\vec{f}_j \right) +
\sum_i \mathcal{P}_{ik}^{\mathrm{bkg}} \left( \vec{x};\vec{\xi}_{ik} \right),
\label{eq:ponshell}
\end{align}
where $\mu_j$ is the ratio of the observed cross section to the SM expectation,
$\vec{f}_j$ is the set of unconstrained parameters describing kinematic distributions in a given process, 
defined in Eqs.~(\ref{eq:fCP_definitions}, \ref{eq:fggH_definitions}, \ref{eq:fa_definitions_hvv}),
and $\vec{\xi}_{jk}$ are the constrained nuisance parameters reflecting the uncertainties in the
above parameterization. 

In the case of the gluon fusion process, $\mu_j=\mu_{\ggH}$ and $\vec{f}_j=\fG$. 
The dependence on the $CP$-sensitive parameter $\fG$ appears only in the \VBF-2jet category 
where correlation of the two associated jets is explored. Therefore, in this category a dedicated
simulation of the \Hboson production with two associated jets with \MGvATNLO is used, and checked 
against the \jhugen and \minlo simulation. 
The interpretation of this process in terms of fermion couplings appearing in the gluon fusion loop is discussed 
in Section~\ref{sec:results_HttHgg}. In cases where the SM fermions are assumed to dominate the gluon fusion loop, 
the $\mu_{\ggH}$ and $\fG$ parameters are correlated to $\mu_{\ttH}$ and $\fT$ in the $\ttH$ process
through Eq.~(\ref{eq:fai-relationship-hgg-tth}).
A more general case, when both SM fermions and heavy BSM particles contribute to the loop, 
is also considered. In all cases the relationship between the \Hff and \Hgg couplings follows \jhugen
with the relative sign of $CP$-odd and $CP$-even coefficients opposite to that assumed in \MGvATNLO, 
as discussed in Ref.~\cite{Gritsan:2020pib}. 

In the case of the $\ttH$ process, $\mu_j=\mu_{\ttH}$ and $\vec{f}_j=\fT$.
The $\tqH$ production is expected to contribute about 5\% of the total $\ttH$ signal in this analysis, 
as shown in Table~\ref{tab:Hffcategory}, and, therefore, the exact treatment of this process is currently not important. 
This production mode depends on both $\HVV$ and $\Hff$ couplings. 
The anticipated small yield may be somewhat larger if the expected destructive interference between the 
$\HVV$ and $\Hff$ couplings in the $\tqH$ process becomes constructive owing to a modification of these couplings. 
In this analysis, we also introduce a possible $CP$-odd Yukawa coupling in the $\tqH$ process.
We fix the sign of the $a_1$ coupling to be positive, and so the sign of the $\kappa_{\PQt}$ coupling 
allows us to float the relative sign of the $\HVV$ and $\Hff$ contributions in this process. 
In Scheme 1, we parameterize the $\tqH$ signal strength with the parameters $\mu_{\ttH}$, $\fT$, and $\muV$, defined below. 
There is a weak dependence of the $\mathcal{D}_{0-}$ distributions on $\fT$ in this channel, 
which we conservatively neglect. We also neglect anomalous $\HVV$ couplings in the $\tqH$ parameterization
because their effect is negligible with the current constraints from analysis of the \VBF and \VH events.
In Scheme 2, where we do not have a dedicated category for the top quark
coupling measurements, we neglect the dependence of the $\tqH$ process on the \HVV couplings because
this would have a negligible effect on the results and unnecessarily complicate the fit parameterization. 

For EW processes (\VBF, \ZH, \WH), the common unconstrained
parameters of interest are $\mu_j=\muV$ and $\vec{f}_j=\left(\fC, \fB, \fL, \fLZg\right)$ in Approach 1
or $\vec{f}_j=\left(\fC, \fB, \fL\right)$ in Approach 2, when using categorization Scheme 2 for fitting 
\HVV couplings. We simplify treatment of the EW processes in categorization Scheme 1 when fitting 
for $CP$-violating $\Hgg$ or $\Htt$ couplings and allow only the $CP$-violating $\fC$ parameter in $\vec{f}_j$.
When all anomalous couplings are set to zero, the signal strength \muV is equal to the ratio of the cross sections
of all EW processes (\VBF, \ZH, \WH) to the SM expectation.
In the case of $f_{ai}=1$, this ratio is corrected by the factor $(\alpha_{ii}/\alpha_{11})$ as quoted 
in Table~\ref{tab:xsec_ratio} for decay cross sections, and the inverse of this factor for the EW production cross sections,
due to the evolution of the cross sections with anomalous couplings. 

It has been shown that there is no sensitivity to the anomalous couplings in the $\bbH$ process~\cite{Gritsan:2016hjl} and 
it is parameterized with the signal strength  $\mu_j=\mu_{\bbH}$. Depending on the fit implementation discussed
in Section~\ref{sec:results}, this signal strength may be correlated with those in other channels, such as $\mu_{\ggH}$
or $\mu_{\ttH}$. The exact treatment of this process is expected to have negligible effect on the results, 
because it cannot be distinguished kinematically from other dominant processes and its relative contribution in 
each kinematic region is expected to be negligibly small, as shown in Tables~\ref{tab:Hffcategory} and~\ref{tab:HVVcategorization}. 

The background processes $i$ in Eq.~(\ref{eq:ponshell}) include the $\qqbar\to4\ell$, $\Pg\Pg\to4\ell$, and EW 
processes, all of which are estimated with simulation, but receive additional constraints from sidebands in data. 
The EW background includes vector boson scattering 
and $\PV\PV\PV$ processes, which are the background counterparts of the \VBF and \VH processes. We also include 
the $\ttbar\PV\PV$ and $\ttbar\PV$ processes in this background contribution, which are important in the study of
the \ttH signal process. Interference of the signal and background processes is negligible in the analysis of the
on-shell \Hboson production. The \ZX background contribution models $\Z$+jets and other related processes
with lepton misidentification and is estimated from the control regions in the data as discussed in Section~\ref{sec:cms_bkg}.
All signal $j$ and background $i$ processes contributing to Eq.~(\ref{eq:ponshell}) and their expected yields 
are shown in Tables~\ref{tab:Hffcategory} and~\ref{tab:HVVcategorization}.

The signal and background probability distributions $\mathcal{P}_{jk}^{\text{sig}}$ and
$\mathcal{P}_{ik}^{\mathrm{bkg}}$ appearing in Eq.~(\ref{eq:ponshell}) are binned 
multidimensional histograms (templates) of observables $\vec{x}$ listed in Table~\ref{tab:Dobservables}.
The binning of these templates has been optimized for memory and speed of computer calculations, 
expected population of events across those bins, and retaining kinematic information. 
In particular, the large number of discriminants used in the Untagged, \VBF-2jet, and  \VH-hadronic categories
in Scheme 2, requires careful optimization of the binning employed in analysis. In these categories,
two bins are used in the two interference discriminants and three bins are used in the other five
discriminants, which corresponds to a total of 972 bins in the seven-dimensional (7D) distribution.  
However,  the bins with a very low yield of expected events for all contributions are merged, and 
the expected symmetry in the distribution of the $\mathcal{D}_{CP}$ observable is enforced.
As a result, the total number of independent bins depends on the category, but does not exceed 
400 in any of the categories. Even though only a limited number of bins
is used in each dimension, the 7D distribution retains substantial kinematic information 
that is nearly optimal for all anomalous couplings targeted in this analysis. 
This has been validated against a dedicated analysis targeting one anomalous 
coupling at a time with a much larger number of bins in each dimension
for a smaller number of discriminants, as employed in Refs.~\cite{Sirunyan:2017tqd,Sirunyan:2019twz}.
This nearly optimal performance is realized in large respect due to the optimal population of events across 
the range of discriminant values by construction of the MELA approach. 

The $\mathcal{P}_{jk}^{\text{sig}}$ and $\mathcal{P}_{ik}^{\mathrm{bkg}}$
probabilities depend on the parameters $\vec{\xi}_{jk}$ and $\vec{f}_j$ and are therefore
interpolated between various templates as a function of these parameters. 
The $\vec{\xi}_{jk}$ reflect systematic uncertainties either in the normalization or shape of
both signal and background templates and an analytical linear interpolation is adopted. 
The $\vec{f}_j$ parameters require nontrivial analytical interpolation of the signal templates,  
which is discussed in more detail below.

\subsection{Signal parameterization}

In the signal production and decay processes, the same anomalous couplings could appear either
on both the production and decay sides simultaneously, as in the case of \HVV couplings in EW production 
and  $\PH\to\VV\to 4\ell$ decay, or only on one side (production or decay). 
This is illustrated in Eq.~(\ref{eq:diff-cross-section2}), with $\bigl(\sum_{il}  \alpha_{il}^{(j)}a_ia_l\bigr)$
appearing on the production side and $\bigl(\sum_{mn} \alpha_{mn}^{(f)}a_ma_n\bigr)$ appearing on the decay side. 
We absorb the width ${\Gamma_{\PH}}$ in Eq.~(\ref{eq:diff-cross-section2}) into the overall signal strength 
and parameterize the kinematic dependence in the signal probability density on the ratio of couplings 
through  $\vec{f}_j$ in the following way. 

In the case of \HVV anomalous couplings, we have either $L=5$ couplings in Approach~1
or $L=4$ couplings in Approach~2, which we can parameterize with four or three components of $\vec{f}_j$, 
defined above, and ${f_{a1}}=(1-\abs{\fB}-\abs{\fC}-\abs{\fL}-\abs{\fLZg})$. Let us denote these as $f_l$ with $l=1,2,3,4,5$. 
For the \ggH and \ttH processes, when we consider anomalous couplings on the production side, we have only
two parameters ${f_l}= (1-\abs{\fG})$ and ${\fG}$, or $(1-\abs{\fT})$ and ${\fT}$, and $L=2$.
When developing the expression in Eq.~(\ref{eq:diff-cross-section2}), one gets a polynomial in the couplings
$a_l\propto\sqrt{\abs{f_{l}}}\, \sign(f_{l})$ following Eq.~(\ref{eq:ai}), 
which is either quartic ($C=4$), in the case of the EW processes,  
or quadratic ($C=2$), when the couplings appear only on the decay or production side. 
Parameterization of the anomalous coupling dependence of the $\mathcal{P}_{jk}^\text{sig}$ probability
density in Eq.~(\ref{eq:ponshell}) is different in these two cases.

This leads to the following general expression for the probability density of the EW processes with~$C=4$
\ifthenelse{\boolean{cms@external}}
{
  \begin{multline}
    \mathcal{P}_{jk}^{\text{sig}}\left(\vec{x};\vec{\xi}_{jk},\vec{f}_j \right) \propto
    \sum_{l\le m\le n\le p =1}^{L}
    \mathcal{P}^{\text{sig}}_{jk,\,lmnp}\left(\vec{x};\vec{\xi}_{jk}\right)\\
    \times
    \sqrt{\abs{f_{l}\, f_{m}\, f_{n}\, f_{p}}} ~ \sign(f_{l}\, f_{m}\, f_{n}\, f_{p})\,.
    \label{eq:psignal}
    \end{multline}  
}
{
\begin{align}
\mathcal{P}_{jk}^{\text{sig}}\left(\vec{x};\vec{\xi}_{jk},\vec{f}_j \right) \propto
\sum_{l\le m\le n\le p =1}^{L}
\mathcal{P}^{\text{sig}}_{jk,\,lmnp}\left(\vec{x};\vec{\xi}_{jk}\right)
\sqrt{\abs{f_{l}\, f_{m}\, f_{n}\, f_{p}}} ~ \sign(f_{l}\, f_{m}\, f_{n}\, f_{p})\,.
\label{eq:psignal}
\end{align}
}
The following general expression applies to the probability density of the processes with~$C=2$
\ifthenelse{\boolean{cms@external}}
{
  \begin{multline}
    \mathcal{P}_{jk}^{\text{sig}}\left(\vec{x};\vec{\xi}_{jk},\vec{f}_j \right) \propto
    \sum_{l\le m=1}^{L}
    \mathcal{P}^{\text{sig}}_{jk,\,lm}\left(\vec{x};\vec{\xi}_{jk}\right)\\
    \times
    \sqrt{\abs{f_{l}\, f_{m}}} ~ \sign(f_{l}\, f_{m})\,.
    \label{eq:psignalshorter}
    \end{multline}
    
}
{
\begin{align}
\mathcal{P}_{jk}^{\text{sig}}\left(\vec{x};\vec{\xi}_{jk},\vec{f}_j \right) \propto
\sum_{l\le m=1}^{L}
\mathcal{P}^{\text{sig}}_{jk,\,lm}\left(\vec{x};\vec{\xi}_{jk}\right)
\sqrt{\abs{f_{l}\, f_{m}}} ~ \sign(f_{l}\, f_{m})\,.
\label{eq:psignalshorter}
\end{align}
}
In both Eqs.~(\ref{eq:psignal}) and~(\ref{eq:psignalshorter}), only the kinematic dependence on $\vec{f}_j$
is expressed, while the overall normalization can be absorbed into $\mu_j$ or accounted for as part of the
cross section measurement. 

In the general case, there are $(C+L-1)!/(C!(L-1)!)$ terms in either Eq.~(\ref{eq:psignal}) or Eq.~(\ref{eq:psignalshorter}).
This leads to 70 terms in Eq.~(\ref{eq:psignal}) when we measure four anomalous \HVV couplings in production and decay  ($L=5$, $C=4$), 
15 terms in Eq.~(\ref{eq:psignalshorter}) when we measure four anomalous \HVV couplings in decay only  ($L=5$, $C=2$), 
and three terms in Eq.~(\ref{eq:psignalshorter}) when we measure $\Hgg$ or $\Htt$ couplings in production only ($L=2$, $C=2$). 
When both \HVV (in decay) and $\Hgg$ or $\Htt$ (in production) couplings are measured at the same time in a given process, 
the number of terms is multiplied ($15\times3=45$), since the two probabilities factorize. 

The $\mathcal{P}^{\text{sig}}_{jk,\,lm(np)}$ templates are extracted from simulation, discussed in Section~\ref{sec:cms_mc},
typically using from three to twelve samples generated with various $\vec{f}_j$ values chosen to map different 
points of phase-space well and reweighted with the \mela package to cover all possibilities with $(C+L-1)!/(C!(L-1)!)$
combinations of couplings. In parameterizing the signal templates $\mathcal{P}^{\text{sig}}_{jk,\,lm(np)}$, 
it is important to ensure that the expected number of events in every bin of the probability densities, defined in 
Eqs.~(\ref{eq:psignal}) and~(\ref{eq:psignalshorter}), remains nonnegative at all possible values of $\vec{f}_j$, 
because a negative yield would cause the likelihood function used for the final fit to become ill-defined. 

To detect a negative event yield, we first minimize Eq.~(\ref{eq:psignal}) or Eq.~(\ref{eq:psignalshorter}),
which are polynomials in $\sqrt{\abs{f_{l}}}$, by finding where the gradient is zero.
For Eq.~(\ref{eq:psignalshorter}), which is quadratic, this is a simple problem in linear algebra.
Equation~(\ref{eq:psignal}) is quartic, so its gradient is a system of cubic equations, which
cannot be solved exactly for $L>1$.  We use the \textsc{Hom4PS} program~\cite{HomotopyContinuation,Hom4PS1,Hom4PS2}
to numerically solve the system of equations.
If we find the minimum to be negative, we adjust the estimated $\mathcal{P}^{\text{sig}}_{jk,\,lmnp}$
until the yield is always positive.  This adjustment is made using the statistical uncertainty on $\mathcal{P}^{\text{sig}}_{jk,\,lmnp}$
through a cutting planes algorithm~\cite{cuttingplanes} implemented using the \textsc{Gurobi} program~\cite{gurobi}.
In all cases, it is found that only small modifications to the initial estimates of $\mathcal{P}^{\text{sig}}_{jk,\,lmnp}$ are needed.

The parameterization in Eqs.~(\ref{eq:psignal}) and~(\ref{eq:psignalshorter}) is written in the self-consistent full-amplitude approach. 
In the EFT interpretation of the amplitude fit, the series in powers of $\abs{f_{ai}}^{1/2}$ corresponds to terms of different dimension,
as discussed in Section~\ref{sec:cross-section}.
For example, in Eq.~(\ref{eq:psignal}) the term with $f_{a1}^2$ corresponds to the SM-like contribution 
with dimension-four operators, while $f_{a1}^{3/2}\abs{f_{ai}}^{1/2}$ corresponds to interference between 
the SM amplitude and the dimension-six contributions in the EFT expansion. 
Assuming that $f_{a1}\sim1$ and all $f_{ai}$ are small, all other terms could in principle be neglected as a test of EFT validity. 
In practice, however, neglecting those terms can easily lead to a negative probability in certain points in the phase space of observables, 
which invalidates the maximum likelihood fit. This does not necessarily mean that the EFT approach is not valid.  
This happens because the sizable interference terms can lead to a negative partial sum in Eq.~(\ref{eq:psignal}), 
especially in the optimized multidimensional space of observables that are sensitive to such interference effects. 
Therefore, in practice, the fits presented in Section~\ref{sec:results} do not allow one to place constraints 
without the full series shown in Eqs.~(\ref{eq:psignal}) and~(\ref{eq:psignalshorter}).

\subsection{Likelihood fit}

The final constraints on parameters $\mu_j$ and $\vec{f}_j$  are placed using the profile likelihood method 
implemented in the \textsc{RooFit} toolkit~\cite{Verkerke:2003ir} within the \textsc{\ROOT}~\cite{Brun:1997pa} framework. 
The extended likelihood function is constructed using the probability densities in
Eq.~(\ref{eq:ponshell}), with each event characterized by the discrete category $k$ and observables $\vec{x}$. 
The likelihood $\mathcal{L}$ is maximized with respect to the nuisance parameters $\vec{\xi}_{jk}$ describing
the systematic uncertainties discussed below, and $\mu_j$ and $\vec{f}_j$ parameters of interest. 
The allowed 68 and 95\% \CL intervals are defined using the profile likelihood function, $-2\Delta\ln\mathcal{L} = 1.00$ 
and 3.84, for which exact coverage is expected in the asymptotic limit~\cite{Wilks:1938dza}.

The reinterpretation of the primary $\mu_j$ and $\vec{f}_j$ results in terms of couplings is performed 
with the help of Eq.~(\ref{eq:diff-cross-section2}) to relate signal strength $\mu_j$ to couplings and 
Eqs.~(\ref{eq:fggH_definitions}--\ref{eq:fa_definitions_hvv}) to relate $\vec{f}_j$ to coupling ratios. 
In this way the couplings directly enter the parameterization in Eq.~(\ref{eq:ponshell}).
However, without further constraints on the \Hboson width, such a fit would not provide useful constraints
on the coupling size.
Therefore, in the total width ${\Gamma_{\PH}}$ parameterization in Eq.~(\ref{eq:diff-cross-section2}),
we assume that there are no unobserved or undetected \Hboson\ decays. We express the width ${\Gamma_{\PH}}$
as the sum of partial decay widths of nine \Hboson decay modes dominant in the SM, $\PH\to\bbar$ being the largest 
and $\PH\to\mu^+\mu^-$ being the smallest. Each partial decay width is scaled as a function of anomalous couplings
following the parameterization in Ref.~\cite{Gritsan:2020pib}.

\subsection{Systematic uncertainties}

Several systematic uncertainties are considered in the set of constrained parameters $\vec{\xi}_{jk}$.
The relative expected yields in different categories and the template shapes describing probability distributions 
in Eq.~(\ref{eq:ponshell}) are varied within either theoretical or experimental uncertainties. 
All results reported at 68 and 95\% \CL are dominated by statistical uncertainties. 
All systematic uncertainties are treated as correlated between different time periods, 
except for the jet-related uncertainties, which originate from statistically independent sources, 
and luminosity uncertainties, which are partially correlated~\cite{CMS-PAS-LUM-17-001, CMS-PAS-LUM-17-004, CMS-PAS-LUM-18-002}. 

The theoretical uncertainties considered are PDF parameterization, factorization and renormalization scales,
the hadronization scale used in \PYTHIA, and the underlying event variations.
The underlying event modeling uncertainty is determined by varying initial- and final-state
radiation scales between 0.25 and 4 times their nominal value. The effects of the modeling of
hadronization are determined by simulating additional events with the variation of the 
nominal \PYTHIA tune described in Section~\ref{sec:cms}.
Experimental uncertainties involve jet energy calibration and \cPqb-quark-tagging efficiency uncertainties, 
which are only relevant when production categories are considered, and lepton efficiency and momentum uncertainties, 
which are similar for the different processes and categories.
In the estimation of the \ZX background, the flavor composition of QCD-evolved jets misidentified as leptons may
be different in the $\PZ+1\ell$ and $\PZ+2\ell$ control regions, and together with the statistical uncertainty
in the $\PZ+2\ell$ region, this uncertainty accounts for about a $\pm$30\% variation in the \ZX background.

The normalization of the background processes derived from the MC simulation is affected 
by a 1.8\% uncertainty in the integrated luminosity. However, all results are found to be insensitive to 
luminosity or theoretical constraints on the nonresonant $\PZ\PZ/\PZ\gamma^*\to4\ell$ background. 
Even though the theoretical cross section is one of the constraints in the fit, the wide sideband included in the range
$105<m_{4\ell}<140\GeV$ constrains this nearly flat background from the data, with nearly identical results when
the theoretical constraints are removed. The main distinguishing feature of this background is the dominant contribution of 
the $\PZ\gamma^*$ intermediate state, which allows effective separation of the already small background from signal
using kinematic information.

\section{Results}
\label{sec:results}

The signal strength $\mu_j$ and the set of parameters $\vec{f}_j$ describing the tensor structure of interactions
are constrained in each production process and decay $\PH\to \ZZ$ / $\PZ\gamma^*$ / $\gamma^*\gamma^*\to 4\ell$. 
In the following, we describe the measurement of $\fG$ in the \ggH process, $\fT$ in the \ttH and \tqH processes, 
and the combination of the two where the top quark contributes to the gluon fusion loop. 
We then report measurements of $\left(\fB, \fC, \fL, \fLZg\right)$ in the \VBF and \VH processes along with the $H\to 4\ell$ 
decay in all production processes, following Approach~1 with the coupling relationship $a_i^{\PW\PW}=a_i^{\PZ\PZ}$.
We also report measurements of $\left(\fB, \fC, \fL\right)$ following Approach~2 within SMEFT, discussed in Section~\ref{sec:pheno}.
These results are interpreted in terms of constraints on the $\Hff$, $\Hgg$, and $\HVV$ operators.
While all operators could potentially be constrained in a joint analysis of all \Hboson decay modes, 
in this paper we analyze only the $\PH\to4\ell$ decay mode, 
and perform a combination with the \tqH and \ttH processes in the $\PH\to\gamma\gamma$ decay mode. 
Therefore, for the purpose of illustration, we make further assumptions on how certain couplings, 
to which this analysis is not sensitive, are related. For example, we must make certain assumptions about the 
relationship between the $\PH\cPqb\cPqb$, $\PH\PQc\PQc$, $\PH\tau\tau$, and $\PH\mu\mu$ couplings
and other couplings. These assumptions are discussed in each of the applications presented below.

\subsection{Constraints on \texorpdfstring{\Hgg}{Hgg} couplings}
\label{sec:results_Hgg}

\begin{figure*}[!tbp]
\centering
\includegraphics[width=0.43\textwidth]{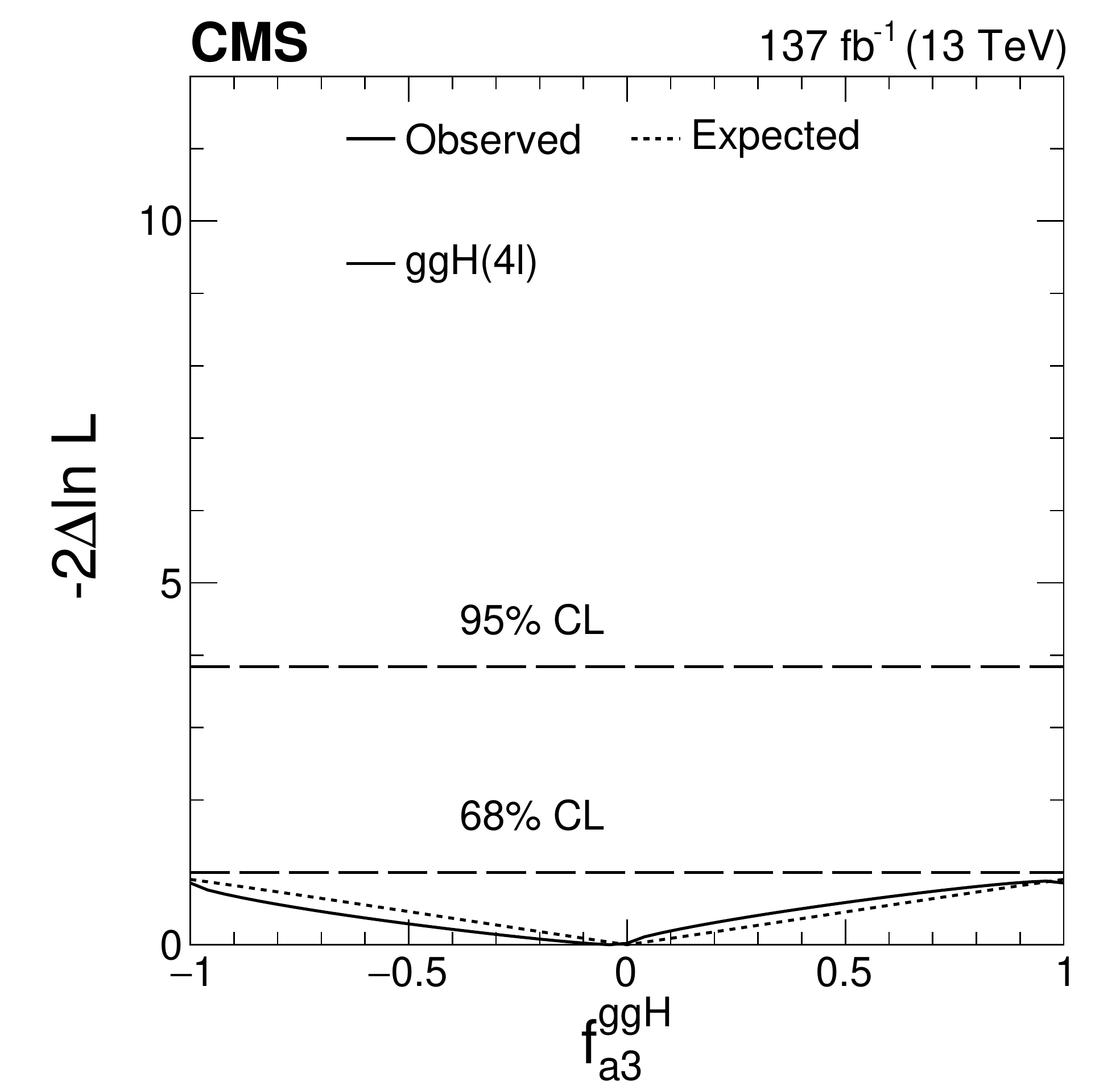}
\includegraphics[width=0.56\textwidth]{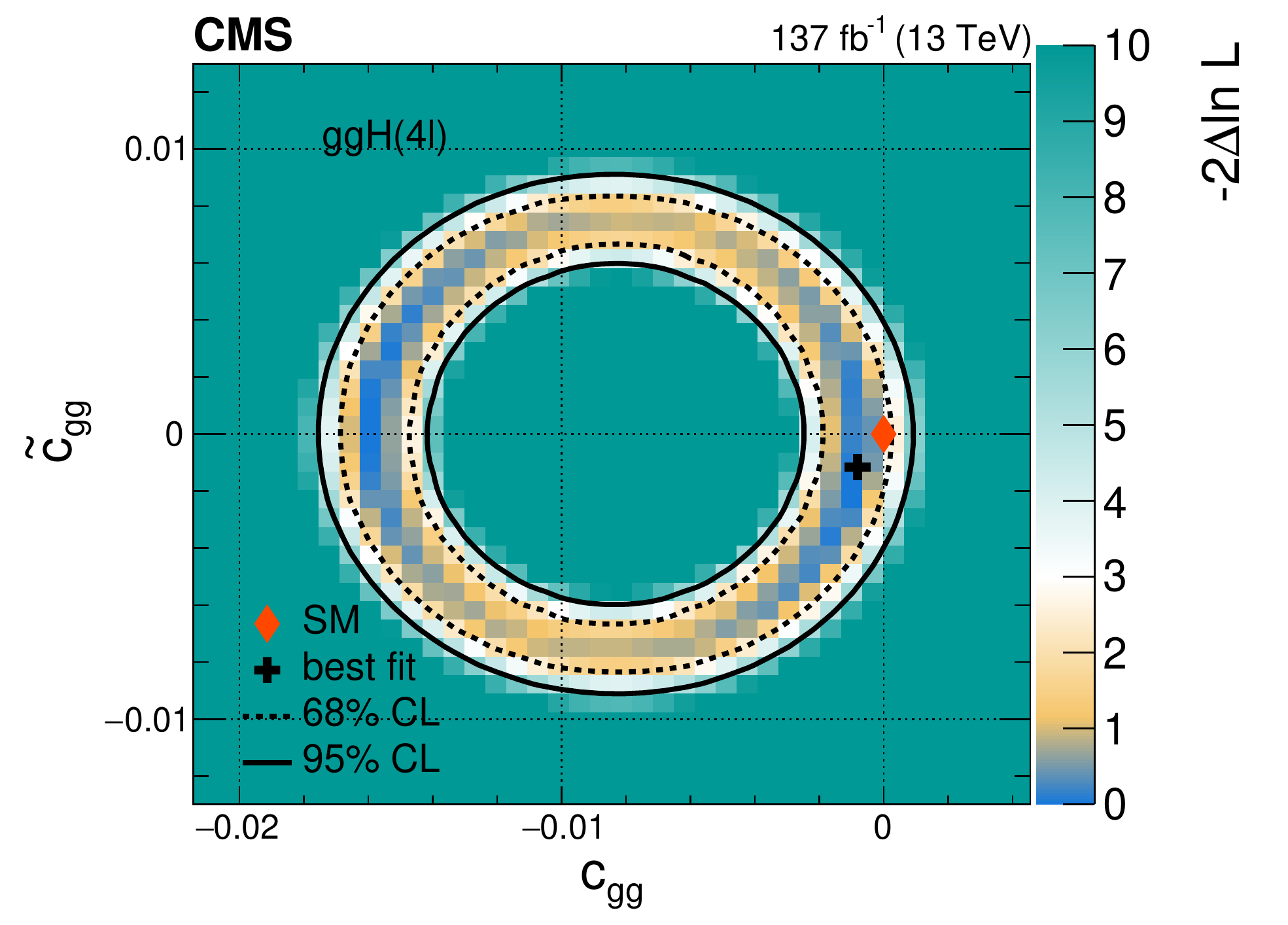}
\caption
{
Constraints on the anomalous \Hboson couplings to gluons in the \ggH process using the $\PH\to 4\ell$ decay. 
Left: Observed (solid) and expected (dashed) likelihood scans of the $CP$-sensitive parameter $\fG$. 
The dashed horizontal lines show $68$ and $95\%$~\CL.
Right: Observed confidence level intervals on the $c_{\Pg\Pg}$ and $\tilde c_{\Pg\Pg}$ couplings reinterpreted
from the $\fG$ and $\mu_{\ggH}$ measurement 
with $\fC$ and $\muV$ profiled, and with $\kappa_{\PQt}=\kappa_{\cPqb}=1$.
The dashed and solid lines show the $68$ and $95\%$~\CL exclusion regions in two dimensions, respectively. 
\label{fig:resultHgg}
}
\end{figure*}

The measurement of anomalous couplings of the \Hboson to gluons is presented in Fig.~\ref{fig:resultHgg}
and Table~\ref{tab:prec_htt}. Since the direct couplings of the \Hboson to SM fermions in the gluon fusion loop 
and to potentially new particles appearing in the loop can not be resolved using this measurement alone, 
both effects are characterized with two parameters, $\fG$ and $\mu_{\ggH}$. The signal strength $\mu_{\ggH}$,
which is the ratio of the measured cross section of the gluon fusion process to that expected in the SM, 
is profiled when the $\fG$ results are reported. The measurement of the $\fG$ is consistent with zero, 
as expected in the SM. This can be clearly seen from the $\mathcal{D}_{\text{CP}}^{\ggH}$ 
and $\mathcal{D}_\text{0-}^{\ggH}$ distributions in Fig.~\ref{fig:ggH_D0mn}.
The measured value of $\mu_{\ggH}=0.86^{+0.13}_{-0.11}$ is consistent with that reported in Ref.~\cite{Sirunyan:2021rug} 
without the fit for the $CP$ structure of interactions. 
The values of $\mu_{\ggH}$ and $\fG$ are uncorrelated.
The signal strength of the \VBF and \VH processes $\muV$ and their $CP$ properties $\fC$ are also profiled 
when this measurement is performed. This measurement is also performed simultaneously in a fit
with the \ttH process with the $\mu_{\ttH}$ and $\fT$ parameters unconstrained, as discussed below.
The \tqH process is always included with the \ttH process with its signal strength expressed through 
the $\mu_{\ttH}$, $\muV$, and $\fT$ parameters.

\begin{table*}[!tb]
\centering
\topcaption{
Constraints on the $\fG$ and $\fT$ parameters with the best fit values and allowed $68\%$~\CL (quoted uncertainties) 
and $95\%$~\CL (within square brackets) intervals, limited to the physical range of $[-1,  1]$.
The $\fT$ constraints obtained in this work are combined with those in the $\PH\to\gamma\gamma$ channel~\cite{Sirunyan:2020sum}. 
The interpretation of the $\fG$ result under the assumption of the top quark dominance in the gluon fusion loop 
are presented in terms of the $\fT$ parameter, where either \ggH or its combination with \tqH and \ttH results are shown. 
\label{tab:prec_htt}
}
\cmsTable{
\begin{scotch}{clrr}
\vspace{-0.3cm} \\
 Parameter   &  Scenario                & Observed                                         & Expected  \\
[\cmsTabSkip]
\hline \\ [-1ex]
$\fG$ &  \ggH ($\PH\to4\ell$)  & $-0.04^{+1.04}_{-0.96}$  $[-1, 1]$  & $0\pm1$ $[-1,  1]$ \\
[\cmsTabSkip]
~~~~~\multirow{6}{*}{$\fT\begin{cases} \\  \\  \\  \\  \\  \end{cases}$ }
  &    \tqH~\&~\ttH ($\PH\to4\ell$)                                                             &                  $\pm(0.88^{+0.12}_{-1.88})$  $[-1, 1]$  & $0\pm1$ $[-1,  1]$ \\
  &    \tqH~\&~\ttH ($\PH\to\gamma\gamma$)  \cite{Sirunyan:2020sum}   &                  $0.00\pm0.33$  $[-0.67, 0.67]$  & $0.00\pm0.49$ $[-0.82,  0.82]$ \\
  &    \tqH~\&~\ttH ($\PH\to4\ell~\&~\gamma\gamma$)                            &                  $0.00\pm 0.33$  $[-0.67, 0.67]$  & $0.00\pm 0.48$ $[-0.81,  0.81]$ \\
  &   \ggH ($\PH\to4\ell$)   & $-0.01^{+1.01}_{-0.99}$ $[-1, 1]$  & $0\pm1$ $[-1,  1]$ \\
  &   \ggH~\&~\tqH~\&~\ttH ($\PH\to4\ell$)    & $-0.56^{+1.56}_{-0.44}$ $[-1, 1]$  & $0.00\pm0.47$ $[-1,  1]$ \\
  &   \ggH~\&~\tqH~\&~\ttH ($\PH\to4\ell~\&~\gamma\gamma$)    &    $-0.04^{+0.38}_{-0.36}$  $[-0.69, 0.68]$  & $0.00\pm 0.30$ $[-0.70,  0.70]$ \\
[\cmsTabSkip]
\end{scotch}
}
\end{table*}

The parameters $\fG$ and $\mu_{\ggH}$ are equivalent to the measurement of the $CP$-even and 
$CP$-odd couplings on the production side, while the $\HVV$ couplings on the decay side are
constrained from the simultaneous measurement of the \VBF and \VH processes with $\fC$ and $\muV$ profiled.  
The $c_{\Pg\Pg}$ and $\tilde{c}_{\Pg\Pg}$ couplings, introduced in Eqs.~(\ref{eq:EFTpar5}) and~(\ref{eq:EFTpar6}), 
can be extracted from the above measurements. 
We follow the parameterization of the cross section and the total width from Ref.~\cite{Gritsan:2020pib},
where $c_{\Pg\Pg}$, $\tilde{c}_{\Pg\Pg}$, $\kappa_{\PQt}$, $\tilde\kappa_{\PQt}$,  $\kappa_{\cPqb}$, and $\tilde\kappa_{\cPqb}$ contribute. 
Since it is not possible to disentangle all these couplings in a single process, we fix $\kappa_{\PQt}=\kappa_{\cPqb}=1$
and $\tilde\kappa_{\PQt}=\tilde\kappa_{\cPqb}=0$ to the SM expectation and leave $c_{\Pg\Pg}$ and $\tilde{c}_{\Pg\Pg}$,
which describe possible BSM contribution in the loop, unconstrained. 
The small contribution of the $\PH\to\gamma\gamma$ and $\PZ\gamma$ decays to the total width is assumed to be SM-like.  

The resulting constraints on $c_{\Pg\Pg}$ and $\tilde{c}_{\Pg\Pg}$ are shown in Fig.~\ref{fig:resultHgg}, right. 
The general features of these constraints are the following. The pure signal strength measurement $\mu_{\ggH}$,
available even without the fit for $\fG$, provides a constraint in the form of a ring on a two-parameter plane 
in Fig.~\ref{fig:resultHgg}, right. The measurement of $\fG$ resolves the areas within this ring. 
Since the sensitivity of the $\fG$ measurement is currently just under $68\%$~\CL, this resolution is not strong.
The \Hboson\ width dependence on $c_{\Pg\Pg}$ and $\tilde{c}_{\Pg\Pg}$ is relatively weak and does not 
alter this logic considerably. The results are consistent with the SM expectation.

As mentioned earlier, it is not possible to resolve the loop contributions from the SM or BSM particles
in this measurement. 
Therefore, the deviations of the SM-like Yukawa couplings $\kappa_{\PQt}$ and $\kappa_{\cPqb}$ 
from unit values are absorbed into the effective $c_{\Pg\Pg}$ measurement, 
and the $CP$-odd Yukawa couplings $\tilde\kappa_{\PQt}$ and $\tilde\kappa_{\cPqb}$
are absorbed into the effective $\tilde{c}_{\Pg\Pg}$ measurement, together with possible 
contributions from BSM particles. 
However, re-interpretation of these results is possible in terms of the independent 
Yukawa couplings and effective point-like gluon couplings in combination with the \ttH and \tqH modes,
as discussed below.

\subsection{Constraints on \texorpdfstring{\Htt}{Htt} couplings}
\label{sec:results_Htt}

\begin{figure*}[!tbp]
\centering
\includegraphics[width=0.43\textwidth]{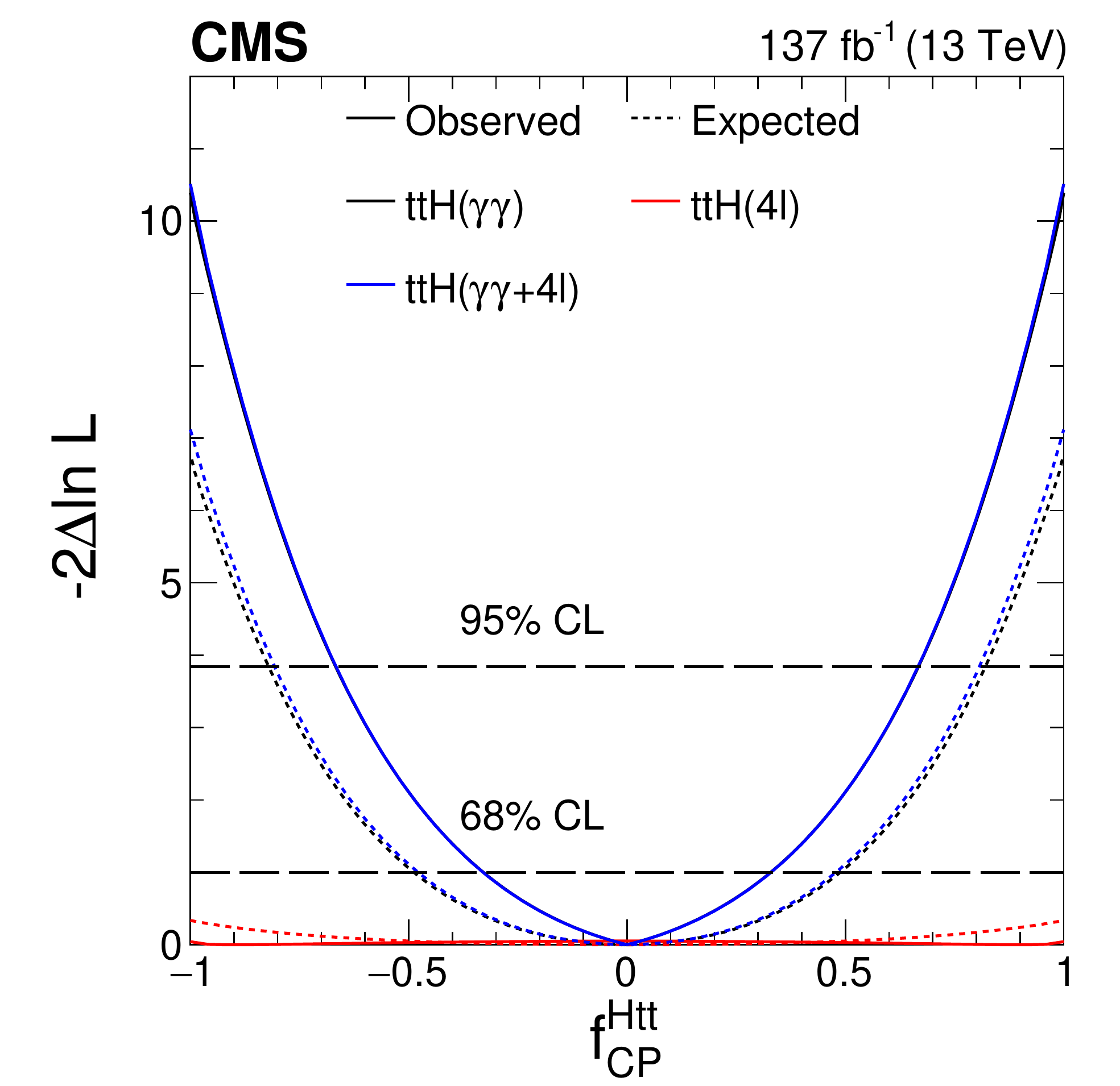}
\includegraphics[width=0.56\textwidth]{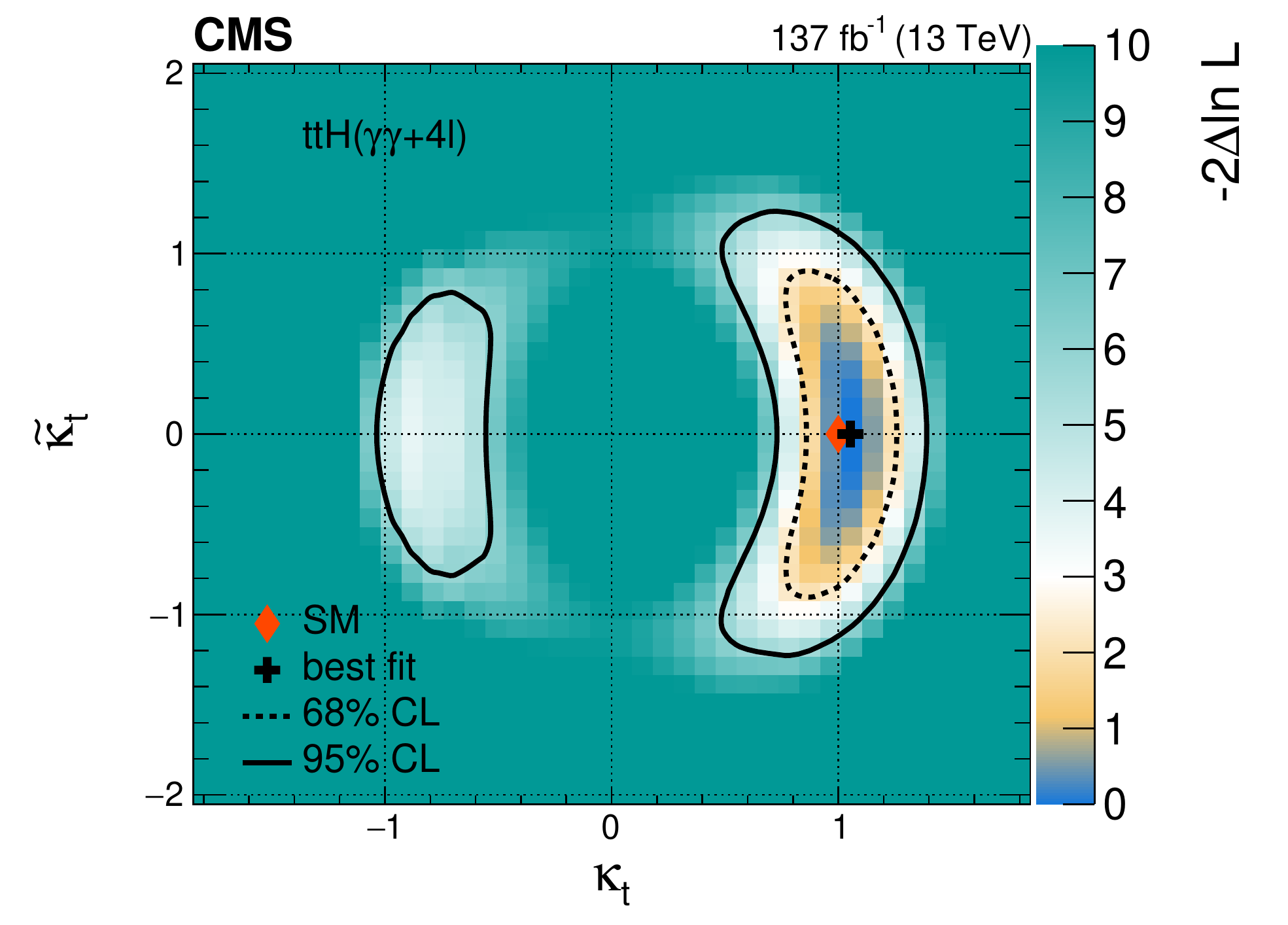} 
\caption
{
Constraints on the anomalous \Hboson couplings to top quarks in the \ttH and \tqH processes 
using the $\PH\to\gamma\gamma$~\cite{Sirunyan:2020sum} and $\PH\to4\ell$ decays. 
Left: Observed (solid) and expected (dashed) likelihood scans of $\fT$ in the \ttH and \tqH processes
in the $\PH\to 4\ell$ (red), $\gamma\gamma$ (black), and combined (blue) channels, where the
combination is done without relating the signal strengths in the two processes. 
The dashed horizontal lines show 68 and 95\%~\CL.
Right: Observed confidence level intervals on the $\kappa_{\PQt}$ and $\tilde\kappa_{\PQt}$ couplings reinterpreted
from the $\fT$, $\mu_{\ttH}$, and $\muV$ measurements in the combined fit of the $\PH\to 4\ell$ and $\gamma\gamma$ channels,
with the signal strengths in the two channels related through the couplings as discussed in text.
The dashed and solid lines show the $68$ and $95\%$~\CL exclusion regions in two dimensions, respectively. 
\label{fig:resultHtt}
}
\end{figure*}

The measurement of anomalous couplings of the \Hboson to top quarks is presented in Fig.~\ref{fig:resultHtt}
and Table~\ref{tab:prec_htt}. First, the measurements of $\fT$ from the \ttH and \tqH processes only are reported. 
The signal strength $\mu_{\ttH}$, which is the ratio of the measured cross section of the $\ttH$ process 
to that expected in the SM, is profiled when the $\fT$ results are reported. 
The measured value of $\mu_{\ttH}=0.17^{+0.70}_{-0.17} $ is consistent with that reported in Ref.~\cite{Sirunyan:2021rug} 
without the fit for the $CP$ structure of interactions. 
In both cases we observe downward fluctuations in the signal yield compared to expectation, but these fluctuations are not statistically significant.
There is no significant linear correlation between $\mu_{\ttH}$ and $\fT$. 
The signal strength of the \VBF and \VH processes $\muV$, \ggH process $\mu_{\ggH}$, 
and their $CP$ properties $\fC$ and $\fG$ are also profiled when this measurement is performed. 
This analysis of the \ttH and \tqH processes is not sensitive to the sign of $\fT$.
However, for later combination with the $\ggH$ measurement, presented above, 
under the assumption of the top quark dominance in the gluon fusion loop, 
symmetric constraints on $\fT$ are reported. 

The observed best fit  $\fT$ value gives preference to the $CP$-odd Yukawa coupling. 
This comes from the negative value of the $\mathcal{D}_\text{0-}^\text{\ttH}$ discriminant for the one 
observed signal-like event in Fig.~\ref{fig:ttH_BDT}. 
However, this result is statistically consistent with the pure $CP$-even Yukawa coupling expected in the SM.
With just about two signal $\ttH$ events and many fewer $\tqH$ events expected to appear in the fit in the $\PH\to4\ell$ channel
under the assumption of the SM cross section, according to Table~\ref{tab:Hffcategory}, the expected confidence sensitivity on the $\fT$ 
constraints is low. Nonetheless, the high signal purity in the $\PH\to4\ell$ channel implies that every observed event candidate 
carries a large statistical weight. 
The importance of including the $CP$ measurements in the $\ttH$ and $\tqH$ production modes also becomes evident when 
combination with $\ggH$ is performed. There is a significant gain in such a combination beyond a simple addition of independent 
measurements, as discussed in Section~\ref{sec:results_HttHgg}. 

The CMS experiment recently reported the measurement of the $\fT$ parameter in the \ttH and \tqH production processes 
with the decay $\PH\to\gamma\gamma$~\cite{Sirunyan:2020sum} (shown also in Table~\ref{tab:prec_htt}). 
In that measurement, the signal strength $\mu_{\ttH}^{\gamma\gamma}$ parameter is profiled,
while the signal strengths in other production processes are fixed to the SM expectation. 
However, there is a very weak correlation of the measurement in the \ttH and \tqH processes 
with parameters in the other production mechanisms. 
Therefore, we proceed with a combination of the $\fT$ measurements in the $\PH\to4\ell$ and $\gamma\gamma$ channels,
where we correlate their common systematic uncertainties, but not the signal strengths of the processes. 
In particular, we do not relate the $\mu_{\ttH}$ and $\mu_{\ttH}^{\gamma\gamma}$ signal strengths 
because they could be affected differently by the particles appearing in the loops responsible 
for the $\PH\to\gamma\gamma$ decay. 
The results of this combination are presented in Fig.~\ref{fig:resultHtt} and Table~\ref{tab:prec_htt}.
The measured signal strength in the $\PH\to4\ell$ channel is $\mu_{\ttH}=0.04^{+0.76}_{-0.04}$, uncorrelated with $\fT$, 
while $\mu_{\ttH}^{\gamma\gamma}=1.23^{+0.33}_{-0.24}$ and the correlation with $\fT$ is $+0.20$.
The pure pseudoscalar hypothesis of the \Hboson corresponding to $\fT=1$ in the case of the $CP$-odd Yukawa 
interaction is excluded at 3.2\,standard deviations, while the expected exclusion is 2.7\,standard deviations. 
Below, we also present an interpretation of these results where the signal strengths in the two \Hboson decay 
channels are related through the couplings. 

In the above measurements, the $\fT$ parameter has the same meaning in both the $\PH\to4\ell$ and 
$\gamma\gamma$ channels. 
In order to make an EFT coupling interpretation of the results, we have to make a further assumption that no BSM particles 
contribute to the loop in the $\PH\to\gamma\gamma$ decay. Without this or a similar assumption, the signal strength 
in the $\PH\to\gamma\gamma$ decay cannot be interpreted without ambiguity. 
We further re-parameterize the cross section following Ref.~\cite{Gritsan:2020pib} with the couplings
$\kappa_{\PQt}$ and $\tilde\kappa_{\PQt}$, and fix $\kappa_{\cPqb}=1$ and $\tilde\kappa_{\cPqb}=0$.
The bottom quark coupling makes a very small contribution to the loop in the $\PH\to\gamma\gamma$ decay,
but it makes a large contribution to the total decay width, where we assume that there are no unobserved or 
undetected \Hboson\ decays.
In order to simplify the fit, we do not allow anomalous $\HVV$ couplings, and the measurement of the signal strength 
$\muV$ constrains the contribution of the $a_1$ coupling in the loop. 
The $\fG$ and $\mu_{\ggH}$ parameters are profiled in this fit. 

The observed confidence level intervals on the $\kappa_{\PQt}$ and $\tilde\kappa_{\PQt}$ couplings from the 
combined fit of the $\PH\to 4\ell$ and $\gamma\gamma$ channels are shown in Fig.~\ref{fig:resultHtt}.
There is no linear correlation between the values of $\kappa_{\PQt}$ and $\tilde\kappa_{\PQt}$.
As was the case for the ($c_{\Pg\Pg}$, $\tilde{c}_{\Pg\Pg}$) measurement in Fig.~\ref{fig:resultHgg}, the pure yield 
measurement in the \ttH process would constrain a ring in the two-dimensional plane. 
However, the $CP$-sensitive measurement of $\fT$ disfavors the values away from $\tilde\kappa_{\PQt}=0$. 
Moreover, the sign ambiguity between the $\kappa_{\PQt}$ and $-\kappa_{\PQt}$ values cannot be resolved in the 
$\ttH$ channel alone. With the inclusion of the $\tqH$ process, the negative values of $\kappa_{\PQt}$
are disfavored because strong constructive interference between the amplitudes induced by the $\HVV$ and $\Htt$
couplings would result in enhanced $\tqH$ yield, inconsistent with the data. 
Therefore, the sign of $\kappa_{\PQt}$ is defined in reference to the tree-level $\HVV$ coupling $a_1$.  
But, the sign ambiguity between the $\tilde\kappa_{\PQt}$ and $-\tilde\kappa_{\PQt}$ values cannot be resolved 
in this fit, unless information from the other channels is incorporated, such as information from the gluon fusion 
loop discussed below.

\subsection{Constraints on \texorpdfstring{\Htt}{Htt} and \texorpdfstring{\Hgg}{Hgg} couplings in combination}
\label{sec:results_HttHgg}

\begin{figure}[!tbp]
\centering
\includegraphics[width=0.43\textwidth]{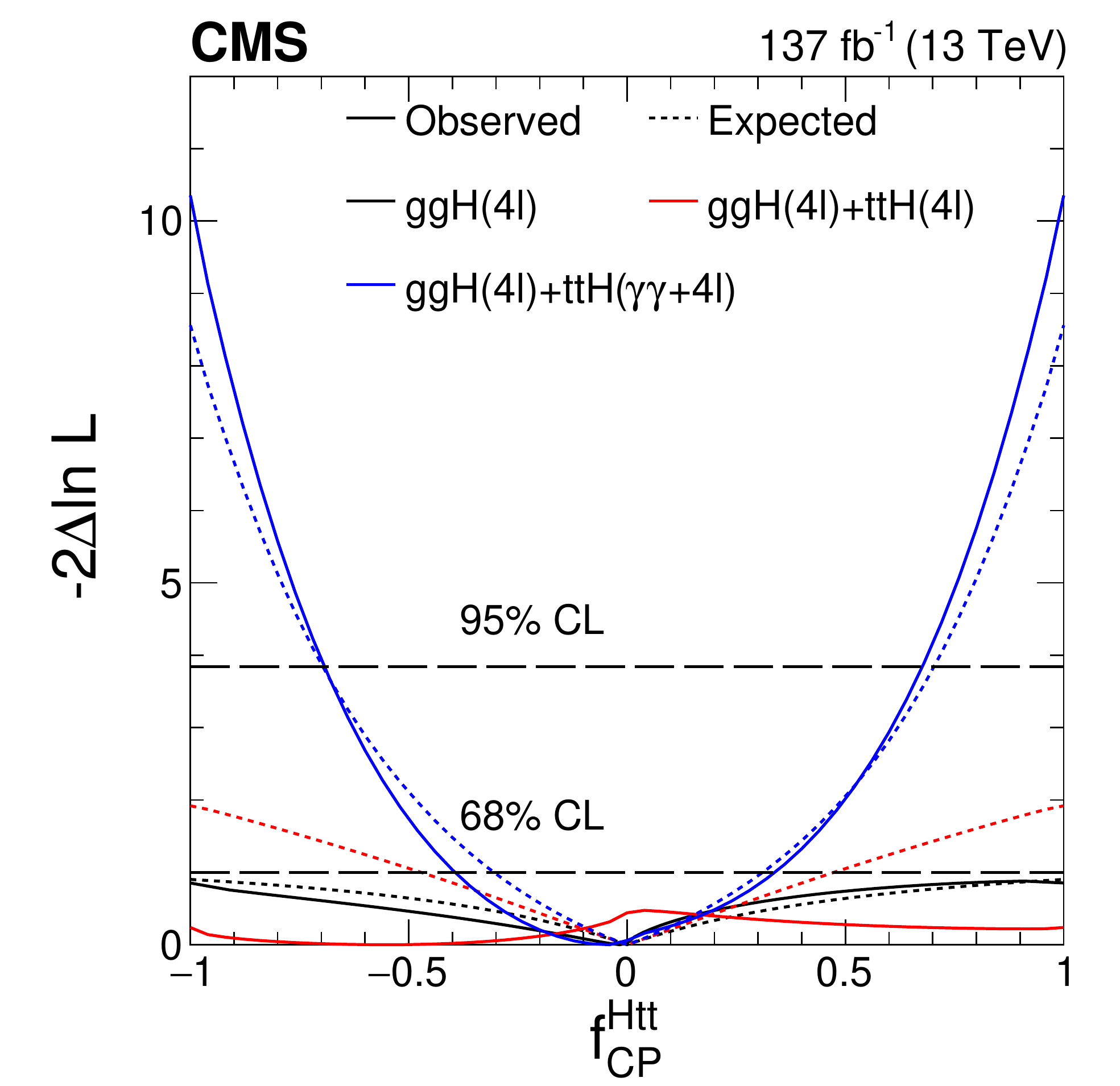}
\caption
{
Constraints on the anomalous \Hboson couplings to top quarks in the \ttH, \tqH, and \ggH processes combined,
assuming top quark dominance in the gluon fusion loop, using the $\PH\to 4\ell$ and $\gamma\gamma$ decays. 
Observed (solid) and expected (dashed) likelihood scans of $\fT$ are shown in the  
\ggH process with $\PH\to 4\ell$ (black), 
\ttH, \tqH, and \ggH processes combined with $\PH\to 4\ell$ (red), 
and in the \ttH,  \tqH, and \ggH processes with $\PH\to 4\ell$ and the \ttH and \tqH processes with $\gamma\gamma$ combined (blue). 
Combination is done by relating the signal strengths in the three processes through the couplings in the loops in both
production and decay, as discussed in the text. The dashed horizontal lines show $68$ and $95\%$~\CL exclusion.
\label{fig:resultHttHgg}
}
\end{figure}

First, we consider the \ggH process under the assumption of top quark dominance in the gluon fusion loop.
The measurement of anomalous couplings of the \Hboson to top quarks for this case is presented
in Fig.~\ref{fig:resultHttHgg} and Table~\ref{tab:prec_htt}. 
Similar to the case of the $\PH\to\gamma\gamma$ loop discussed above, 
the cross section of the \ggH process, normalized to the SM expectation, 
is parameterized following Ref.~\cite{Gritsan:2020pib} 
to account for $CP$-odd Yukawa couplings as follows:
\begin{align}
	& \frac{\sigma(\Pg\Pg\PH)}{\sigma_{\mathrm{ SM}}} 
	= \kappa_\mathrm{f}^2 + 2.38 \tilde\kappa_\mathrm{f}^2,
	\label{eq:sigma_ggH}
\end{align}
where we set $\kappa_\mathrm{f}= \kappa_{\PQt}=\kappa_{\cPqb}$ and $\tilde\kappa_\mathrm{f}= \tilde\kappa_{\PQt}=\tilde\kappa_{\cPqb}$.
Equation~(\ref{eq:sigma_ggH}) sets the relationship between $\fT$ and $\fG$, reported in Fig.~\ref{fig:resultHgg} 
and Table~\ref{tab:prec_htt}, according to Eq.~(\ref{eq:fai-relationship-hgg-tth}). 

\begin{figure*}[!tbp]
\centering
\includegraphics[width=0.49\textwidth]{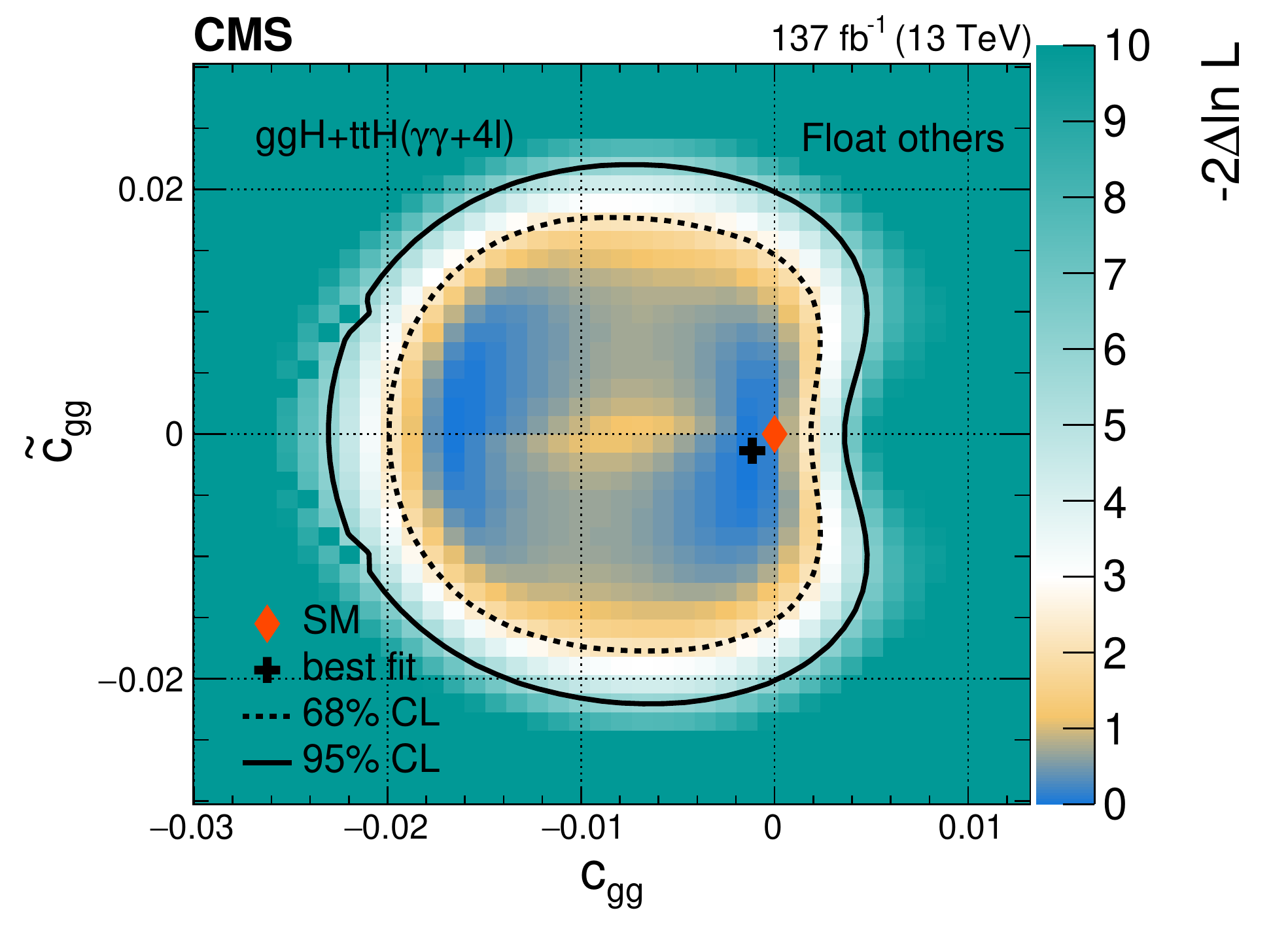}
\includegraphics[width=0.49\textwidth]{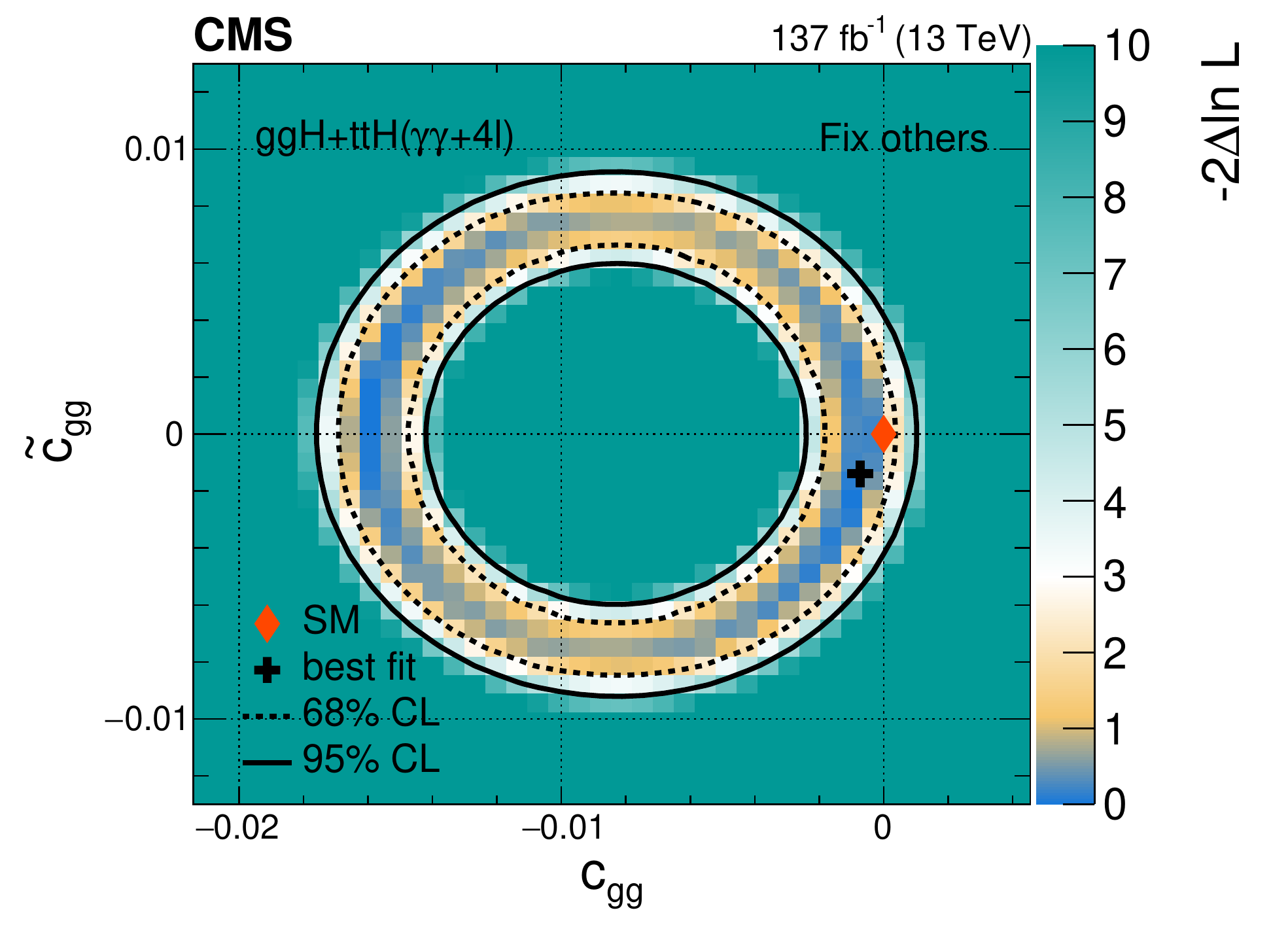}\\
\includegraphics[width=0.49\textwidth]{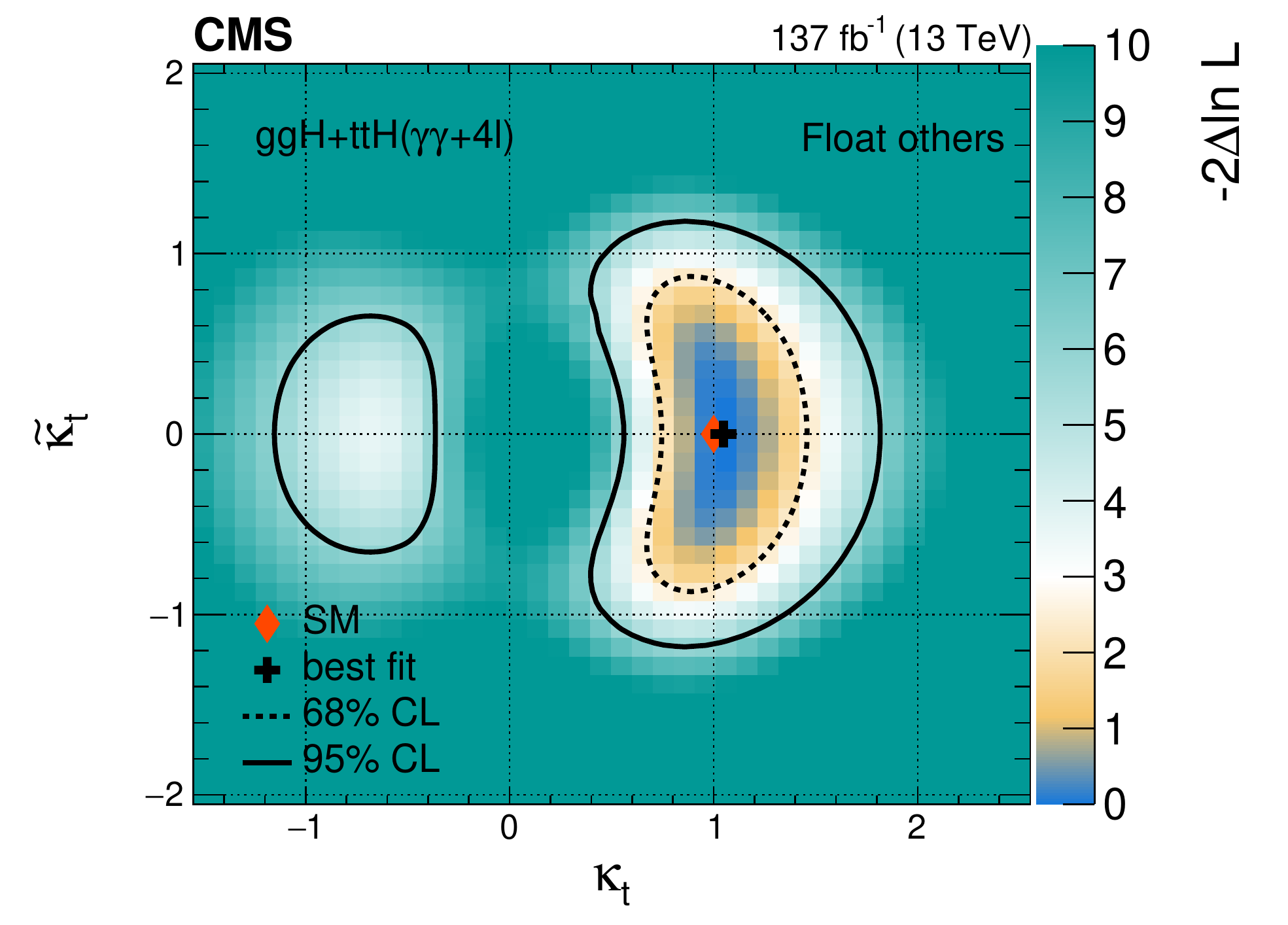}
\includegraphics[width=0.49\textwidth]{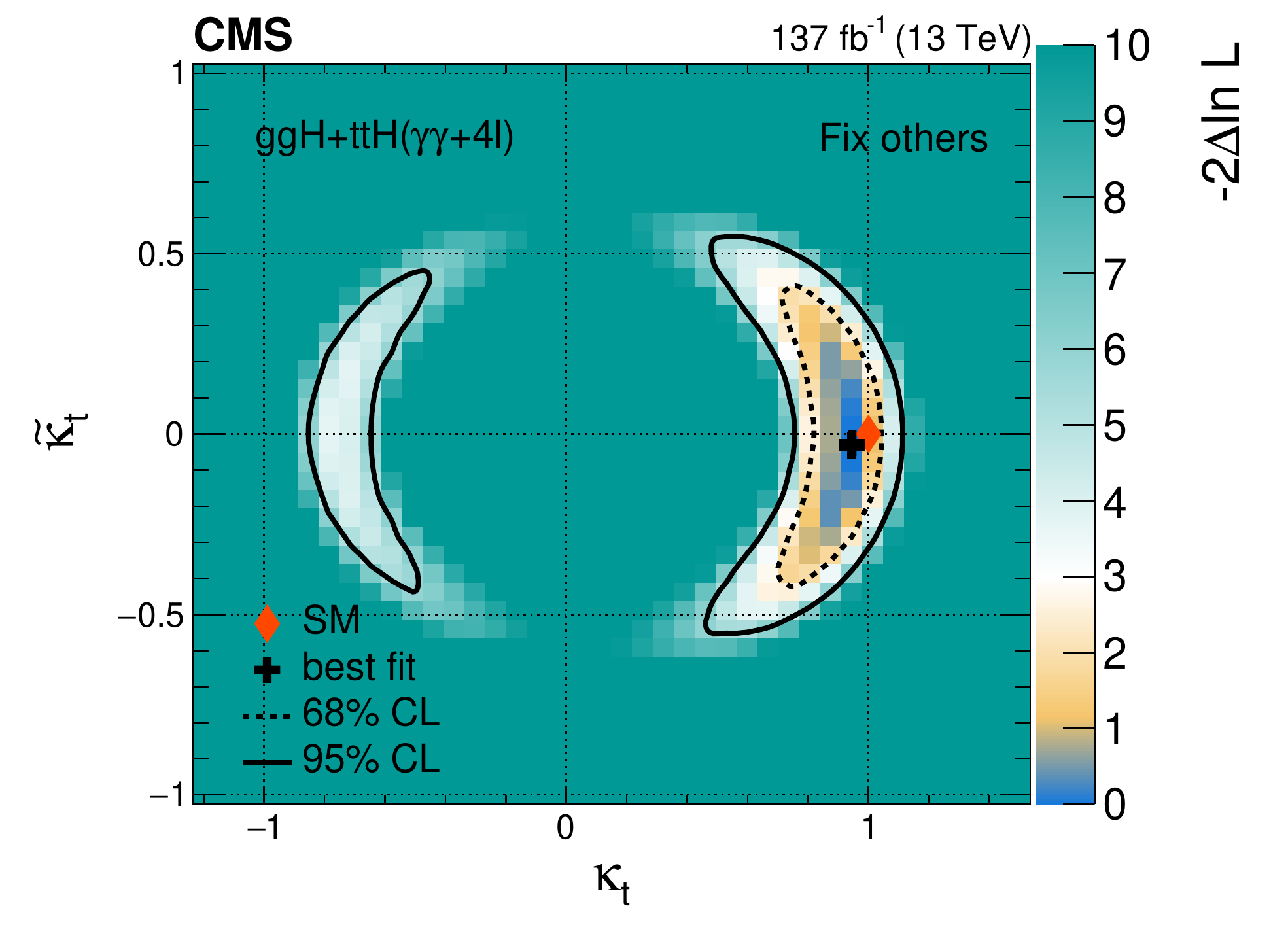}
\includegraphics[width=0.49\textwidth]{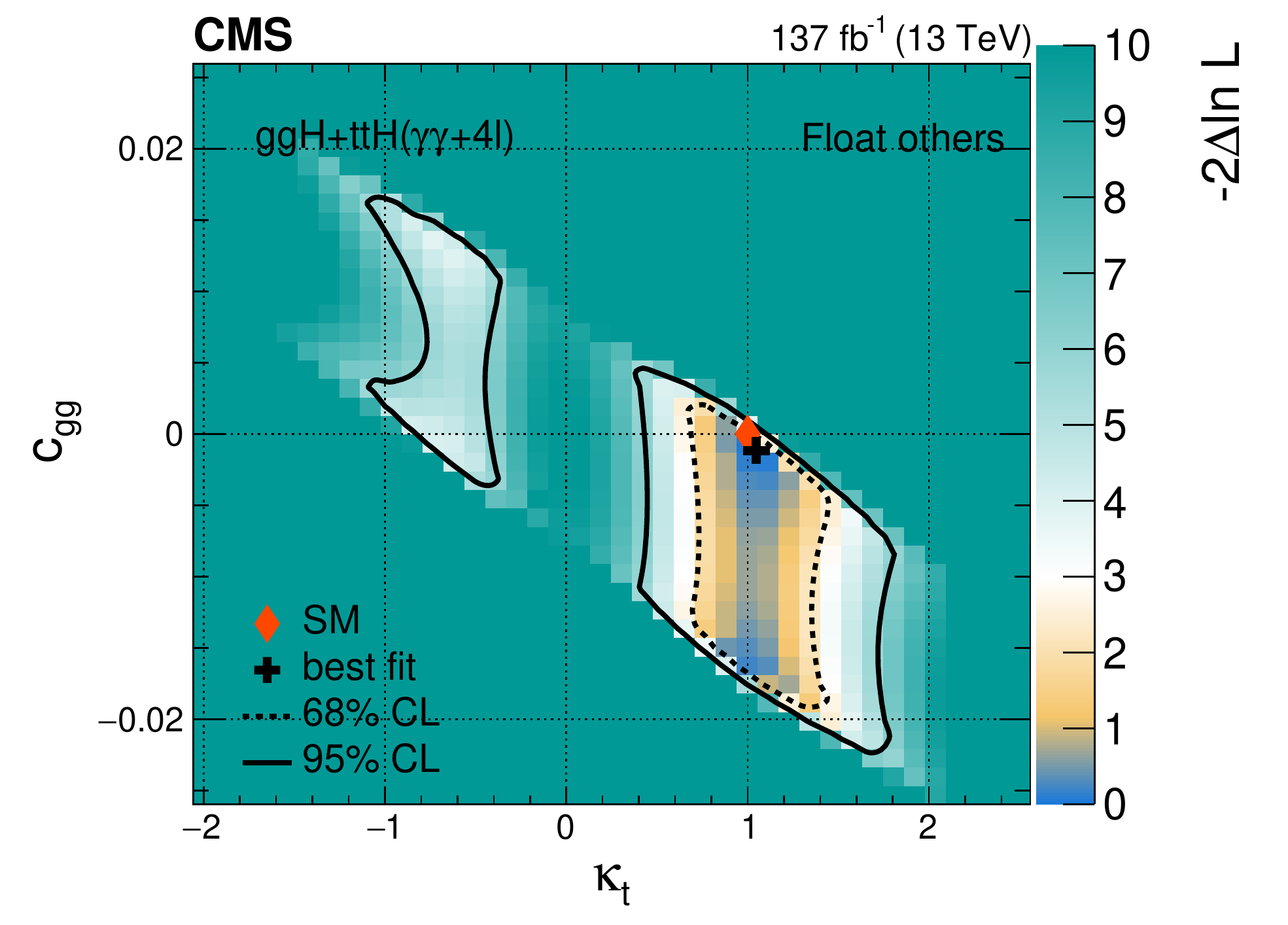}
\includegraphics[width=0.49\textwidth]{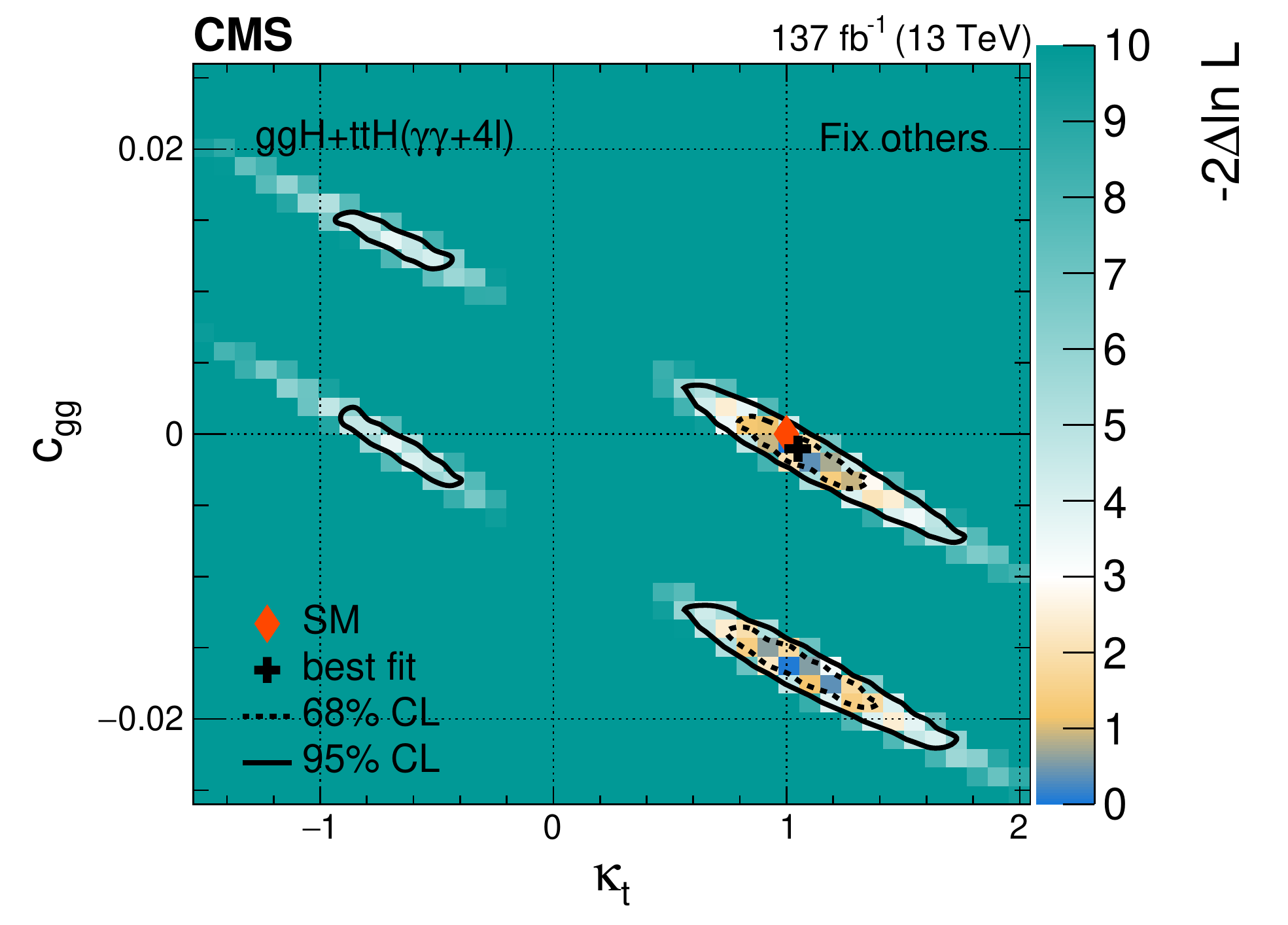}
\caption
{
Constraints on the anomalous \Hboson couplings
 $c_{\Pg\Pg}$, $\tilde{c}_{\Pg\Pg}$, $\kappa_\cPqt$, and $\tilde\kappa_\cPqt$
in the \ttH, \tqH, and \ggH processes combined, using the $\PH\to 4\ell$ and $\gamma\gamma$ decays. 
The constraints are shown for the pairs of parameters: 
$c_{\Pg\Pg}$ and $\tilde{c}_{\Pg\Pg}$ (upper), 
$\kappa_\cPqt$ and $\tilde\kappa_\cPqt$ (middle),
$\kappa_\cPqt$ and $c_{\Pg\Pg}$ (lower), and 
with the other two parameters either profiled (left) or fixed to the SM expectation (right). 
The dashed and solid lines show the 68 and 95\%~\CL exclusion regions in two dimensions, respectively. 
\label{fig:resultHttHgg_2D}
}
\end{figure*}

Constraints on $\fT$ are also shown for the combination of the \ttH, \tqH, and \ggH processes with $\PH\to 4\ell$
in Fig.~\ref{fig:resultHttHgg} and Table~\ref{tab:prec_htt}. 
The combination of the $\PH\to 4\ell$ and $\gamma\gamma$ channels with \ttH, \tqH, and \ggH processes 
proceeds in a similar manner and is also shown in Fig.~\ref{fig:resultHttHgg} and Table~\ref{tab:prec_htt}. 
In this case, we do not allow anomalous $\HVV$ couplings, 
and the measurement of the signal strength $\muV$ constrains the contribution of the $a_1$ coupling in the 
$\PH\to\gamma\gamma$ loop. 
 
The gain in this combination of the \ggH and \tqH~\&~\ttH processes
is beyond the simple addition of the two constraints. While in the \ggH and \ttH
analyses the signal strengths of the two processes are independent, they could be related under the 
assumption of top quark dominance in the loop using Eq.~(\ref{eq:sigma_ggH}).
As discussed in Section~\ref{sec:pheno}, $CP$-odd coupling predicts rather different cross sections in the two processes: 
$\sigma(\tilde\kappa_\mathrm{f}=1)/ \sigma(\kappa_\mathrm{f}=1)$ is 2.38 in the gluon fusion 
process dominated by the top quark loop and $0.391$ in the \ttH process. 
This means that the ratio differs by a factor of 6.09 for $\fT=1$ when compared to SM ($\fT=0$).
This correlation enhances the sensitivity in the $\fT$ measurement. 
For example, the combined sensitivity from $\tqH$ and $\ttH$ (with either $\PH\to4\ell$ 
alone or together with $\PH\to\gamma\gamma$), and $\ggH$ is significantly improved compared to separate analyses,
and the result is not just a simple addition of two independent results, 
as shown in Fig.~\ref{fig:resultHttHgg} and Table~\ref{tab:prec_htt}.
This effect also enhances correlation between $\fT$ and the yield parameters. 
In the full combination, the measured signal strength is $\mu_{\ttH}=0.70^{+0.30}_{-0.25} $ 
and the correlation with $\fT$ is $+0.96$. 

Finally, we present the reinterpretation of the $\fG$, $\fT$, and signal strength measurements in terms of
constraints on $c_{\Pg\Pg}$, $\tilde{c}_{\Pg\Pg}$, $\kappa_\cPqt$, and $\tilde\kappa_\cPqt$.
In this fit, it is assumed that $\kappa_{\cPqb}=\kappa_{\PQc}=\kappa_\mu=1$ and 
$\tilde\kappa_{\cPqb}=\tilde\kappa_{\PQc}=\tilde\kappa_\mu=0$ in the fermion coupling contribution 
to the loops and in the decay width parameterization~\cite{Gritsan:2020pib}.
The gluon fusion loop is parameterized in terms of point-like couplings 
$c_{\Pg\Pg}$ and $\tilde{c}_{\Pg\Pg}$, and the top and bottom quark contributions. 
These point-like and top couplings can be resolved in combination with the \ttH and \tqH production processes. 
The $\PH\to\gamma\gamma$ loop is parameterized with the top and bottom quark and with the $\PW$ contributions.
One cannot generally relate the point-like couplings in this loop and in the gluon fusion loop, and they
are assumed to be zero in $\PH\to\gamma\gamma$. The measurement of the signal strength $\muV$ constrains 
the contributions of the $a_1$ coupling, affecting the $\PW$ contribution to the $\PH\to\gamma\gamma$ loop 
and \tqH process, and anomalous $\HVV$ couplings are not allowed.
By convention, the $a_1$ coupling is constrained to be positive, which sets the relative sign of $\kappa_{\PQt}$. 
It is assumed that there are no unobserved or undetected \Hboson\ decays. 

The constraints are shown in Fig.~\ref{fig:resultHttHgg_2D}, where the likelihood scans
are plotted for the pairs of parameters: ($c_{\Pg\Pg}$, $\tilde{c}_{\Pg\Pg}$), ($\kappa_{\PQt}$, $\tilde\kappa_{\PQt}$),
and ($\kappa_{\PQt}$, $c_{\Pg\Pg}$), with the other two parameters either profiled or fixed to the SM expectation. 
On the likelihood scan of ($c_{\Pg\Pg}$, $\tilde{c}_{\Pg\Pg}$) with the other parameters fixed, 
the appearance is generally similar to the scan shown in Fig.~\ref{fig:resultHgg}. 
This is because the addition of the \ttH and \tqH processes does not alter the results much with their couplings fixed to the SM. 
However, with the other parameters affecting the gluon fusion loop left floating, the contours are washed out,
as one would expect with more degrees of freedom in a fit. 

On the likelihood scan of ($\kappa_{\PQt}$, $\tilde\kappa_{\PQt}$) with the other parameters profiled, 
one can observe the effects similar to that in Fig.~\ref{fig:resultHtt}. This is because the \ggH process does not
bring additional constraints due to uncertainties with the point-like interactions, but may introduce additional 
modifications to the fit parameters. However, with the point-like interactions $c_{\Pg\Pg}$ and $\tilde{c}_{\Pg\Pg}$ set to zero, 
constraints on the $\kappa_{\PQt}$ and $\tilde\kappa_{\PQt}$ couplings appear tighter as a result of the combination
of information from the \ttH, \tqH, and \ggH processes. In both cases, the general features remain similar 
to Fig.~\ref{fig:resultHtt}, such as pure yield measurements leading to constraints within a ring, 
$CP$-sensitive measurements resolving areas within this ring, 
and the $\tqH$ process leading to the exclusion of negative values of $\kappa_{\PQt}$.
However, in this case, ambiguity between the positive and negative values of $\tilde\kappa_{\PQt}$
can be resolved with the inclusion of the \ggH process, where the $\mathcal{D}_{CP}^{\ggH}$
discriminant carries information sensitive to the sign. 

On the likelihood scan of ($\kappa_{\PQt}$, $c_{\Pg\Pg}$) with the other parameters fixed, 
we observe a resolved four-fold ambiguity of the best fit ranges. Within each range, there is a 
large correlation between the two parameters. This happens because the point-like interaction 
$c_{\Pg\Pg}$ is equivalent to a BSM heavy quark \PQQ contribution to the loop. 
It is hard to distinguish between such a new heavy quark and the heavy top quark. 
The two amplitudes add constructively, leading to a large anticorrelation. 
The rate of the gluon fusion process would be roughly proportional to 
$(\kappa_{\PQt}+\alpha\,\kappa_{\PQQ})^2$ with $\alpha\sim1$. 
Given that the \ttH rate constrains $\kappa_{\PQt}\sim\pm1$, the \ggH rate would 
constrain ($\kappa_{\PQt}$, $\kappa_{\PQQ}$) to four discrete sets of values around
$(+1,0)$, $(+1,-2)$, $(-1,0)$, and $(-1,+2)$. The presence of the $\tqH$ process 
shifts the preferred negative $\kappa_{\PQt}$ solutions away from $-1$ and makes 
it less likely than the $+1$ value for the reasons discussed above. 
However, the local minimum near $\kappa_{\PQQ}\sim-2$, corresponding to $c_{\Pg\Pg}\sim -0.017$, 
cannot be excluded, even though the global minimum is at $c_{\Pg\Pg}=-0.001$, close to the null SM expectation. 
In the case with the other parameters profiled, the constraints on the ($\kappa_{\PQt}$, $c_{\Pg\Pg}$) 
plane get washed out further, as expected in a fit with more degrees of freedom. 
In this case, the $CP$-odd amplitudes can compensate for some effects of the $CP$-even ones. 
However, some sensitivity is retained because $CP$-sensitive measurements 
constrain the relative contribution of $CP$-odd amplitudes. 

\begin{table*}[!tbh]
\centering
\topcaption{Summary of constraints on the anomalous \HVV coupling parameters 
with the best fit values and allowed $68$ and $95\%$~\CL intervals.
Three scenarios are shown for each parameter: 
with three other anomalous \HVV couplings set to zero (first), 
with three other anomalous \HVV couplings left unconstrained (second), 
in Approach~1 with the relationship $a_i^{WW}=a_i^{ZZ}$ in both cases; 
and with two other anomalous \HVV couplings left unconstrained (third), 
in Approach~2 within SMEFT with the symmetry relationship of couplings set in Eqs.~(\ref{eq:EFT1}--\ref{eq:EFT5}).
The $\fLZg$ parameter is not independent in the latter scenario. 
\label{tab:prec}
}
\cmsTable{
\begin{scotch}{cllll}
\vspace{-0.3cm} \\
 Parameter   &  Scenario~~~~~~~~~~~~~~~~~~~~~~~  &          & Observed~~~~~~~~~~~~~~~~~~~~~~~~            & Expected~~~~~~~~~~~~~~~~~~~~~~~~  \\
[\cmsTabSkip]
\hline \\ [-1ex]
\multirow{9}{*}{$\fC\begin{cases} \\  \\  \\  \\  \\  \\  \\  \\ \end{cases}$ }
 &   Approach~1                   & best fit & $0.00004$ &  $0.00000$ \\
 &     {\fB=\fL=\fLZg=0}                             & 68\% \CL &    $[-0.00007, 0.00044]$         &             $[-0.00081, 0.00081]$                          \\
 &                                                             & 95\% \CL & $[-0.00055,  0.00168 ]$     & $[-0.00412,  0.00412 ]$ \\
 &    Approach~1                   & best fit & $-0.00805$                                        & $0.00000$  \\
 &      {float \fB,\fL,\fLZg}                         & 68\% \CL & $[-0.02656, 0.00034]$                                & $[-0.00086, 0.00086]$ \\
 &                                                            & 95\% \CL & $[-0.07191, 0.00990]$                                  & $[-0.00423,  0.00422 ]$ \\
 &     Approach~2                  &  best fit &   $0.00005$  & $0.0000$  \\
 &       {float \fB,\fL}                                & 68\% \CL & $[-0.00010,0.00061]$ & $[-0.0012,0.0012]$ \\
 &                                                          & 95\% \CL & $[-0.00072,0.00218]$ & $[-0.0057,0.0057]$ \\
[\cmsTabSkip]
\multirow{10}{*}{$\fB\begin{cases} \\  \\  \\  \\  \\  \\  \\  \\  \\ \end{cases}$ }
 & Approach~1   & best fit & $0.00020$ & $0.0000$ \\
 & {\fC=\fL=\fLZg=0}    & 68\% \CL &           $[{-0.00010},{0.00109}]$                                      & $[{-0.0012},{0.0014 }]$ \\
 &                               & 95\% \CL & $[-0.00078,  0.00368 ]$                            & $[-0.0075,  0.0073 ]$ \\
 &Approach~1  & best fit & $-0.24679$                                          & $0.0000$ \\
 & {float \fC,\fL,\fLZg} & 68\% \CL & \specialcell{$[-0.41087,    -0.15149  ]$ \\ $\cup [-0.00008, 0.00065]$}   &  $[{-0.0017 }, {0.0014 }]$ \\
 &                               & 95\% \CL & \specialcell{$[-0.66842,    -0.08754  ]$ \\ $\cup [-0.00091,  0.00309 ]$} & $[-0.0082,  0.0073 ]$ \\
 &    Approach~2 &  best fit &   $-0.00002$  & $0.0000$  \\
 &       {float \fC,\fL}                                & 68\% \CL & $[-0.00178,0.00103]$ & $[-0.0060,0.0033]$ \\
 &                                                          & 95\% \CL & $[-0.00694,0.00536]$ & $[-0.0206,0.0131]$ \\
[\cmsTabSkip]
\multirow{10}{*}{$\fL\begin{cases} \\  \\  \\  \\  \\  \\  \\  \\ \end{cases}$ }
 & Approach~1    & best fit & {$0.00004$} & {$0.00000$} \\
 & {\fC=\fB=\fLZg=0}  & 68\% \CL &      $[-0.00002,0.00022]$                         & $[-0.00016,0.00026]$  \\
 &                               & 95\% \CL & $[-0.00014, 0.00060]$     & $[-0.00069, 0.00110]$ \\
 & Approach~1 & best fit & $0.18629$                                           &  {$0.00000$} \\
 & {float \fC,\fB,\fLZg} & 68\% \CL & \specialcell{$[-0.00002, 0.00019]$ \\ $\cup [0.07631,     0.27515   ]$}   &   $[-0.00017,0.00036]$ \\
 &                               & 95\% \CL & $[-0.00523,    0.35567  ]$   & $[-0.00076, 0.00134]$ \\
 &  Approach~2 &  best fit &   $0.00012$  & $0.0000$  \\
 &       {float \fC,\fB}                                & 68\% \CL & $[-0.00021,0.00141]$ & $[-0.0013,0.0030]$ \\
 &                                                          & 95\% \CL & $[-0.00184,  0.00443]$ & $[-0.0056,0.0102]$ \\
[\cmsTabSkip]
\multirow{7}{*}{$\fLZg\begin{cases} \\  \\  \\  \\  \\  \\ \end{cases}$ }
 & Approach~1   & best fit &  {$-0.00001$} & {$0.0000$} \\
 & {\fC=\fB=\fL=0}      & 68\% \CL &           $[-0.00099,0.00057]$       &   $[-0.0026,0.0020]$ \\
 &                               & 95\% \CL & $[-0.00387,  0.00301 ]$     & $[-0.0096,  0.0082 ]$ \\
 &  Approach~1 & best fit & $-0.02884$                                          &   {$0.0000$} \\
 & {float \fC,\fB,\fL}     & 68\% \CL & \specialcell{$[-0.09000,    -0.00534  ]$ \\ $\cup [-0.00068, 0.00078]$}   &    $[-0.0027, 0.0026]$  \\
 &                               & 95\% \CL & $[-0.29091,    0.03034   ]$                            & $[-0.0099,   0.0096  ]$ \\
[\cmsTabSkip]
\end{scotch}
}
\end{table*}

\subsection{Constraints on \texorpdfstring{\HVV}{HVV} couplings}
\label{sec:results_HVV}

\begin{figure*}[!tbp]
\centering
\includegraphics[width=0.49\textwidth]{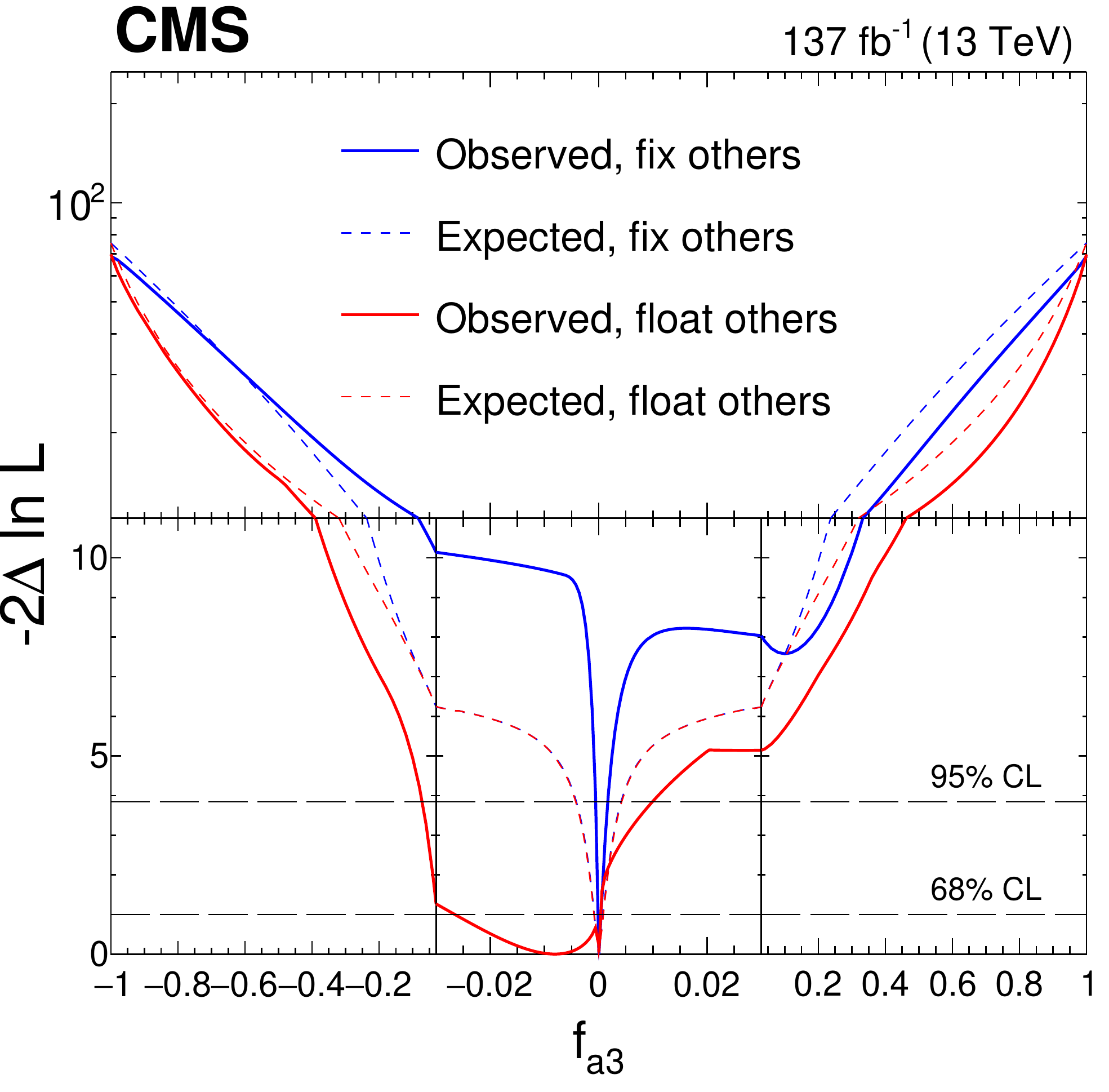}
\includegraphics[width=0.49\textwidth]{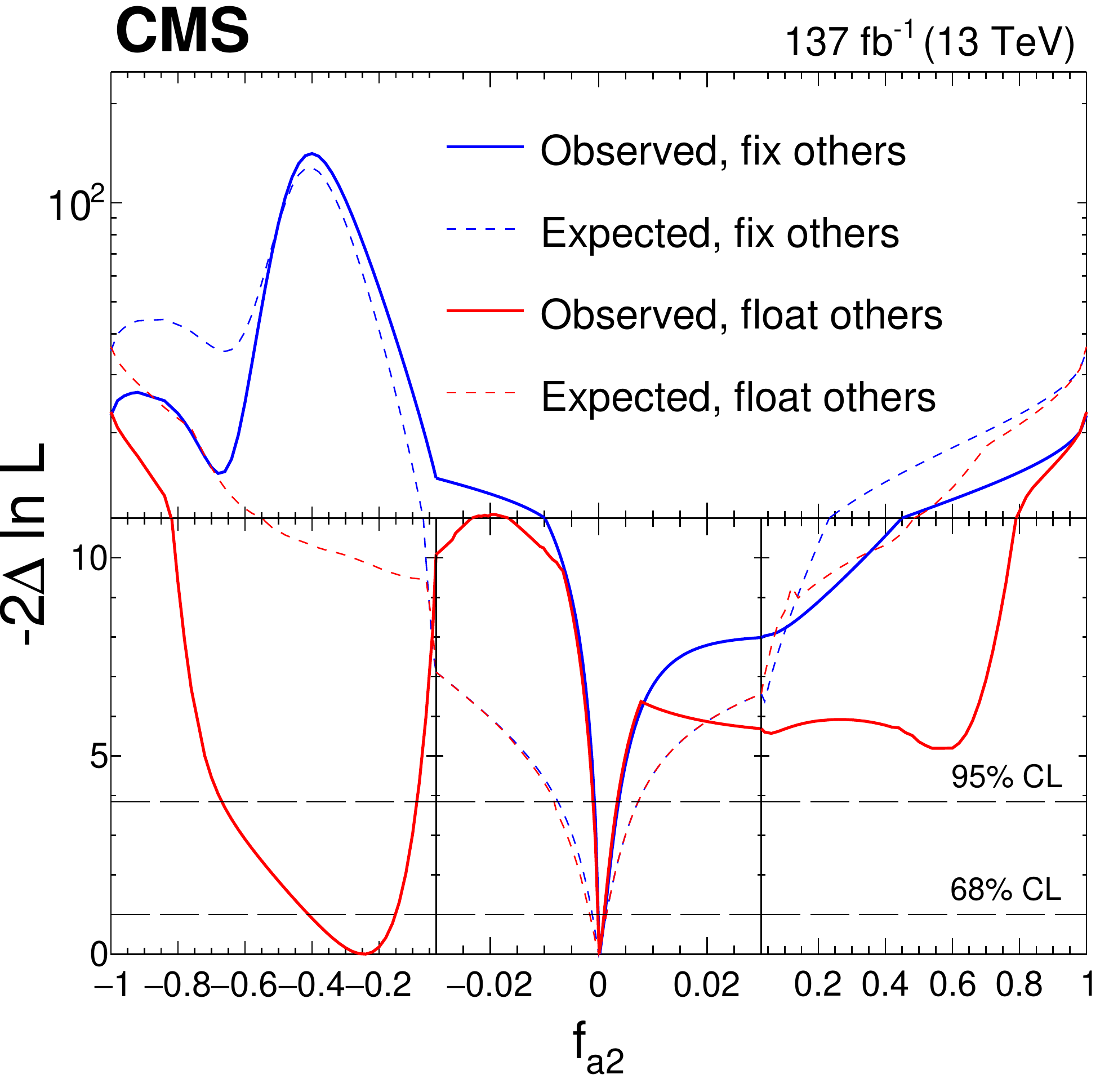} \\
\includegraphics[width=0.49\textwidth]{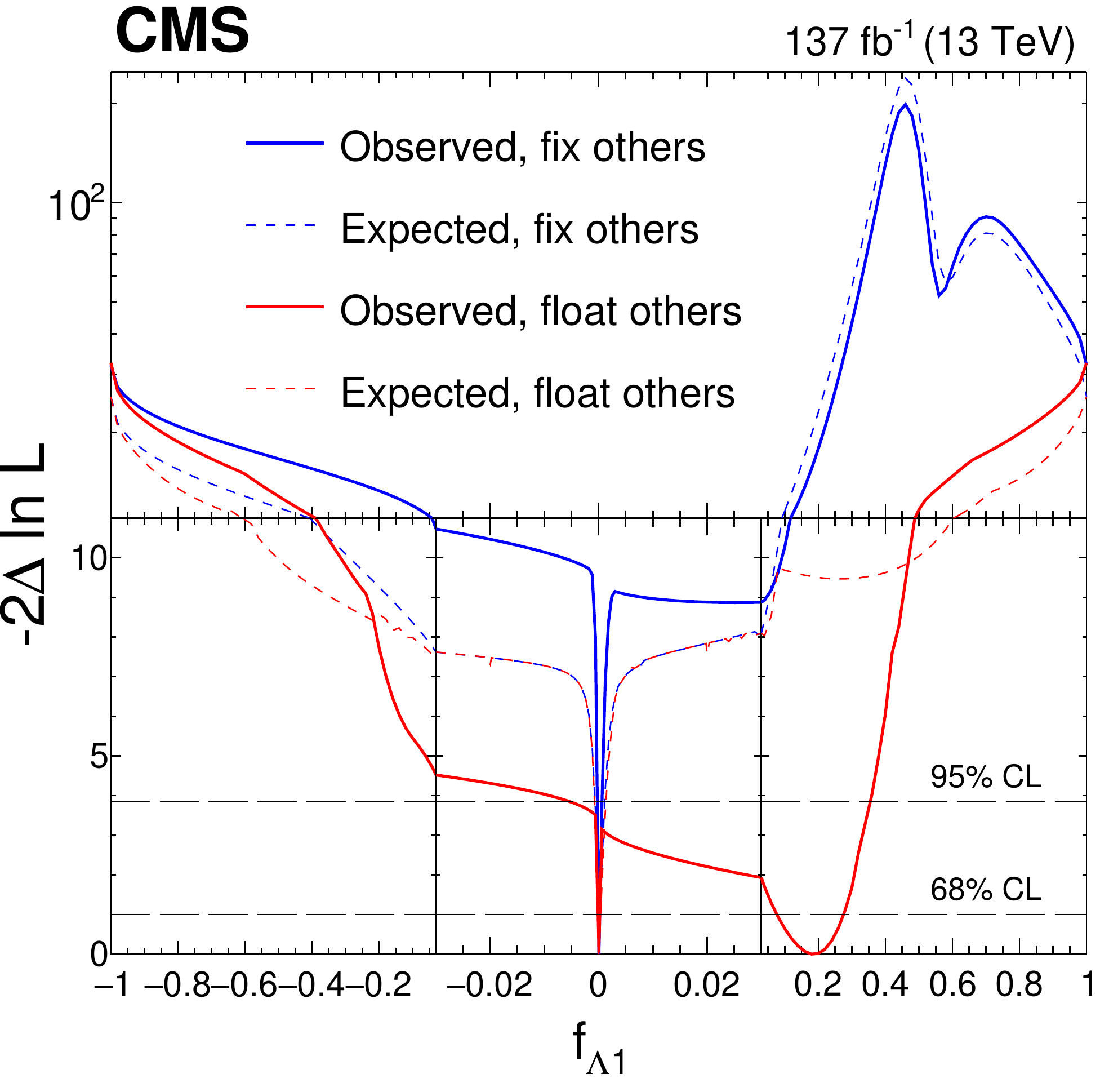}
\includegraphics[width=0.49\textwidth]{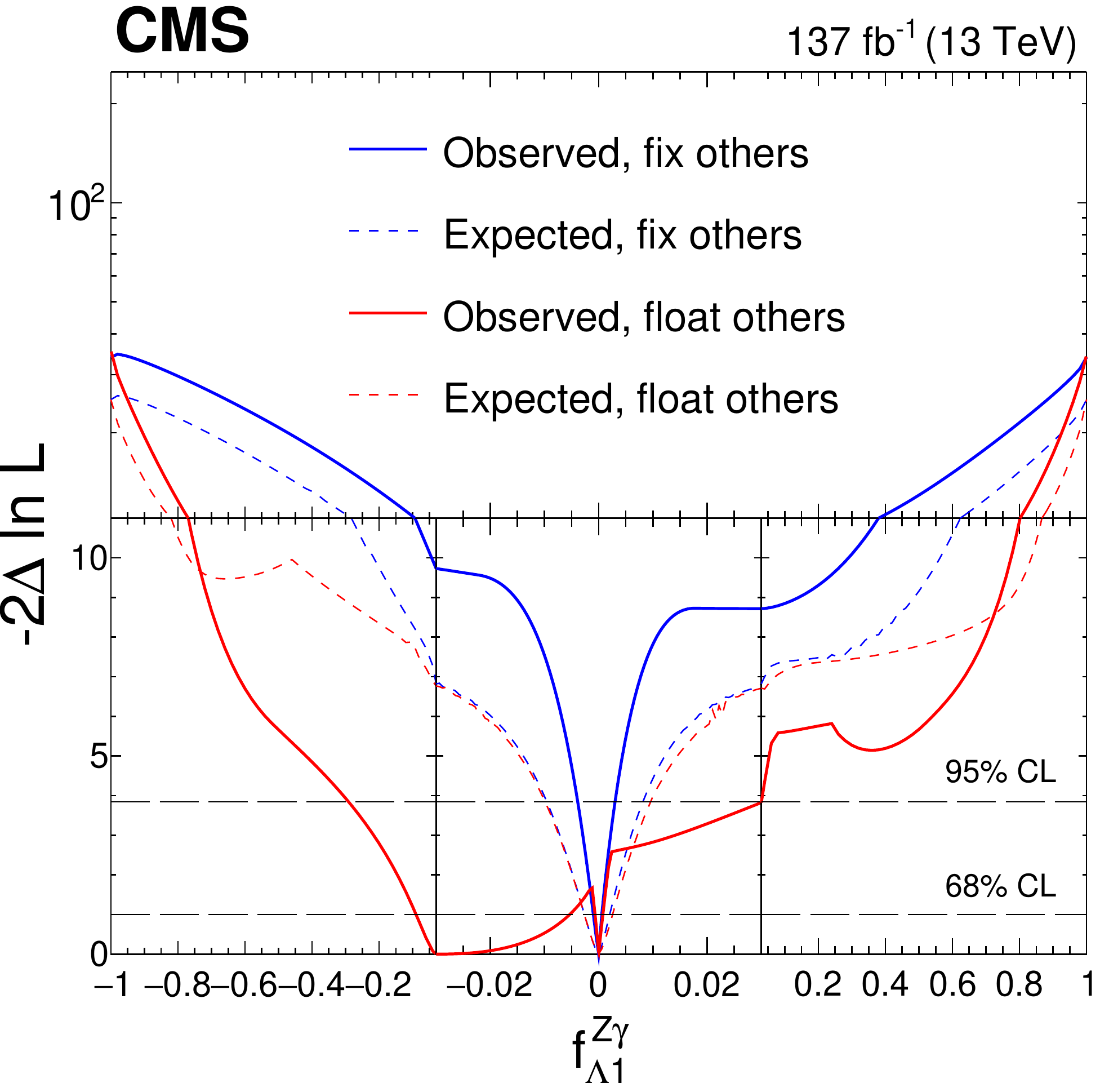}
\caption{
Observed (solid) and expected (dashed) likelihood scans of
\fC (upper left),  \fB (upper right), \fL (lower left), and \fLZg (lower right)
in Approach~1 with the coupling relationship $a_i^{WW}=a_i^{ZZ}$.
The results are shown for each coupling fraction separately with the other three anomalous coupling fractions
either set to zero or left unconstrained in the fit. In all cases, the signal strength parameters have been left unconstrained.
The dashed horizontal lines show the $68$ and $95\%$~\CL regions.
For better visibility of all features, the $x$ and $y$ axes are presented with variable scales.
On the linear-scale $x$ axis, an enlargement is applied in the range $-0.03$ to $0.03$. 
The $y$ axis is shown in linear or logarithmic scale for values of $-2\ln\mathcal{L}$ below or above 11, respectively. 
}
\label{fig:resultHVV}
\end{figure*}

\begin{figure*}[!tbp]
\centering
\includegraphics[width=0.49\textwidth]{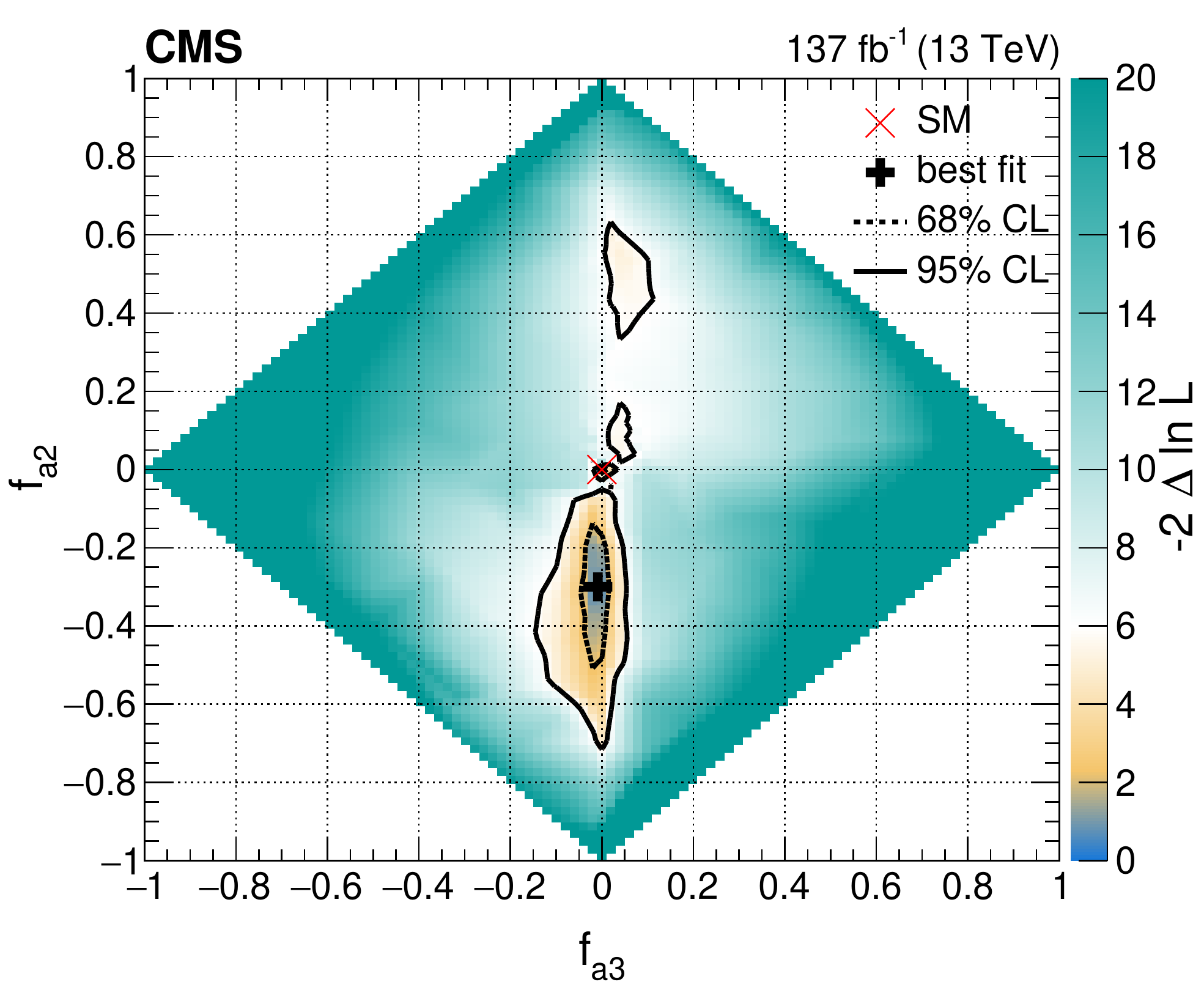}
\includegraphics[width=0.49\textwidth]{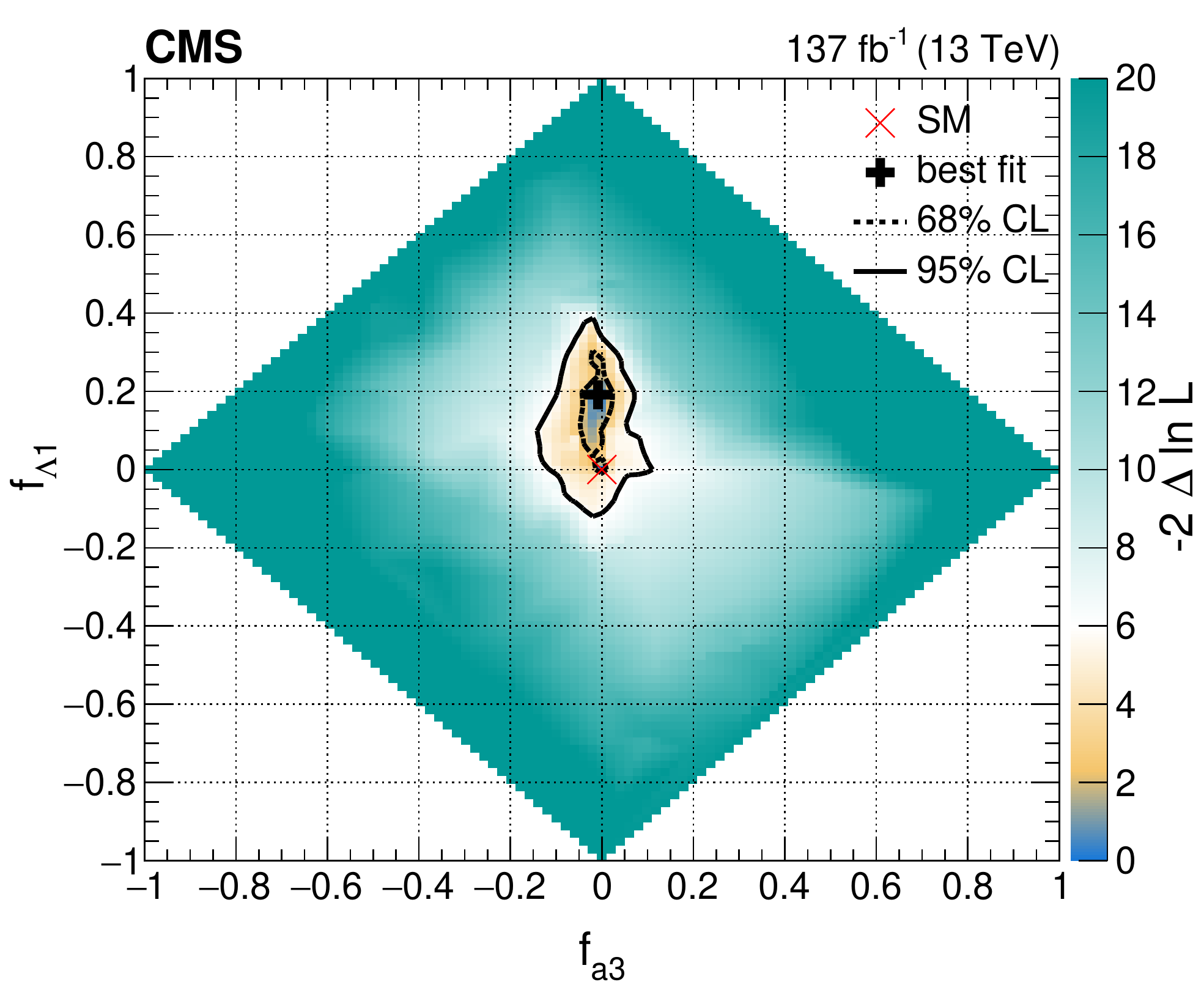}
\includegraphics[width=0.49\textwidth]{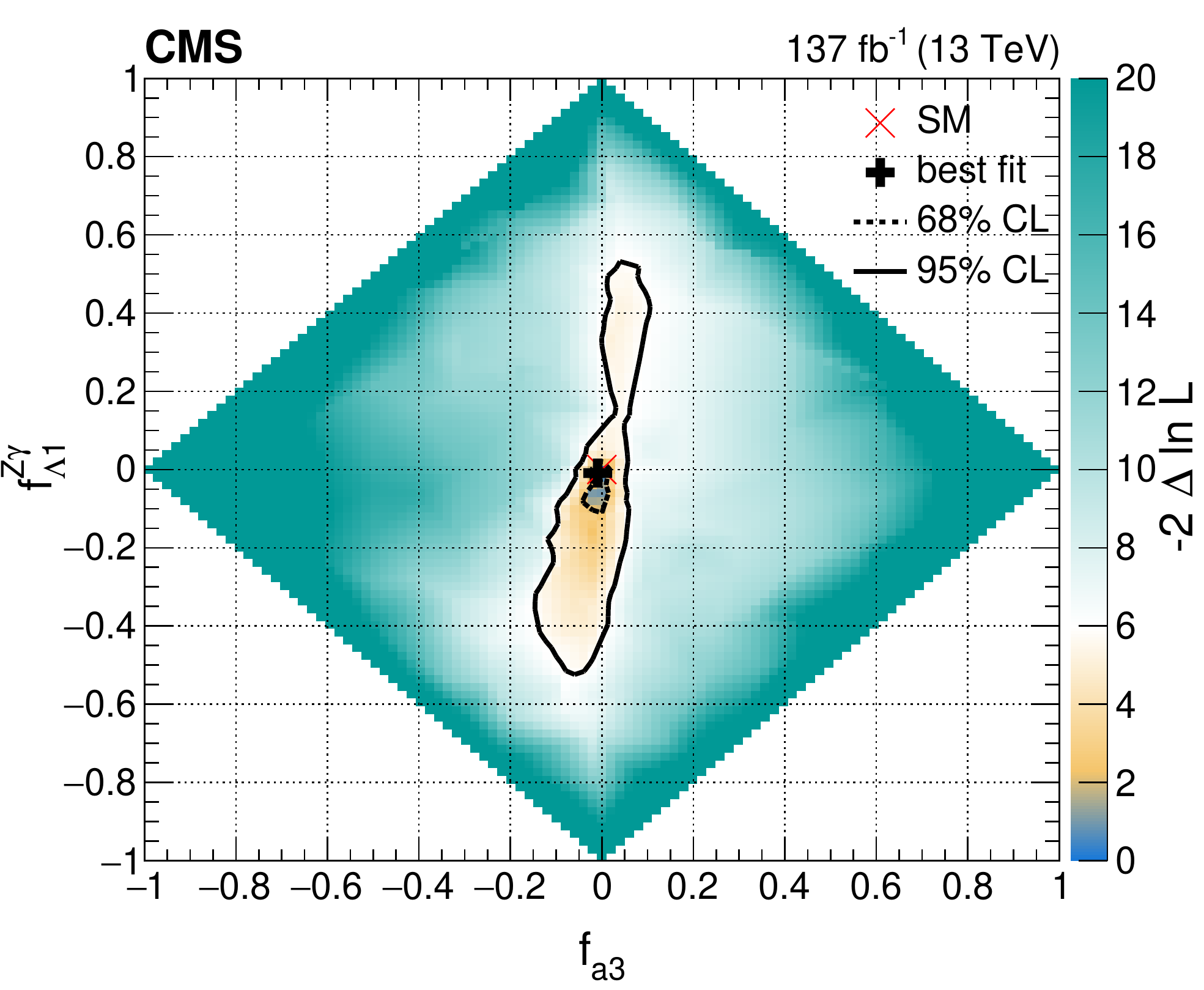}
\includegraphics[width=0.49\textwidth]{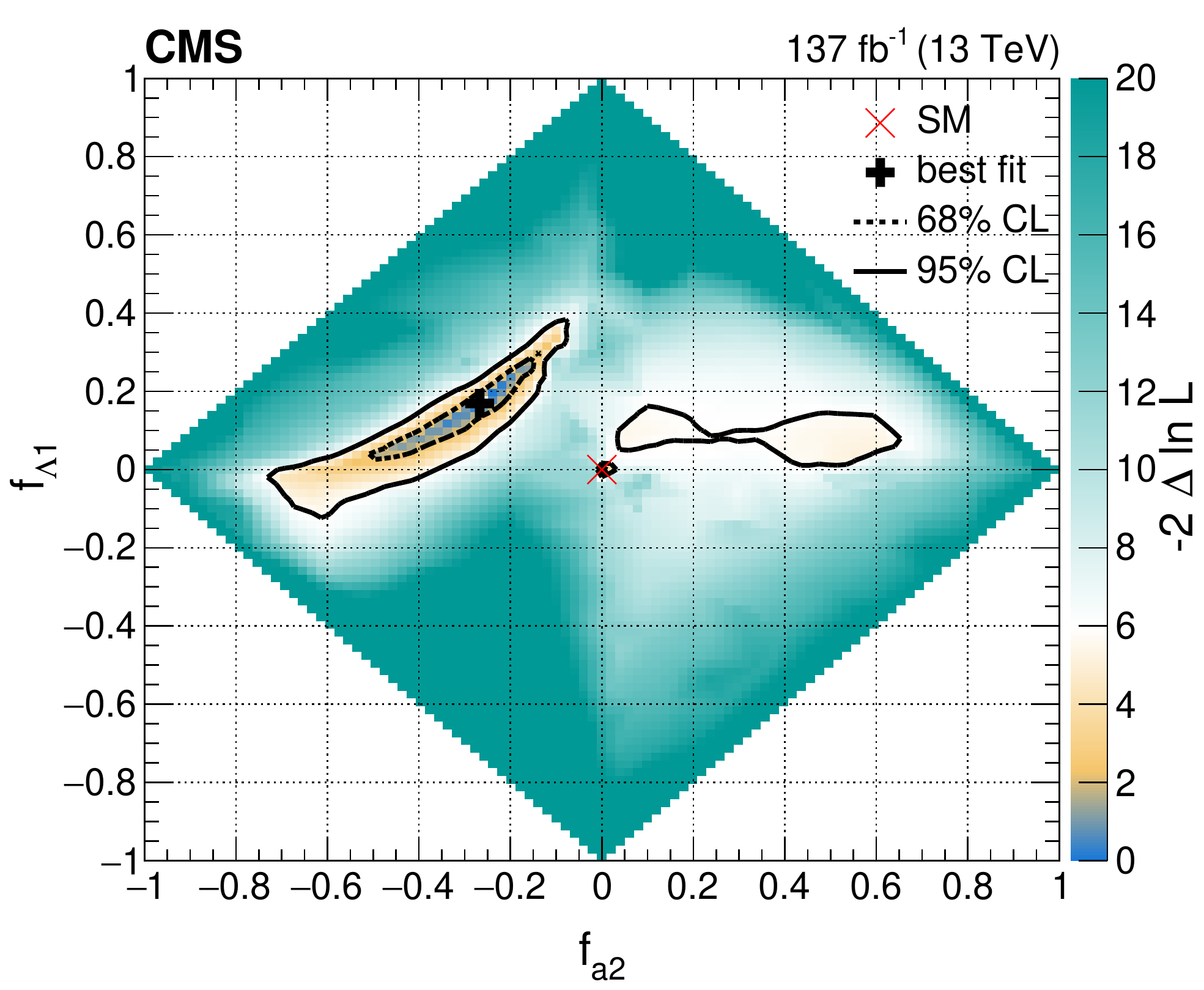}
\includegraphics[width=0.49\textwidth]{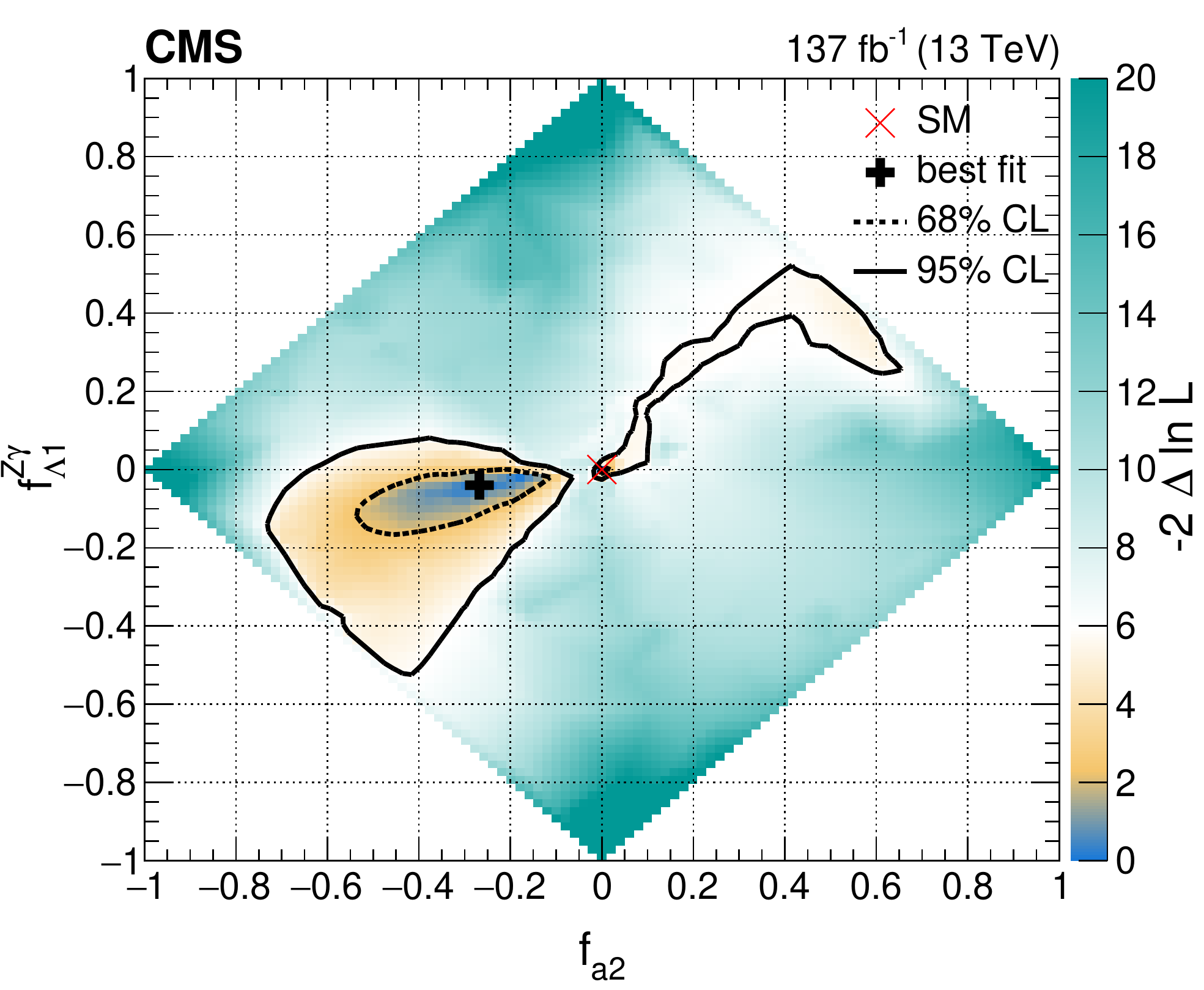}
\includegraphics[width=0.49\textwidth]{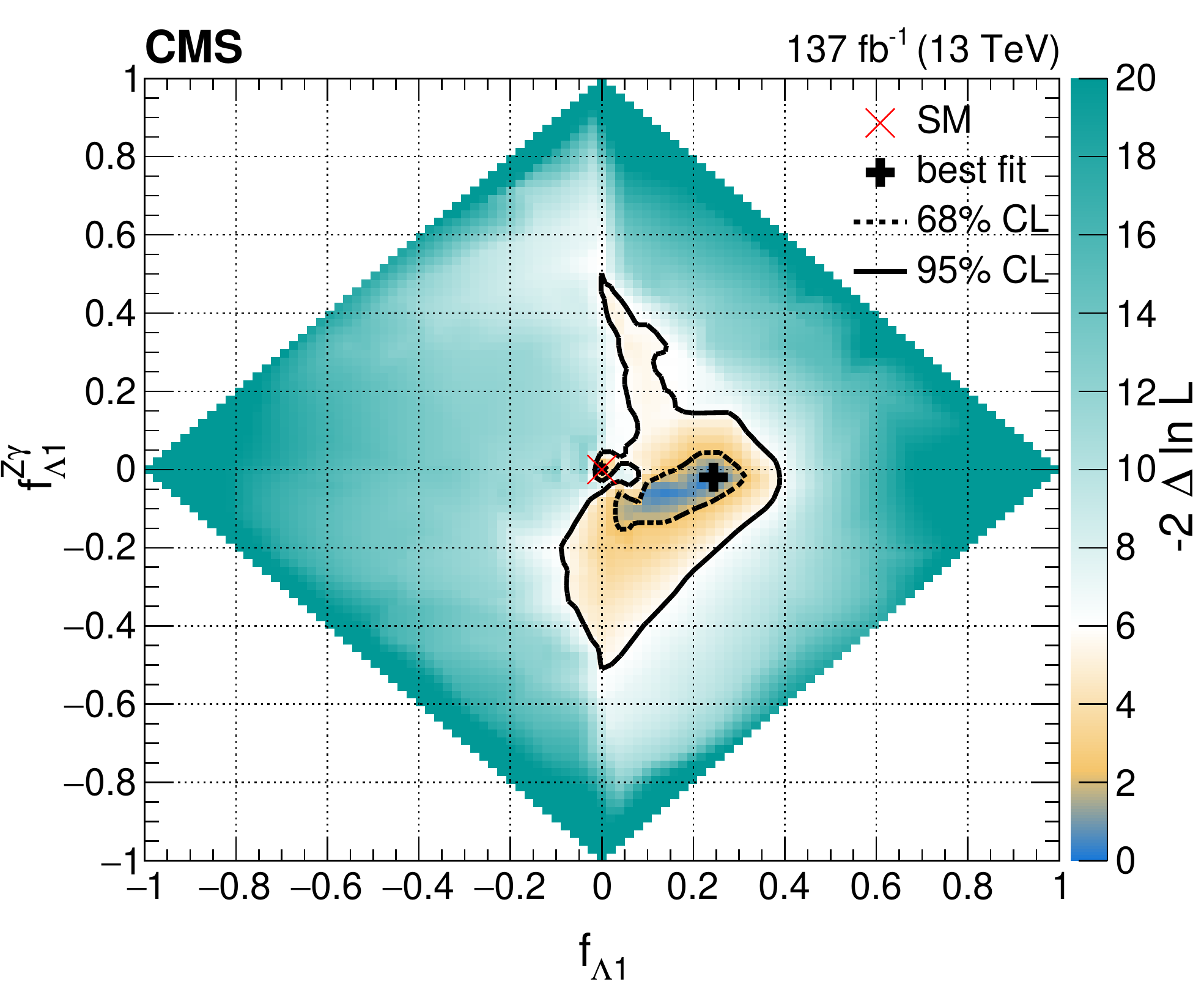}
\caption{Observed two-dimensional likelihood scans of the four coupling parameters \fC,  \fB, \fL, and \fLZg
in Approach~1 with the coupling relationship $a_i^{WW}=a_i^{ZZ}$.
In each case, the other two anomalous couplings along with the signal strength parameters have been left unconstrained.
The $68$ and $95\%$~\CL regions are presented as contours with dashed and solid black lines, respectively.
The best fit values and the SM expectations are indicated by markers.
}
\label{fig:resultHVV_2Dscans_obs}
\end{figure*}

The measurement of anomalous couplings of the \Hboson to EW vector bosons in Approach~1 
with the relationship $a_i^{WW}=a_i^{ZZ}$ is presented in Fig.~\ref{fig:resultHVV} and Table~\ref{tab:prec}. 
Figure~\ref{fig:resultHVV} shows the observed and expected likelihood scans in the simultaneous 
measurement of \fC,  \fB, \fL, and \fLZg, where the $CP$-sensitive parameter $\fG$ and
the signal strength parameters $\muV$ and $\mu_{\ggH}$ are profiled, 
and where we relate $\mu_{\ttH}$ and $\fT$ to $\mu_{\ggH}$ and $\fG$ assuming top quark dominance in the loop. 
The results are shown for each coupling separately, with the other three anomalous couplings either set to zero 
or left unconstrained in the fit. Figure~\ref{fig:resultHVV_2Dscans_obs} shows the same results presented 
as two-dimensional contours, where all couplings discussed above are left unconstrained. 
In all cases, the likelihood scans are limited to the physical range of $\abs{\fC}+\abs{\fB}+\abs{\fL}+\abs{\fLZg}<1$.

There are several features visible on these plots. First, the results with all other couplings constrained
to zero exhibit narrow minima near $\fai=0$ in both the expected and the observed scans. 
This effect comes from utilizing production information. 
The anomalous coupling terms in Eq.~(\ref{eq:formfact-fullampl-spin0}) are multiplied by a factor
of $q_i^2$, which is larger in \VBF and \VH production than in \Hell decay.
As a result, the cross section in \VBF and \VH production increases quickly with \fai.
At the same time, the constraints above $\fai\sim0.02$
are dominated by the decay information from \Hell.

However, when all four anomalous couplings are allowed to float independently, the best fit
value is $(\fC ,  \fB, \fL, \fLZg)=(-0.00805, -0.24679, 0.18629, -0.02884)$. This global minimum
is driven by the decay information from \Hell and is only slightly preferred to the local minimum
at $(0,0,0,0)$, with a difference in $-2\ln\left(\mathcal{L}\right)$ of $0.05$ between the SM value and the global minimum.
The local minimum at $(0,0,0,0)$ is still evident in the four-dimensional distribution and its
projections on each parameter, and is driven by the production information, as discussed above for
the fits with one parameter. Owing to what appears to be a statistical fluctuation in the observed
data when the $-2\ln\left(\mathcal{L}\right)$ minima obtained from the decay and from the 
production kinematics differ, the observed constraints appear weaker than expected. 
However, the results are still statistically consistent with the SM and with the expected 
constraints in the SM. 
Should the global minimum nonetheless persist away from $(0,0,0,0)$ with more data, 
it will be interesting to study consistency of the constraints from the \VBF and \VH production 
and from the \Hell decay. The production and decay test different ranges of $q_i^2$,
as discussed above. If the $q_i^2$ growth is truncated in the \VBF and \VH production 
due to lower-energy BSM effects, the decay information becomes more important. 

\subsection{Constraints on \texorpdfstring{\HVV}{HVV} couplings within \texorpdfstring{$SU(2)\times U(1)$}{SU(2) x U(1)} symmetry}
\label{sec:results_HVVEFT}

\begin{figure*}[!tbp]
\centering
\includegraphics[width=0.49\textwidth]{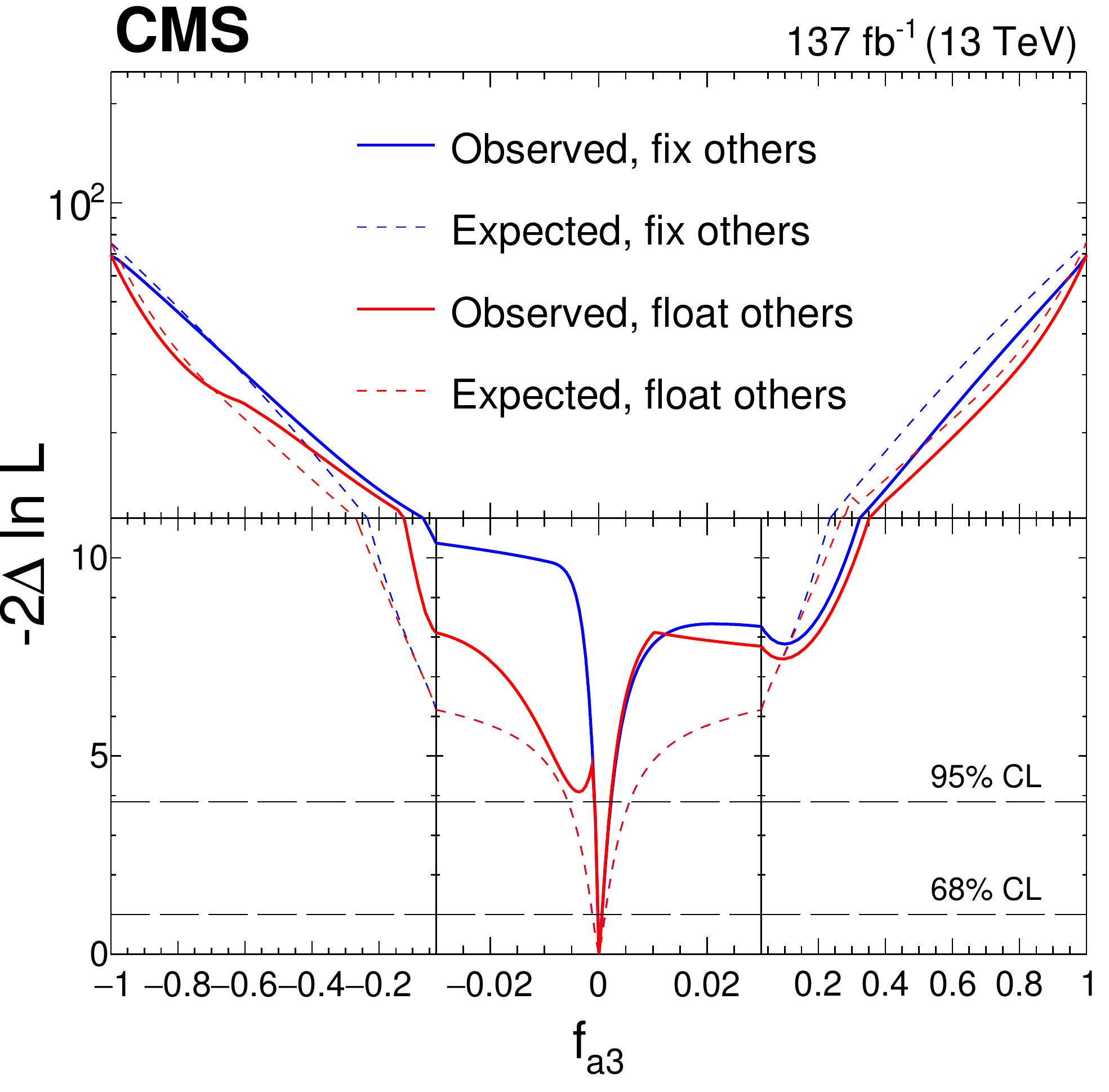}
\includegraphics[width=0.49\textwidth]{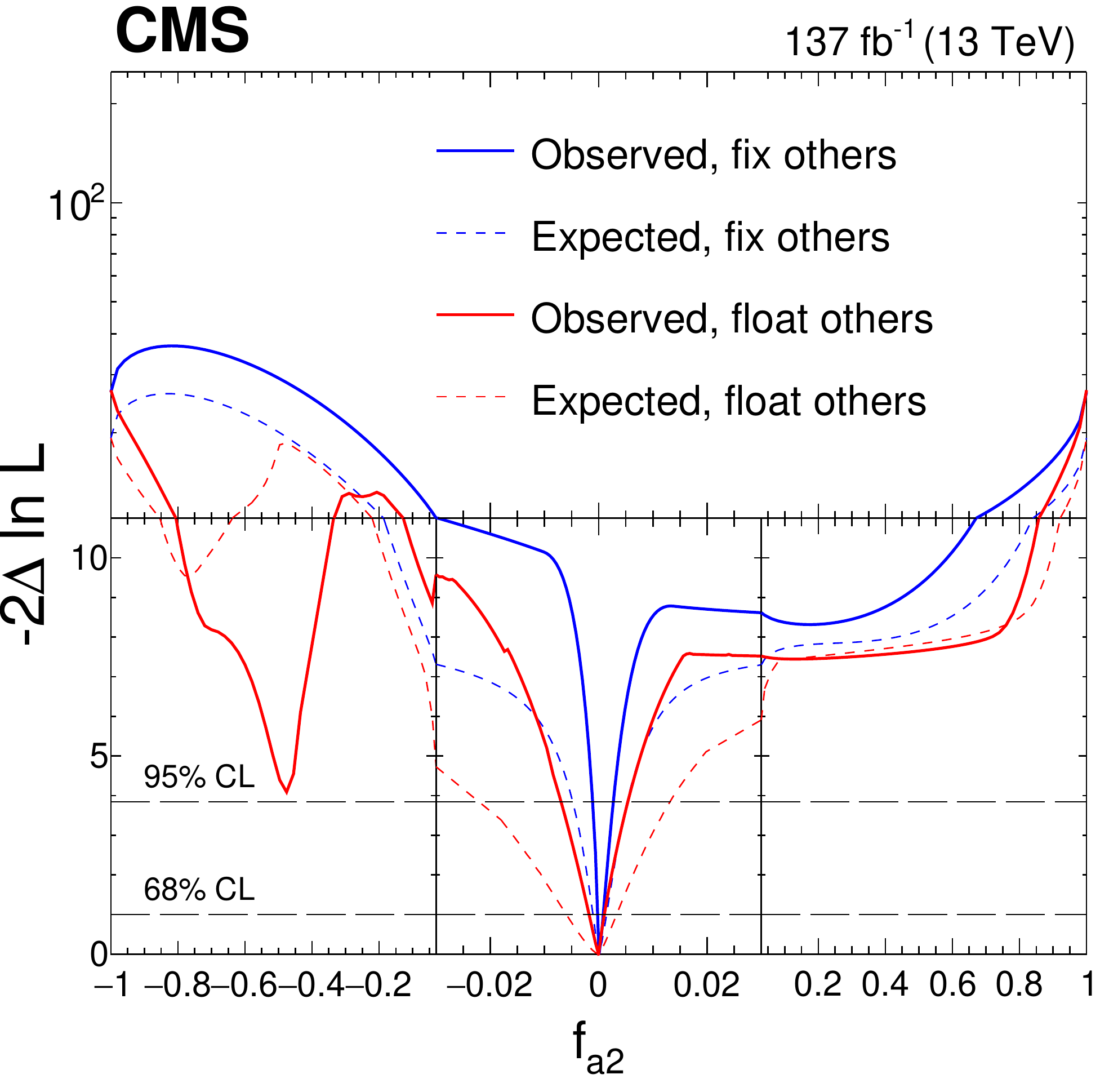} \\
\includegraphics[width=0.49\textwidth]{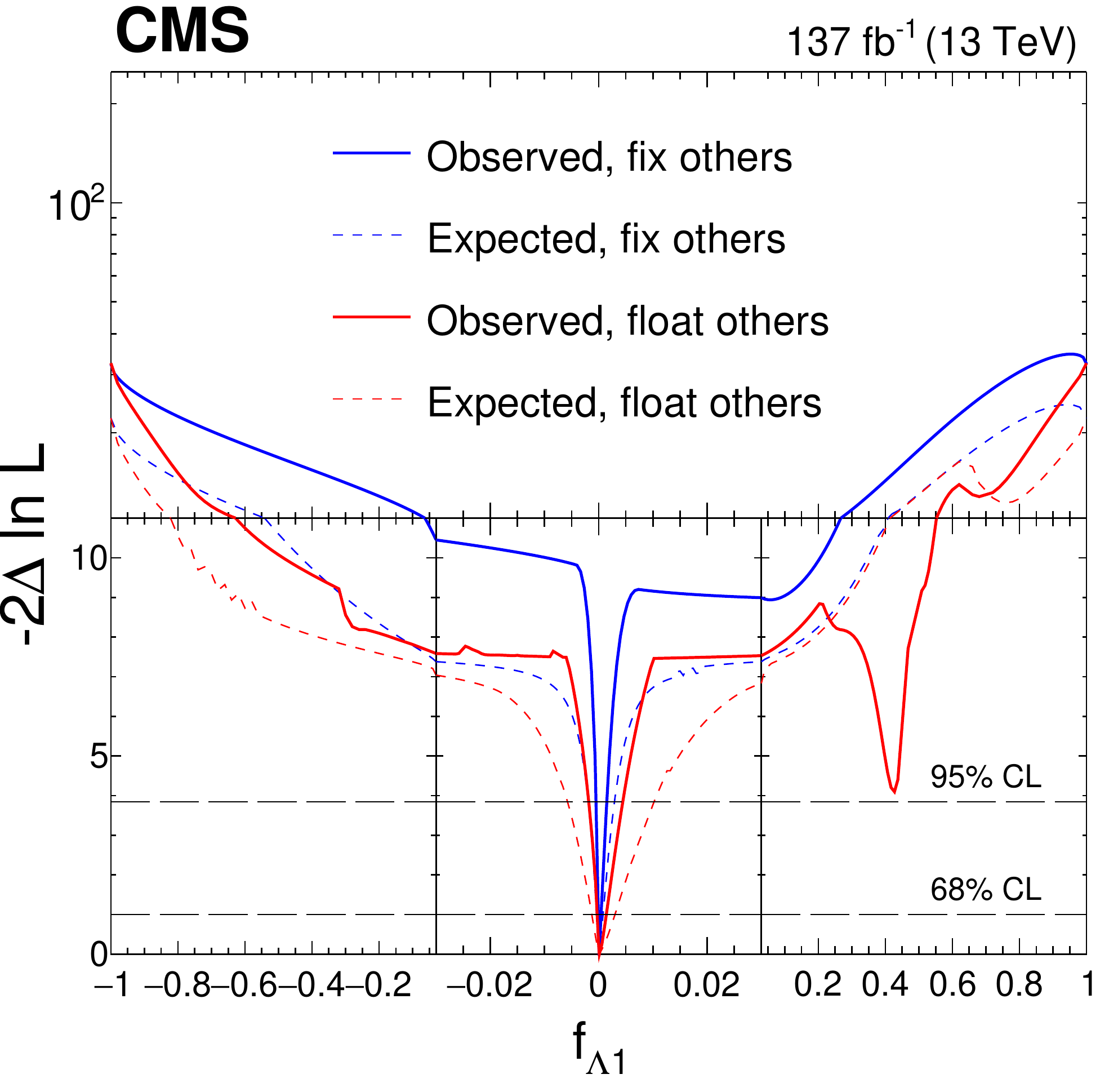}
\caption{
Observed (solid) and expected (dashed) likelihood scans of \fC (upper left),  \fB (upper right), and \fL (lower)
in Approach~2 within SMEFT with the symmetry relationship of couplings set in Eqs.~(\ref{eq:EFT1}--\ref{eq:EFT5}).
The results are shown for each coupling separately with the other anomalous coupling fractions
either set to zero or left unconstrained in the fit. 
In all cases, the signal strength parameters have been left unconstrained.
The dashed horizontal lines show the $68$ and $95\%$~\CL regions.
For better visibility of all features, the $x$ and $y$ axes are presented with variable scales.
On the linear-scale $x$ axis, an enlargement is applied in the range $-0.03$ to $0.03$. 
The $y$ axis is shown in linear or logarithmic scale for values of $-2\ln\mathcal{L}$ below or above 11, respectively. 
}
\label{fig:resultHVV_EFT}
\end{figure*}

The above studies of the \Hboson couplings to EW vector bosons are repeated following Approach~2 
within SMEFT with the symmetries in Eqs.~(\ref{eq:EFT1}--\ref{eq:EFT5}). 
In this case, the $\fLZg$ parameter is not independent.
Therefore, constraints on the three parameters \fC, \fB, \fL, and the signal strength are obtained in this scenario following
the same approach as above. These constraints are shown in Fig.~\ref{fig:resultHVV_EFT} and Table~\ref{tab:prec}. 
The measured signal strength is $\muV=1.10^{+0.50}_{-0.42} $.
The observed correlation coefficients are shown in Table~\ref{tab:correlation_faiEFT}.
Keeping only linear terms and dropping terms with order greater than one for anomalous couplings
does not allow us to make a reasonable likelihood scan, since the probability density goes negative,
as discussed in Section~\ref{sec:cross-section}. 

\begin{table}[!bpt]
\centering
\topcaption{The observed correlation coefficients of the signal strength \muV and the \fC, \fB, \fL parameters 
in Approach~2 within SMEFT with the symmetry relationship of couplings set in Eqs.~(\ref{eq:EFT1}--\ref{eq:EFT5}).
\label{tab:correlation_faiEFT}
}
\begin{scotch}{crrrr}
\vspace{-0.3cm} \\
Parameter  &   \multicolumn{4}{c}{Observed correlation}  \\
[\cmsTabSkip]
\hline \\ [-1ex]
    & \muV & \fC & \fB & \fL \\
[\cmsTabSkip]
\muV &  1       &   $-0.242$ &  $-0.060$  &  $-0.025$     \\
\fC  &              & 1                 &  $-0.082$  &   $+0.008$       \\
\fB &               &                    & 1                 &   $-0.763$  \\
\fL &               &                    &                  & 1 \\
[\cmsTabSkip]
\end{scotch}
\end{table}

Since the relationship of the $\HWW$ and $\HZZ$ couplings does not affect the measurement
of the \fC parameter in the $\PH\to4\ell$ decay, the constraints from the decay information in the wider range of \fC
in Approach~2 are unaffected compared to Approach~1, when other couplings are fixed to zero. 
However, with one less parameter to float, 
the constraints are modified somewhat when all other couplings are left unconstrained. The modified relationship between 
the $\HWW$ and $\HZZ$ couplings also leads to some modification of constraints using production information
in the narrow range of \fC. On the other hand, the \fB and \fL parameters are modified substantially because
the $\fLZg$ information gets absorbed into these measurements through symmetry relationships. 

\begin{figure*}[!tbp]
\centering
\includegraphics[width=0.49\textwidth]{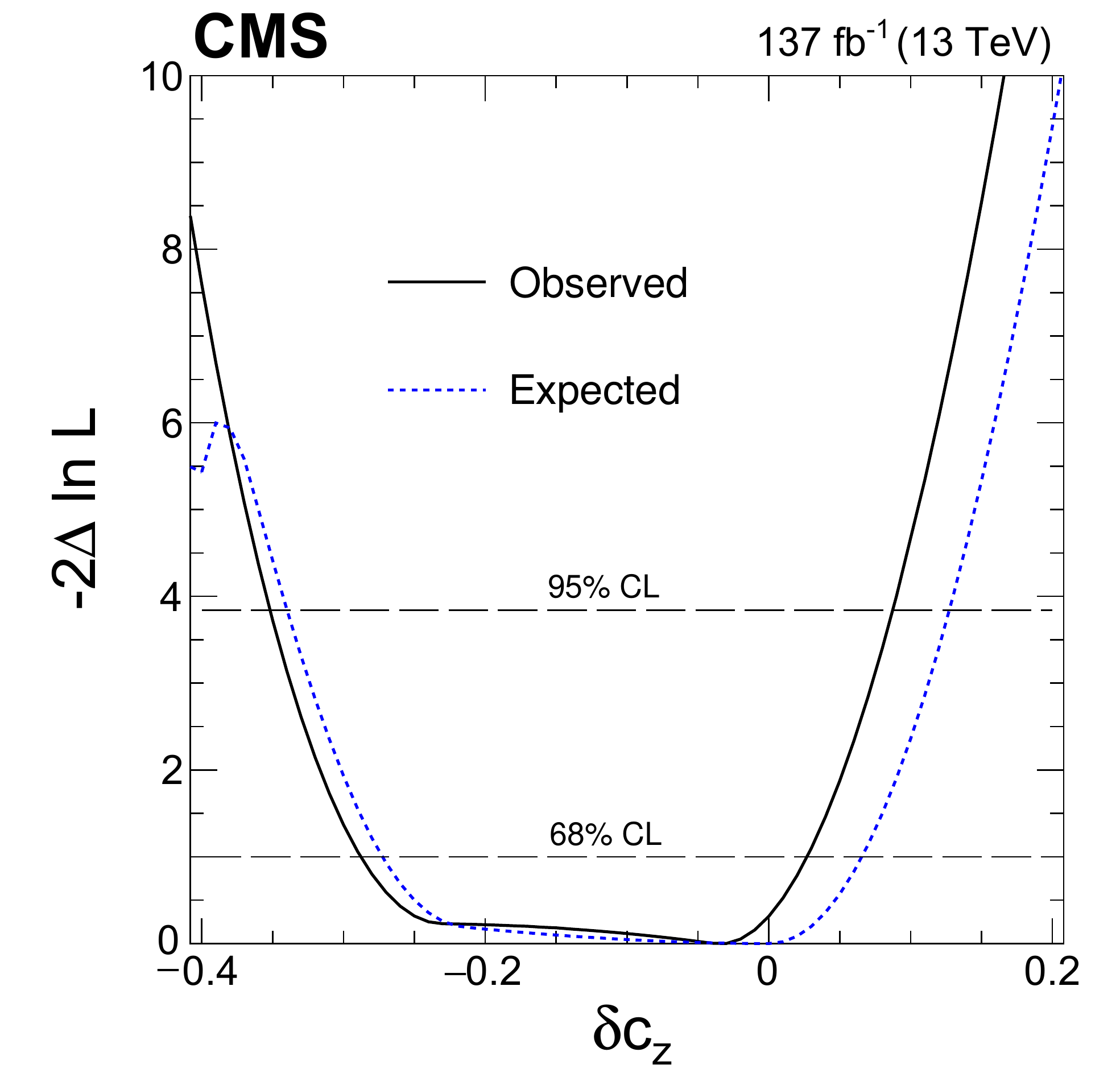}
\includegraphics[width=0.49\textwidth]{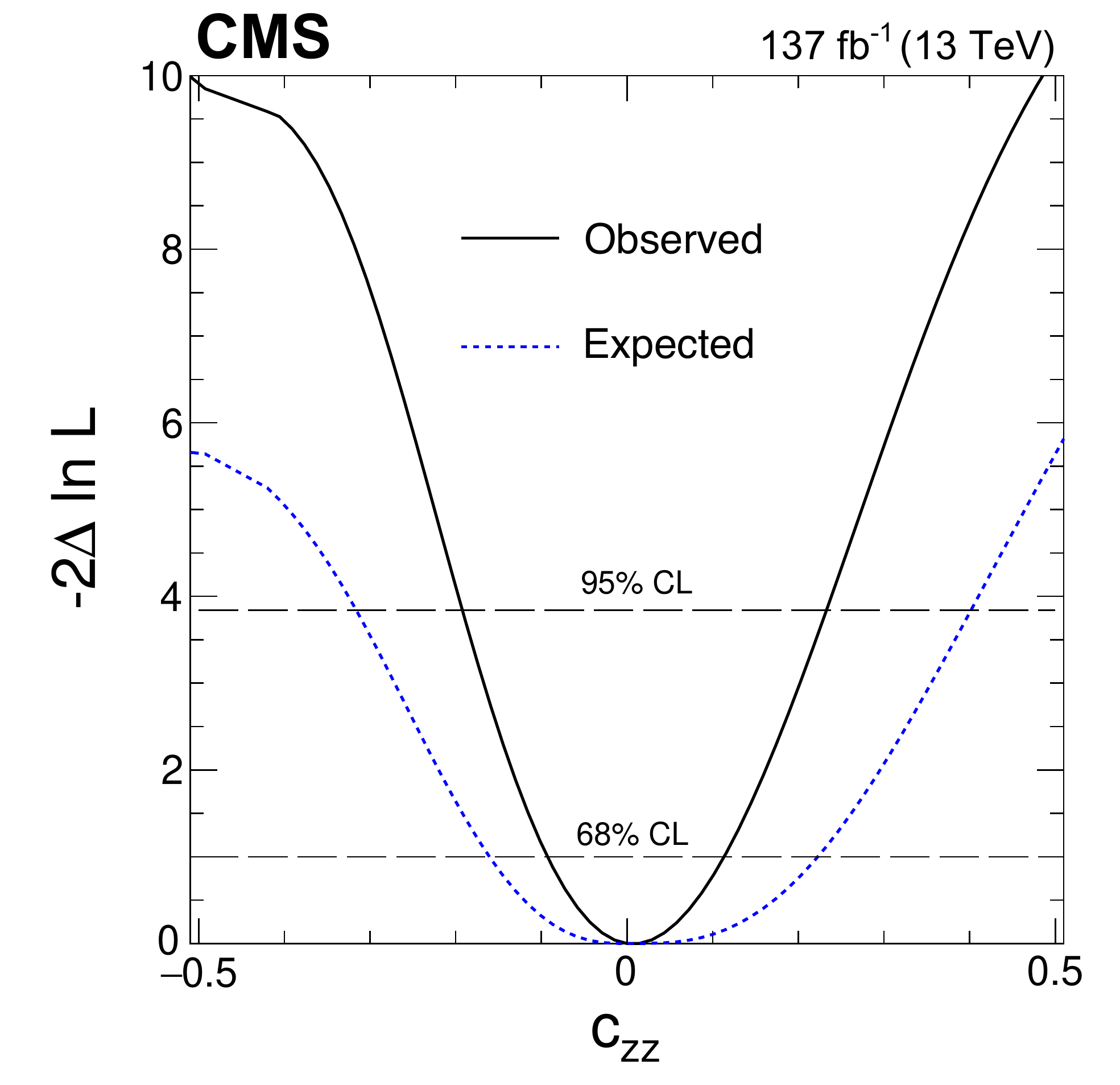}
\includegraphics[width=0.49\textwidth]{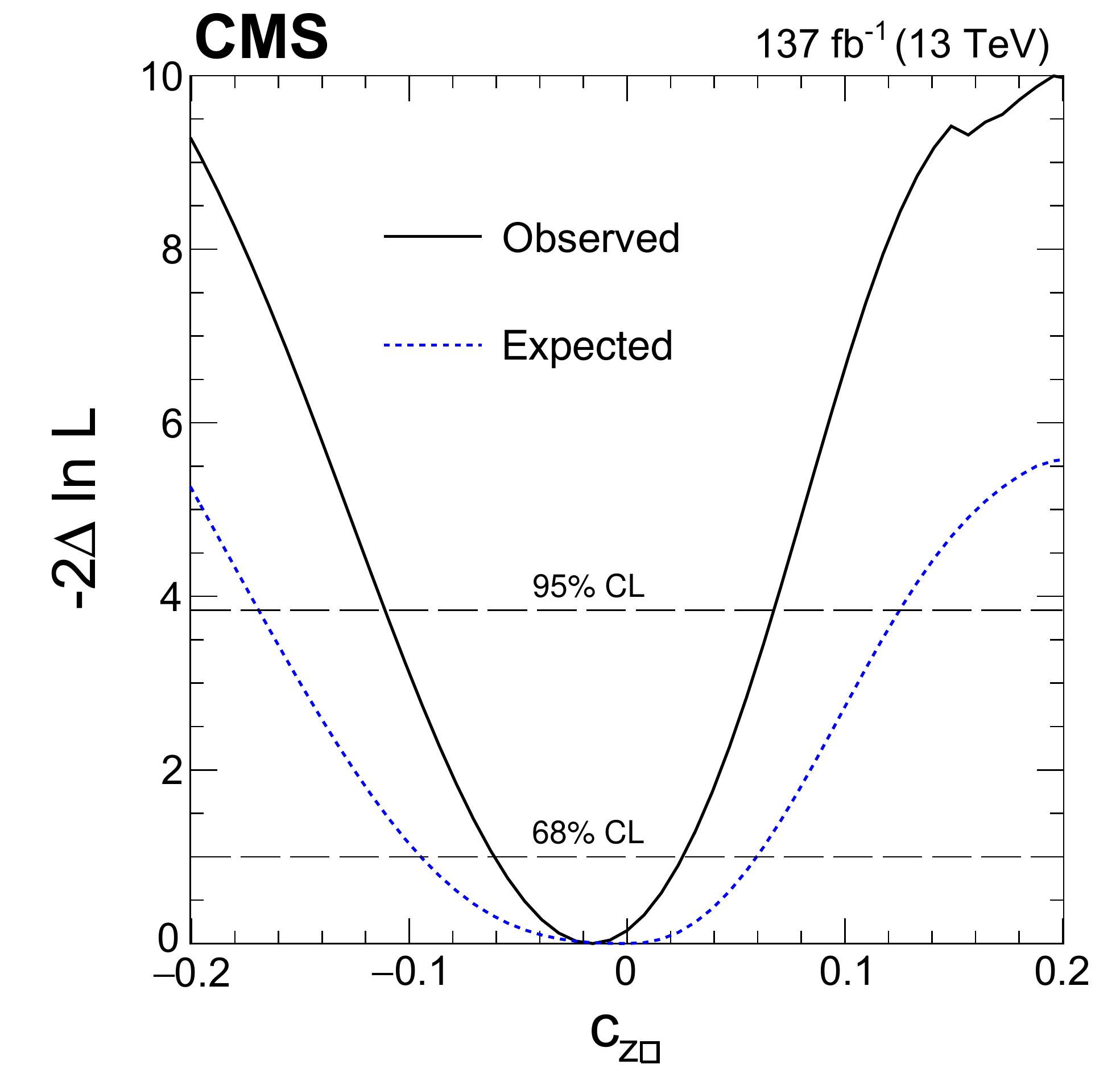}
\includegraphics[width=0.49\textwidth]{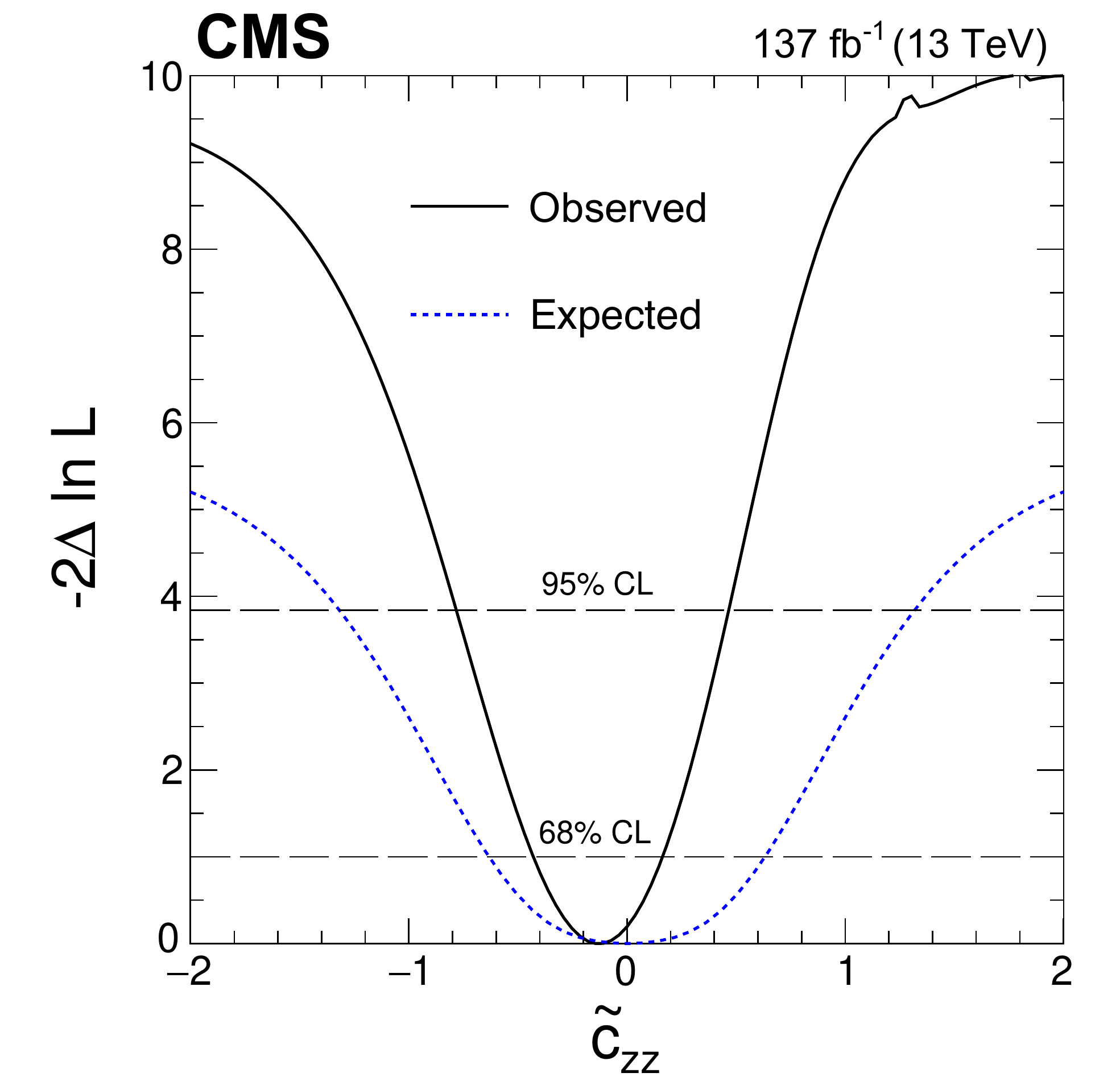}
\caption{
Observed (solid) and expected (dashed) constraints from a simultaneous fit of the SMEFT parameters 
$\delta c_\mathrm{z}$ (upper left), $c_\mathrm{zz}$ (upper right), $c_{\mathrm{z} \Box}$ (lower left), and $\tilde c_\mathrm{zz}$ (lower right)
with the $c_{\Pg\Pg}$ and $\tilde c_{\Pg\Pg}$ couplings left unconstrained. 
}
\label{fig:resultHVV_1DEFT}
\end{figure*}

\begin{figure*}[!tbp]
\centering
\includegraphics[width=0.49\textwidth]{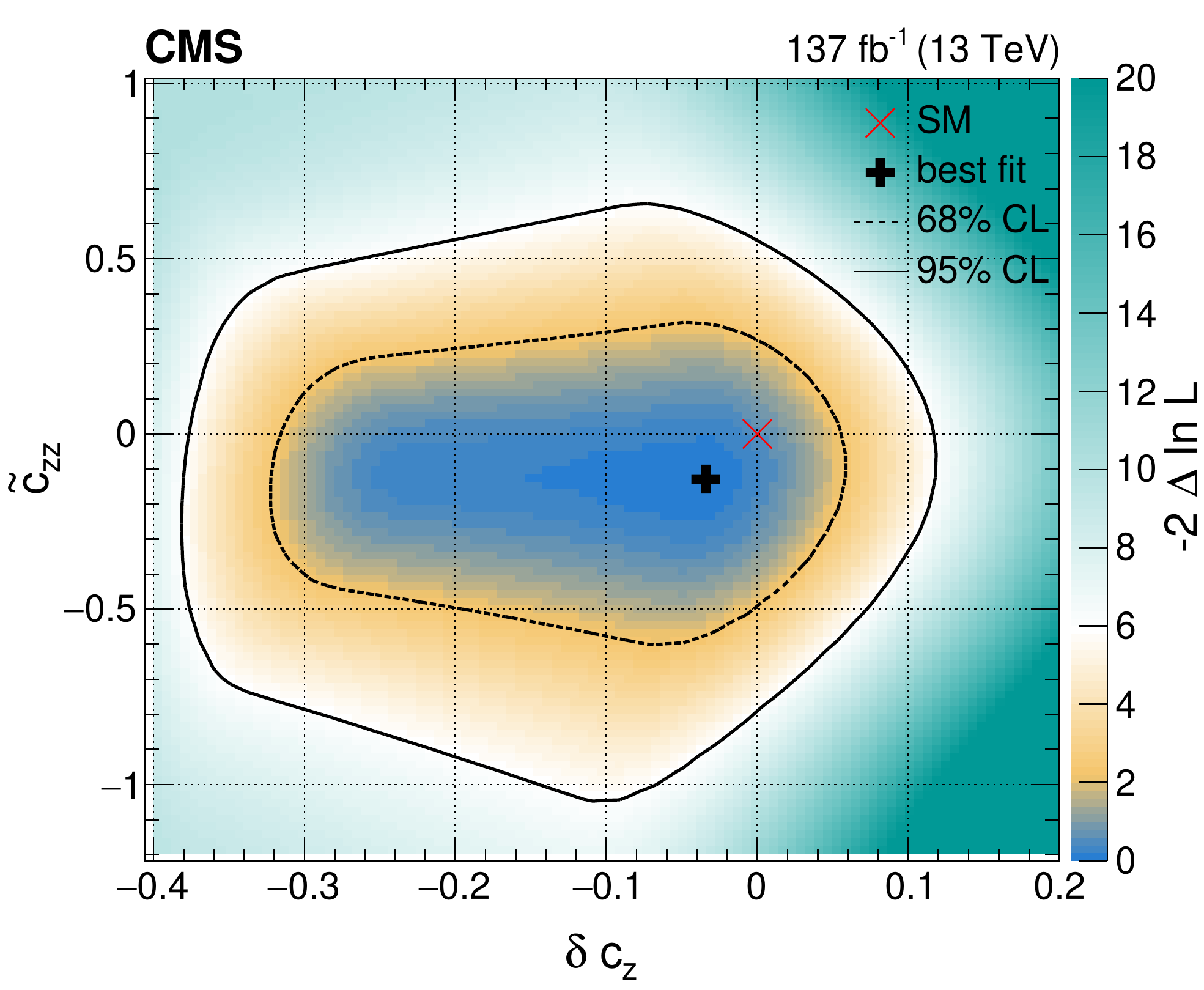}
\includegraphics[width=0.49\textwidth]{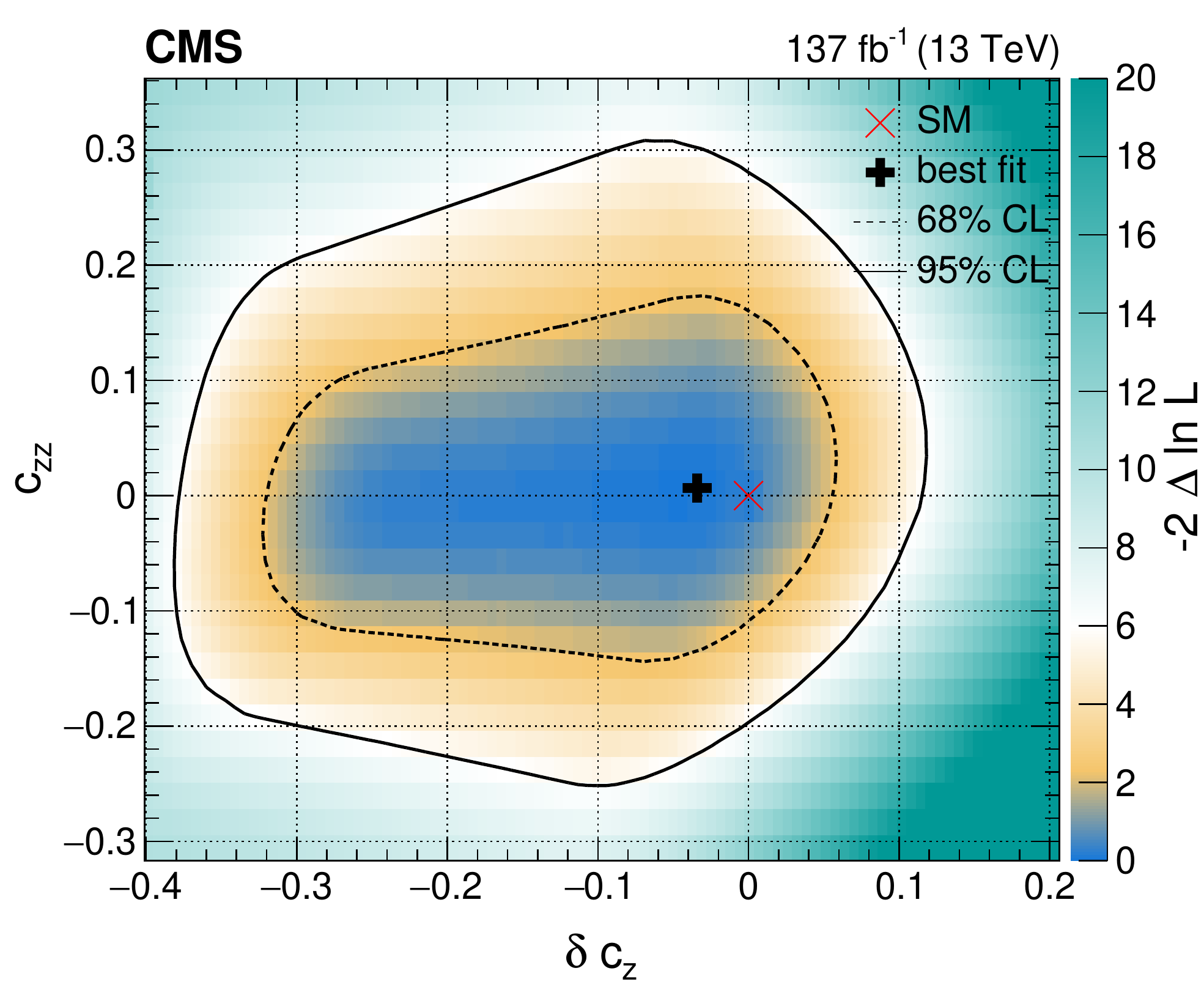}
\includegraphics[width=0.49\textwidth]{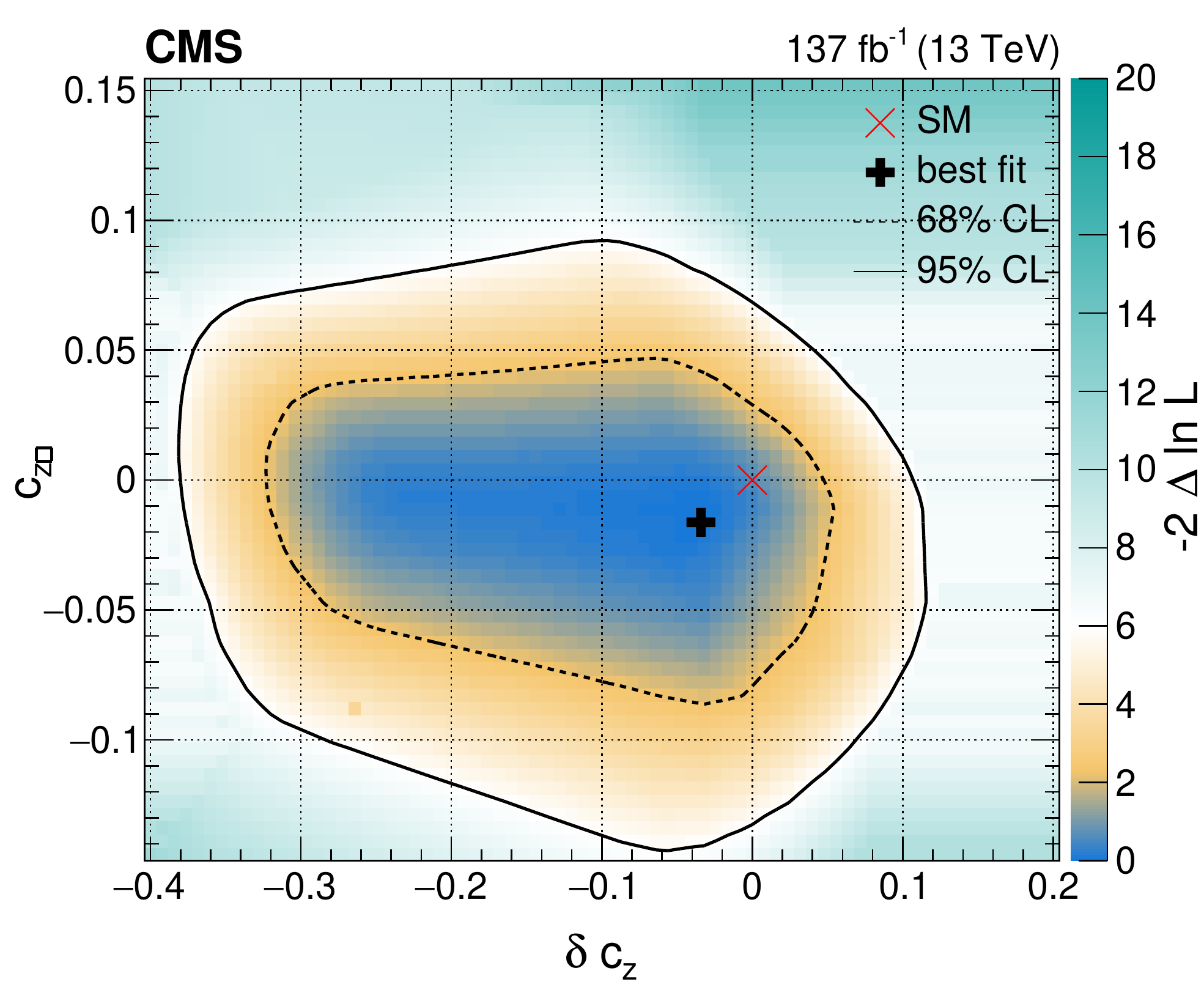}
\includegraphics[width=0.49\textwidth]{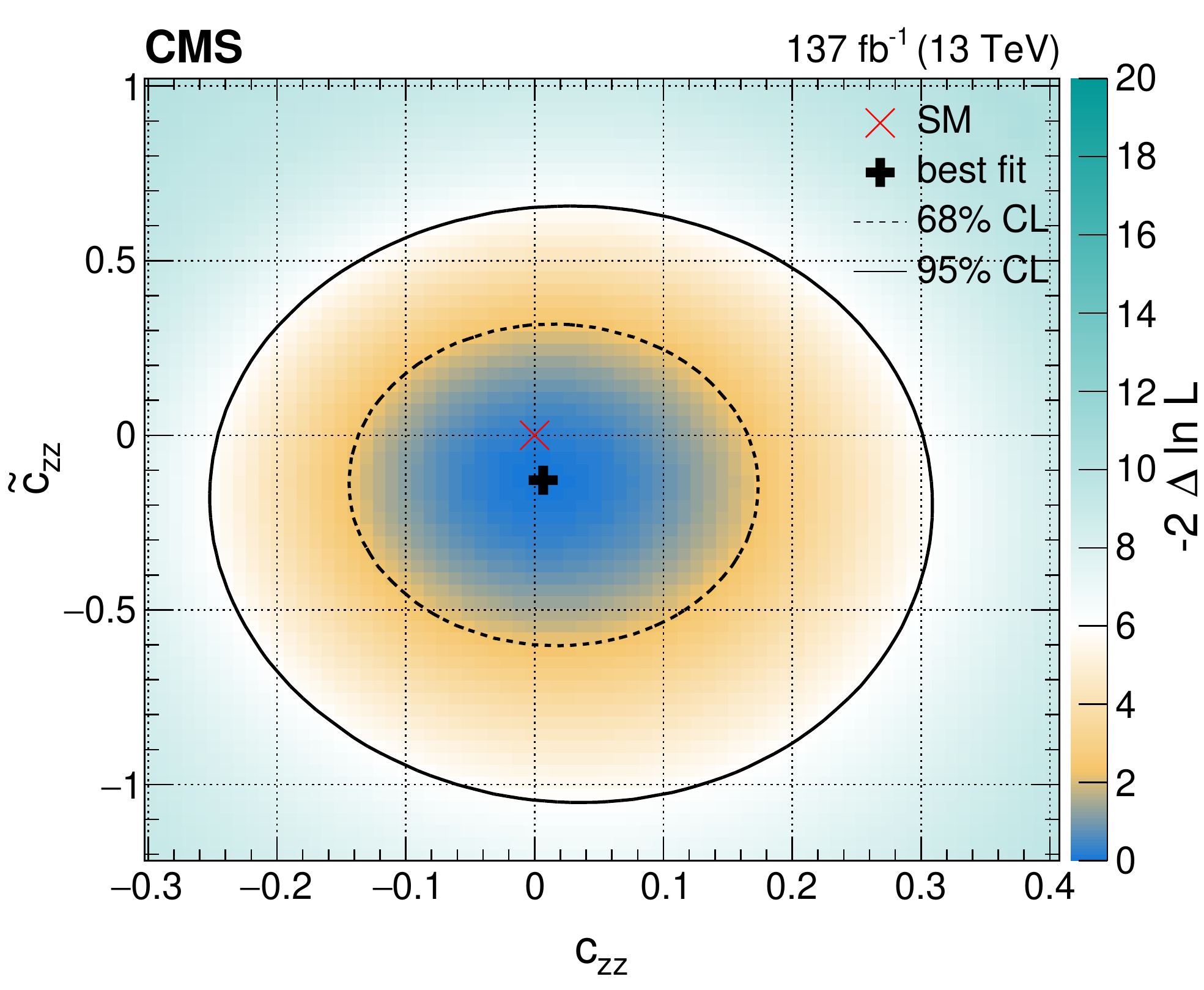}
\includegraphics[width=0.49\textwidth]{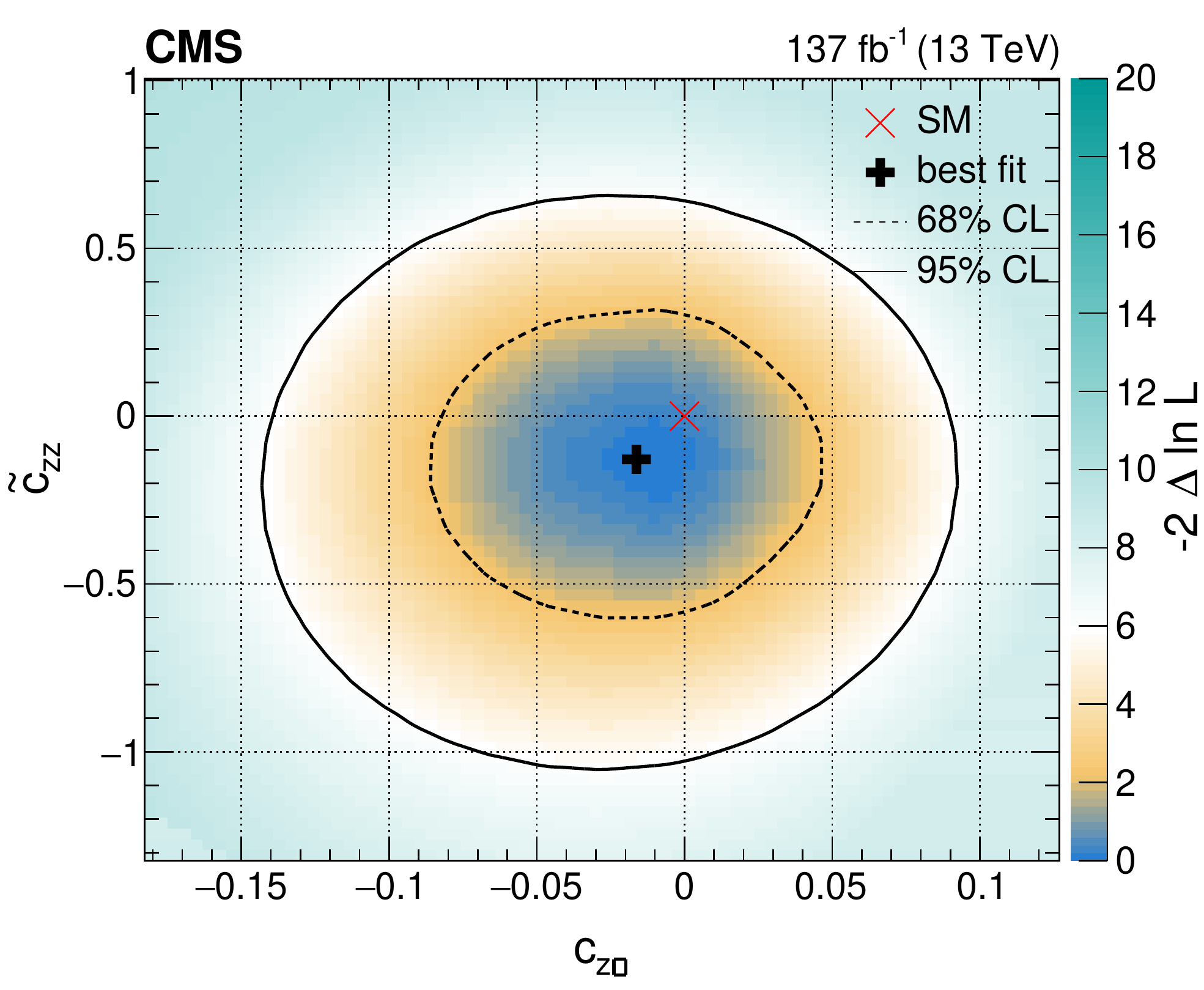}
\includegraphics[width=0.49\textwidth]{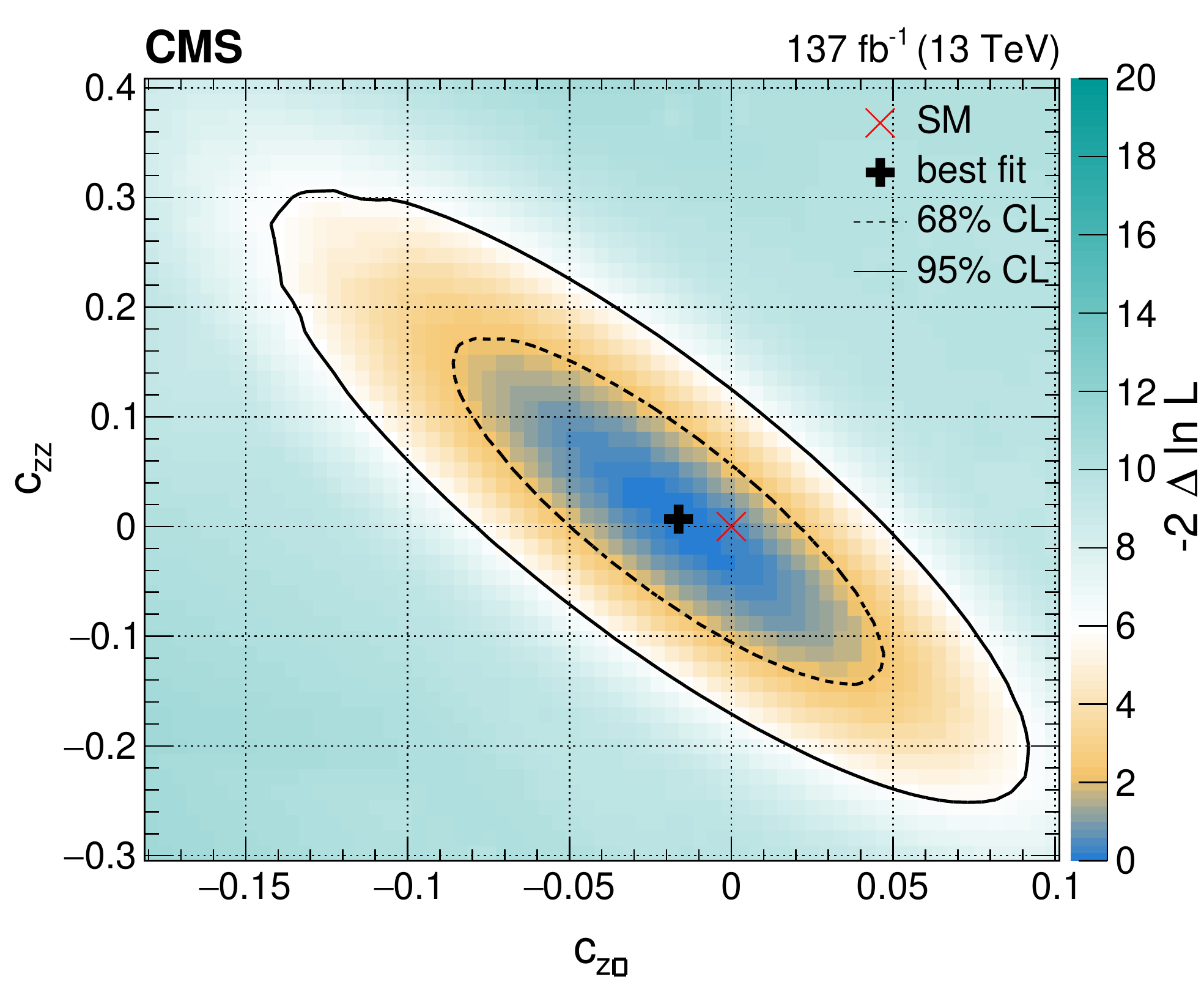}
\caption{
Observed two-dimensional constraints from a simultaneous fit of the SMEFT parameters 
$\delta c_\mathrm{z}$, $c_\mathrm{zz}$, $c_{\mathrm{z} \Box}$, and $\tilde c_\mathrm{zz}$ with the $c_{\Pg\Pg}$ and $\tilde c_{\Pg\Pg}$ couplings
left unconstrained. 
}
\label{fig:resultHVV_2DEFT}
\end{figure*}

The measurement of the signal strength $\muV$ and the \fC , \fB, \fL parameters can be reinterpreted in 
terms of the $\delta c_\mathrm{z}$, $c_\mathrm{zz}$, $c_{\mathrm{z} \Box}$, and $\tilde c_\mathrm{zz}$ coupling strength parameters. 
Observed one- and two-dimensional constraints from a simultaneous fit of SMEFT parameters are shown
in Figs.~\ref{fig:resultHVV_1DEFT} and~\ref{fig:resultHVV_2DEFT}.
The $c_{\Pg\Pg}$ and $\tilde c_{\Pg\Pg}$ couplings are left unconstrained. 
A summary of all constraints on the \Htt, \Hgg, and \HVV coupling parameters in the Higgs basis of SMEFT,
including the correlation coefficients, is shown in Table~\ref{tab:summary_EFT}. 
The results in this table are taken from Sections~\ref{sec:results_HttHgg} and~\ref{sec:results_HVVEFT},
as measured in the \tqH, \ttH, \ggH, and EW processes. 

The above interpretation of $\HVV$ results in terms of the $\delta c_\mathrm{z}$, $c_\mathrm{zz}$, $c_{\mathrm{z} \Box}$, and $\tilde c_\mathrm{zz}$ 
couplings can be extended into an interpretation in terms of the couplings in the Warsaw SMEFT basis~\cite{deFlorian:2016spz}. 
In this basis, nine operators are considered:
$c_{\PH\Box}$, $c_{\PH\mathrm{D}}$, $c_{\PH\PW}$ , $c_{\PH\PW\mathrm{B}}$, $c_{\PH\mathrm{B}}$, 
$c_{\PH\tilde{\PW}}$, $c_{\PH\tilde{\PW}\mathrm{B}}$, $c_{\PH\tilde{\mathrm{B}}}$, and $\delta_v$,
where the latter is a linear combination of additional Warsaw basis operators~\cite{Falkowski:2015wza}.
However, not all nine of these operators are independent. First of all, consideration of Eq.~(\ref{eq:EFT1})
leads to $\delta_v$ expression as a linear combination of $c_{\PH\mathrm{D}}$ and $c_{\PH\PW\mathrm{B}}$. 
Four constraints on the couplings $a_2^{\gamma\gamma,\PZ\gamma}$ and $a_3^{\gamma\gamma,\PZ\gamma}$
lead to only one of the three operators $c_{\PH\PW}$ , $c_{\PH\PW\mathrm{B}}$, and $c_{\PH\mathrm{B}}$ being independent, 
and only one of $c_{\PH\tilde{\PW}}$, $c_{\PH\tilde{\PW}\mathrm{B}}$, and $c_{\PH\tilde{\mathrm{B}}}$ being independent. 
Therefore, we obtain only four independent constraints, the same number as in the Higgs basis. 
We note that the couplings of the $\PZ$ boson to fermions are fixed to those expected in the SM
because those are well constrained from prior measurements and this constraint is already 
included in our primary measurements. Even though some of the above EFT operators may affect
couplings of the $\PZ$ boson, their effect must be compensated by the other EFT operators not
affecting the \Hboson couplings directly. 
With the above constraints, we use the tools in Refs.~\cite{Falkowski:2015wza,Gritsan:2020pib}
to relate operators in the Higgs and Warsaw bases. Since it is arbitrary which one of the three 
operators is considered to be independent, we present results with all three choices in each case. 
The results can be found in Table~\ref{tab:summary_Warsaw}, where three other independent 
couplings are left unconstrained for each measurement. 

\begin{table*}[!tbh]
\centering
\topcaption{Summary of constraints on the \Htt, \Hgg, and \HVV coupling parameters in the Higgs basis of SMEFT. 
The observed correlation coefficients are presented for the \Htt$\&\,$\Hgg and \HVV couplings in the fit configurations discussed 
in text and shown in Figs.~\ref{fig:resultHttHgg_2D} and~\ref{fig:resultHVV_2DEFT}, respectively. 
\label{tab:summary_EFT}
}
\cmsTable{
\begin{scotch}{ccrrrrrr}
\vspace{-0.3cm} \\
 Channels & Coupling   &  Observed                                         & Expected & \multicolumn{4}{c}{Observed correlation} \\
[\cmsTabSkip]
\hline \\ [-1ex]
   &                                     &                                                                  &                            & $c_{\Pg\Pg}$ & $\tilde c_{\Pg\Pg}$ & $\kappa_{\PQt}$ & $\tilde\kappa_{\PQt}$ \\
[\cmsTabSkip]
\multirow{4}{*}{\tqH\,\&\,\ttH\,\&\,\ggH} 
   & $c_{\Pg\Pg}$                     &      $-0.0012_{-0.0174}^{+0.0022}$          &         $0.0000_{-0.0196}^{+0.0019}$       &  1        &  $-0.050$ &  $-0.941$  &   $+0.029$     \\
   & $\tilde c_{\Pg\Pg}$             &       $-0.0017_{-0.0130}^{+0.0160}$          &         $0.0000_{-0.0138}^{+0.0138}$     &                  & 1        &  $+0.046$  &    $-0.568$      \\
  & $\kappa_{\PQt}$            &      $1.05_{-0.20}^{+0.25}$           &          $1.00_{-0.26}^{+0.34}$                             &                    &         & 1        &   $+0.168$  \\
  & $\tilde\kappa_{\PQt}$    &      $-0.01_{-0.67}^{+0.69}$           &          $0.00_{-0.71}^{+0.71}$                              &                 &         &            & 1     \\     
[\cmsTabSkip]
&                                     &                                                                  &                            & $\delta c_\mathrm{z}$ & $c_\mathrm{zz}$ & $c_{\mathrm{z} \Box}$  & $\tilde c_\mathrm{zz}$ \\
[\cmsTabSkip]
\multirow{4}{*}{\VBF\,\&\,\VH\,\&~$\PH\to4\ell$} 
   & $\delta c_\mathrm{z}$                  &      $-0.03^{+0.06}_{-0.25}$         &     $0.00^{+0.07}_{-0.27}$      &  1       & $+0.241$ & $-0.060$  &  $-0.009$       \\
   & $c_\mathrm{zz}$                       &       $0.01^{+0.11}_{-0.10}$          &     $0.00^{+0.22}_{-0.16}$      &           & 1        &  $-0.884$ &   $+0.058$       \\
   & $c_{\mathrm{z} \Box}$                 &       $-0.02^{+0.04}_{-0.04}$         &     $0.00^{+0.06}_{-0.09}$     &           &            & 1        &   $+0.020$ \\
   & $\tilde c_\mathrm{zz}$               &       $-0.11^{+0.30}_{-0.31}$         &      $0.00^{+0.63}_{-0.63}$    &           &           &            & 1     \\     
[\cmsTabSkip]
\end{scotch}
}
\end{table*}

\begin{table}[!bth]
\centering
\topcaption{Summary of constraints on the \HVV coupling parameters in the Warsaw basis of SMEFT. 
For each coupling constraint reported, three other independent operators are left unconstrained, where
only one of the three operators $c_{\PH\PW}$, $c_{\PH\PW\mathrm{B}}$, and $c_{\PH\mathrm{B}}$ is independent, 
and only one of $c_{\PH\tilde{\PW}}$, $c_{\PH\tilde{\PW}\mathrm{B}}$, and $c_{\PH\tilde{\mathrm{B}}}$ is independent. 
\label{tab:summary_Warsaw}
}
\begin{scotch}{ccrr}
\vspace{-0.3cm} \\
 Channels & Coupling   &  Observed                                         & Expected \\
[\cmsTabSkip]
\hline \\ [-1ex]
\multirow{8}{*}{\VBF\,\&\,\VH\,\&~$\PH\to4\ell$} 
   & $c_{\PH\Box}$                  &      $0.04^{+0.43}_{-0.45}$         &     $0.00^{+0.75}_{-0.93}$             \\
   & $c_{\PH\mathrm{D}}$                       &       $-0.73^{+0.97}_{-4.21}$          &     $0.00^{+1.06}_{-4.60}$         \\
   & $c_{\PH\PW}$                 &       $0.01^{+0.18}_{-0.17}$         &     $0.00^{+0.39}_{-0.28}$         \\
   & $c_{\PH\PW\mathrm{B}}$               &       $0.01^{+0.20}_{-0.18}$         &      $0.00^{+0.42}_{-0.31}$        \\
   & $c_{\PH\mathrm{B}}$               &       $0.00^{+0.05}_{-0.05}$         &      $0.00^{+0.03}_{-0.08}$       \\
   & $c_{\PH\tilde{\PW}}$                 &       $-0.23^{+0.51}_{-0.52}$         &     $0.00^{+1.11}_{-1.11}$         \\
   & $c_{\PH\tilde{\PW}\mathrm{B}}$               &       $-0.25^{+0.56}_{-0.57}$         &      $0.00^{+1.21}_{-1.21}$        \\
   & $c_{\PH\tilde{\mathrm{B}}}$               &       $-0.06^{+0.15}_{-0.16}$         &      $0.00^{+0.33}_{-0.33}$        \\
[\cmsTabSkip]
\end{scotch}
\end{table}

\section{Summary}
\label{sec:Summary}

In this paper, a comprehensive study of $CP$-violation, anomalous couplings, and the tensor structure of \Hboson interactions 
with electroweak gauge bosons, gluons, and fermions, using all accessible production mechanisms and
the $\Hell$ decay mode, is presented. The results are based on the 2016--2018 data from $\Pp\Pp$ collisions recorded with the CMS 
detector during Run~2 of the LHC, corresponding to an integrated luminosity of $137\fbinv$ at a center-of-mass 
energy of 13\TeV. These results significantly surpass our results from Run~1~\cite{Khachatryan:2014kca}
in both precision and coverage. The improvements result not only from a significantly increased sample of $\PH$ bosons,
but also from a detailed analysis of kinematic distributions of the particles associated with the \Hboson production 
in addition to kinematic distributions in its decay. 
These results also surpass our earlier studies of \onshell production of the \Hboson in this decay channel
with a partial Run~2 dataset~\cite{Sirunyan:2017tqd,Sirunyan:2019twz}.

The parameterization of the \Hboson production and decay processes is based on a scattering amplitude or, equivalently, 
an effective field theory Lagrangian, with operators up to dimension six. Additional symmetries and prior measurements 
allow us to reduce the number of independent parameters and make a connection to the standard model effective field theory 
(SMEFT) formulation. Dedicated Monte Carlo programs and matrix-element re-weighting techniques provide modeling of all 
kinematic effects in the production and decay of the \Hboson, with any variation of parameters of the scattering amplitude and 
with full simulation of detector effects. Each production process of the $\PH$ boson is identified using the kinematic
features of its associated particles.  The matrix element likelihood approach (MELA) is employed to construct observables 
that are optimal for the measurement of the targeted anomalous couplings in each process, including $CP$-sensitive observables.
A maximum likelihood fit allows a simultaneous measurement of up to five \HVV, two \Hgg, and two \Htt couplings.

For the first time, we present a complete and dedicated study of $CP$ properties in the \Hboson coupling to gluons 
through a loop of heavy particles using the $CP$-sensitive observables, while separating the electroweak and strong boson 
fusion processes. An interpretation of the loop contribution is made both with and without an assumption of top quark dominance, 
which allows for a new heavy particle to contribute. In both cases, combination with the 
$CP$-sensitive measurement of the \Htt coupling in the \ttH and \tqH processes allows either simultaneous 
or separate measurements of the two effective point-like \Hgg couplings and the two \Htt couplings, both $CP$-odd and $CP$-even. 
For the \ttH and \tqH processes, results in the $\Hell$ channel are combined with those in the
 $\PH\to\gamma\gamma$ channel~\cite{Sirunyan:2020sum}. This is the first comprehensive study of $CP$ properties
 in the \Hgg and \Htt couplings from a simultaneous measurement of the \ggH, \ttH, and \tqH processes. 

Also for the first time, we present the measurement of $CP$ properties and the tensor structure of the \Hboson's interactions 
with two electroweak bosons with up to five parameters measured simultaneously. 
The $\HVV$ coupling is analyzed in \VBF and \VH production and in $\PH\to\VV\to4\ell$ decay. The measurements 
are performed with two approaches. In the first approach, more lenient symmetry considerations are applied, 
which allow a less restrictive interpretation of results. In the second approach, SU(2)$\times$U(1) symmetry is invoked 
and the formulation becomes equivalent to SMEFT. The operator basis is chosen to coincide with the couplings of 
the mass eigenstates, which allows us to minimize the number of independent parameters. A translation of the SMEFT
results to the Warsaw basis is also presented for easier comparison with other results. 

In all cases, we first present 
results in terms of the total cross section of a process and the fractional contribution of each anomalous coupling.
These results are further re-interpreted in terms of direct constraints on the couplings by applying certain assumptions about 
the \Hboson total width. 
Each of the measurements presented here is limited by statistical precision and is consistent with the expectations 
for the standard model Higgs boson. 
\ifthenelse{\boolean{cms@external}}{\clearpage}{}

\begin{acknowledgments}
\hyphenation{Bundes-ministerium Forschungs-gemeinschaft Forschungs-zentern Rachada-pisek} 
We thank Markus Schulze for optimizing the \textsc{JHUGen} Monte Carlo simulation program and  matrix element library for this analysis. We thank Amitabh Basu for guidance on implementing the cutting planes algorithm and  Tianran Chen for updating the \textsc{Hom4PS} program for this analysis and providing support. We are grateful to the members of the LHC Higgs and EFT Working Groups for stimulating the development of several phenomenological aspects of this work, among which is relating the EFT operator bases for the Higgs boson couplings.

We congratulate our colleagues in the CERN accelerator departments for the excellent performance of the LHC and thank the technical and administrative staffs at CERN and at other CMS institutes for their contributions to the success of the CMS effort. In addition, we gratefully acknowledge the computing centers and personnel of the Worldwide LHC Computing Grid and other centers for delivering so effectively the computing infrastructure essential to our analyses.
We also acknowledge the Maryland Advanced Research Computing Center (MARCC) for providing computing resources essential for this analysis. Finally, we acknowledge the enduring support for the construction and operation of the LHC, the CMS detector, and the supporting computing infrastructure provided by the following funding agencies: the Austrian Federal Ministry of Education, Science and Research and the Austrian Science Fund; the Belgian Fonds de la Recherche Scientifique, and Fonds voor Wetenschappelijk Onderzoek; the Brazilian Funding Agencies (CNPq, CAPES, FAPERJ, FAPERGS, and FAPESP); the Bulgarian Ministry of Education and Science; CERN; the Chinese Academy of Sciences, Ministry of Science and Technology, and National Natural Science Foundation of China; the Ministerio de Ciencia Tecnolog\'ia e Innovaci\'on (MINCIENCIAS), Colombia; the Croatian Ministry of Science, Education and Sport, and the Croatian Science Foundation; the Research and Innovation Foundation, Cyprus; the Secretariat for Higher Education, Science, Technology and Innovation, Ecuador; the Ministry of Education and Research, Estonian Research Council via PRG780, PRG803 and PRG445 and European Regional Development Fund, Estonia; the Academy of Finland, Finnish Ministry of Education and Culture, and Helsinki Institute of Physics; the Institut National de Physique Nucl\'eaire et de Physique des Particules~/~CNRS, and Commissariat \`a l'\'Energie Atomique et aux \'Energies Alternatives~/~CEA, France; the Bundesministerium f\"ur Bildung und Forschung, the Deutsche Forschungsgemeinschaft (DFG), under Germany's Excellence Strategy -- EXC 2121 ``Quantum Universe" -- 390833306, and under project number 400140256 - GRK2497, and Helmholtz-Gemeinschaft Deutscher Forschungszentren, Germany; the General Secretariat for Research and Technology, Greece; the National Research, Development and Innovation Fund, Hungary; the Department of Atomic Energy and the Department of Science and Technology, India; the Institute for Studies in Theoretical Physics and Mathematics, Iran; the Science Foundation, Ireland; the Istituto Nazionale di Fisica Nucleare, Italy; the Ministry of Science, ICT and Future Planning, and National Research Foundation (NRF), Republic of Korea; the Ministry of Education and Science of the Republic of Latvia; the Lithuanian Academy of Sciences; the Ministry of Education, and University of Malaya (Malaysia); the Ministry of Science of Montenegro; the Mexican Funding Agencies (BUAP, CINVESTAV, CONACYT, LNS, SEP, and UASLP-FAI); the Ministry of Business, Innovation and Employment, New Zealand; the Pakistan Atomic Energy Commission; the Ministry of Science and Higher Education and the National Science Center, Poland; the Funda\c{c}\~ao para a Ci\^encia e a Tecnologia, Portugal; JINR, Dubna; the Ministry of Education and Science of the Russian Federation, the Federal Agency of Atomic Energy of the Russian Federation, Russian Academy of Sciences, the Russian Foundation for Basic Research, and the National Research Center ``Kurchatov Institute"; the Ministry of Education, Science and Technological Development of Serbia; the Secretar\'{\i}a de Estado de Investigaci\'on, Desarrollo e Innovaci\'on, Programa Consolider-Ingenio 2010, Plan Estatal de Investigaci\'on Cient\'{\i}fica y T\'ecnica y de Innovaci\'on 2017--2020, research project IDI-2018-000174 del Principado de Asturias, and Fondo Europeo de Desarrollo Regional, Spain; the Ministry of Science, Technology and Research, Sri Lanka; the Swiss Funding Agencies (ETH Board, ETH Zurich, PSI, SNF, UniZH, Canton Zurich, and SER); the Ministry of Science and Technology, Taipei; the Thailand Center of Excellence in Physics, the Institute for the Promotion of Teaching Science and Technology of Thailand, Special Task Force for Activating Research and the National Science and Technology Development Agency of Thailand; the Scientific and Technical Research Council of Turkey, and Turkish Atomic Energy Authority; the National Academy of Sciences of Ukraine; the Science and Technology Facilities Council, UK; the US Department of Energy, and the US National Science Foundation.

Individuals have received support from the Marie-Curie program and the European Research Council and Horizon 2020 Grant, contract Nos.\ 675440, 724704, 752730, 765710, and 824093 (European Union); the Leventis Foundation; the Alfred P.\ Sloan Foundation; the Alexander von Humboldt Foundation; the Belgian Federal Science Policy Office; the Fonds pour la Formation \`a la Recherche dans l'Industrie et dans l'Agriculture (FRIA-Belgium); the Agentschap voor Innovatie door Wetenschap en Technologie (IWT-Belgium); the F.R.S.-FNRS and FWO (Belgium) under the ``Excellence of Science -- EOS" -- be.h project n.\ 30820817; the Beijing Municipal Science \& Technology Commission, No. Z191100007219010; the Ministry of Education, Youth and Sports (MEYS) of the Czech Republic; the Lend\"ulet (``Momentum") Program and the J\'anos Bolyai Research Scholarship of the Hungarian Academy of Sciences, the New National Excellence Program \'UNKP, the NKFIA research grants 123842, 123959, 124845, 124850, 125105, 128713, 128786, and 129058 (Hungary); the Council of Scientific and Industrial Research, India; the National Science Center (Poland), contracts Opus 2014/15/B/ST2/03998 and 2015/19/B/ST2/02861; the National Priorities Research Program by Qatar National Research Fund; the Ministry of Science and Higher Education, project no. 0723-2020-0041 (Russia); the Programa de Excelencia Mar\'{i}a de Maeztu, and the Programa Severo Ochoa del Principado de Asturias; the Thalis and Aristeia programs cofinanced by EU-ESF, and the Greek NSRF; the Rachadapisek Sompot Fund for Postdoctoral Fellowship, Chulalongkorn University, and the Chulalongkorn Academic into Its 2nd Century Project Advancement Project (Thailand); the Kavli Foundation; the Nvidia Corporation; the SuperMicro Corporation; the Welch Foundation, contract C-1845; and the Weston Havens Foundation (USA).

\end{acknowledgments}

\bibliography{auto_generated}

\cleardoublepage \appendix\section{The CMS Collaboration \label{app:collab}}\begin{sloppypar}\hyphenpenalty=5000\widowpenalty=500\clubpenalty=5000\input{HIG-19-009-authorlist.tex}\end{sloppypar}
%%% END EDITABLE REGION %%%
% skeleton_end
\end{document}

%% file: HIG-19-009-authorlist.tex
\vskip\cmsinstskip
\textbf{Yerevan Physics Institute, Yerevan, Armenia}\\*[0pt]
A.M.~Sirunyan$^{\textrm{\dag}}$, A.~Tumasyan
\vskip\cmsinstskip
\textbf{Institut f\"{u}r Hochenergiephysik, Wien, Austria}\\*[0pt]
W.~Adam, J.W.~Andrejkovic, T.~Bergauer, S.~Chatterjee, M.~Dragicevic, A.~Escalante~Del~Valle, R.~Fr\"{u}hwirth\cmsAuthorMark{1}, M.~Jeitler\cmsAuthorMark{1}, N.~Krammer, L.~Lechner, D.~Liko, I.~Mikulec, P.~Paulitsch, F.M.~Pitters, J.~Schieck\cmsAuthorMark{1}, R.~Sch\"{o}fbeck, M.~Spanring, S.~Templ, W.~Waltenberger, C.-E.~Wulz\cmsAuthorMark{1}
% ...
\vskip\cmsinstskip
\dag: Deceased\\
1:  Also at TU Wien, Wien, Austria\\
% ...
